\documentclass[a4paper,11pt,twoside]{book}

\pdfoutput 1

\usepackage{graphicx} 
\usepackage[english]{babel}
\usepackage{amsmath,amssymb}
\usepackage{natbib}
\usepackage[utf8x]{inputenc}
\usepackage{times}
\usepackage{makeidx}
\usepackage[dvips]{epsfig}
\usepackage[dvips]{graphicx}
\usepackage{textcomp}
\usepackage{rotating}
\usepackage{longtable}
\usepackage{pdfpages}
\usepackage{import}
\usepackage{appendix}
\usepackage{verbatim}
\usepackage{ulem}

\usepackage{multirow}
\usepackage{hhline}

\usepackage{deluxetable} 

\def\ltsim{\raise 2pt \hbox {$<$} \kern-1.1em \lower 4pt \hbox {$\sim$}}
\def\ltapprox{\raise 2pt \hbox {$<$} \kern-1.1em \lower 5pt \hbox {$\approx$}}
\def\gtsim{\raise 2pt \hbox {$>$} \kern-1.1em \lower 4pt \hbox {$\sim$}}
\def\gtapprox{\raise 2pt \hbox {$>$} \kern-1.1em \lower 5pt \hbox {$\approx$}}

\def\lapp{\ifmmode\stackrel{<}{_{\sim}}\else$\stackrel{<}{_{\sim}}$\fi}
\def\gapp{\ifmmode\stackrel{>}{_{\sim}}\else$\stackrel{<}{_{\sim}}$\fi}

\def\eps@scaling{.95}
\def\epsscale#1{\gdef\eps@scaling{#1}}
\def\plotone#1{\centering \leavevmode
    \epsfxsize=\eps@scaling\columnwidth \epsfbox{#1}}
\def\plottwo#1#2{\centering \leavevmode
    \epsfxsize=.45\columnwidth \hbox{\epsfbox{#1}} \hfil
    \epsfxsize=.45\columnwidth \epsfbox{#2}}

\def\acom{a_{\rm COM}}

\def\Msun{M_{\odot}}
\def\Lsun{L_{\odot}}
\def\Rsun{R_{\odot}}

\def\mcom{M_{\rm COM}}

\def\Pspin{P}
\def\Pdot{\dot P}

\def\psrA{PSR J1740$-$5340A}
\def\psrC{PSR J1518$+$0204C}
\def\psrH{PSR J1824$-$2452H}
\def\psrI{PSR J1824$-$2452I}
\def\psrIM{PSR J1439$-$5501}
\def\psrBW{PSR J0610$-$2100}

\def\ltsima{$\; \buildrel < \over \sim \;$}
\def\lsim{\lower.5ex\hbox{\ltsima}}

\def\comBW{COM\, J0610$-$2100}
\def\rrl{R_{\rm RL}}
\def\mpsr{M_{\rm PSR}}
\def\rcom{R_{\rm COM}}
\def\dpsr{d_{\rm PSR}}

\def\apsr{a_{\rm PSR}}

\def\com{ the companion}

\def\igr{IGR J18245$-$2452}
\def\halpha{H$\alpha$}
\def\B{F435W}
\def\R{F625W}
\def\U{F390W}
\def\V{F606W}
\def\I{F814W}
\def\H{F656N}
\def\VV{F555W}
\def\UU{F336W}
\def\Ha{F658N}
\def\nUV{F255W}
\def\fUV{F170W}

\def\comA{COM J1740$-$5340A}

\def\comC{COM-M5C}
\def\psrC{PSR J1518+0204C}
\def\Uc{\rm m_{F390W}}
\def\Vc{\rm m_{F606W}}
\def\Ic{\rm m_{F814W}}
\def\Hc{\rm m_{F656N}}

\usepackage[top=4cm, bottom=3cm, left=3.5cm, right=3.5cm]{geometry}

\usepackage{fancyhdr}
\newcommand\arcsec{\mbox{$^{\prime\prime}$}}%
\def\aj{AJ}%
\def\araa{ARA\&A}%
\def\apj{ApJ}%
\def\apjl{ApJ}%
\def\apjs{ApJS}%
\def\aap{A\&A}%
\def\aapr{A\&A~Rev.}%
\def\aaps{A\&AS}%
\def\mnras{MNRAS}%
\def\na{New A}%
\def\pasp{PASP}%
\def\pasj{PASJ}%
\def\nat{Nature}%
\def\aplett{Astrophys.~Lett.}%
\def\physrep{Phys.~Rep.}%
\begin{document}

  \thispagestyle{empty}   \hsize 140 mm
  \topmargin 0mm
  \enlargethispage*{3cm} 
\begin{center}
    
    {\Large \textbf{Alma Mater Studiorum}}\\\vspace{5mm}
    {\Large \textbf{Universit\`{a} degli Studi di Bologna}}\\ 
    \vspace{1mm}
    \rule{140mm}{0.2mm}\\
    \rule[2ex]{140mm}{0.5mm}
    \vspace{5mm}
    \large{DIPARTIMENTO DI FISICA E ASTRONOMIA\\      
           Dottorato di ricerca in Astronomia\\ 
            Ciclo XXVI\\

           \vspace{2.5cm}}
           
                  {\huge\textbf{Cosmic-Lab: Optical companions\\
                  \vspace{5mm} 
                  to binary Millisecond Pulsars}}\\
                   		     
\end{center}

  \vspace{2.5  cm}
 
  \hspace{-6mm}
  \parbox[t]{50mm}{
    \begin{flushleft}
      {\large Dottoranda: \vspace{1mm}\\
        \textbf{Cristina Pallanca}}\\
    \end{flushleft}
  } \ \         
  \hspace{-5mm}
  \parbox[t]{90mm}{
    \vspace{0cm}
    \begin{flushright}
      {\large Relatore:\\ \vspace{1mm}         \textbf{Chiar.mo Prof. Francesco R. Ferraro}\\
        \vspace{0.3 cm}
        Co--Relatore: \\ \vspace{1mm}
        \textbf{Dott. Emanuele Dalessandro} \\
        \vspace{0.3 cm}
        Co--Relatrice: \\ \vspace{1mm}
        \textbf{Dott.sa Barbara Lanzoni} \\
        \vspace{1 cm}
        Coordinatore: \\ \vspace{2mm}
        \textbf{Chiar.mo Prof. Lauro Moscardini}} \\
    \end{flushright}
  } \\

\begin{center}

    \vspace{1.5cm}
      {\large Esame finale anno 2014}\\
    \rule{140mm}{0.2mm}\\
    \rule[2ex]{140mm}{0.5mm}\\

       {\small         Settore Concorsuale: 02/C1 -- Astronomia, Astrofisica, Fisica della Terra e dei Pianeti\\
     Settore Scientifico-Disciplinare: FIS/05 -- Astronomia e Astrofisica\\}
         
          \end{center}

 \hsize 140 mm

\clearpage{\pagestyle{empty}\cleardoublepage}

\newpage
\mbox{ }
\thispagestyle{empty}

\vspace{4cm}
\begin{flushright}
{\sl ``I do not seek. I find.''}\\ 
   \vspace{2mm}
               { Pablo Picasso\ \ \  } 
\\
\end{flushright}

\newpage
\mbox{ }
\thispagestyle{empty}

\baselineskip 4ex

\newcommand{\ltae}{\raisebox{-0.6ex}{$\,\stackrel
{\raisebox{-.2ex}{$\textstyle <$}}{\sim}\,$}}
\newcommand{\gtae}{\raisebox{-0.6ex}{$\,\stackrel
{\raisebox{-.2ex}{$\textstyle >$}}{\sim}\,$}}

\baselineskip 4ex

\renewcommand{\thepage}{  }
\newpage

\pagenumbering{roman}

\markboth{\sc \ }{\sc Contents}
\tableofcontents
\clearpage{\pagestyle{empty}\cleardoublepage}

\listoffigures
\clearpage{\pagestyle{empty}\cleardoublepage}

\clearpage
\markboth{ACRONYMS}{ACRONYMS}
\newpage
\chapter*{List of Acronyms}

\begin{table}[!h]
\begin{tabular}{ll}
BB & Black body\\
BD &  Brown dwarf\\
BW  &  Black widow\\
CMD  & Color magnitude diagram\\
EW  & Equivalent width\\
FOV  & Field of view\\
FWHM  & Full width half maximum\\
GC  & Globular Cluster\\
GF  & Galactic Field\\
IMM  & Intermediate mass Millisecond Pulsar\\
LMXB  & Low mass X-ray Binary\\
MS  &  Main sequence\\
MSP   & Millisecond Pulsar\\
NS  &  Neutron star\\
PSF &  Point spread function\\
PSR &  Pulsar\\
RB  & Red Back\\
RL  & Roche Lobe\\
TO  & Turn Off\\
WD   & White dwarf\\
\end{tabular}
\end{table}

\clearpage{\pagestyle{empty}\cleardoublepage}

\setcounter{page}{1}

\renewcommand{\thepage}{\arabic{page}}

\clearpage
\markboth{ABSTRACT}{ABSTRACT}
\addcontentsline{toc}{chapter}{Abstract}
\newpage
\chapter*{Abstract}

Millisecond Pulsars (MSPs) are  fast rotating, highly  magnetized  neutron stars.
They are thought to form  in binary systems  containing a slowly rotating neutron star that, during a phase of heavy mass accretion from an evolving companion, is spun up to millisecond spin period.  
The final stage consists in a binary made of a fast rotating  Pulsar (a MSP) and a deeply peeled or even an exhausted star (as a white dwarf).
This theoretical scenario of MSP formation is known as {\it ``canonical recycling scenario''}.
However, in the last years an increasing  number of systems deviating from the expectations of the {\it canonical recycling scenario} has been discovered and the formation and evolution of MSPs still remains  unclear.

In this framework, the identification of companion stars to MSPs is not only useful  to fully characterize the binary systems, allowing for example, to infer the masses of the two stars, but it also represents a powerful tool to constrain the formation and evolution of these objects.
Moreover, in dense environments such as Globular Clusters (where most MSPs reside), the identification of the companion stars is helpful also to understand the interplay between dynamics and the evolution of binary systems.
Despite the paramount importance of identifying the companion stars to binary MSPs, before this Thesis work started, only six companions were identified in 5 Globular Clusters and a few in the Galactic Field.

This Thesis is devoted to the search for companions to MSPs in Galactic Globular Clusters and in the Galactic Field, with the aim  of shading new light on their formation mechanism.
We report on the identification of a total of 5 new companions.
The main results can be summarized as follows:
\begin{itemize}
\item We identified three new companions in two  Globular Clusters (two in M28 and one in M5), thus increasing by 50\% the number of MSP optical counterparts in Globular Clusters  known before this work (6 in total).
They are  non-degenerate objects, in contrast with the prediction of the {\it canonical recycling scenario} and in support to the hypothesis that exchange interactions are quite efficient in modifying the expected evolutionary path of these objects.
\item We detected the optical counterpart to the INTEGRAL transient \igr, both in a quiescent and in an outburst state.
\item We identified  the first companion to a black widow in a Globular Cluster: this is the case of \psrC\ in M5.
\item We spectroscopically confirmed that  the companion to \psrA\ in NGC 6397 is a deeply peeled star descending from a $\sim0.8\Msun$ progenitor, which lost $\sim75\%$ of its original material.
\item The two identifications in the Galactic Field concern the companion to a black widow  recently detected by the Fermi Gamma-ray Telescope in the Galactic plane, and the white dwarf companion to the intermediate mass binary pulsar J1439-5501, suggesting that the spin-down age  derived from radio properties could be overestimated by a factor of 10.
\end{itemize}

The Thesis is organized in 9 Chapters:
\begin{itemize}
\item[-] In Chapter 1, I give a brief introduction to binary MSPs and to the expected companions.
\item[-] In Chapter 2, I describe the used method of data analysis and the adopted procedures to discover and characterize the companions to MSPs.
\item[-] In Chapter 3, I report on the identification of the companion to \psrH\ in the  Globular Cluster M28.
\item[-] In Chapter 4, I present the detection of the companion to \igr/\psrI\ in the Globular Cluster M28.
\item[-] In Chapter 5, I describe the identification of the companion to \psrC\ in the Globular Cluster M5.
\item[-] In Chapter 6, I report on the main results from the spectroscopic follow-up of the previously identified companion to \psrA\ in the Globular Cluster NGC 6397.
\item[-] In Chapter 7, I present the identification of the companion to \psrBW\ in the Galactic Field.
\item[-] In Chapter 8, I discuss the detection of the companion to \psrIM\ in the Galactic Field.
\item[-] Finally, the Conclusions summarize the main results of this Thesis and future perspectives of this work.
\end{itemize}

\clearpage{\pagestyle{empty}\cleardoublepage}

 \chapter{Millisecond Pulsars}\label{Chap:introPSR}

Pulsars (PSRs) are  rapidly spinning, strongly magnetized neutron stars (NSs) that emit beams of radiation crossing the line of sight of the observer once per rotation and thus producing a detectable pulse \citep{stairs04}. PSRs are commonly classified on the basis of their observed properties: the spin period ($\Pspin$) and the period derivative ($\Pdot$). 

\begin{figure*}[!hbt]
\begin{center}
\includegraphics[height=110mm]{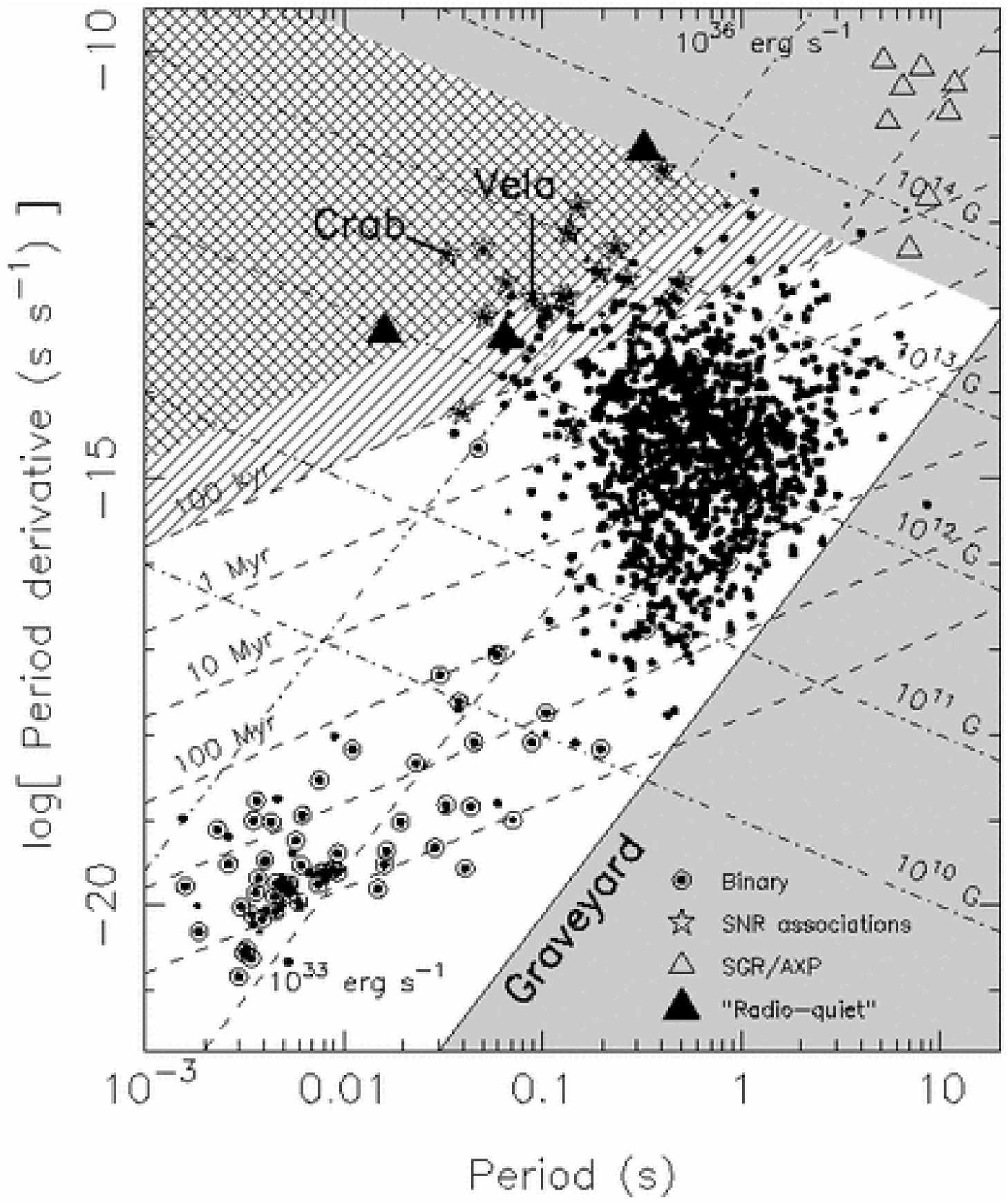}
  \caption[$\Pspin$ - $\dot P$ diagram.]{$\Pspin$ - $\dot P$ plot. MSPs are located in the lower left corner of the plot \citep{handbook}.}\label{PPdot}
\end{center}
\end{figure*}

In the framework of this classification it is possible to distinguish two main groups \citep[see Figure \ref{PPdot} and][]{handbook}. The larger group is composed of ``normal PSRs'' which have $\Pspin\sim0.5$ s and $\dot P \lapp 10^{-15}$ s s$^{-1}$. A second significant group clearly separated and located in the lower left side of the diagram in Figure \ref{PPdot} consists of  PSRs characterized by $\Pspin\sim3$ ms and $\dot P \sim 10^{-20}$ s s$^{-1}$. These systems are called Millisecond Pulsars (MSPs) because of their very short spin periods.
Lines of constant magnetic field and characteristic age (see Appendix) are drawn in Figure \ref{PPdot}, showing that  typical magnetic fields and ages are of the order of $10^{12}$ G and $10^7$ yr for normal PSRs, and $10^8$ G and and $10^9$ yr for MSPs.

The first MSP was discovered in 1982 by \citet{backer82} and \citet{alpar82} who finally unveiled the nature of the source 4C21.53.
\citet{backer82} suggested that the observed radio properties of 4C21.53 could be interpreted assuming that this system is a PSR with  $\Pspin$ of 1.558 ms, hence rotating near the maximum rate possible for a NS.
They also proposed that  it was not a young system, despite its large rotational velocity and energy.
At the same time \citet{alpar82} suggested that NS with sufficiently low magnetic field,  accreting from a surrounding Keplerian disk for long time, can be spin up to millisecond periods. 
Therefore they proposed that 4C21.53 belongs to a new class of objects, thereafter defined as MSPs. They have short $\Pspin$, long apparent ages, and pulsed optical, X-ray and $\gamma$-ray fluxes significantly below those expected for canonical PSRs with similar $\Pspin$.

\section{The formation of MSPs}
\begin{figure*}[t]
\begin{center}
\includegraphics[width=140mm]{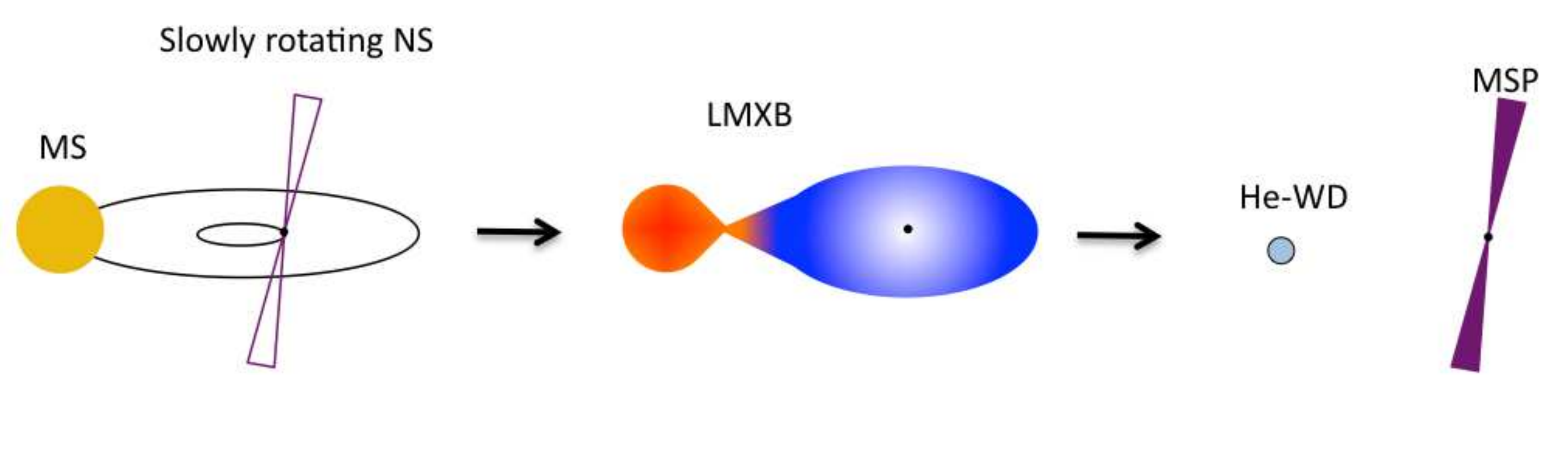}
  \caption[Schematic view of the {\it canonical recycling scenario}.]{Schematic view of the main phases of the {\it canonical recycling scenario}.}\label{recycling}
\end{center}
\end{figure*}

The commonly accepted formation scenario of MSPs \citep[the {\it canonical recycling scenario};][]{alpar82, bhattvan91} states that MSPs form  in binary systems containing a slowly rotating NS that, during a phase of heavy mass accretion from an evolving, non-degenerate, Roche Lobe (RL) filled companion, is eventually spun up to millisecond spin periods. 
The binary system, at its final stage, will be formed by  a deeply peeled or even exhausted star (as a white dwarf; WD) orbiting a rapidly spinning PSR \citep[e.g.][see Figure \ref{recycling} for a schematic representation of such scenario]{lyne87,alpar82,bhattvan91,handbook}.
Moreover, if the companion has been completely evaporated, the final configuration could even consists of an isolated MSP.

MSPs are thought to be the descendants of Low Mass X-ray Binaries (LMXBs;  binary systems with an accreting NS and characterized by X-ray luminosities larger than $\sim 10^{35}$ erg s$^{-1}$).
Strong support to this evolutionary scenario is given by the recently discovered transient source  \igr / \psrI\ \citep{papitto13} in the Globular Cluster (GC) M28, which has been found to swing between rotation-powered (``radio dominated'')   and accretion-powered  (``X-ray dominated'') states on short timescales, providing conclusive evidence for the evolutionary link between LMXBs and radio MSPs.

\section{Demography of MSPs}\label{Demography}

\begin{figure*}[t]
\begin{center}
\includegraphics[width=140mm]{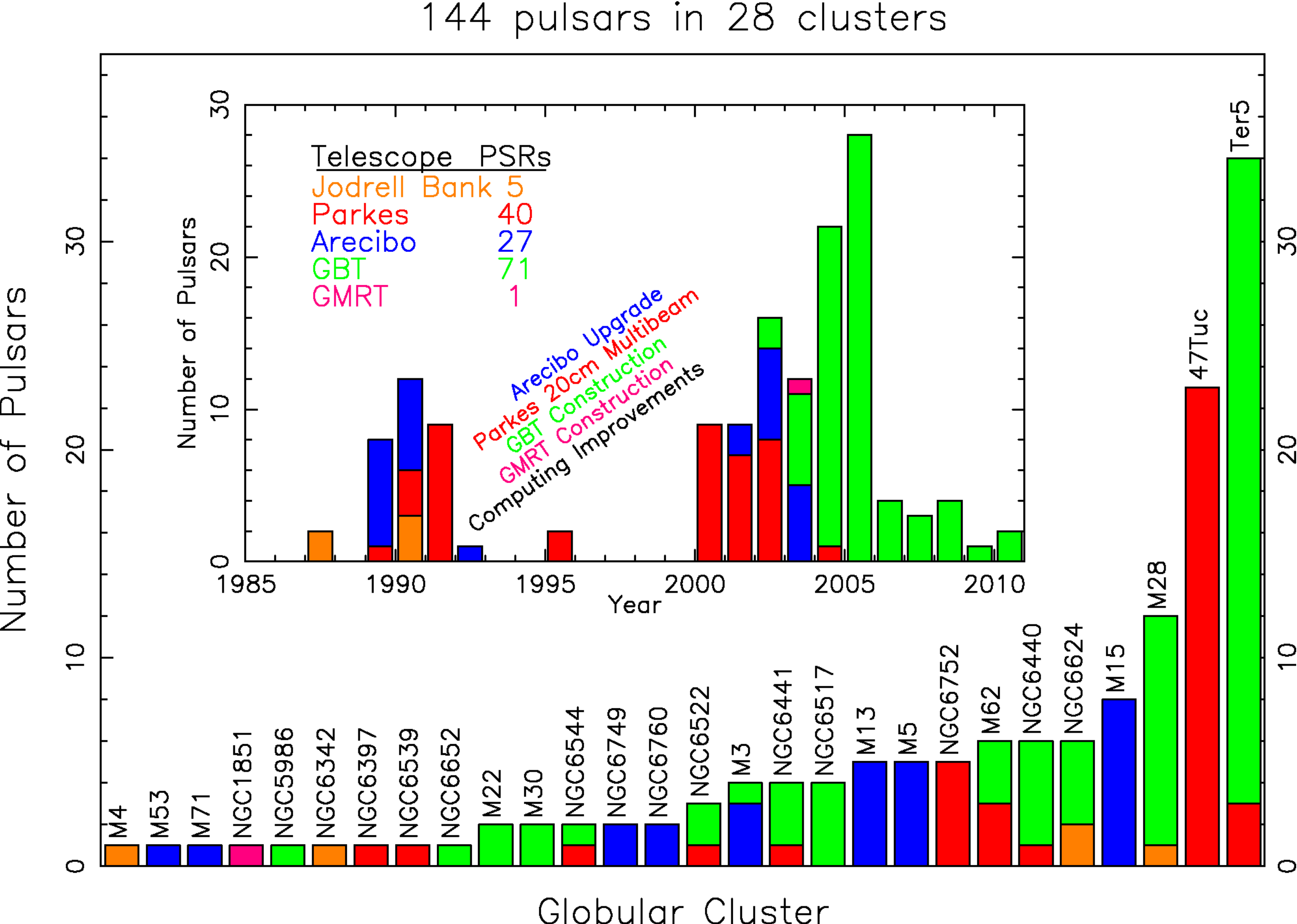}
  \caption[Histogram of the  discovered MSPs per year and GCs.]{Updated version of the plot of discoveries per year and GCs \citep{ransom07}.}\label{GCpsrs}
\end{center}
\end{figure*}

The second date to remember in the history of MSP science is 1987, when the first MSP in a GC has been discovered  \citep{lyne87}.
Currently, 132 of the 324 known MSPs are located in GCs (Patruno catalog\footnote{A recent compilation of MSPs both in  GCs and in the GF is available at the following web site: http://apatruno.wordpress.com/about/millisecond-pulsar-catalogue/}).
Finding these objects has required high-performance computing, sophisticated algorithms, state-of-the-art instrumentation, and deep observations with some of the largest radio telescopes on the world, primarily Parkes, Arecibo, and the Green Bank telescope \citep[][see Figure{\ref{GCpsrs}}]{ransom07}.  

As shown in Figure \ref{GCpsrs}, 144 PSRs\footnote{see {\it http://www.naic.edu/$\sim$pfreire/GCpsr.html} for an updated list}, out of which  $\sim 90\%$ are canonical MSPs ($P_{spin}\lapp20$ ms),  have been detected in 28 GCs and most of them are hosted in three clusters: Terzan 5 with 34 objects, 47 Tucanae with 23, and M28 with 12.

Indeed Terzan 5 deserves particular attention, since it contains the largest known population of MSPs \citep[][and Freire list]{ransom05}. 
Interestingly, it has been recently shown that Terzan 5  is not a genuine GC, but it more likely is the relic of a primordial building block that contributed to the formation of the Galactic bulge \citep{fe09ter,lanzoni10,origlia11,origlia13}. 
Two main populations with metallicities [Fe/H]$=-0.3$ and [Fe/H]$=+0.3$ have been discovered in this system\footnote{Note that only another GC-like stellar system has been found to show a similar behavior: $\omega$ Centauri in the Galactic halo. Indeed the CMD of $\omega$ Centauri shows a variety of sub populations \citep{lee99,pancino02,ferraro04,ferraro06omega,sollima07} with different iron content \citep{norris95,sollima05,johnson10,origlia03}, and for this reason it is considered to be the remnant of a disrupted dwarf galaxy  accreted by the Milky way \citep{bekkifreeman03, romano07}.} \citep{fe09ter, origlia11}. Their iron and $\alpha$-element content indicates that the metal-poorer component formed from gas enriched by a huge amount of  Supernovae II, while the metal-richer one further requires a significant contribution of Supernovae Ia.
In order to account for its remarkably high metallicity, it has been suggested that Terzan 5 was likely 100 times more massive 
in the past than observed today  \citep[its current mass being $\sim10^6\Msun$,][]{lanzoni10}.
Such a scenario would naturally explain also the  extraordinary population of MSPs currently observed in Terzan 5. 
In fact,  because of its large primordial mass and the huge amount of  Supernovae II, most of 
the produced NSs were likely retained within  the deep potential well of the  system.
In addition, its extremely large  collision rate \citep{verbunthut87,lanzoni10} could have favored the formation of binary systems containing NSs and
promoted the recycling process that finally generated the large population of MSPs now observed in Terzan 5.

\section{MSP preferred habitats}

Although the absolute number of known MSPs is larger in the Galactic Field (GF) than in GCs (192 MSPs and 132, resepctively), its value per unit mass is significantly larger in GCs. In fact, about  40\%  of known MSPs are found in GCs, although the Galaxy mass 
\citep[$2.4\times10^{11}\le M/ \Msun \le 1.2\times 10^{12}$;][]{little87,kochanek96} is 100 times larger than that of the whole GC system.

\begin{figure*}[t]
\begin{center}
\includegraphics[width=100mm]{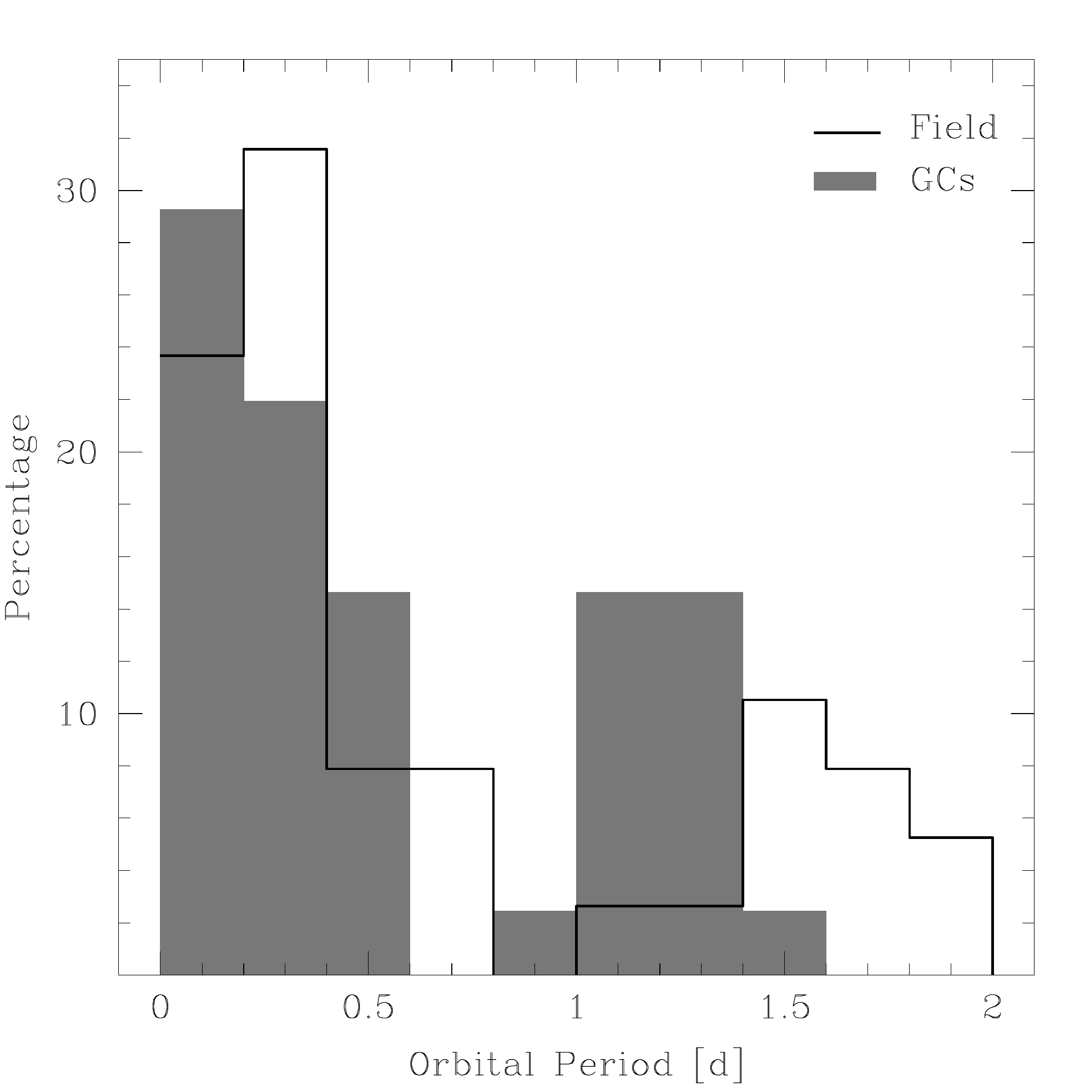}
  \caption[Number of MSPs as a function of  orbital period.]{Number of MSPs as a function of orbital period, for $P_b$ smaller than two days,  in  the GF (black histogram) and in GCs (gray filled histogram). Each population is normalized to the total number of binaries with orbital period smaller than 2 days.  Data are from the Patruno catalog.  
}\label{Pbshort}
\end{center}
\end{figure*}

Such a peculiarity can be explained by considering that,  in the
ultra-dense GC cores, dynamical interactions can promote the formation of binaries suitable for recycling NSs into MSPs \citep[e.g.][] {davieshansen98}.
Indeed GCs are very efficient ``furnaces" for generating exotic objects, which are  thought to be the result of the evolution of various kinds of binary systems originated and/or hardened by stellar interactions \citep[e.g.][]{clark75,hillsday76,bailyn92,ferraro95,fe09m30,ferraro12,ivanova08,bellazzini95}.
In fact, LMXBs, cataclysmic variables and blue stragglers  are preferentially found in GCs \citep[e.g.][]{paresce92, bailyn95,verbunt97,ferraro0147tuc, fe09m30, pooley03,fregeau08}.
An additional observed property thought to be connected to the dynamical origin of MSPs in GC is 
the distribution of orbital periods ($P_b$). In fact GC MSPs have on average shorter orbital periods than the GF population (see Figure \ref{Pbshort}).
This is likely due to the fact that,   at least in the densest GCs, most wide orbit binary systems  are disrupted\footnote{In this respect the empirical determination of the fraction of primordial binaries in GCs is of paramount importance. Some progresses have been made in this issue in the recent past (see the work by \citealp{sollima07bin} and \citealp{milone12}) also providing interesting relations as that found by \citet{sollima08} between the blue straggler star specific frequency and the binary fraction in low density clusters.}. 

\begin{figure*}[t]
\begin{center}
\includegraphics[width=100mm]{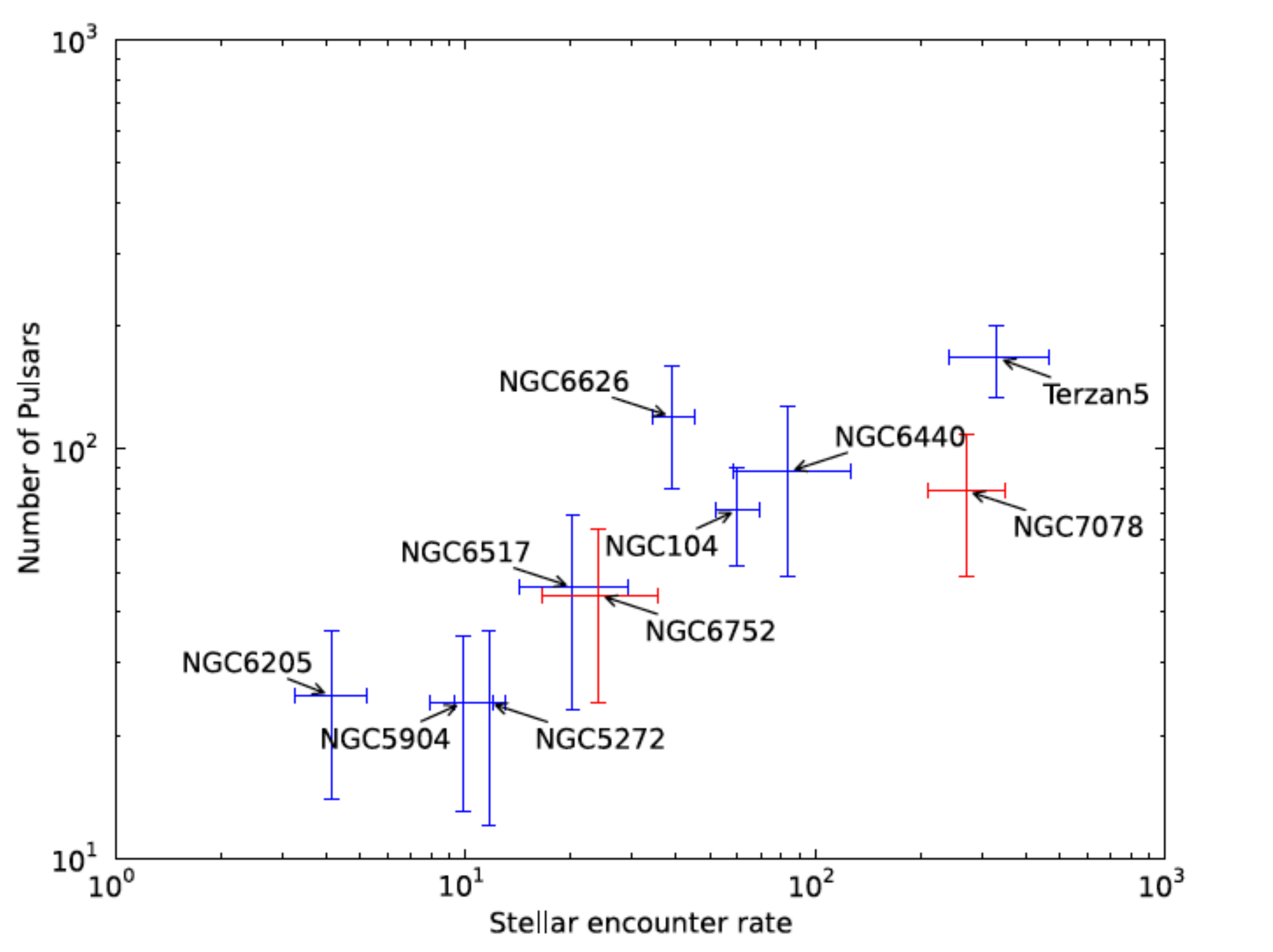}
\caption[Number of MSPs as a function of the collisional parameter.]{Number of MSPs within a GC \citep[from][]{bagchi11} compared to the calculated $\Gamma$ values (from \citealt{bahramian13}). Core-collapsed clusters are shown in red. A correlation is clearly seen. The normalization of $\Gamma$ is chosen to be similar to the choice of normalization in \citet[][$\Gamma_{\rm NGC\ 6266} = 100$]{bagchi11}.}\label{NGamma}
\end{center}
\end{figure*}

Hence, the formation and evolution of MSPs in GCs are highly affected by dynamical interactions, while in the GF they are thought to follow the natural evolutionary paths of primordial binaries.
In order to quantify the role of dynamics  in GCs, and in analogy with what done for the LMXBs population \citep{katz75, clark75}, several studies have been performed to 
check whether the number of  MSPs correlates  with some GC dynamical parameter.
It has been estimated \citep[][and references therein]{verbfrei13} that the total two-body encounter rate within a GC core of mean density $\rho_c$, velocity dispersion $\nu$ and radius $r_c$ is given by $\Gamma \propto \rho_c^2 r_c^3/\nu$. 
Indeed, the existence of a  correlation between the number of MSPs and the $\Gamma$ parameter has been investigated by many authors    \citep[see e.g.][]{pooley03, hui10, bahramian13}. The  strong correlation with $\Gamma$ is shown in Figure \ref{NGamma}.

In addition, \citet{verbfrei13} proposed a second order classification. According to these authors, despite the successful use of $\Gamma$ as indicator of the MSPs population, there are a few aspects which $\Gamma$ does not describe.
In fact, in some GCs the MSP population is mainly contributed by binaries, in others by single and exotic systems thought to be the result of exchange interactions.
In order to better characterize different populations, they proposed to use also the 
rate of encounters of a binary with a single star, given by $\gamma \propto \rho_c/\nu$, as useful indicator.
In particular they found that the higher is the value of $\gamma$, the higher is the number of  isolated and  of exchange interaction products. 
Hence, \citet{verbfrei13} concluded that $\Gamma$ is a reasonable indicator of the number of MSPs, while $\gamma$ provides a first indication of the characteristics of the PSR populations in GCs. In particular, the larger $\gamma$ is, the more deviations from the canonical scenario are expected.

\section{Deviations from the canonical scenario}
As described above,  the formation and evolution of GC MSPs are thought to be highly affected by dynamical interactions, while in the GF they are thought to follow the natural evolutionary paths of primordial binaries.
In addition, as suggested by  \citet{clark75}, given the extremely high stellar densities in the cores of some GCs, it occasionally happens that many old, dead NSs lurking in the core may ``collide'' with a binary, disrupt it and acquire a (new) companion (the so-called ``exchange-interactions''). The latter then evolves, fills its RL and starts transferring matter to the NS, forming a LMXB.
In such a scenario it is likely to observe a non-degenerate companion, bloated because of the influence of the NS.
Hence, a non-degenerate  nature of the companion  seems to be strongly related to the cluster core dynamics and in particular to the efficiency of exchange interactions. 

Therefore the characterization of MSPs and in particular  the relative frequency of  non-degenerate and  canonical companions could be used  as a powerful diagnostics of GC 
internal dynamical processes \citep{ferraro95, goodmanhut89,hut92, meylanheggie97, pooley03, fregeau08, ferraro12}. 
Of course understanding the interplay between stellar dynamics and stellar evolution is extremely complex, and many details of the involved processes are not well understood yet \citep{hut03,sills03}.   

Moreover,  thanks both to dedicated surveys of $\gamma$-ray sources and to new blind searches \citep[see][and references therein]{roberts11} the number of MSPs in the GF
has  significantly increased.
In particular, several MSPs with non-canonical companions have been identified.
Since exchange interactions are expected to be negligible in the GF,  many open questions still remain about their formation and evolutionary paths.

\section{A ``mass/type'' Classification of MSPs}\label{masstype}
The MSP population is mainly divided in isolated and binary MSPs.
Of the known MSPs, about $40\%$ are isolated and $60\%$ are members of binary systems. 
The latter group  can be also divided in different classes related to the mass range and the evolutionary stadium of the companion:
\begin{itemize}
\item High Mass MSPs: characterized  by a massive companion with mass $M \gapp 1 \Msun$ (likely a NS) ;
\item Intermediate Mass MSPs (IMMs): they have a relatively massive companion with mass $0.5\lapp M \lapp 1 \Msun$ (likely a CO/ONe-WD) ; 
\item Low Mass MSPs: characterized by a companion with mass $M \lapp 0.5 \Msun$. 
\end{itemize}
This classification  clearly reflects the differences in the latest stages of  evolution of stars with low, intermediate and high mass, that evolve to Helium (He), Carbon-Oxygen (CO) or Oxigen-Neon-Magnesium (ONeMg) WDs, and NS, respectively \citep[][]{vanKerkKulk95}.

\subsection{High mass MSPs}
Double NS binary PSRs consist of two NSs, at least one of which is pulsating at radio frequencies \citep{stairs04}.
A few high mass MSPs with  candidate NS companions are known \citep[e.g. see][]{anderson90, jacoby06, lynch12}, the most secure detection being  the double PSR system J0737$-$3039 \citep{burgaypsr2,lyne04}.
In this case  a pulsating radio signal has been detected from both NSs.
As proposed by \citet{vankerk05}, high mass MSPs can form in binaries with a NS and a massive star. Unless the orbit is very wide, the massive star will overflow its RL  during its evolution and an unstable mass transfer to the NS will occur.  Mass accretion on the NS spins it up and  reduces its magnetic field by still unknown mechanisms. Finally, if the remaining core is massive enough to experience a supernovae explosion, the resulting ``recycled'' PSR is left in an eccentric  binary with another NS.

\subsection{Intermediate mass MSPs}   
IMMs are thought to be generated from binaries with donor masses typically of $3-6~\Msun$. 
After that  the supernovae explosion generates the primary  (a NS),  the secondary evolves and recycles the PSR through mass transfer, eventually forming a CO- or ONeMg-WD with a He envelope \citep{vandenheuvel94,tauris00,taam00}.
The latter scenario may be characterized by a prolonged accretion phase, which will likely lead to spinning at shorter periods
 and with a more strongly reduced magnetic field with respect to the case of a shorter accretion phase. The WD  is expected to have still a hydrogen envelope \citep{vankerk05}. 
 Moreover, the PSR spin periods are usually a few tens of milliseconds, meaning that the PSRs are less recycled that the fastest spinning ones. 
Seventeen candidate IMM systems are currently known \citep{vankerk05, jacoby06,tauris12, burgay12}.
 
\subsection{Low mass MSPs} \label{lowmass}

The Low mass MSPs category is slightly more complex and a further classification, likely connected to  the type of the companion, is needed:
\begin{itemize}
\item[-] ``Canonical'' binary MSPs likely hosting  He-WD companions with masses   $0.1\lapp M \lapp 0.5 \Msun$, as expected in the context of the {\it canonical recycling scenario};
\item[-] ``Red back'' (RB) systems in which the companion likely  is a Main Sequence (MS) star with mass $0.1\lapp M \lapp 0.5 \Msun$;
\item[-] Black widow (BW) systems in which the companion is an almost exhausted MS star or a brown dwarf (BD) with a very low mass, $M \lapp 0.1 \Msun$.
\end{itemize} 

\begin{figure*}[t]
\begin{center}
\includegraphics[width=130mm]{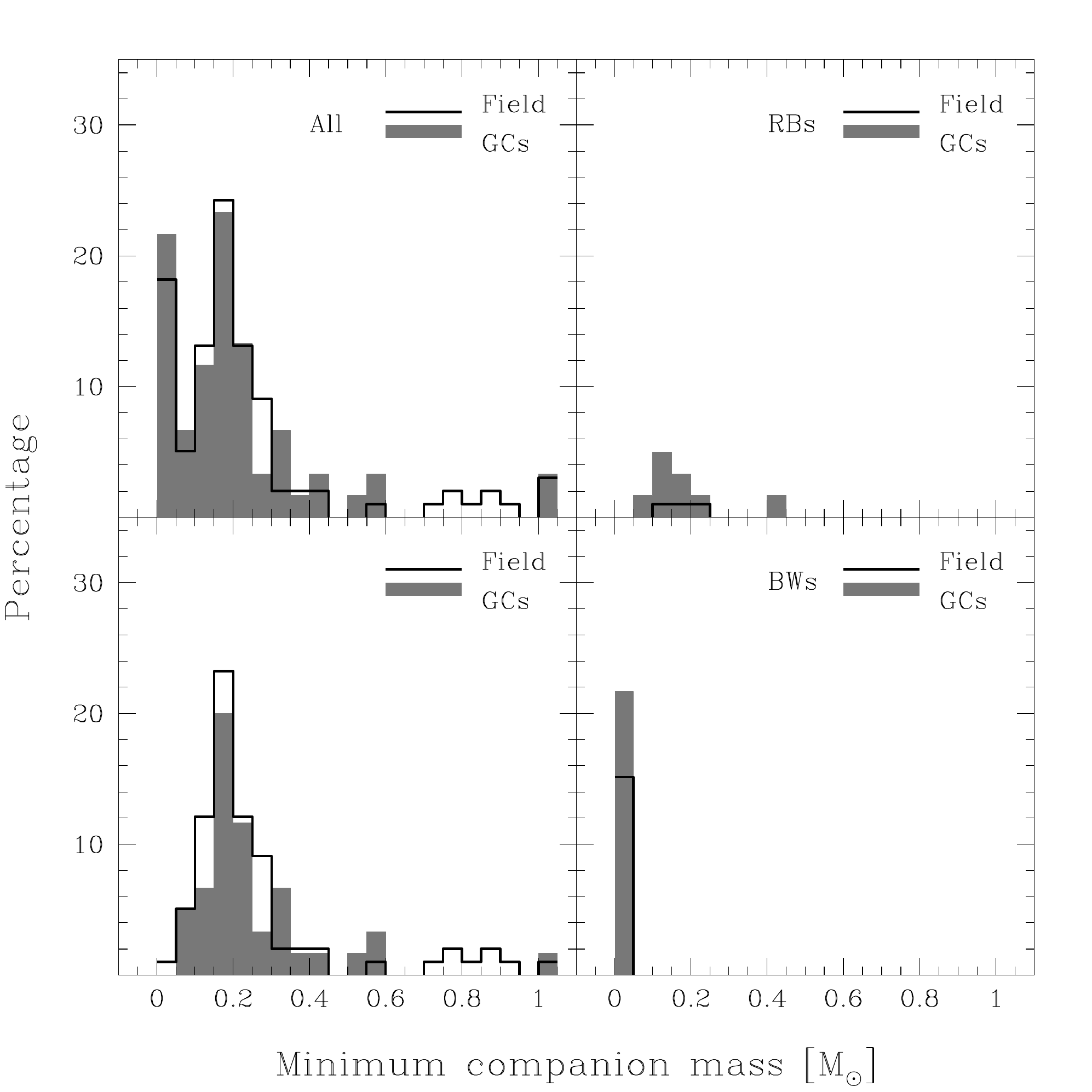}
 \caption[Number of MSPs as function of the minimum companion mass.]{Number of MSPs in bins of minimum masses of the companion (as measured from the PSR mass function assuming a NS mass of $1.4\Msun$) in both the GF (black histogram) and in GCs (gray filled histogram). Each population is normalized to the total number of binaries, for which a measure of the minimum mass is available. {\it Top-Left panel:} the whole population. {\it Top-right panel}: the RB population.  {\it Bottom-left panel}: the BW population.  {\it Bottom-right panel}: the  population complementary to RBs and BWs. Data are from the Patruno catalog.}\label{mmin}
\end{center}
\end{figure*}

This classification is well depicted in Figure \ref{mmin}.

\subsubsection{``Canonicals''}
According to the {\it canonical recycling scenario} the low mass  companion to the NS starts a stable mass transfer when it evolves and overfills its RL.
A lot of mass is transferred leading to fast spin periods (ten milliseconds),  low magnetic field, and, presumably,  greatly increased mass. If mass transfer started before the He flash, a He core WD will be left with a H envelope. If it started afterward, a WD with a CO core will be formed and the atmosphere may be of either H, or He. In both cases circular orbits are expected \citep{vankerk05}. 
``Canonical'' low mass binary MSPs have  magnetic field strengths of about $10^8$ G and likely have typical He-WD companions of  mass $\mcom\sim0.1-0.2 \Msun$. 

\begin{figure*}[t]
\begin{center}
\includegraphics[width=140mm]{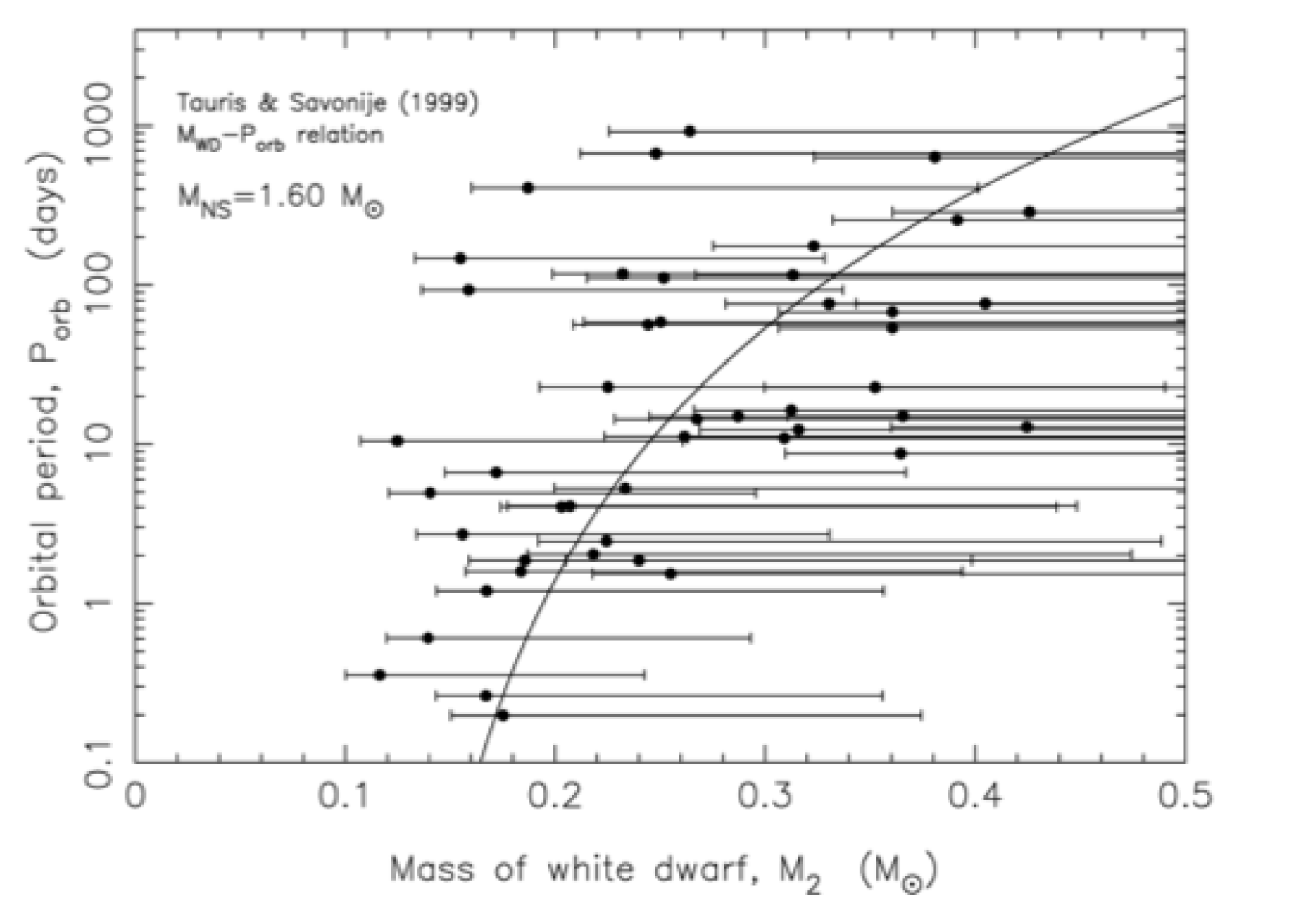}
\caption[Relation between the orbital period and the WD mass.]{ Orbital period as a function of WD mass for ``canonical'' MSPs (from \citealt{tauris11}). The error bars of the WD masses reflect the unknown orbital inclination angle: the left end corresponds to an inclination of $90^\circ$ and the right end marks the $90\%$ probability limit. The additional assumption here is that the NS mass is  $1.6\Msun$ in all cases. The theoretical  correlation is shown as  solid line \citep{taurissav99}.  }\label{relcan}
\end{center}
\end{figure*}

Moreover, in the case of canonical He-WD companions the evolutionary scenario leads to a few predictions about the properties of the binary pulsars: there should be a relation between the companion mass and the orbital period, one between eccentricity and orbital period, and the neutron star mass should have increased \citep{phinneykulk94, taurissav99, vankerk05, tauris11}.
As an example, in Figure \ref{relcan} are reported the observed orbital periods as function of companion masses for MSPs with He-WD companions \citep[see also Figure 2 in] []{vankerk05}.
The expected relation between these two parameters \citep[solid line in Figure \ref{relcan},][]{taurissav99}  is in good agreement with the observations.  In particular more massive companions tend to be members of binary systems with longer periods. 

\subsubsection{Red Backs and Black Widows}\label{introbw}

RBs show irregular eclipses and erratic timing, often have hard X-ray counterparts, and they likely have non-degenerate (e.g. MS) companions with a few tenths of a solar mass \citep{ransom08}.
Until a few years ago they have been observed only in GCs.  For this reason they were thought to be the result of exchange   interactions, where a MSP acquires a new MS companion. However, in the last few years, beyond the 10 observed in GCs, at least 7 RBs have been discovered  in the GF \citep{roberts13} thus requiring some mechanisms able to form these objects also by the evolution of isolated binaries. 

BWs  have very low mass companions ($\mcom \sim 0.04 \Msun$) and they are often characterized by eclipses of the radio signal  occurring when  the PSR passes behind its companion, thus indicating that the low mass companion is considerably bloated or is evaporating.
In some cases the eclipse of the radio signal is so extended in time that it implies a size of the companion larger than its RL,   suggesting that the obscuring material is plasma released by the companion because of the energy injected by the PSR. However since the size of the eclipse depends on the inclination angle \citep{king05}, not all BWs are expected to show eclipses.

\citet{king03} suggested that the formation of BWs needs two phases: a first one in which the companion spins-up the NS to millisecond periods and a second one where the companion is ablated by the PSR. 
In general it is difficult to describe the two phases using the same star as a companion. However, in GCs, where encounters and exchange interactions are frequent, the WD companion responsible for the PSR spinning-up can be replaced by a MS star  via an exchange interaction.
BWs are frequently found with BDs companions  \citep{vankerk11, pallanca12,romani12,breton13,kaplan13,archibald13,bellm13,stappers01}.
However these stars were likely captured by the PSR when still having  larger masses, but then they suffered a progressive vaporization in a long X-ray binary phase because of the strong irradiation flux by the PSR.
While this scenario could be applied in GCs, it is less likely in the GF where dynamical interactions are less probable.
Indeed, until recently only few BWs were known in the GF and they were thought to be formed in GCs and then ejected. However the increasing number of BWs discovered in the last years \citep[to date 17 BWs are known in the GF, while 18 are observed in GCs; Patruno catalog;][]{roberts13}  asks for a new interpretative scenario able to efficiently create BWs in the GF.

As noticed above, the formation of RBs and BWs is not well understood. Both populations have circular orbits and periods $\lapp 1.5$  days, but they have companion masses of the order of  $\sim 0.2$ and $\sim 0.04\Msun$, respectively \citep{chen13}. 
\citet{chen13} have recently proposed a common formation scenario for these two classes of objects. They found that the determining factor for producing either BWs or RBs is the different efficiency of the irradiation process.
In particular this bimodality is likely related to the geometric effect of beaming that depends on the orientation of the PSR  magnetic field axis with respect to the direction of the companion star.

\section{Eccentric MSPs}\label{Sec:ecc}
\begin{figure*}[b]
\begin{center}
\includegraphics[width=120mm]{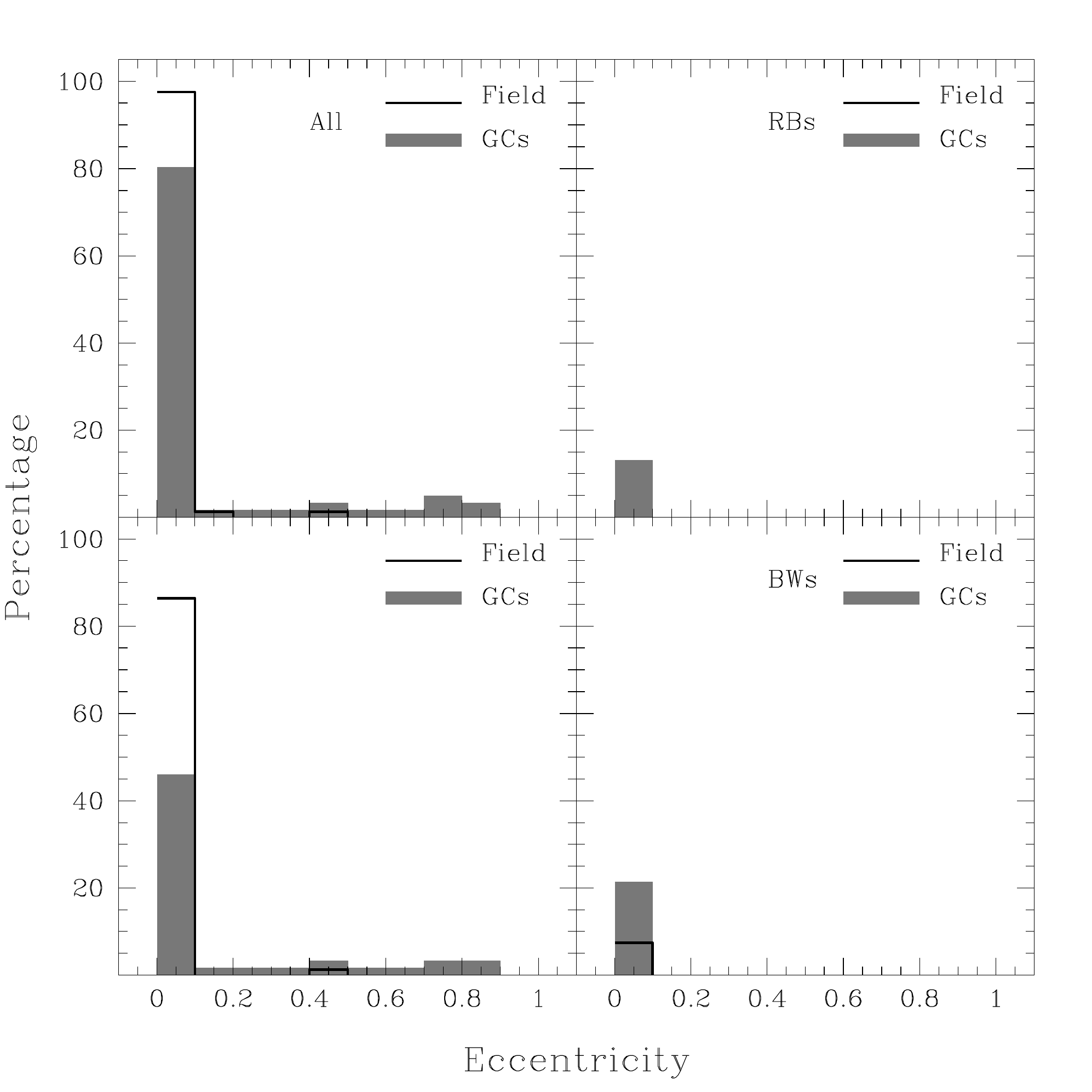}
  \caption[Number of MSPs as function of eccentricity.]{Number of MSPs in bins of eccentricity  both in the GF (black histogram) and in GCs (gray filled histogram). Each population is normalized to the total number of binaries, for which a measure of the eccentricity is available. {\it Top-Left panel:} the whole population. {\it Top-right panel}: the RB population.  {\it Bottom-right panel}: the BW population.  {\it Bottom-left panel}: the complementary populations to RBs and BWs. Data are from the  Patruno catalog.}\label{ecc}
\end{center}
\end{figure*}

In addition to the ``mass/type classification'' a bimodality  in the eccentricity values could also be used as an element of classification.
In fact, most of MSPs are in circular orbits ($e<0.1$), but a few having a  significantly high eccentricity value (see Figure \ref{ecc}). 

Eccentric binary MSPs are important because in these systems it is possible to measure at least the rate of advance of periastron.
If this is totally due to the effects of general relativity, it yields an estimate of the total mass of the binary. Moreover, when more post-Keplerian measurements become available, it is possible to derive the masses of both components and in some cases to test general relativity \citep[][see also Appendix]{lynch12, jacoby06}.

In principle  these systems should be rare because eccentricities should be reduced by tidal effects in case of a bloated companion (e.g. during the LMXB phase).
The main scenario proposed to explain the existence of eccentric binary systems requires either a recent  dynamical interaction (likely scenario in GCs), or a double NS system.
This is in agreement with the observed properties of MSPs. In fact, in the GF the fraction of MSPs in circular orbits correspond to about $100 \%$ of the total \citep[with only a few, recently detected, exceptions,][]{deneva13,barr13}, while in GCs it is reduced to $80 \%$ (see Figure \ref{ecc}).  It is also interesting to note that almost all the RBs and BWs are in circular orbits, likely because of circularization due to tidal effects from their non-degenerate companion.

However, \citet{deneva13} and \citet{barr13} recently discovered two binary  MSPs with unusual  eccentricities ($e\gapp0.1$)  in the GF.
To explain the existence of such unexpected objects, \citet{freiretauris13} proposed an alternative mechanism to the {\it canonical recycling scenario}.
They proposed that MSPs can also be formed directly through a ``rotationally-delayed accretion-induced collapse'' of a super-Chandrasekhar mass WD.
In such a way the accretion ceases completely before the formation of the MSP and there will be no re-circularization.

\clearpage{\pagestyle{empty}\cleardoublepage}

\chapter{Optical study of companions to MSPs}\label{Chap:introCOM}

The identification of companions to MSPs and the study of their nature  is a fundamental step to investigate the formation and evolution of binary MSPs, and  to clarify the recycling processes occurred.
In the case of GCs   \citep{ferraro01com6397, ferraro03com6752,cocozza08, pallanca10, pallanca12,pallanca13com1439,pallanca13comM28I}, it also represents a crucial tool for quantifying the occurrence of dynamical interactions, understanding the effects of crowded stellar environments on the evolution of binaries, determining the shape of the GC potential well, and estimating the mass-to-light ratio in  GC cores \citep [e.g.,][]{phinney92, bellazzini95, ferraro03dyn6752, possenti03}.

Typically   the companion star to a MSP  is expected to be  a MS or a WD (see Chapter \ref{Chap:introPSR} for details). Therefore its emission should peak in the Ultraviolet-optical region of the electromagnetic spectrum.
The study of these targets essentially consists of two steps: first, the photometric identification and characterization of the companion and then, if it is bright enough, a  spectroscopic follow-up to constrain the orbital parameters of the system and its (likely anomalous) chemical composition.

In this Chapter I will describe the needed data-set and the adopted analysis procedure to perform the search and a thorough study of MSP companions.

\section{Photometric analysis}

\subsection{The data-set}

\begin{figure*}[b]
\plottwo{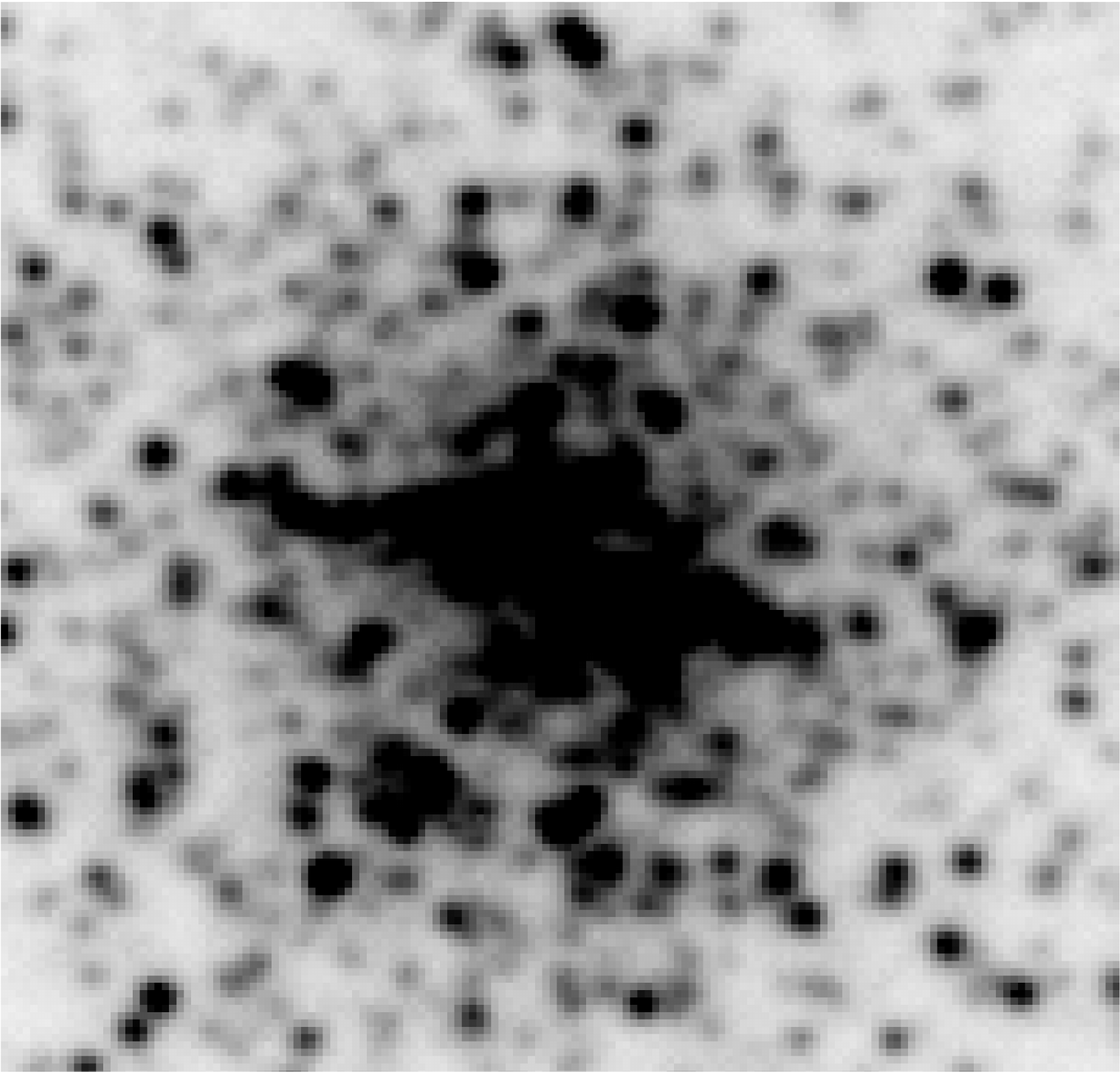}{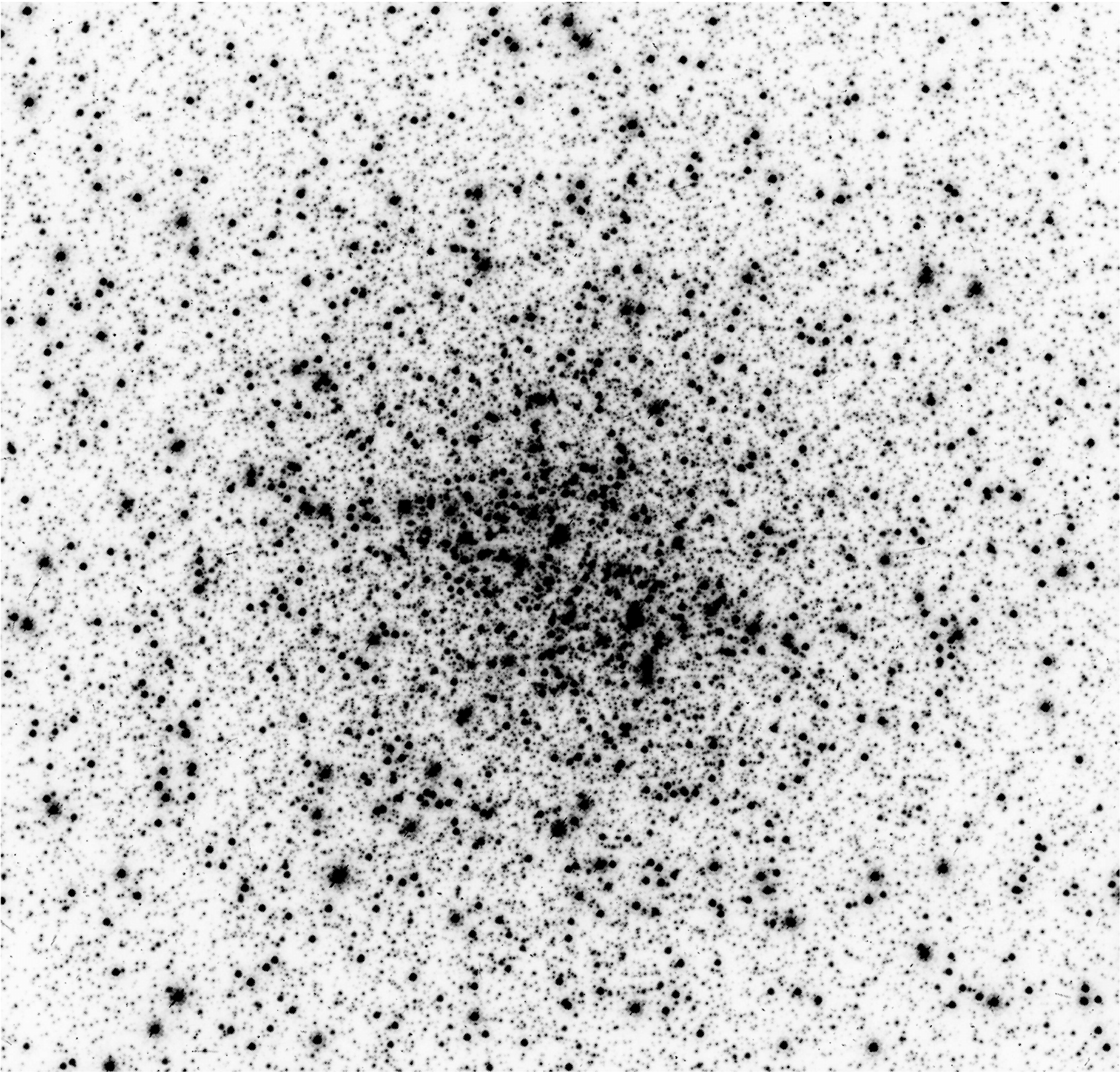}
\caption[Comparison between a ground-based and a HST space image.]{Comparison between a ground-based image (left panel) and a space image (right panel) of the central region of the GC NGC 6440. In the space image the population is clearly much better resolved.}\label{groundspace}
\end{figure*}

The photometric data-sets used in this Thesis have been obtained through both space and ground-based telescopes (see Figure \ref{groundspace} for a comparison).
Since  MSPs in GCs are typically found in the innermost and highly crowded regions, where they sink because of their large mass, their search requires the use of high resolution space data.  At odd with the GF, where crowding does not represent an issue, the use of ground-based facilities guarantees a  good  enough  quality for this kind of studies.

The space data used in this Thesis have been  obtained with the Wide Field Camera 3 (WFC3), the Advanced Camera for Survey (ACS) and the Wide Field Planetary Camera 2 (WFPC2) on board the Hubble Space Telescope (HST).
All these detectors are characterized by a pixel scale of the order of $\sim 0.04-0.05\arcsec$/pixel and a field of view (FOV) ranging from $\sim 80\arcsec\times 80\arcsec$ for  the WFPC2, to  $\sim 162\arcsec\times 162\arcsec$ for the ultraviolet visible (UVIS) channel of the WFC3 and $\sim  202\arcsec\times 202\arcsec$ for the Wide Field Camera of ACS.  
Such characteristics allowed to resolve the crowded stellar population in the central region of GCs.
Note that the typical Full Width Half Maximum (FWHM) of the stellar brightness profile is about  0.06-0.08$\arcsec$ for these data (see Section \ref{psf-fitting} for more details).

The data-sets used to study MSPs in the GF consist of ground-based images.
Unlike the space observations, the ground-based images are  limited by  the perturbative effects of the atmosphere and the FWHM is at most  of the order of $0.6-0.8\arcsec$.  
In Figure \ref{groundspace} we show a comparison between a ground-based image (obtained with the Wide Field Imager mounted at the MPG/ESO telescope at La Silla) and a space image of the core of the Galactic GC NGC 6440. Clearly, stars can be properly resolved only by using the latter.

The ground-based data-sets used in this Thesis, have been acquired with the FOcal Reducer/low dispersion Spectrograph 2 (FORS2) mounted at the ESO Very Large Telescope (VLT).
Two resolution modes and two binnings are available. The combination of these settings leads to different pixel scales ($0.126\arcsec$/pixel or $0.25\arcsec$/pixel). Moreover, in order to obtain deep images free from the blooming due to heavy saturation of bright stars, which can significantly limit the search for faint objects, all the brightest stars in the FOV have been  covered with occulting masks.

In order to perform a thorough analysis in the context of the study of MSPs, ad hoc data-sets made of deep, multi-wavelength and multi-epoch exposures  are required.
In fact  multi-band observations allow to derive colors and hence to build CMDs and color-color diagrams, that are useful diagnostics to study MSP companions (e.g. to constrain their degenerate or non-degenerate nature). 
Also observations in a narrow filter, corresponding to the \halpha\ transition, are a useful diagnostic to select objects with \halpha\ excess (likely due to some ionized material).

 On the other hand also a multi-epoch data-set is an important requirement, because it allows  to detect both  luminosity modulation on time-scales of the orbital period, likely related to the orbital motion,  and possible significant changes of magnitude between different epochs.

\subsection{The photometric analysis}\label{psf-fitting}

In order to fully exploit the telescope capabilities it is important to adopt an accurate method for the data analysis.

In this Thesis, the photometric  analysis has been carried out by using the Point Spread Function (PSF)-fitting method, which consists in reproducing the star brightness profile with  an analytical model.
The most commonly used analytical model for resolved stars is a Moffat function \citep{moffat69}, which can be expressed in the following way:
\begin{displaymath}
f(x,y)=\left [1+\frac{(x-x_0)^2+(y-y_0)^2}{\sigma^2}\right]^{-\beta}
\end{displaymath}  
where $\sigma$ and $\beta$ are two parameters related to the curve width and  the height of the wings of the brightness profile, respectively. 
The FWHM depends on these two quantities following the relation FWHM=$2\sigma \sqrt{2^{1/\beta}-1}$.

\begin{figure*}[b]
\begin{center}
\includegraphics[width=140mm]{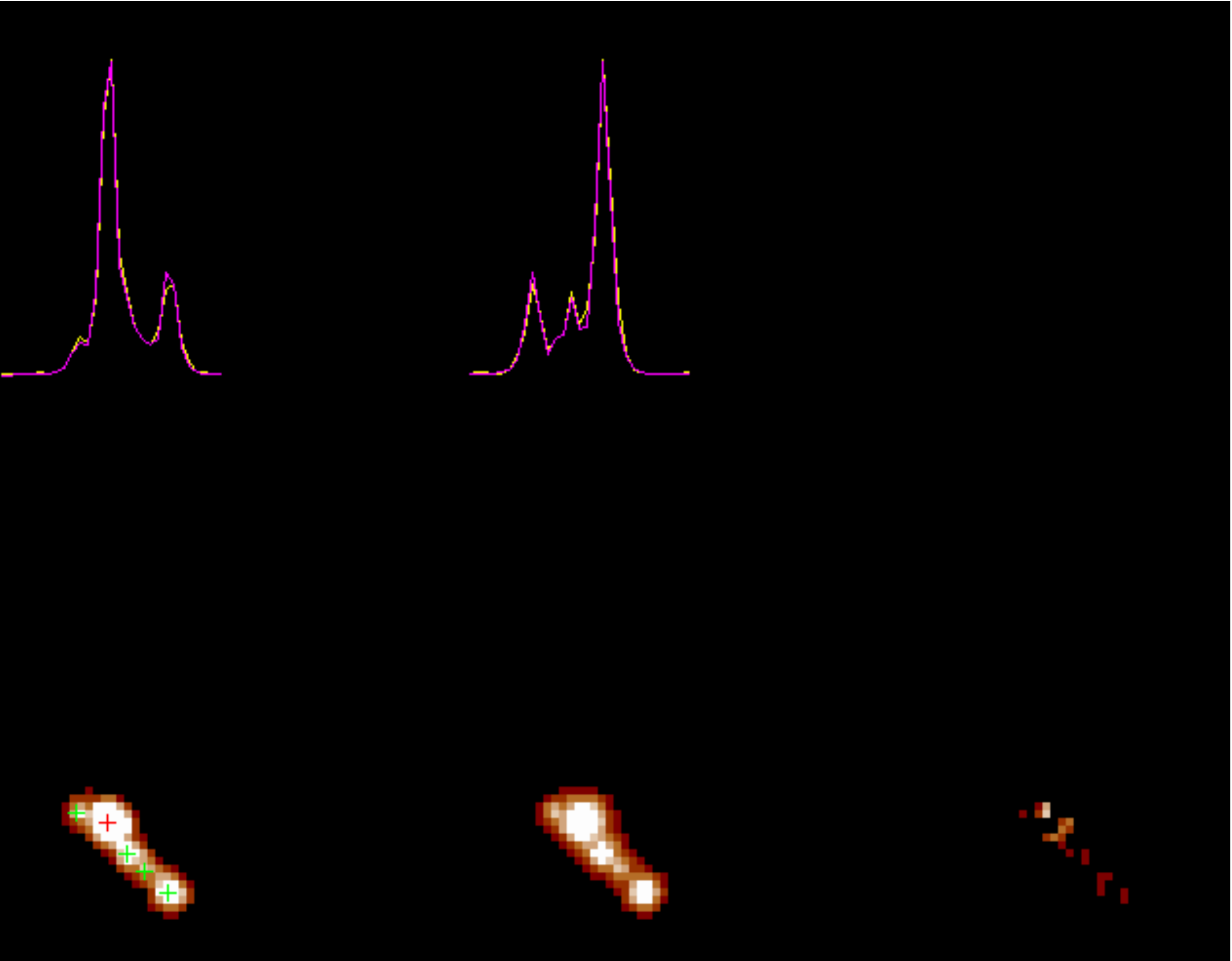}
  \caption[Graphical scheme  of the PSF-fitting method.]{Graphical example of the PSF-fitting method performed with the software {\sc romafot} \citep{buonanno83}. The lower three panels from left to right  are the observed image, the reconstructed image and the residual image, respectively. The top two panels are the projections of the observed image along the horizontal and the vertical directions. The yellow curve is the observed total light profile and the magenta is its reproduction obtained with PSF-fitting method. Each fitted star marked by a cross is reproduced through a PSF-model previously  built on properly selected isolated stars. With this method we are able to reproduce both the peak of the brightest stars and
the light increase due to faint stars on the wings of  brighter neighbors.}\label{romapsf}
\end{center}
\end{figure*}

The crucial advantage of using the PSF-fitting method is that, once derived a proper PSF-model, it allows to estimate the magnitude of each star independently of the surrounding region. In particular, it allows to reproduce the intrinsic brightness profile  of each star by avoiding the contamination from close (maybe brighter) stars
and by accurately estimating the background level.
Such qualities are crucial in crowded environments (as GCs). In fact,  as shown in Figure \ref{romapsf},   even in the case of partially overlapped stars, the PSF-fitting allows to 
deconvolve the light contribution of each component, thus recovering the effect of mutual contamination. 

To apply this photometric method we used the {\sc daophot} package \citep[][]{stetson87} and the software {\sc romafot} \citep{buonanno83}.

To model the PSF, it is important to use a large number ($\sim 50 - 200$) of bright, isolated and not
saturated stars homogeneously  distributed in the FOV.  
Starting from a family of PSF-models, the most appropriate for a given image is chosen on the basis of a $\chi^2$-test performed on the selected stars.
This procedure is automatically performed by the {\tt PSF} routine of  {\sc daophot}.
Once the best-fit PSF-model has been selected, it is applied to all the sources detected in the images.

In order to obtain a list of stars as much complete as possible we typically used the following approach.
We combined all the images in the sample by using  the {\sc daophot} {\tt MONTAGE2} routine, thus to obtain a master-frame with a S/N larger that in single images. Since the master-frame is the result of a combination of images obtained with different filters, it  also has a higher sensitivity to different stellar spectral types.
On the master-frame we performed by using the {\sc daophot} {\tt FIND} routine a search for sources at  $3\sigma$ from the local background (where  $\sigma$ is the standard deviation of the measured background), thus obtaining a master-list of stars.

Finally, starting from the master-list we forced the PSF-fitting on each single image by
 using the {\sc daophot} packages {\tt ALLSTAR} and {\tt ALLFRAME} \citep{stetson87, stetson94}.
This procedure allows us to achieve an improved determination of the  star centroids and a better reconstruction of the star  intensity profiles.

In the final catalog both single image instrumental magnitudes and the mean magnitudes in each filter are reported.
In particular, for each star the magnitudes estimated in different images are homogenized \citep[see][]{ferraro91,ferraro92} and their weighted mean and standard deviation are adopted as the star magnitude and its photometric error, respectively.

\subsection{Calibration}

We reported the instrumental magnitudes to  a standard photometric system. HST data were reported to the VEGAMAG \citep[e.g.][]{sirianni05} photometric system, while ground-based data to the classical Johnson one \citep{johnson53, johnson66}.

In the case of observations made with HST, the PSF and the instrumentation performances are very stable. Therefore the calibration is performed by adopting very simple equations. In particular. 
the calibrated magnitude ($m_{cal}$) is given by \citet{holtzman95}:
\begin{displaymath}
m_{cal}=m_{instr}+2.5 \lg(t_{exp})+ZP+AC
\end{displaymath}
where $m_{instr}$ is the instrumental magnitude measured as $-2.5\lg(counts)$ and
$t_{exp}$ is the exposure time of a given observation.
The Zero Point (ZP)  is the term that reports the magnitude to a particular photometric system and it depends also on the wavelength. Its value is available at  the instrumentation web pages. 
The Aperture Correction (AC) quantifies the missed light in the wings of a truncated (for computational reasons) PSF-model.

In the case of the ground-based observations, since the PSF depends on the photometric conditions during the observations (mainly the seeing), proper calibration equations have to be derived every time.
This can be achieved by observing, with the same configuration of the scientific target observations and during the same night, a few photometric standard stars.
Then, by comparing the measured instrumental magnitudes in the standard field with those tabulated, for example in the {\it standard  Stetson catalog} available on the  CADC web  site\footnote{http://cadcwww.dao.nrc.ca/community/STETSON/standards/} \citep{stetson00}, it is possible to derive the proper transformation to report the instrumental magnitudes to the Johnson system. Finally, the derived equations are applied to the instrumental magnitudes obtained for the scientific targets.

\subsection{Astrometry}
The astrometry is the procedure that consists in the conversion  from instrumental coordinates relative to the detector, to the absolute coordinate system, e.g. in terms of  {\it Right Ascension} and as {\it Declination}, commonly cited both as ($R.A.$, $Dec$) and ($\alpha$,$\delta$).

Since the PSR radio positions are known with a very high precision ($<0.1\arcsec$), obtaining  accurate astrometry is a critical requirement for searching for 
the optical companions to binary MSPs.

Instrumental coordinates (x,y) are reported to the absolute coordinate system  ($\alpha$,$\delta$) through the cross-correlation with catalogs of astrometric standards \citep[as the GSCII\footnote{The Guide Star Catalogue-II is a joint project of the Space Telescope Science Institute and the Osservatorio Astronomico di Torino. Space Telescope Science Institute is operated by the Association of Universities for Research in Astronomy, for the National Aeronautics and Space Administration under contract NAS5-26555. The participation of the Osservatorio Astronomico di Torino is supported by the Italian Council for Research in Astronomy. Additional support is provided by European Southern Observatory, Space Telescope European Coordinating Facility, the International GEMINI project and the European Space Agency Astrophysics Division.}, 2MASS and PPMXL;][]{2mass06,roeser10}.
To this aim we use CataXcorr\footnote{CataXcorr is a code aimed at cross-correlating catalogs and finding astrometric solutions, developed by P. Montegriffo at INAF - Osservatorio Astronomico di Bologna. This package has been used in a large number of papers of our group in the past 10 years.}, a software that  finds the best roto-translation between the (x,y)  and  the ($\alpha$,$\delta$) coordinates by using the stars in common with the publicly  available catalogs of astrometric standards.
Then the best astrometric solution is applied to all the stars in a given catalog.
Usually, given the current instrumentations and standard catalogs, the root mean squares of the transformations are  of the order  of $0.2\arcsec$ in both $\alpha$ and $\delta$. Hence the  astrometric solution  typically has an accuracy $\lapp0.3\arcsec$.

ACS and WFC3 images heavily suffer from geometric distortions within their FOVs.
Hence, before reporting the instrumental coordinates to the absolute system they have to be properly corrected.
To this aim it is possible to use equations quoted by \citet{anderson03}, \citet{bellini09} and \citet{bellini11}.

There are cases, however, for which it is not possible to find astrometric standards. This is typically the case for the cores of GCs and, therefore, for the HST data-sets. In these cases, the astrometric procedure requires an additional step. We complement the HST sample with  ground-based wide-field catalogs.
We then place the latter onto the absolute coordinate system by using the stars in common with the available astrometric standard catalogs. Then we use the stars in common between the HST and the  ground-based data as secondary astrometric standards. At this point we re-run the same procedure described above and finally obtain the absolute astrometry for the HST sample. 

The centre of gravity of each cluster ($C_{grav}$) has been determined by averaging $\alpha$ and $\delta$ coordinates of a subsample of stars (typically the high-resolution sample) following the iterative procedure described in \citet[][see also \citealt{ferraro0147tuc,ferraro03dyn6752}]{montegriffo95}.

Another important point to deal with when GF stars have to be put in the absolute reference frame, is proper motions. 
Indeed, proper motions  in the GF may not be negligible, thus strongly affecting the internal astrometric accuracy.
In order to obtain a precise astrometric solution also in these cases it is necessary to use catalogs of astrometric standards also listing the stellar proper motions \citep[e.g][]{roeser10},
and compute the absolute position of the standard stars at the same epoch of the target observations. Then, a standard procedure can be applied to cross-correlate the observed list of positions with the proper motion-corrected reference catalog.

\section{Photometric search for MSP companions}

Starting from the information obtained with radio observations, the identification of the optical companions to MSPs is based on three signatures: positional coincidence, location in the CMD and optical variability.

\subsubsection{Positional coincidence}
\begin{figure*}[!t]
\begin{center}
\includegraphics[width =140mm]{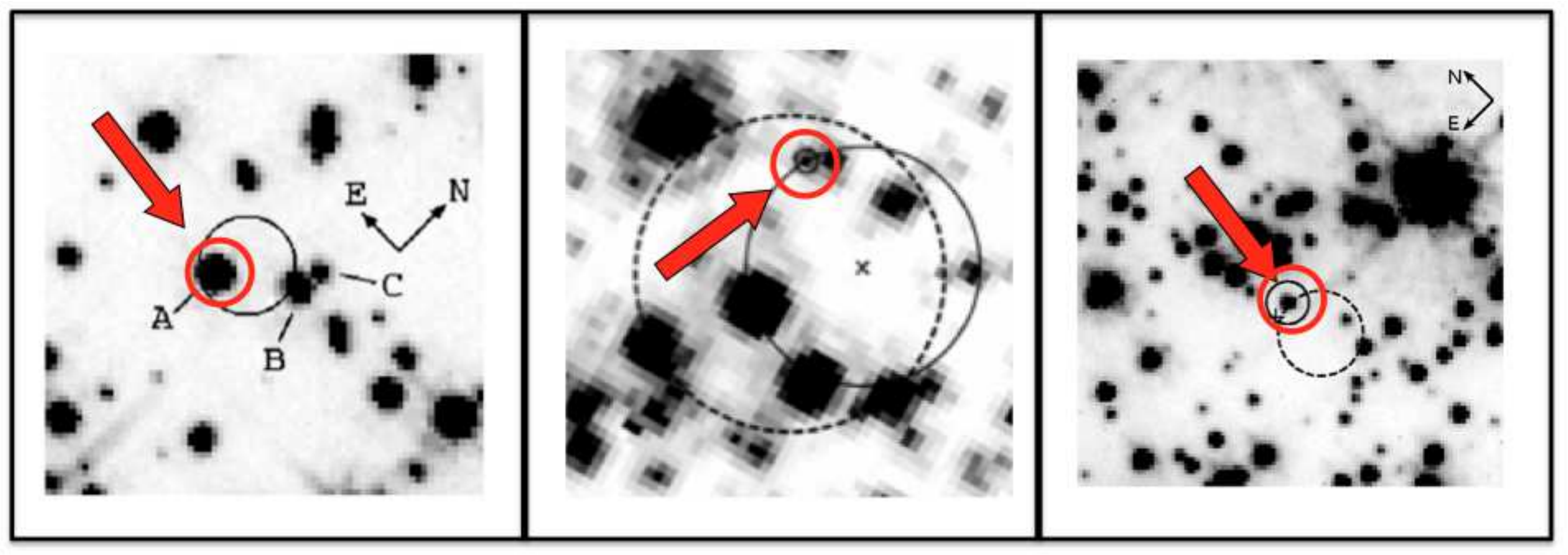}
  \caption[Example of positional coincidence of companions to MSPs.]{Maps of the regions around three identified companions to MSPs in GCs. From left to right: PSR J1740$-$5340A \citep{ferraro01com6397}, PSR J1701$-$3006B \citep{cocozza08} and PSR J1824$-$2452H \citep{pallanca10}. In each panel the identified companion is marked by a red circle. Each of them is located within the astrometric uncertainty of the nominal radio position (black circle in the left panel and cross in the middle and right panels).
   }\label{poscoin}
\end{center}
\end{figure*}

To identify a companion to MSP a necessary (but not sufficient) condition is the positional coincidence:
the candidate companion has to lie within the astrometric error-box, as shown in Figure  \ref{poscoin}.
The PSR position is accurately ($<0.1\arcsec$) known from the radio observations. Therefore it is extremely important to have a good astrometric solution for the optical catalog. This is particularly relevant for the case of  crowded GCs, where  more than one candidate within the positional error is  typically found.

\subsubsection{CMD position}
\begin{figure*}[!t]
\begin{center}
\includegraphics[width =99mm]{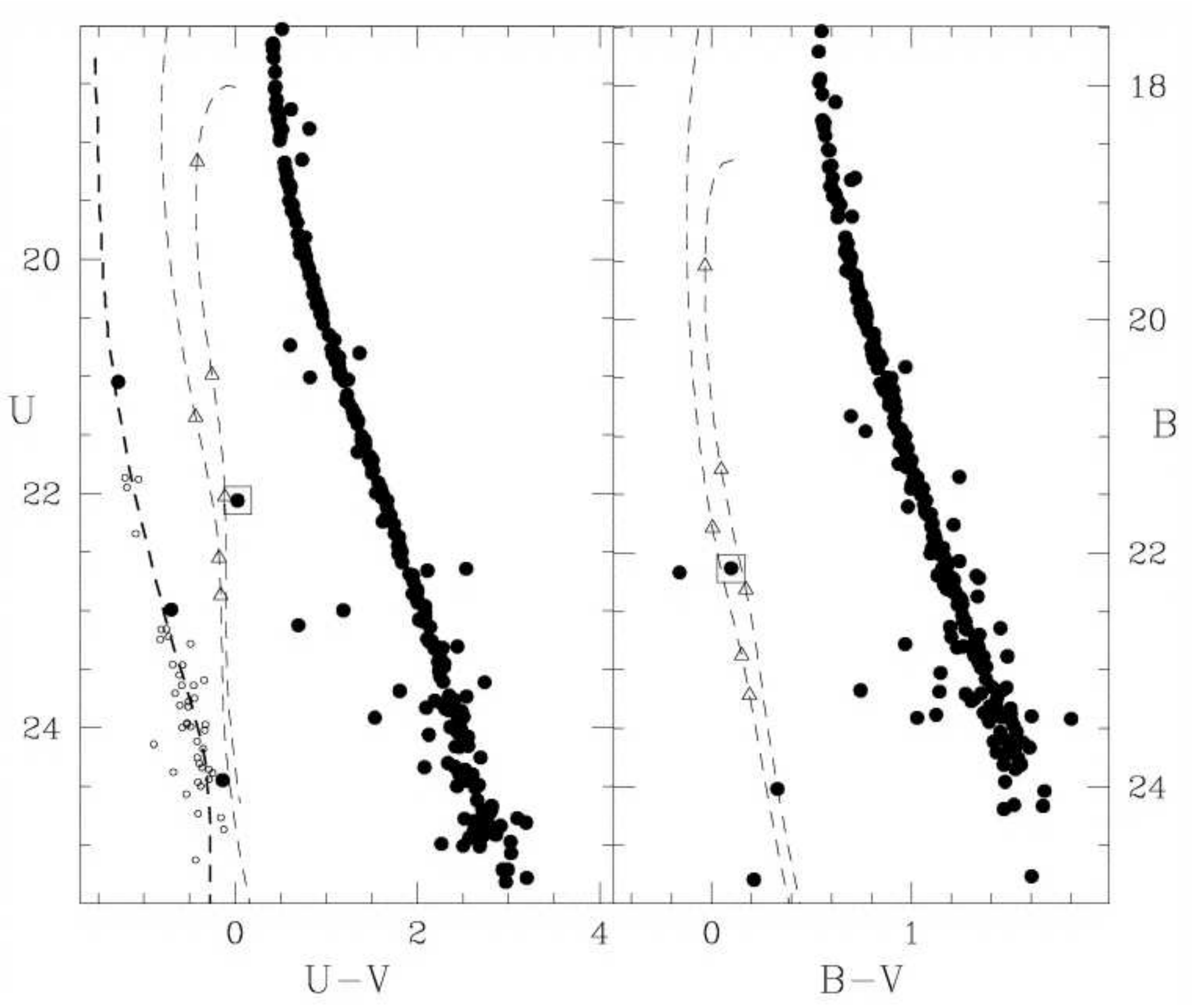}
  \caption[Example of the CMD position of a He-WD companion.]{ CMD position of the companion to PSR J1911$-$5958A in the GC NGC 6752. Its location is in agreement with the He-WD cooling sequences \citep[from][]{ferraro03com6752}.}\label{cmdWD}
\end{center}
\end{figure*}

\begin{figure*}[!b]
\begin{center}
\includegraphics[width =99mm]{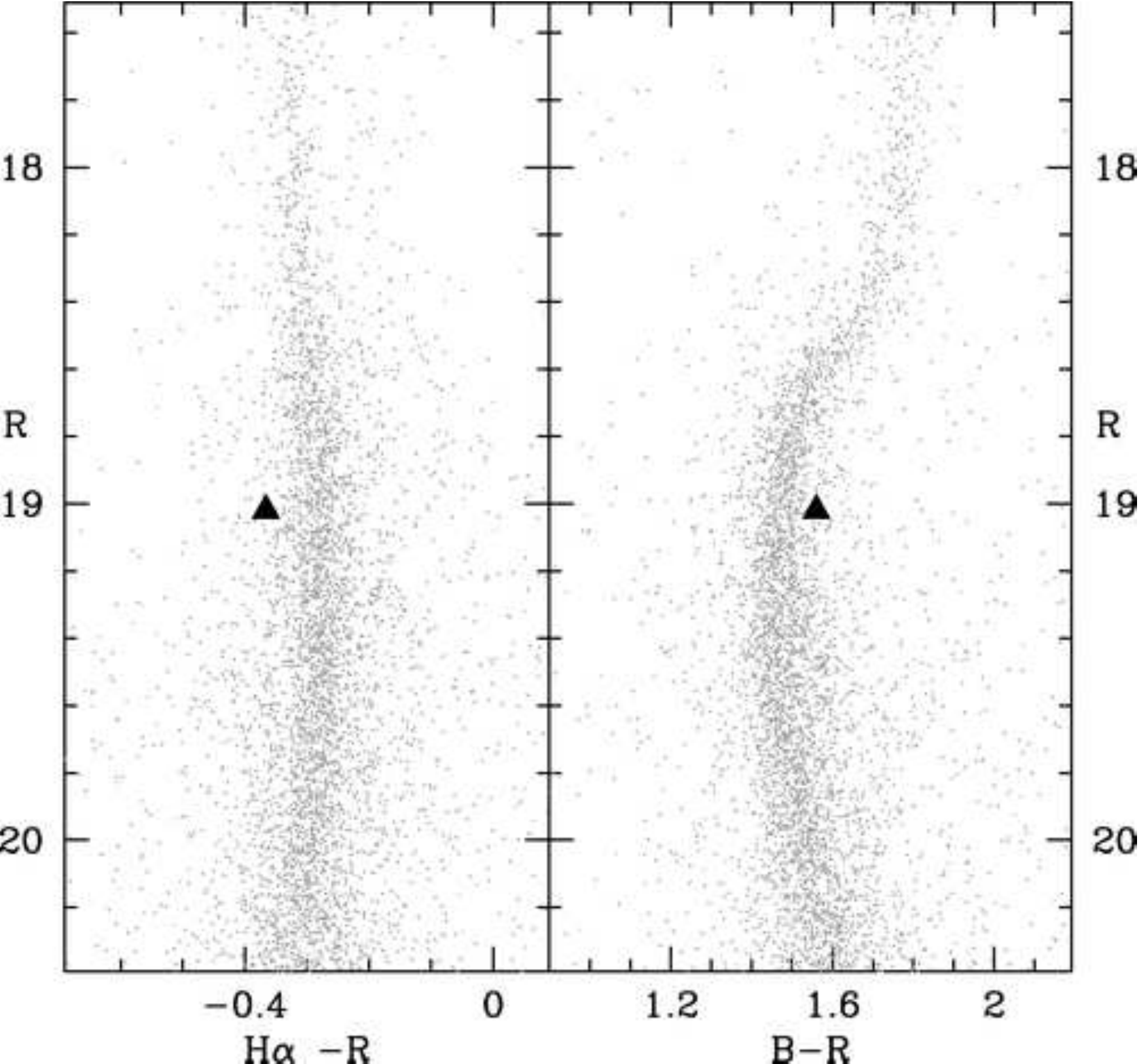}
  \caption[Example of the CMD position of a non-degenerate companion.]{ CMD position of the companion to PSR J1701$-$3006B in the GC NGC 6266. Its location  denotes a non-degenerate nature of the companion  \cite[from][]{cocozza08}.}\label{cmdMS}
\end{center}
\end{figure*}

A  data-set  made of multi-band (to derive colors) and deep (to reach the faintest objects) observations, allows us to use the CMD as an important tool to identify and characterize MSP companions.
In fact, since different kinds of companions (likely related also to different radio properties) are expected,
the CMD position is a good indicator of the nature (degenerate or not) of the companion.
In particular it can allow us to distinguish if the companion is a WD as predicted by the {\it canonical recycling scenario}, or if it is  a non- or semi-degenerate companion, like a MS or a BD (see Figures \ref{cmdWD} and \ref{cmdMS}).


In addition, from the CMD position,   it is possible to estimate the companion mass through the comparison with theoretical stellar evolutionary models.
In the particular case of a WD companion the position along WD cooling sequences is a good indicator of mass and age\footnote{This is important because by comparing the derived cooling age with the spin down age  is possible to put some constrain on the spin-down theory (see Appendix).}. 
It is important to note, however, that in the case  of a non-degenerate star that filled its RL, 
the mass estimate obtained by projecting its position on a theoretical model,  might be overestimated because the stellar structure might be not in 
hydrostatic  equilibrium.

In addition, when three observational bands are available the color-color diagram can be investigated.
This  is particularly advantageous in the case of MSPs in the GF because it offers the possibility to accurately estimate the value of the stellar  extinction E(B-V).

\begin{figure*}[b]
\begin{center}
\includegraphics[height=130mm]{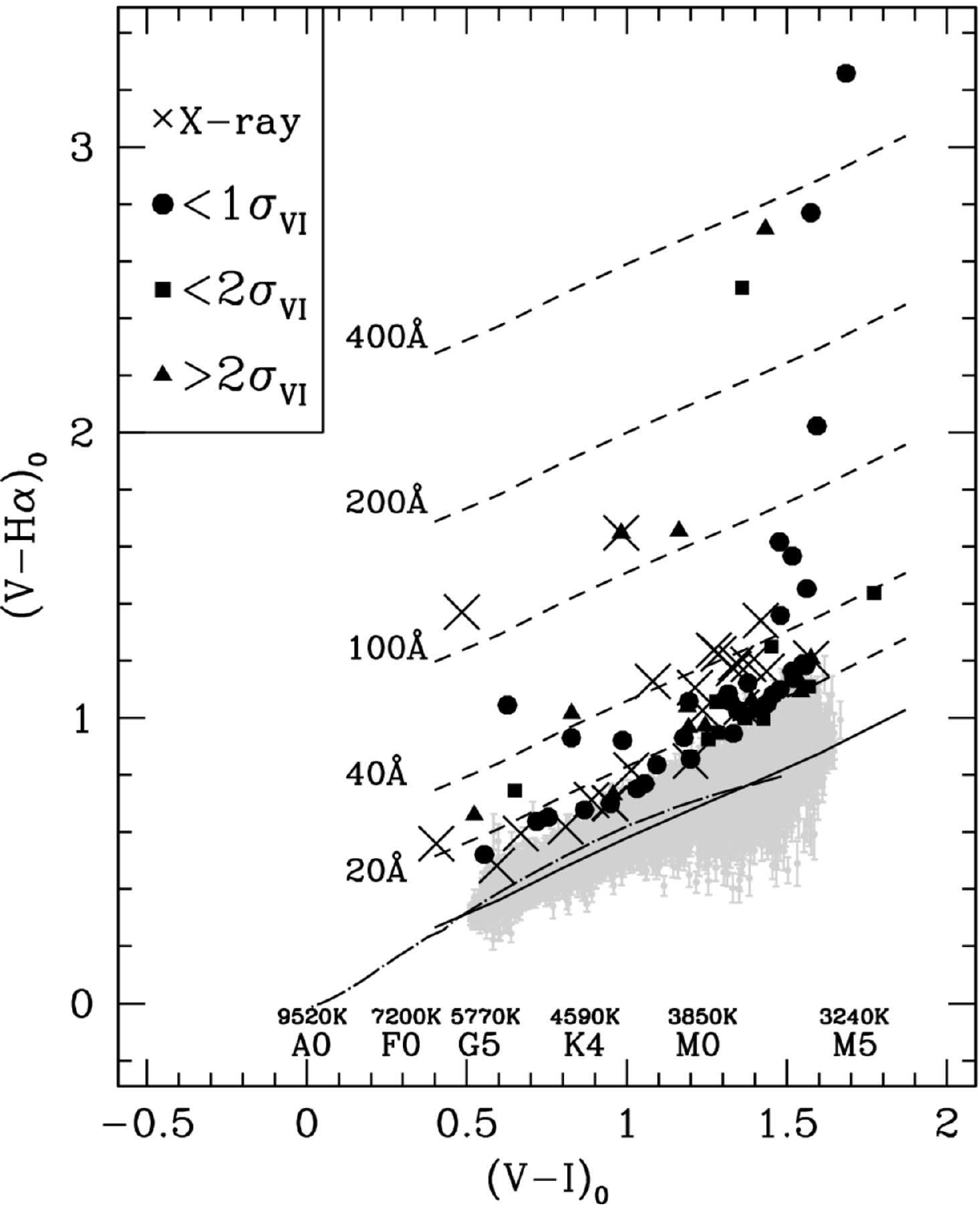}
  \caption[Color-color diagram for distinguishing  objects with \halpha\ excess.]{Example of a color-color diagram of the GC 47 Tucanae in which objects with \halpha\ excess (black symbols) have been selected \citep[from][]{beccari13}.  The solid line represents  the locus of stars without \halpha\ excess emission and hence the location of stars with ${\rm EW}({\rm H}\alpha)=0$. The dashed lines show the position of stars at increasing levels of \halpha\ emission. The dot-dashed line shows the location of the color relationship derived for these bands using the \citet{bessel98} atmospheric models.
 For more details  see \citet{beccari13}.
 }\label{47tuc}
\end{center}
\end{figure*}

Finally, if narrow band observations around the  wavelength of the \halpha\ transition are available, it is possible to photometrically constrain the presence of  \halpha\ excess, likely due to the presence of ionized  material (see Figure \ref{47tuc} for an application of this method).

\subsubsection{Variability}
Beside the positional coincidence,  the main signature of a  companion is the presence of a magnitude modulation with a period  compatible with the PSR orbital motion  (known from radio observations).

A method to identify variable stars possibly related to a given PSR consists in  using radio orbital parameters (see Appendix) to phase  the magnitude measures  of stars located at the PSR nominal position and then  check if the light curve  is consistent with the PSR orbital motion. 
Of course a better  approach would be to blindly derive orbital parameters from optical observations and then compare them with those known from radio.
However, this procedure would require a large number of images properly distributed  in time, which often, in particular for GCs, are not available.

Two main  light curve shapes are commonly observed:
\begin{itemize}
\item[-] {\it ``single-hill'' shape}: characterized by one maximum and  one minimum,  typically associated to  heating processes;
\item[-]{\it ``double-hill'' shape}: characterized by two maxima and two (asymmetrical) minima, typically associated to tidal distortions.   
\end{itemize}

\begin{figure*}[!t]
\begin{center}
\includegraphics[width=115mm]{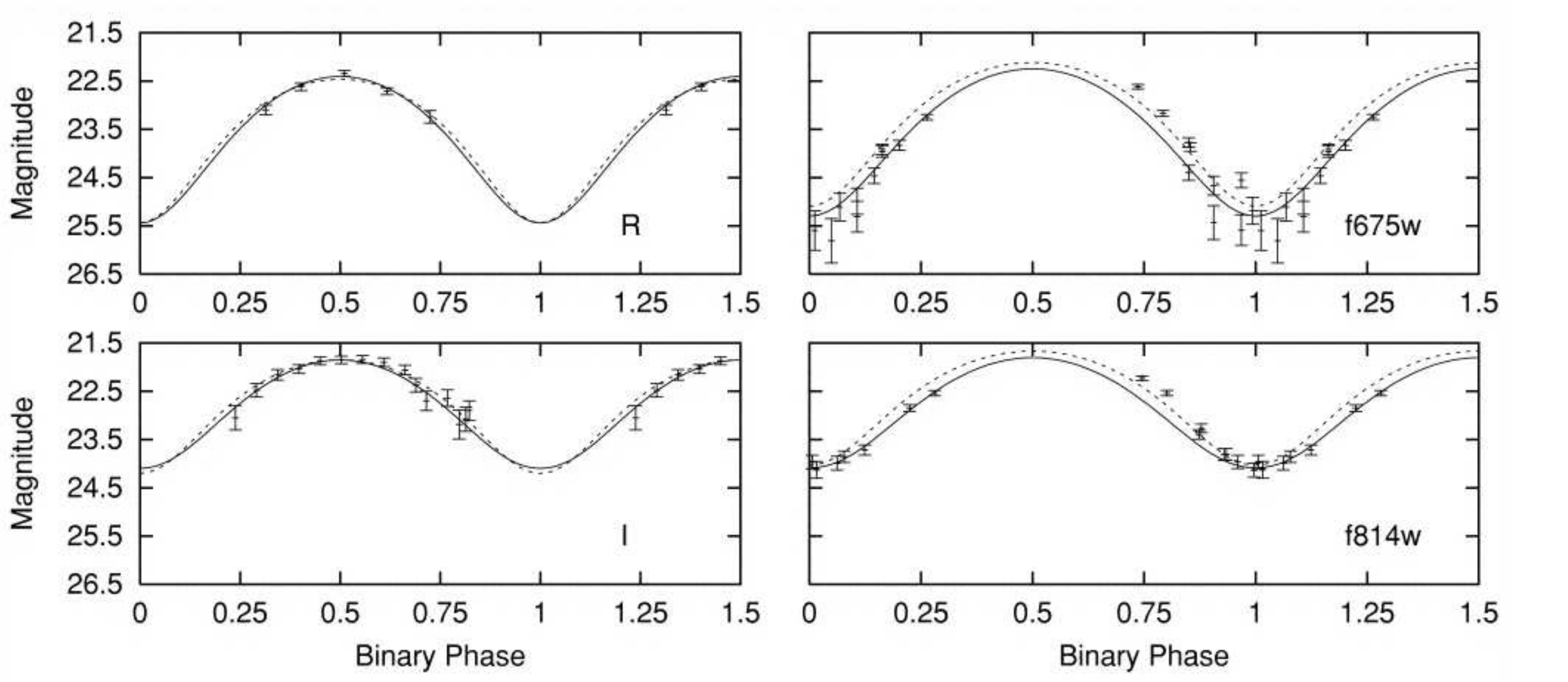}
  \caption[Example of a ``single-hill'' shape light curve.]{ {\it ``Single-hill'' shape} light curve due to heating of the companion to PSR J2051$-$0827 \citep[from][]{stappers01}. Note that the phase definition is different to that used in this Thesis.}\label{single}
\end{center}
\end{figure*}

The ``single-hill'' light curve shape (see an example in Figure \ref{single}) is characterized by a minimum at the PSR superior conjunction\footnote{Note that several phase definitions (depending on the reference $\Phi=0$ configuration) can be adopted. However, in this Thesis we adopted the radio convention. In particular, in the case of circular orbits, $\Phi=0$ corresponds to the PSR ascending node (see the Appendix for more details).} ($\Phi=0.25$) and a maximum at the PSR inferior conjunction ($\Phi=0.75$). In particular the luminosity minimum corresponds to the phase in which the observer is looking the back side of the companion, while the luminosity maximum corresponds to the phase in which the observer is looking the side of the companion facing the PSR. The explanation to this sinusoidal shape is that the increase of luminosity  is due to the PSR flux reprocessed by the companion and, of course, this effect is maximized at the phase during which the observer sees the larger fraction of the surface  illuminated by the PSR, while it is almost absent when observing the companion back side.

\begin{figure*}[!b]
\begin{center}
\includegraphics[height=105mm]{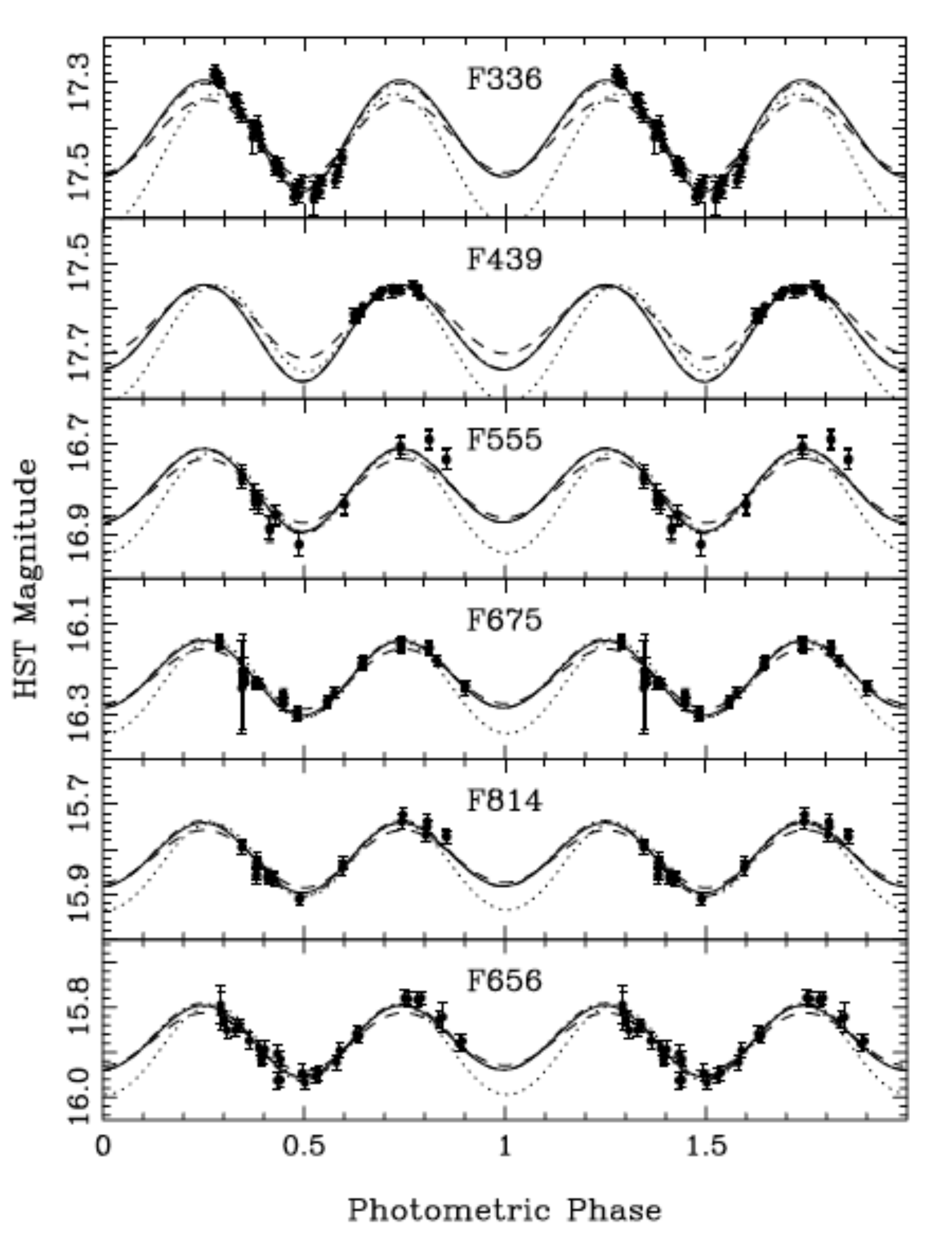}
  \caption[Example of a ``double-hill'' shape light curve.]{ {\it ``Double-hill'' shape} light curve due to tidal distortion of the companion to PSR J1740$-$5340A \citep[from][]{orosz03}. The solid, dash-dotted, dashed and dotted  lines are different models (with  different inclination angles, RL filling factors and the presence or not of heating) built to reproduce the light curve.  Note that the phase definition is different to that used in this Thesis.}\label{double}
\end{center}
\end{figure*}

Moreover, also the inclination angle of the binary system plays an important role. In fact in case of a face-on ($i=0^\circ$) configuration
the illuminated surface visible to the observer corresponds to a quarter of the star surface and it remains constant during the entire orbit, thus not producing any magnitude modulation. Conversely, in the case of an edge-on ($i=90^\circ$) configuration the illuminated surface visible to the observer ranges between zero at the PSR superior conjunction (minimum in the light curve), and half of the star surface at the PSR inferior conjunction (maximum in the light curve).
Hence, the larger is the inclination angle the larger is the amplitude of the light curve.
Note that such a mechanism is able to produce  variation amplitudes  up to several magnitudes, in particular when the companion is a faint cold object.

\begin{figure*}[!b]
\begin{center}
\includegraphics[width=140mm]{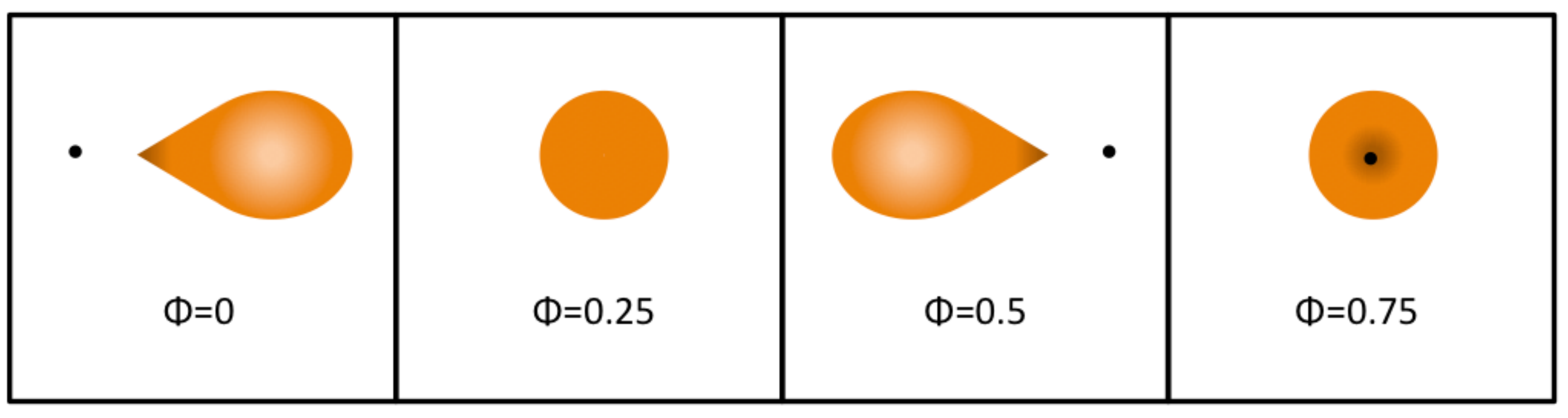}
  \caption[Schematic view of tidal distortion effects.]{Schematic representation of the effect of tidal distortions for a system seen in an edge-on configuration. The PSR and the companion are drawn     in black and orange, respectively. As evident the 
 surface of the companion visible to an observer changes along the orbital period. In particular, at phases 0.25 and 0.75 the lower gravity zones, corresponding to lower fluxes (darker orange), are facing the observer and two minima in the ``double-hill'' shape light curve are detected. Instead, at phases 0 and 0.5 
the visible surface with higher gravity, and hence higher flux (lighter orange), is maximized and two maxima are therefore detected in the light curve.
   }\label{tidaldist}
\end{center}
\end{figure*}

The ``double-hill'' light curve shape (see an example in Figure \ref{double}) is characterized by two maxima at the two quadratures ($\Phi=0$ and $\Phi=0.5$) and two minima at the PSR superior and inferior conjunctions ($\Phi=0.25$ and $\Phi=0.75$). 
In particular, the two (asymmetrical) minima correspond to the phases in which the observer is looking the back and the front side of the companion, while  the luminosity maxima correspond to the phases in which the observer is looking the companion from the lateral sides.
To understand the origin of such a shape, we should first consider that the companion is probably  bloated up to fill its RL and it is
suffering from the strong gravitational field of the massive NS. 
As a result, tidal distortions induce a ``drop-like'' shape to the companion and the surface gravity changes along the star, with the lower gravity zones corresponding to the back and the tip of the drop (see Figure \ref{tidaldist}).
Consequently, also stellar radiative flux, which is proportional to the local gravity  \citep{vonzeipel24}, changes along the star surface.
Therefore, when the system is observed from the side, the visible surface with higher gravity is maximized, and hence the detected flux is higher and the light curve shows the two maxima. Conversely, when the lower gravity zones (the tip and the back of the companion) are facing the observer, the flux is lower and the light curve shows the two luminosity minima.
In particular, the relative depth of the two minima is slightly different:  the deepest minimum occurs when the tip of the star is visible. However, this minimum could be less deep if an 
extra source of flux (the reprocessed PSR flux) is present.

The light curve is also very helpful to constrain some parameters of the system as the mass ratio,   the RL filling factor  and the inclination angle (see Figure \ref{double}). Note that, despite a weak  degeneracy between the effects of different mass ratios and  inclination angles,  the amplitude of the light curve is  mainly dependent  on the inclination angle and hence this is the most suitable parameter to be constrained. Also a completely filled RL is often required to justify the observed light curve.

Although the light curve modeling is a powerful tool to study these objects, not all companions to MSPs show variability.
In particular the absence of variability could be related either  to a selection effect (small inclination angles), or to an intrinsic lack of variability (due to negligible heating and tidal distortion).

\section{Spectroscopic follow-up of  a MSP companion}\label{specana}
Because of the current instrumental limits, a spectroscopic follow-up can be performed only for bright  companions.
In general, for very bright companions ($V\lapp17$) it is possible to obtain high spectral resolution data,    while for fainter stars ($17\lapp V \lapp 21$) only low spectral resolution observations can be performed.
However,  the situation is much more complicated in GCs, where the  companion stars are usually located in very crowded regions. In fact, to the detriment of the S/N, the observations must be planned as to avoid the contamination by close bright stars  (e.g. by using a thinner slit). 
Depending on the spectral resolution, only a kinematical (for low-resolution spectra), or both a kinematical and a chemical analysis (for high resolution spectra) can be performed.

\subsubsection{Kinematical analysis}
The kinematical analysis consists in the measure of the line of sight velocity of the target star through the Doppler shift of the central wavelength of the detected spectral lines.
To measure this shift is possible to use the cross-correlation method \citep{tonry79}, as implemented for instance in the {\tt fxcor} IRAF\footnote{ IRAF is distributed by the National Optical  Astronomy Observatory, which is operated by the Association of Universities for Research in Astronomy, Inc., under the cooperative agreement with the National Science Foundation.} task.
Basically, it consists in the cross-correlatation of the observed spectrum  with a template of known velocity computed in the same wavelength range and with the same spectral resolution.
The radial velocity is then calculated from the value corresponding to the maximum correlation probability.

\begin{figure*}[t]
\begin{center}
\includegraphics[height=95mm]{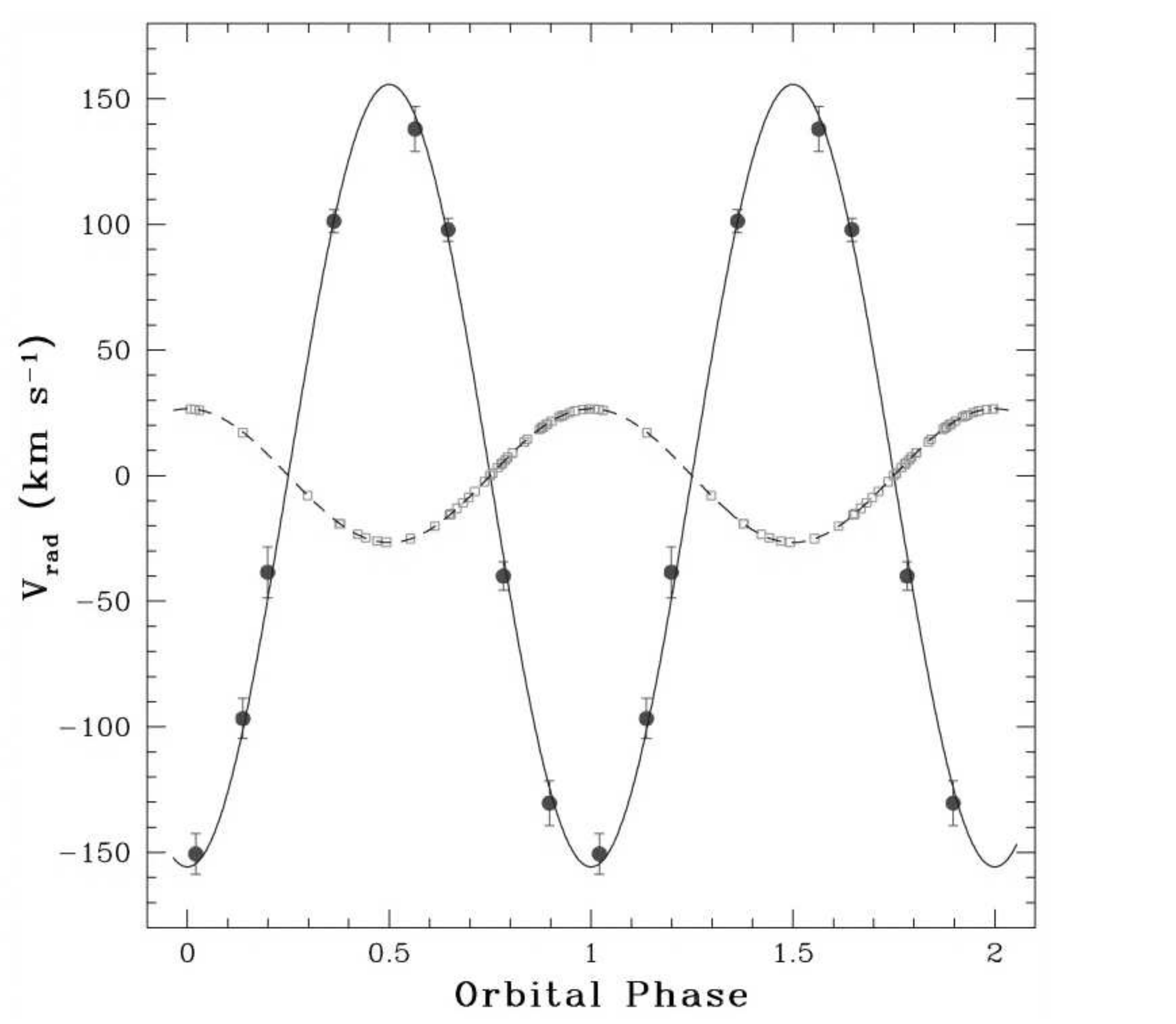}
  \caption[Radial velocity curve of the companion to PSR J1740-5340A.]{Radial velocity curve of the companion to PSR J1740-5340A (black filled points and solid line) and of the PSR (open squares and dashed line). Note that negative velocities mean that the object is approaching, while positive values mean that it is receding \citep[from][]{ferraro03rv6397}. }\label{rv}
\end{center}
\end{figure*}

The knowledge of the radial velocity is crucial to confirm the cluster membership for binary MSPs in GCs. Moreover, phase resolved observations allow us to build the radial velocity curve due to the companion star orbiting the MSP. 
Hence, the radial velocity curve of a MSP companion is expected to be periodically variable with the same orbital period of the radio PSR.  Of course, with respect to the observer line of sight, the companion star and the PSR move in  opposite directions. Thus in correspondence of the PSR ascending node ($\Phi=0$, when the PSR is receding) the companion is approaching the observer, while in the opposite configuration ($\Phi=0.5$, when the PSR is approaching) the companion star recedes. In the two quadratures both stars are moving tangentially to the line of sight and hence no Doppler effect due to the binary motion is detected.
The resulting radial velocity curve  (for circular orbits) is  a sinusoidal curve with a mean velocity corresponding to the systemic velocity of the object (representative of the mean motion of the binary system from/toward the Earth) and an amplitude corresponding to the companion motion in the binary system (see Figure \ref{rv}).

Hence, first of all, if the radial velocity curve has the same orbital period measured for the PSR and if the epoch of the companion descending node corresponds to that of the PSR ascending node, it is possible to confirm the membership to the same binary system.
Second, the amplitude of the light curve gives an indication of the  system projected dimension and in case of a GC the mean velocity is important to confirm the membership to the cluster.
Finally, combining the two amplitudes of the companion and PSR radial velocity curves, the dependence on the inclination angle is removed and it is possible to directly estimate the mass ratio as the inverse ratio of the amplitudes of the two curves (see Appendix).

\subsubsection{Chemical analysis}
The chemical analysis consists in the measure of the surface abundances of chemical elements.
To derive them, the traditional method consists in the measure of the equivalent widths. However  in the case of low-resolution spectra affected by   severe line blanketing conditions, such a method makes  difficult to accurately determine the continuum (see the procedure described in \citealt{m12b} for a method to reduce such a problem).

\begin{figure*}[!t]
\begin{center}
\includegraphics[width=90mm]{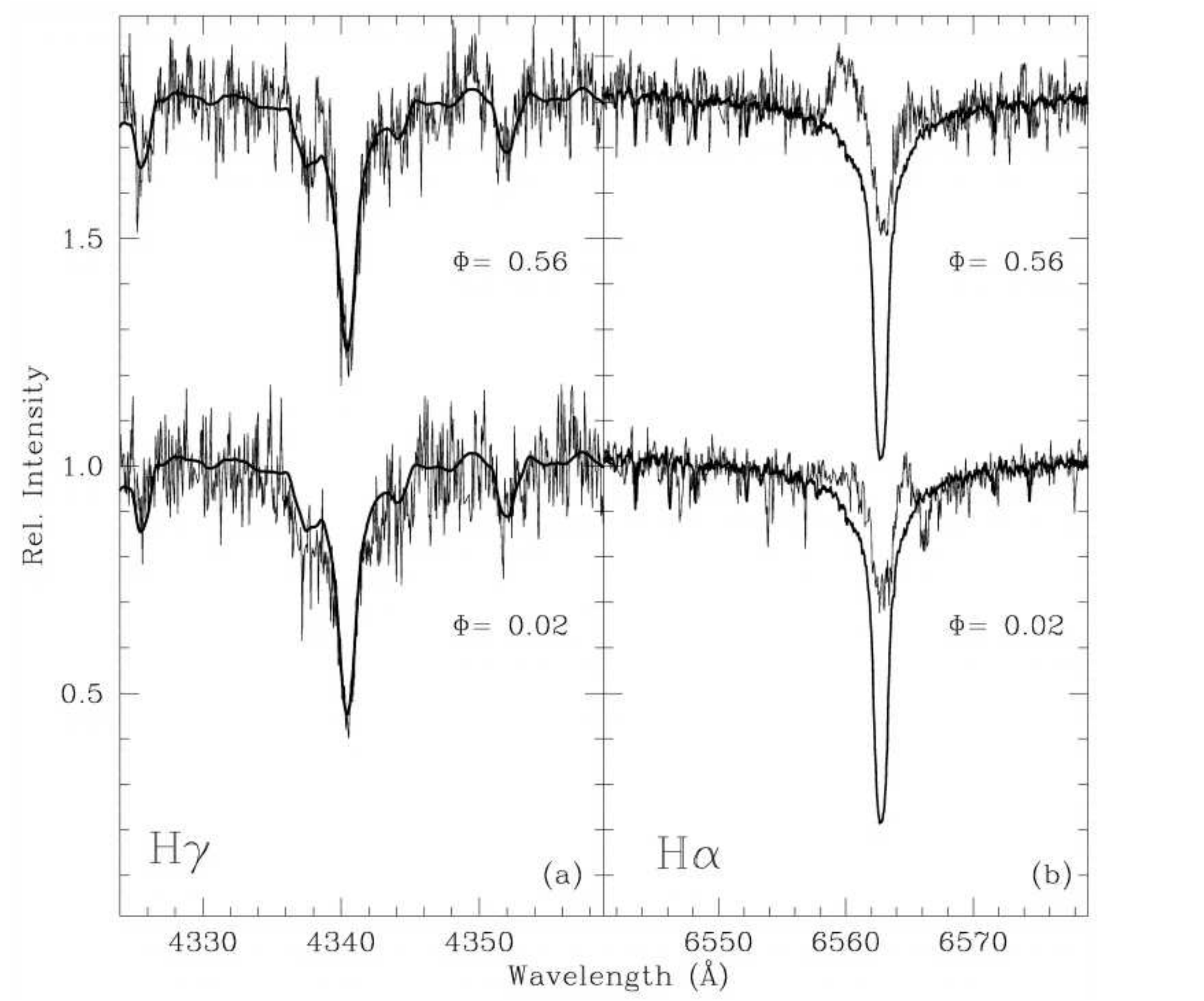}
  \caption[Balmer lines of the companion to PSR J1740$-$5340A.]{ Observed spectra of the companion star to PSR J1740$-$5340A in the $H\gamma$ (left panel) and \halpha\ (right panel) regions, at two orbital phases \citep[from][]{sabbi03Ha}. The thick solid line is a reference template derived by averaging three normal sub giant stars in the GC. The \halpha\ line of the PSR companion  is shallower and it shows anomalous wings suggesting the existence of a complex emitting structure.}\label{halpha1}
\end{center}
\end{figure*}

\begin{figure*}[!b]
\begin{center}
\includegraphics[width=90mm]{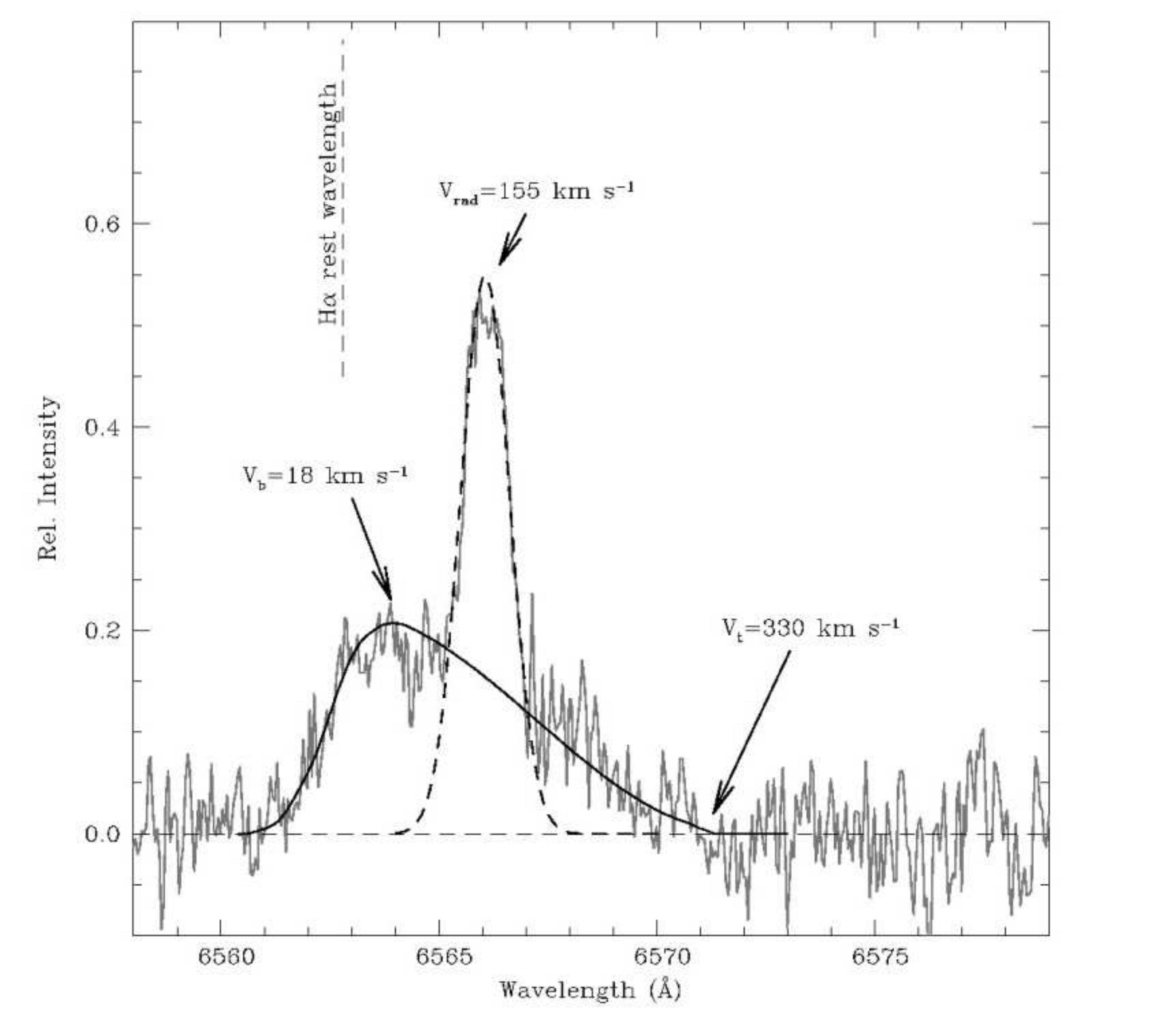}
  \caption[\halpha\ line of the companion to PSR J1740$-$5340A.]{Subtracted spectrum of the companion to PSR J1740$-$5340A in the GC NGC 6397 in the \halpha\ region \citep[from][]{sabbi03Ha}. Two components of  the \halpha\ emission were detected (solid and dashed lines). See \citet{sabbi03Ha} for more details.}\label{halpha2}
\end{center}
\end{figure*}

From the detailed chemical analysis of spectral  lines it is possible to deeply look into  MSP systems.
In  the case of low-resolution spectra and for high $T_{eff}$, Balmer lines are the main available diagnostics to infer the stellar parameters.
For example, in the case of  WD companions, by comparing the photometric properties and  H spectral lines with WD models, it is possible to constrain the temperature, the gravity, the radius and the mass \citep{antoniadis13}.
The latter is probably the most important parameter. In fact, together with the knowledge of the mass ratio (from radial velocity curve), it furnishes the NS mass.  
Moreover, the characterization of the \halpha\ line is important also to detect \halpha\ excess due to some ionized material \citep[see][and Figures \ref{halpha1} and \ref{halpha2}]{sabbi03Ha}.


\begin{figure*}[t]
\begin{center}
\includegraphics[width =140mm]{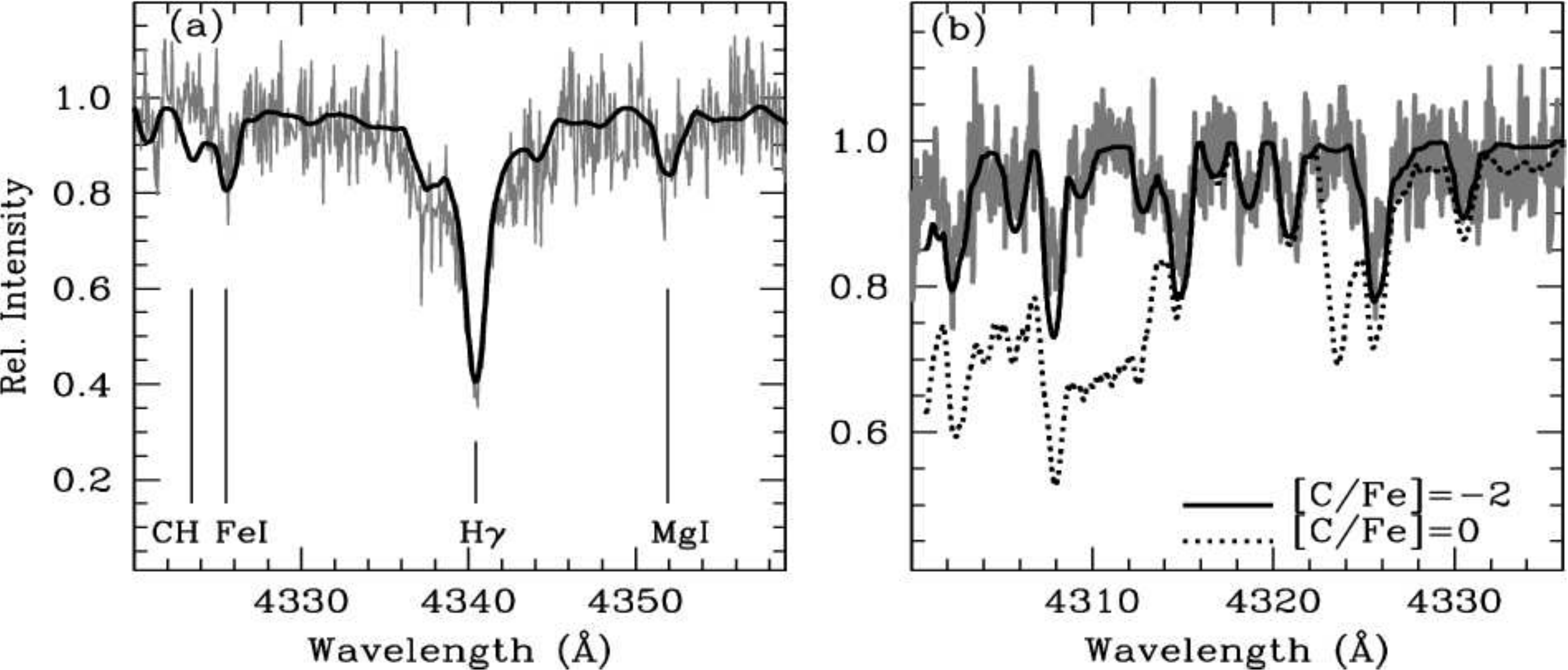}
  \caption[Carbon abundance of the companion to PSR J1740$-$5340A.]{ Evidence  of the Carbon abundance depletion  from the CH band for the companion to PSR J1740$-$5340A \citep[from][]{sabbi03C}. }\label{carbon}
\end{center}
\end{figure*}

Finally, also the chemical abundances of heavier elements are interesting. For example in the case of GCs,  chemical abundances in agreement with that of the parent cluster population are a good indicator of the membership to the GC. In addition, some elements are expected to show anomalies.
In fact, the companion is expected to lose a significant quantity of material (partially accreted by the NS) and to be therefore deprived of its outermost layers. 
In the case of a strongly peeled star the residual material is what remains of the original inner  part of the star, where the chemical composition has been previously modified by the thermonuclear reactions. Hence the measure of chemical abundances can give precious information about the nature of the companion. 
 For instance, in case of material partially processed by the CNO cycle, Carbon depletion and  Nitrogen enhancement are expected (see Figure \ref{carbon}).

\section{The state of the art}
\begin{figure*}[t]
\begin{center}
\includegraphics[height=120mm]{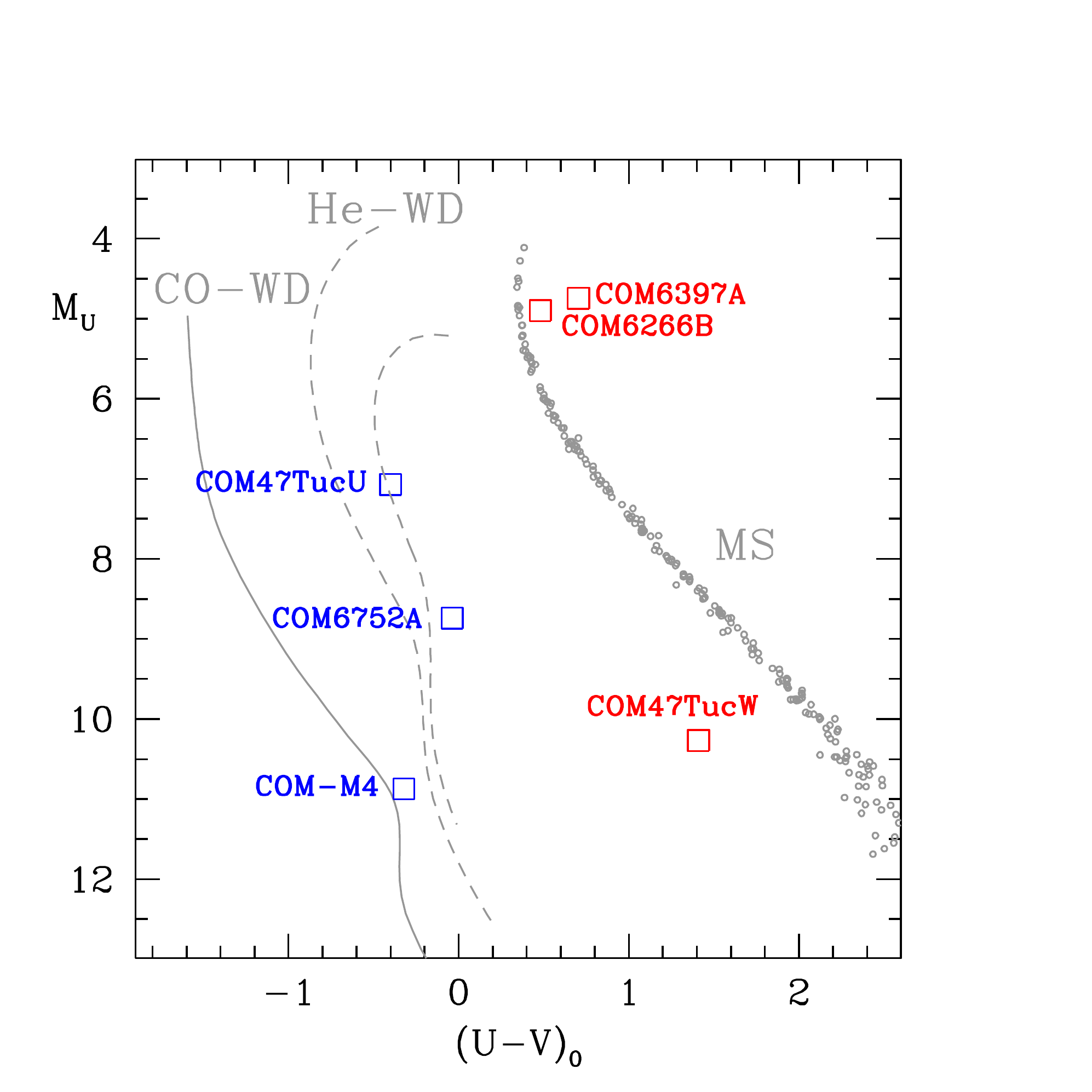}
  \caption[CMD position of the known MSP companions in GCs discovered until 2009.]{Location in the ($\rm M_U, (U-V)_0$) absolute plane of the six optical companions (squares) to binary MSPs detected in GCs before 2009.  The blue and red squares mark the companion to ``canonical'' and red backs systems, respectively.}\label{stateArt}
\end{center}  
\end{figure*}  

Despite the paramount importance of identifying the companion stars to binary MSPs, before this Thesis work started,  only 6 companions were identified in 5 GCs
(out of the known 132 GC MSPs).

Three of them had positions in the CMD consistent with the cooling sequences of He-WDs (see Figure \ref{stateArt}), in agreement with the expectations of the {\it canonical recycling scenario}. These are the companions
to MSP-U in 47 Tucanae \citep{edmonds01}, to MSP-A in NGC 6752 \citep{ferraro03com6752}, and to PSR B1620$-$26 in M4 \citep{sigurdsson03}. The other identified companions showed, instead, quite peculiar properties and positions in the CMD not compatible with He-WD cooling sequences (see Figure \ref{stateArt}).
The luminosity and colors of the optical companion to MSP-A in NGC 6397 are totally incompatible with those of a WD. Indeed, this is a relatively bright, tidally deformed star, suggesting that the system either harbors a newly born MSP, or is the result of an exchange interaction \citep{ferraro01com6397}.  
The companion star to MSP-B in NGC 6266 is a similarly bright object, with luminosity comparable to the cluster MS turn-off (TO), an anomalous red color and an optical variability suggestive of a tidally deformed star which filled its RL \citep{cocozza08}. This object is also a Chandra X-ray source, thus supporting the hypothesis that some interaction is occurring between the PSR wind and the gas streaming off the companion. Finally the companion to MSP-W in 47 Tucanae has been identified to be a faint MS star, showing large-amplitude, sinusoidal luminosity variations probably due to the heating effect by the PSR \citep{edmonds02}.
The relative small number of known companions to GC MSPs is due to the observational difficulties to identify and to study them (see previous Sections). 

On the other hand, the situation in the GF is quickly evolving with an increasing number of new detections of MSPs by radio and $\gamma$-ray searches, and  the contribution to the optical identification of companions by several  ground-based telescopes.
In the particular case of exotic MSPs,  the number of detected objects is significantly increased in the last years and  7 BWs (out of the known 17) and 4 RBs (out of 7) have identified optical companions \citep{kulkarni88,stappers96b,stappers01,  vankerk11, pallanca12,romani12,breton13,kaplan13,bellm13}.

\section{The targets of this Thesis}
This work  is part of a large project (Cosmic-Lab) aimed at exploiting Galactic GCs as natural laboratories where to study the complex interplay between dynamics and stellar evolution by using Blue Straggler stars, MSPs and intermediate-mass black holes as probe particles. This Thesis deals with MSPs.

In particular we report the identification of  the companions to three MSPs in GCs, namely 
\begin{itemize}
\item[-] the RB \psrH\ in M28 (see Chapter \ref{Chap:M28H});
\item[-] the RB \psrI\ in M28 (see Chapter \ref{Chap:M28I});
\item[-] the BW \psrC\ in M5 (see Chapter \ref{Chap:M5C});
\end{itemize}
and we show the results of a spectroscopic follow up of the companion to 
\begin{itemize}
\item[-] \psrA\ in NGC 6397 (see Chapter \ref{Chap:6397}).
\end{itemize}
Moreover, we also report  the  identification of the companions to two MSPs in the GF, namely
\begin{itemize}
\item[-] the BW \psrBW\ (see Chapter \ref{Chap:0610});
\item[-] the IMM \psrIM\ (see Chapter \ref{Chap:1439}).
\end{itemize}

\clearpage{\pagestyle{empty}\cleardoublepage}

\chapter[The companion to \psrH\ in M28]{The optical companion to the binary Millisecond Pulsar J1824$-$2452H in the Globular Cluster M28}\label{Chap:M28H}

In this Chapter we describe the optical identification of the companion star to the RB eclipsing MSP J1824$-$2452H in the Galactic GC M28 (NGC 6626). This star is located at only $0.2\arcsec$ from the
nominal position of the PSR and it shows optical variability ($\sim 0.25$ mag) that nicely correlates with the PSR orbital period.  It is positioned on the blue side of the cluster MS, $\sim 1.5$ mag fainter than the TO point. The observed light curve shows two distinct and asymmetric minima, suggesting that the companion star is suffering tidal distortion from the PSR. 
This discovery increases the number of non-degenerate MSP companions  in GCs, suggesting that these systems could be a common outcome of the PSR recycling process, at least in dense environments where they can be originated by exchange interactions. - {\it This Chapter is mainly based on \citealt{pallanca10}, ApJ, 725, 1165.}

\section{Introduction}

 In this Chapter we focus our attention on M28 (NGC 6626),  a Galactic GC with intermediate central density \citep[$\log \rho_0=4.9$ in units of $M_\odot/$pc$^3$;][]{pryormeylan93}. 
It is the first GC where a MSP was discovered \citep{lyne87} and to date it is known to harbor a total of twelve PSRs \citep{begin06}. This is the third largest population of known PSRs in GCs, after that of Terzan 5 (with 33 objects; \citealp{ransom05}, but see the recent results by \citealp{fe09ter} suggesting that Terzan 5 is not a genuine GC) 
and that of 47 Tuc \citep[with 23 MSPs;][]{camilo00, freire03}.

Among the binary MSPs harbored in M28, J1824$-$2452H (hereafter M28H) deserves special attention since it is an eclipsing system showing a number of timing irregularities, possibly due to the tidal effect on the companion star \citep{begin06, stairs06}.  
It is located at $\alpha_{2000}=18^{\rm h} 24^{\rm m} 31.61^{\rm s}$ and $\delta_{2000}=-24^\circ 52' 17.2\arcsec$, it has an orbital period $P_b=0.43502743$ days and it shows eclipses for $\sim 20\%$ of it (S. B\'egin et al., in preparation).  There is also an associated hard X-ray source, possibly variable at the binary period \citep{bogdanov11}. Such X-ray emission is likely due to the shock between the MSP magnetospheric radiation and the matter released by the companion, similar to that detected in the case of MSP-W in 47 Tuc \citep{bogdanov05}. This further suggests that the companion to M28H is a non-degenerate star.  In this  Chapter we present its optical identification, based on high-quality, phase-resolved photometry obtained with the WFC3 on board the HST.
 
\section{Observations and data analysis}
The photometric data-set used for this work consists of HST high-resolution images obtained with the UVIS channel of the WFC3.  A set of supplementary HST WFPC2 images, and ground-based wide-field images obtained at the ESO have been retrieved from
the Science Archive and used for variability and astrometric purposes.

The WFC3 images have been obtained on 2009 August 8 (Prop. 11615, PI: Ferraro) in four different bands.
The data-set consists of: 6 images obtained through the F390W filter ($\sim U$) with exposure times  $t_{\rm exp}=800-850$ s each; 7 images in F606W ($\sim V$) with $t_{\rm exp}=200$ s; 7 images in F814W ($\sim I$) with $t_{\rm exp}=200$ s; and 7 images in F656N (a narrow filter corresponding to H$\alpha$) with exposure time
ranging from $t_{\rm exp}=935$ s, up to $t_{\rm exp}=1100$ s.  
All the images are aligned and the cluster is almost centered in CHIP1.

Additional public WFPC2 images have been retrieved from the archive.
The first data-set (hereafter WFPC2-A) was obtained in 1997 (Prop. 6625, PI: Buonanno) and consists of 8 images in F555W ($\sim V$) with $t_{\rm   exp}=140$ s and 9 images in F814W ($6\times t_{\rm exp}=160$ s and $3\times t_{\rm exp}=180$ s).  
The second sample (hereafter WFPC2-B) consists of 13 images in F675W ($\sim R$), with $t_{\rm   exp}=100$ s each, secured in 2008 (Prop. 11340, PI: Grindlay).

Finally, the wide-field data-set consists of 6 images in the $V$ and $I$ filters, obtained in August 2000 with the Wide Field Imager (WFI) at the ESO-MPI 2.2 m telescope (La Silla, Chile). The WFI consists of a mosaic of eight chips, for a global FOV of $34'\times34'$.

The data reduction procedure 
has been performed on the WFC3 ``flat fielded"  (flt) images, once corrected for ``Pixel-Area-Map"  (PAM) by using standard IRAF procedures. The photometric analysis has been carried out by using the {\sc daophot} package (Stetson 1987). For each image we modeled the 
PSF by using a large number ($\sim 200$) of bright and nearly isolated stars. 
Then we performed the PSF fitting by using the {\sc daophot} packages {\tt ALLSTAR} and {\tt ALLFRAME} \citep{stetson87, stetson94}.  
The final star list consists of all the sources detected in at least 14 frames on a total number of 27. A similar procedure has been adopted to reduce the WFPC2 images. 
For the WFPC2-A data-set we demanded that sources were in at least 9 frames out of 17, whereas for the WFPC2-B data-set in at least 7 frames out of 13. Since the WFC3 images heavily suffer from geometric distortions within the FOV, we corrected the instrumental positions of stars by applying the equations reported by \citet{bellini09} for the filter F336W, neglecting any possible dependence on the wavelengths. Standard procedure \citep[see e.g.,][]{lanzoni07M5, lanzoni071904, lanzoni076388,dalessandro086388,dalessandro082419} has been adopted to analyze the WFI data.  
Here we use this data-set only for astrometric purposes.  
In fact, a very accurate astrometry is the most critical task in searching for the optical counterparts to MSPs, especially in crowded fields such as the central regions of GCs, where primary astrometric standards are lacking \citep[e.g.,][]{ferraro0147tuc, ferraro03com6752}.  
For this reason we first placed the wide-field catalog obtained from the WFI images on the absolute astrometric system, and we then used the stars in common with the high
resolution data-sets as secondary standards.  
In particular, the WFI catalogue has been reported, through cross-correlation, onto the coordinate system defined by the GSCII. 
Then, we placed the WFC3 and the WFPC2 catalogs on the same system through cross-correlation with the WFI data-set.  Each transformation has been performed by using several thousand stars in common, and at the end of the procedure the typical accuracy of the astrometric solution was $\sim0.2\arcsec$ in both $\alpha$ and $\delta$.

Finally, the WFC3 instrumental magnitudes have been calibrated to the VEGAMAG system by using the photometric zero-points and the procedures reported on the WFC3 web
page.\footnote{http://www.stsci.edu/hst/wfc3/phot\_zp\_lbn} 
The WFPC2  magnitudes have been reported to the same photometric system by using the procedure described in \citet{holtzman95}, with the gain settings and zero-points listed in Tab. 28.1 of the HST data handbook.

\section{The optical companion to M28H}
In order to search for the optical counterpart to the M28H companion, we carefully re-analyzed a set of $4\arcsec\times 4\arcsec$ WFC3 sub-images centered on the nominal radio position of the MSP.  For these sub-images the photometric reduction has been re-performed by using both {\sc daophot} \citep{stetson87} and {\sc romafot} \citep{buonanno83}. 
In both cases, in order to optimize the identification of faint objects, we performed the source detection on the median image in the $U$ band, thus obtaining a master-list. 
The master-list was then applied to all the single images in each band and we performed the PSF-fitting by using appropriate PSF models obtained in each image.
The resulting instrumental magnitudes were reported to those of a reference image in each filter, and from the frame-to-frame scatter, a mean magnitude and a standard deviation have been obtained for all the objects.

\begin{figure}[b]
\begin{center}
\includegraphics[height=105mm]{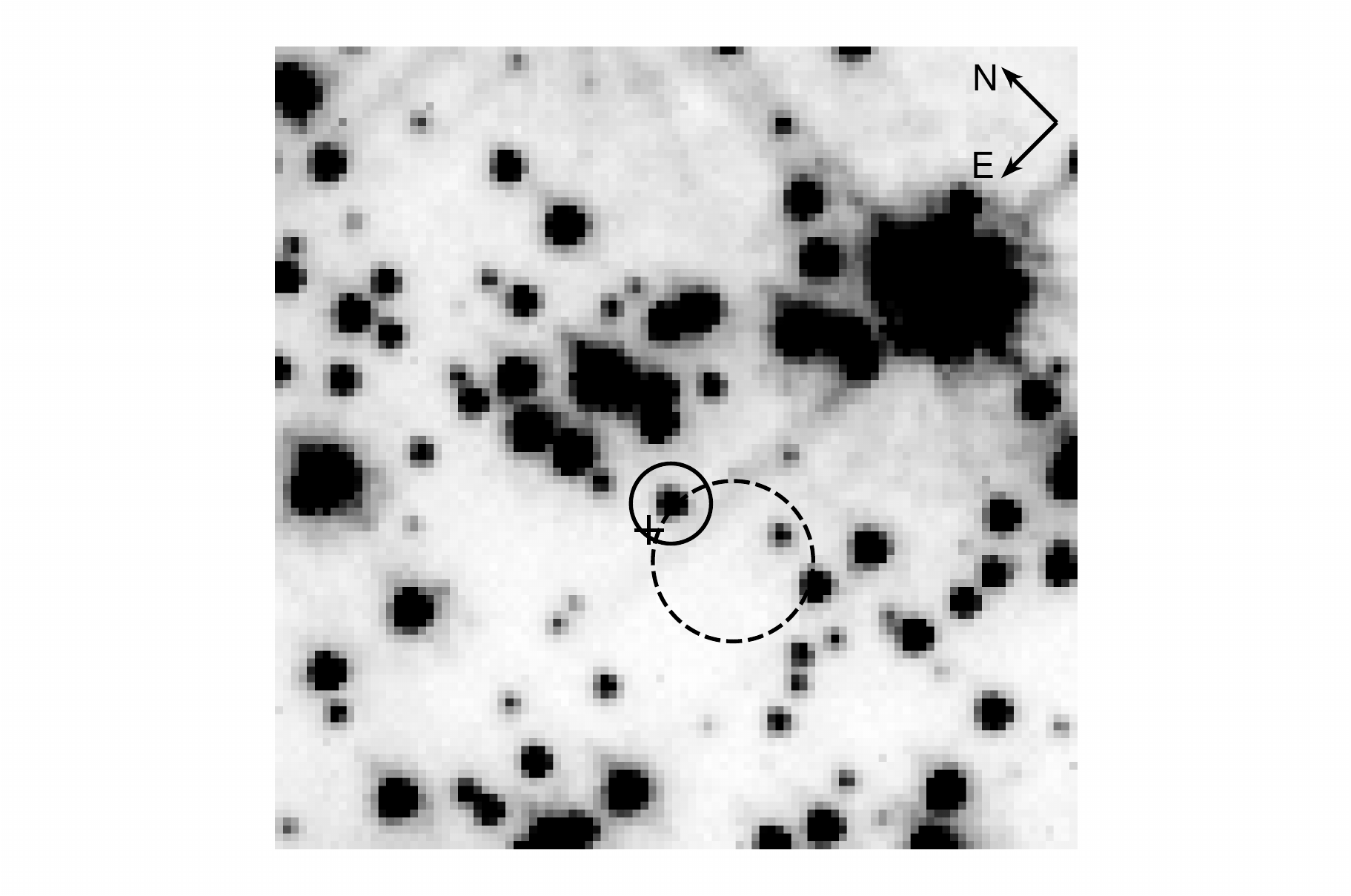}
\caption[Map of the region around \psrH.]{$U$-band, $4\arcsec\times4\arcsec$ WFC3 sub-image of M28   centered on the position of COM-M28H, identified as the   companion star to the MSP M28H.  The solid circle has a radius of   $0.2\arcsec$, corresponding to the estimated astrometric accuracy of   our analysis. The position of the radio source M28H is marked with   the cross.  The dashed circle ($0.4\arcsec$ radius) marks the   position and estimated uncertainty of the X-ray source.}
\label{Fig:map}
\end{center}
\end{figure}
  
We then selected the stars that showed significant variations in the $U$ band, looking for those that have a periodic variability compatible with the orbital period of M28H.  Only one object has been found to match such requirements. 
This star is located at $\alpha_{2000}=18^{\rm h} 24^{\rm m} 31.60^{\rm s}$ and $\delta_{2000}=-24^\circ 52' 17.2\arcsec$, just $0.17\arcsec$ from the radio position of M28H, and $\sim 0.4\arcsec$ from the X-ray source (Figure \ref{Fig:map}).

\begin{figure}[b]
\begin{center}
\includegraphics[width=130mm]{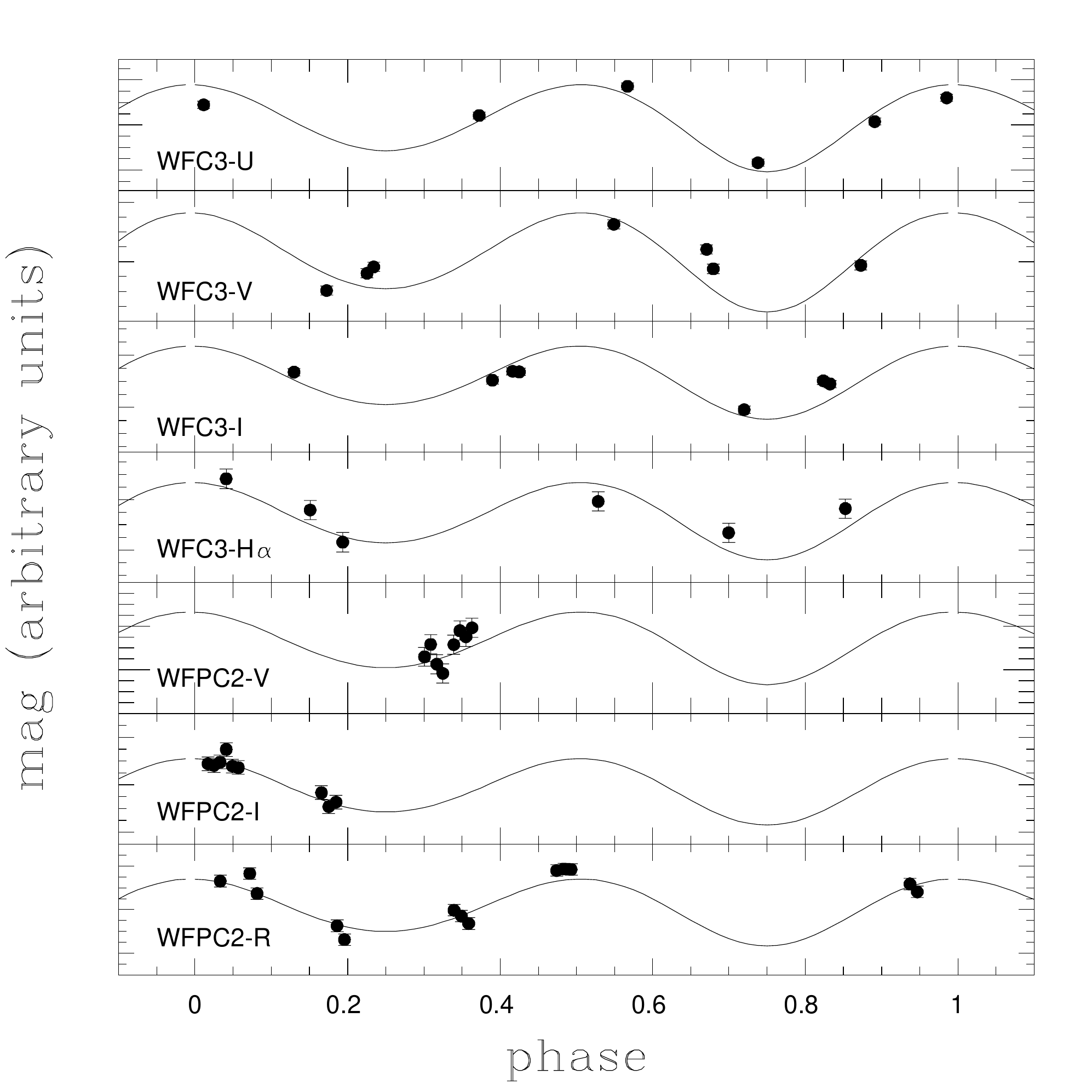}
\caption[Light curves of \psrH.]{The observed light curves obtained from the WFC3 and WFPC2 data. The best fit model is shown as a solid line in each panel.}
\label{Fig:lc_filters}
\end{center}
\end{figure}

\begin{figure}[b]
\begin{center}
\includegraphics[width=145mm]{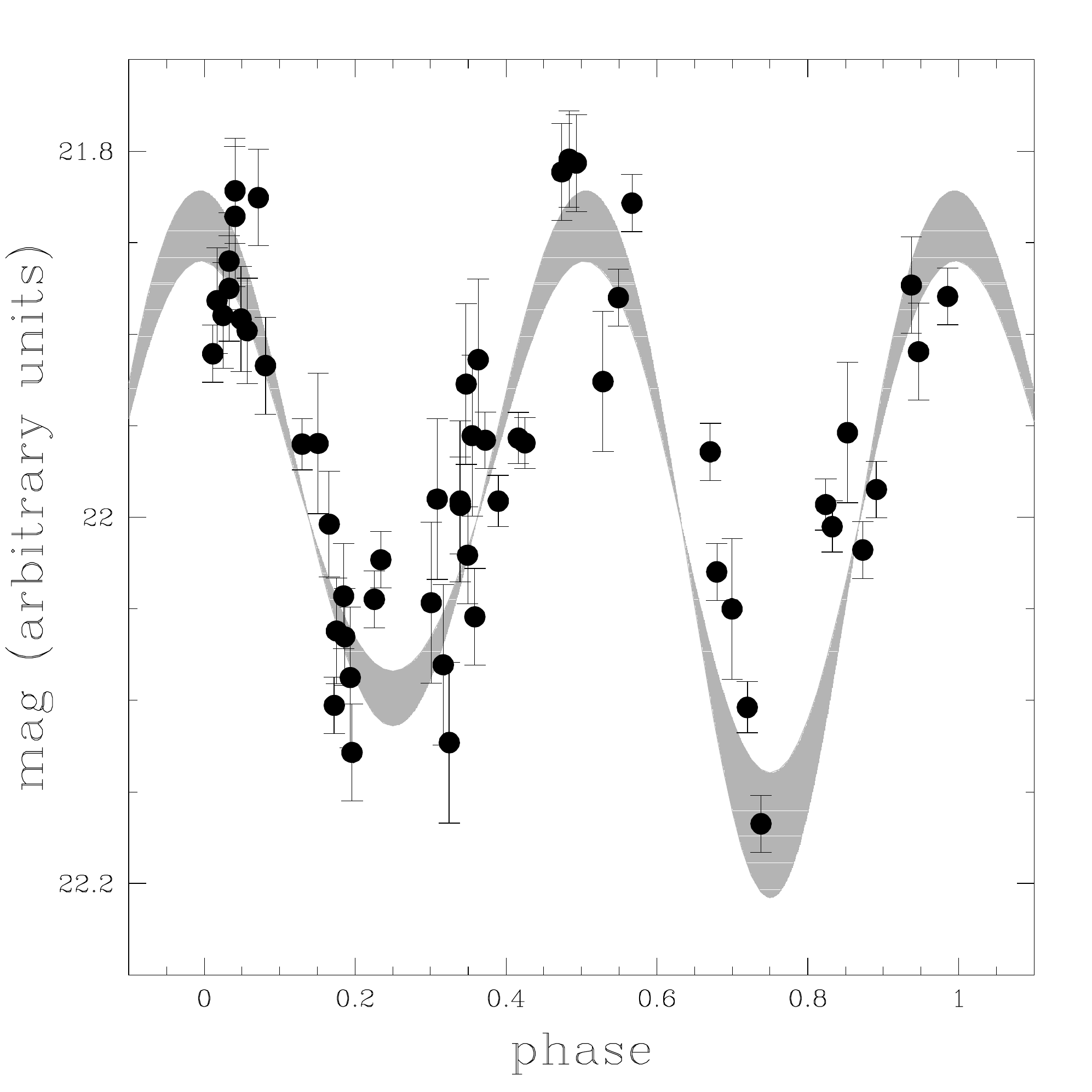}
\caption[Combined light curve of \psrH.]{The global light curve of COM-M28H, obtained by combining the   data points shown in Figure \ref{Fig:lc_filters}. The gray area   represents the region spanned by the best-fit light curves in each   photometric band.}
\label{Fig:lc_combined}
\end{center}
\end{figure}

In order to carefully investigate the optical modulation of this star, we first folded all the measured data points obtained in each filter by using the orbital period $P_b$ and the reference epoch ($T_0=53755.226397291$ MJD; B\'egin et al. in preparation) of the M28H radio ephemeris.  
Since the few measured data points in the single bands are not sufficient to properly cover the entire period (see dots in Figure \ref{Fig:lc_filters}), we computed the average magnitude in each filter and the shift needed to make it match the mean $U$-band magnitude, which we adopted as a reference. 
We thus obtained the combined light curve shown in Figure \ref{Fig:lc_combined}, which well samples the entire period of the PSR. Indeed, the optical modulation of the identified star nicely agrees with the orbital period of the MSP, thus fully  confirming that the variability is associated with the PSR binary motion. 
Hence we conclude that the identified star (hereafter COM-M28H) is the optical companion to M28H.

\begin{figure}[b]
\begin{center}
\includegraphics[width=140mm]{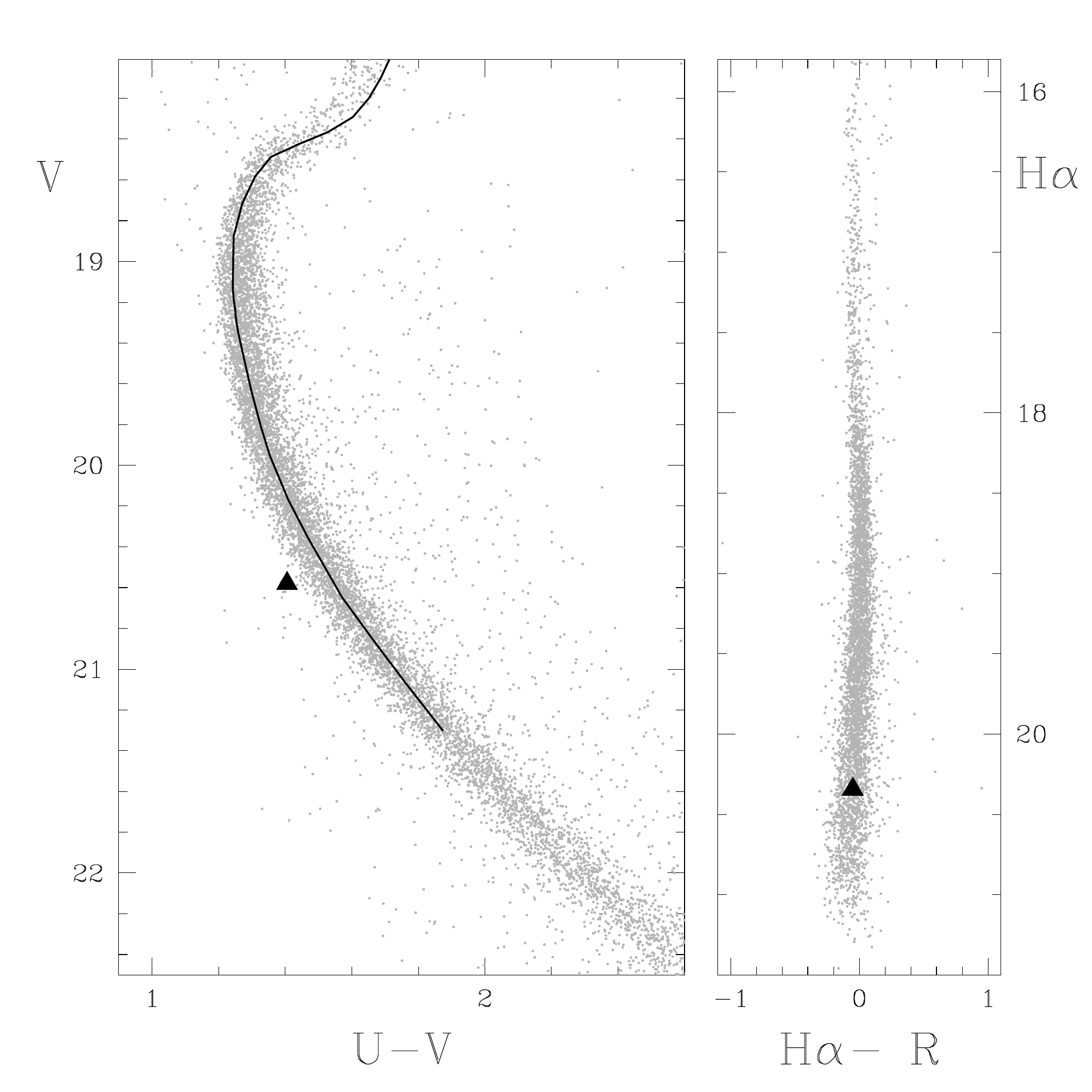}
\caption[CMD position of the companion to \psrH.]{CMDs of M28  in a circular region of   $\sim 30\arcsec$ radius centered on the position of COM-M28H derived from WFC3 data (left   panel), and from a combination of WFC3 and WFPC2 data (right   panel). The solid triangle marks the position of COM-M28H.  The   plotted isochrone \citep[from][]{marigo08} has been obtained for   $t=13$ Gyr, [Fe/H]$=-1.27$, $E(B-V)=0.4$ and (m-M)$_V=14.97$.  }
\label{Fig:cmd}
\end{center}
\end{figure}

This object (with average $U=21.99,~V=20.58,~ I=19.49$) is $\sim 1.5$ mag fainter than the cluster TO and slightly bluer than the MS (Figure \ref{Fig:cmd}). 
Clearly, it is far too red and bright to be compatible with a WD (the typical product expected from the {\it canonical recycling scenario}, see Chapter \ref{Chap:introPSR}).  
Instead, its position in the CMD is marginally   consistent with a MS star. 
In particular, the cluster MS is well   reproduced by a $t=13$ Gyr isochrone \citep{marigo08},   with metallicity [Fe/H]$=-1.27$ (from \citealt{zinn80}, after calibration   to the scale of \citealt{carrettagratton97}, following \citealt{ferraro99}) and assuming a color excess $E(B-V)=0.4$ and a distance   modulus (m-M)$_V=14.97$ \citep{harris96}.  
By projecting the observed   magnitudes and colours of COM-M28H onto this isochrone, the   resulting mass, temperature and radius would be $M_{\rm COM}\sim0.68   M_\odot$, $T\sim 6000$ K, $R\sim 0.64 R_\odot$, respectively.
  However these quantities should be considered as just an indication,   since the observational properties of this object (see below)   strongly suggest that it is a highly perturbed star.

\begin{figure}[b]
\begin{center}
\includegraphics[width=100mm]{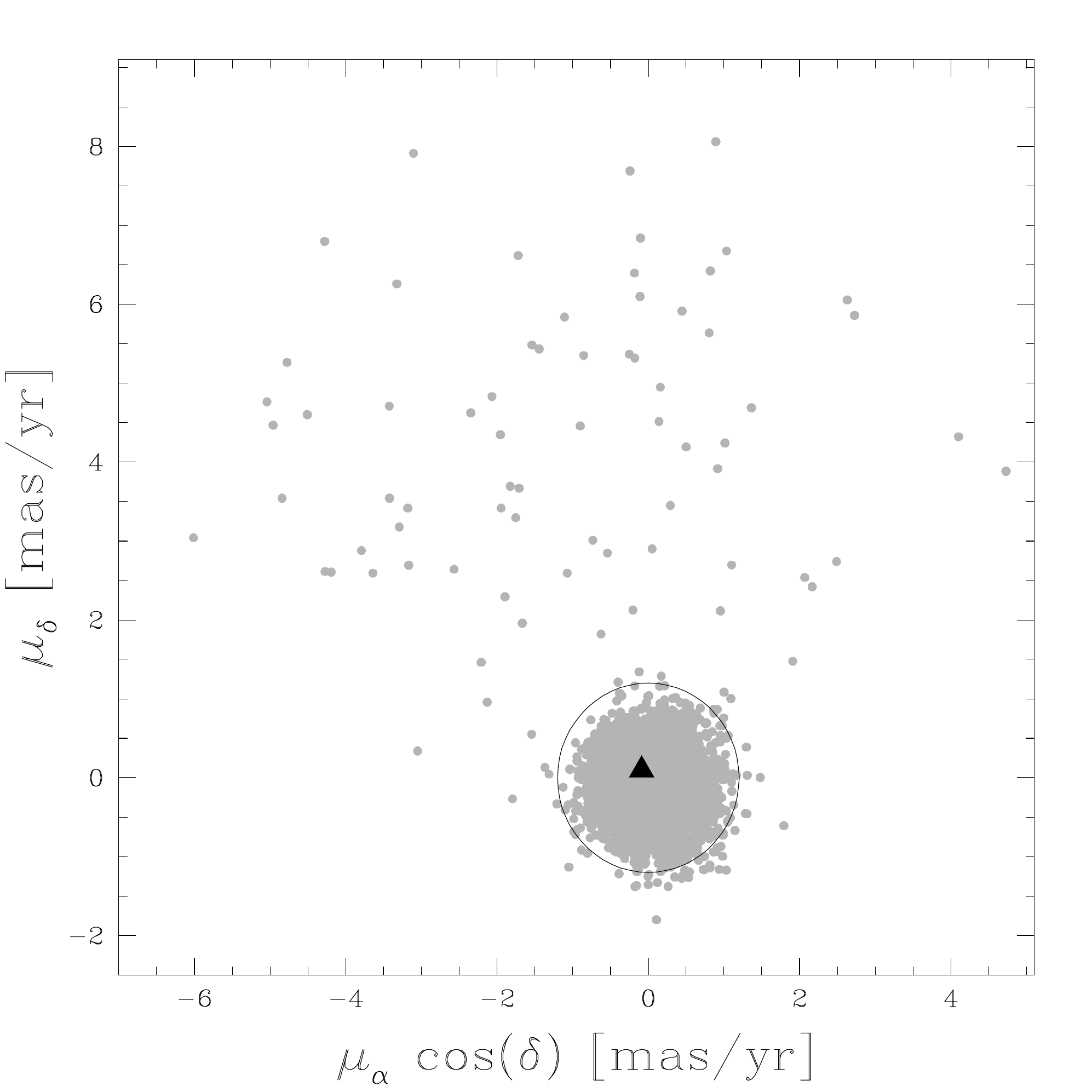}
\caption[Relative proper motion of the companion to \psrH.]{Relative proper motion diagram of M28 in equatorial coordinates. The solid triangle represents the position of COM-M28H. }
\label{Fig:proper_motion}
\end{center}
\end{figure}  

Since the two WFPC2 data-sets have been obtained in two different epochs with a time baseline of more than 10 years, and given the relative small distance of M28 \citep[d=5.6 Kpc;][]{harris96}, we have been able to perform a proper-motion analysis.  
As shown in Figure \ref{Fig:proper_motion}, the bulk of stars lie around the position ($\mu_\alpha cos(\delta)=0$, $\mu_\delta=0$) [mas/yr], within a radius $\sigma_{\mu} \sim1$.  These most likely are members of the cluster, while field stars are clearly separated (the position in the CMD of these two classes of stars further confirms such a conclusion).
 Since no extra-Galactic source can be identified in the FOV adopted for this analysis, no absolute proper motion determination can be obtained. However a rough estimate can be derived by averaging the positions of field stars: we obtain $\mu_\alpha cos(\delta)=-1.40$ and $\mu_\delta=3.50$ [mas/yr], in agreement with previous results \citep{cudworth93}. 
COM-M28H lies at $\mu_\alpha cos(\delta)=-0.09 \pm 0.15$, $\mu_\delta=0.09 \pm 0.15$ [mas/yr], thus fully behaving as a member of the cluster.

\section{Discussion} 
The observed light curve of COM-M28H (Figure \ref{Fig:lc_combined}) clearly shows two distinct and asymmetric minima, at phases $\Phi\sim 0.25$ and $\Phi\sim 0.75$, quite similar to what is observed for two other MSP companions \citep{ferraro01com6397,cocozza08}.
Such a shape is a clear signature of ellipsoidal variations induced by the NS tidal field on a highly perturbed, bloated star. Moreover, the relative depth of the two minima (consistent with a light curve purely due to ellipsoidal variations) suggests that only a marginal (if any) over-heating is affecting the side of the companion facing the PSR.  
This is also supported by the non-detection of \halpha\ emission from the system (see the right panel of Figure \ref{Fig:cmd}).

Given the mass function derived from the radio observations ($f_{\rm PSR}=0.00211277 M_\odot$; B\'egin 2006), and assuming $1.4 M_\odot$ for the MSP mass and $0.68 M_\odot$ for the mass of COM-M28H (as derived from the cluster best-fit isochrone), the resulting orbital inclination of the system would be $i\sim 18^\circ$.  
Such a low value for the inclination angle would not produce any optical modulation.  
Indeed both the light curve shape and the occurrence of eclipses in the radio signal point toward a significantly higher value of the orbital inclination (which corresponds to a lower companion mass for a given mass function).  
By assuming $i=60^\circ$ (the median of all possible inclination angles) a companion mass of $\sim 0.2 M_\odot$ is obtained.  
In this configuration the corresponding total mass of the system would be $M_{\rm TOT}=1.6 M_\odot$ and the physical orbital separation of the system is $a\sim 2.8 R_\odot$.  
In order to check whether such a configuration reproduces the observed light curve, we employed the publicly available software NIGHTFALL\footnote{Available at   http://www.hs.uni-hamburg.de/DE/Ins/Per/Wichmann/Nightfall.html.}.
We fixed the orbital period of the PSR to the radio value and the surface temperature of COM-M28H to $T_{eff}=6000$ K (as inferred from the position in the CMD). 
We then used an iterative procedure letting the orbital inclination, the mass ratio ($M_{\rm NS}/M_{\rm COM}$) and the RL filling factor vary respectively in the ranges
$0^\circ-90^\circ$, $1-20$, and $0.1-1$.  
By using as selection criterion a $\chi^2$-test, the best-fit model (Figures \ref{Fig:lc_filters} and \ref{Fig:lc_combined}) was obtained for an inclination $i\simeq 65^\circ$, a mass ratio $M_{\rm   NS}/M_{\rm COM}\simeq 7$ and a RL filling factor equal to 1. 
These results confirm that a configuration with a highly distorted companion of about $0.2 M_\odot$, orbiting a $1.4 M_\odot$ MSP, in a plane with an orbital inclination of $\sim 60-70^\circ$ well reproduces both the mass function of the system derived from the radio observations, and the optical light curve of COM-M28H.  
It is worth noticing that a good fit can be obtained only if COM-M28H completely filled its RL, which, following \citet{paczynski71}, we estimate to be $\sim 0.65 R_\odot$. While such a large value of the stellar radius allows to account for the observed luminosity of COM-M28H,\footnote{Under the assumption of black body radiation (with   the luminosity $L$ given by $L\propto R^2 T_{eff}^4$), a $0.2   M_\odot$ star heated to the observed temperature ($T_{eff}\sim 6000$   K) and bloated to a radius of $\sim 0.65 R_\odot$, has a luminosity   which is fully consistent with the observed one. 
This reinforces the   hypothesis that COM-M28H completely filled its RL. 
For sake   of comparison, the same object with a radius $R\sim 0.2 R_\odot$   \citep[the value expected for a $0.2 M_\odot$ star on the MS;][]{marigo08},   would be a factor of $\sim 10$ too faint.} it is far too small to cause the observed radio eclipse. 
In fact an eclipse lasting for $\sim 20\%$ of the orbital period corresponds to an eclipsing region of $\sim 3.3 R_\odot$ size. 
This suggests that the eclipsing material is extending well beyond the RL and that it is probably constantly replenished (see also B\'egin 2006). 
Indeed, under the influence of the MSP intense radiation field, a (otherwise normal) MS star may expand to fill its RL \citep{dantona93} and even start to lose mass, while the accretion on the PSR is inhibited by its magnetic pressure (as in the case of MSP-A in NGC 6397; \citealt{ferraro01com6397};  see also \citealt{archibald09}). 
Moreover B\'egin (2006) found a large orbital period derivative for this system, suggesting that the binary is losing material and is spiraling out to longer orbital period.

All these considerations indicate that COM-M28H is a highly-perturbed star which is currently losing mass, and that the system is surrounded by large clouds of gas. Whether or not part of the lost mass was accreted by the NS and served to reaccelerate it in the past \citep[as in the case of J1023+0038;][]{archibald09} cannot be inferred from the available data. We note however that, while from the natural cluster dynamical evolution massive objects are expected to be concentrated close to the centre, M28H is the second most off-centered (after M28F)
and it is located outside the cluster core. 
Hence, such an offset position may suggest the following scenario: the NS was recycled by another companion (that eventually became a very low mass, exhausted star, because of the heavy mass transfer); then an exchange interaction occurred in the cluster core between the MSP binary and a MS star, thus causing the ejection of the lightest star and kicking the newly-formed system away from the centre; the new companion started to suffer heavy perturbations (bloating, mass loss, etc.)  induced by the MSP and we currently observe it as COM-M28H; then it eventually will become a He-WD.  
If the MSP was ejected from the cluster core in an exchange encounter, its current position suggests a relatively recent epoch for the formation of this system, since the expected time for such a heavy system to sink back into the cluster centre is lower than a few Gyr.

While a spectroscopic follow-up may help to better clarify this scenario, the identification of COM-M28H further increases the number of MSPs with a non-degenerate companion in GCs.
This likely indicates that exchange interactions are common events in the dense environments of GC cores and they are quite effective in modifying the ``natural" outcomes of the PSR recycling processes \citep{freire05}.

\clearpage{\pagestyle{empty}\cleardoublepage}

\chapter[The companion to \igr\ in M28]{The optical counterpart to the X-ray transient  \igr\ in the Globular Cluster M28}\label{Chap:M28I}

In this Chapter we describe the identification of the optical counterpart to the recently detected INTEGRAL transient \igr\ in the GC M28.  From the analysis of a multi-epoch HST data-set we have identified a strongly variable star positionally coincident with the radio and Chandra X-ray sources associated to the INTEGRAL transient.  The star has been detected during both a quiescent and an outburst state.  In
the former case it appears as a faint, unperturbed MS star, while in the latter state it is about two magnitudes brighter and slightly bluer than MS stars.  We also detected \halpha\ excess during the outburst state, suggestive of active accretion processes by the NS.  - {\it This Chapter is mainly based on \citealt{pallanca13comM28I}, ApJ, 773, 122.}

\section{Introduction}\label{M28I_intro}

When a close binary system contains a compact object, mass transfer processes can take place. The streaming gas, its impact on the compact star, or the presence of an accretion disk can produce significant X-ray and UV radiation, together with emission lines (such as the \halpha) or rapid luminosity variations.  The first evidence of interacting binaries in Galactic GCs was indeed obtained through the discovery of X-ray sources.  In particular, LMXBs are thought to be binary systems with an accreting NS and are characterized by X-ray luminosities larger than $\sim 10^{35}$ erg s$^{-1}$. Their final stage is thought to be a binary system containing a MSP.
Moreover, during their life  some LMXBs,  usually called X-ray transients \citep{white84}, show a few outbursts and during the quiescent states their millisecond pulsations can become detectable \citep[][]{chak98}.
 
As described in Chapter \ref{Chap:introCOM}, the identification of the optical counterparts is a fundamental step for characterizing these exotic binary systems, both in quiescent and in outburst state, and for clarifying their formation and evolutionary processes \citep[][]{testa12}.  Determining the nature and the properties of the companion (which dominates the optical emission in the quiescent state) is also very useful to tightly constrain the orbital parameters of the system \citep[e.g.][]{d'avanzo09, engel12}.

A total of 46 X-ray sources, 12 of which lie within one core radius \citep[$r_c=14.4\arcsec$;][]{harris96} from the centre, has been detected with Chandra \citep{becker03} in the GC M28.

During the observations of the Galactic center performed on 2013 March 28 with INTEGRAL, a new hard X-ray transient (IGR J18245-2452) has been revealed in the direction of M28 \citep{eckertAtel}. Subsequent observations with SWIFT/XRT confirmed the detection of the transient source and its location within the core of the cluster, at $\alpha_{2000}=18^{\rm h} 24^{\rm m} 32.20^{\rm s}$ and $\delta_{2000}=-24^\circ 52' 05.5\arcsec$, with error radius $3.5\arcsec$ \citep[at 90\% confidence;][]{heinkeAtel, romanoAtel}.  SWIFT/XRT time-resolved spectroscopy performed on 2013 April 7 revealed a thermal spectrum with a ``cooling tail'', unambiguously identifying the burst as thermonuclear and suggesting that the source is a low-luminosity LMXB in the hard state, where a NS is accreting matter from a companion (\citealp{linaresAtel}; see also \citealp{serinoAtel}).  A radio follow-up has been performed with ATCA on 2013 April 5, for a total of 6 hours, at two different frequencies (9 and 5.5 Hz). A single source has been identified at $\alpha_{2000}=18^{\rm h} 24^{\rm m} 32.51^{\rm s}$ and $\delta_{2000}=-24^\circ 52' 07.9\arcsec$, with a 90\% confidence error of $0.5\arcsec$ \citep{pavanAtel}.  This position is only marginally consistent with that derived from the SWIFT/XRT data, but the detected strong variability (up to 2.5 times the mean flux density during the first 90 minutes of observations) suggests a possible association with the X-ray  transient. Its position well corresponds  to the location of the X-ray source \#23 identified by \citet{becker03} from Chandra observations and associated to \igr\ by \citet{homanAtel}.

In this Chapter we report on the identification of the optical counterpart to IGR J18245-2452, obtained from the analysis of high resolution HST data acquired with the WFPC2, WFC3 and ACS/WFC in three different epochs (see also \citealp{pallancaAtel}, and \citealp{cohnAtel}).

\section{Observations and data analysis}\label{M28I_obs} For this work we adopted the same catalog used to identify the companion to PSR J1824-2452H and fully described in  Chapter \ref{Chap:M28H}.  In order to unveil luminosity variations among different epochs, two additional sets of HST data acquired with the WFPC2 and the ACS have been analyzed. In particular, because we were interested only in the GC core, we limited the analysis to the Planetary Camera (PC) of the WFPC2 and  CHIP2 of the ACS/WFC mosaic.  The available samples have been acquired through various filters, at three different epochs (see Table \ref{M28I_dataset}): the WFPC2 dataset was collected on 2009, April 7 (epoch 1, hereafter EP1), WFC3 observations were performed on 2009, August 9 (epoch 2, EP2) and the ACS data-set was acquired on 2010, April 26 (epoch 3, EP3).

\begin{table}[b]
\centering\begin{tabular}{ c  c  l  c  l  c  l }
\hline
\hline
Epoch & Date & Instrument & Filter & ${\rm t}_{{\rm exp}}$ [s] & State & Proposal ID/PI\\
\hline
\multirow{4}{*}{{ \small EP1}}   & \multirow{4}{*}{{ \small 09/04/07}}   & \multirow{4}{*}{{ \small WFPC2/PC}}    & { \scriptsize F170W} & { \scriptsize $2 \times1700$}  &\multirow{4}{*}{{ \small Q}}  & \multirow{4}{*}{{ \scriptsize GO11975/Ferraro}}\\
\multirow{4}{*} {}          & \multirow{4}{*} {}                      &  \multirow{4}{*} {}                       & { \scriptsize F255W} &  { \scriptsize $3 \times 1200$}   &  \multirow{4}{*}{}   & \multirow{4}{*}{}\\
\multirow{4}{*} {}          & \multirow{4}{*} {}                      &  \multirow{4}{*} {}                       & { \scriptsize F336W} &  { \scriptsize $3 \times 800$}     &  \multirow{4}{*}{}   & \multirow{4}{*}{}\\
\multirow{4}{*} {}          & \multirow{4}{*} {}                      &  \multirow{4}{*} {}                       & { \scriptsize F555W} &  { \scriptsize $2 \times 80$}       &  \multirow{4}{*}{}   & \multirow{4}{*}{}\\
\hline
\multirow{5}{*} {{ \small EP2}}  & \multirow{5}{*} {{ \small 09/08/09}}  &  \multirow{5}{*} {{ \small WFC3/UVIS}} & { \scriptsize F390W} & { \scriptsize $5\times 850+1\times800$}    &  \multirow{5}{*}{{ \small B}} &  \multirow{5}{*}{{ \scriptsize GO11615/Ferraro}}\\
\multirow{5}{*} {}          & \multirow{5}{*} {}                      &  \multirow{5}{*} {}                       &{ \scriptsize  F606W} & { \scriptsize  $7 \times 200$}     &  \multirow{5}{*}{}    & \multirow{5}{*}{}\\
\multirow{5}{*} {}          & \multirow{5}{*} {}                      &  \multirow{5}{*} {}                       & { \scriptsize F814W} & { \scriptsize $7 \times 200$}      &  \multirow{5}{*}{}    & \multirow{5}{*}{}\\
\multirow{5}{*} {}          & \multirow{5}{*} {}                      &  \multirow{5}{*} {}                       & { \scriptsize F656N}  &  { \scriptsize $2\times1100+1\times1070$}&  \multirow{5}{*}{}    & \multirow{5}{*}{}\\
\multirow{5}{*} {}          & \multirow{5}{*} {}                      &  \multirow{5}{*} {}                       &  \multirow{5}{*} {}      &  { \scriptsize $3\times1020+1\times935$}&  \multirow{5}{*}{}    & \multirow{5}{*}{}\\
\hline
\multirow{3}{*}{{ \small EP3}}   & \multirow{3}{*}{{ \small 10/04/26}}   & \multirow{3}{*}{{ \small ACS/WFC}}      & { \scriptsize F435W}  & { \scriptsize $4 \times 464$}      &\multirow{3}{*}{{ \small Q}}  & \multirow{3}{*}{{ \scriptsize GO11340/Grindlay}}\\ 
\multirow{3}{*} {}          & \multirow{3}{*} {}                      &  \multirow{3}{*} {}                       & { \scriptsize F625W}  & { \scriptsize $4 \times 60$}        &  \multirow{3}{*}{}   & \multirow{3}{*}{}\\
\multirow{3}{*} {}          & \multirow{3}{*} {}                      &  \multirow{3}{*} {}                       & { \scriptsize F658N}   & { \scriptsize $6\times724 + 3\times717$}   &  \multirow{3}{*}{}   & \multirow{3}{*}{}\\
\hline
\end{tabular}
\caption{Summary of the multi-epoch data-set used in this work.  The quiescent and outburst states (see Section \ref{M28I_identification}) are marked by letters Q and B, respectively.}
\label{M28I_dataset}
\end{table}

The data reduction procedure for the ACS sample has been performed on the charge transfer efficiency corrected (flc) images, once corrected for Pixel-Area-Map (PAM) by using standard IRAF procedures.  The photometric analysis has been carried out by 
following the general outline described in Chapter \ref{Chap:introCOM}. In particular we used the {\sc daophot} package \citep{stetson87}. For each image we modeled the PSF by using a large number ($\sim 100$) of bright and nearly isolated stars.  Then, all \B\ and \R\ images have been combined with {\tt MONTAGE2} and used to produce a master-frame on which we optimized a master-list of stars.
Finally we performed the PSF fitting on this master-list by using the {\sc daophot} packages {\tt ALLSTAR} and {\tt ALLFRAME} \citep{stetson87, stetson94}.  A similar  procedure has been adopted to reduce the flat-fielded (c0m) WFPC2 images.

Since the ACS images heavily suffer from geometric distortions within the FOV, we corrected the instrumental positions of stars by
applying the equations reported by \citet{anderson03}.  We then placed, through cross-correlation, the ACS and the WFPC2 data sets on
the same astrometric system of the WFC3 sample, for which the astrometric solution has an accuracy of $\sim0.2''$ in both $\alpha$ and $\delta$  \citep{pallanca10}.

Finally, the instrumental magnitudes have been calibrated to the VEGAMAG system by using the photometric zero-points reported on the
instrument web pages\footnote{www.stsci.edu/hst/acs/analysis/zeropoints/zpt.py and   www.stsci.edu/documents/dhb/web/c32$\_$wfpc2 dataanal.fm1.html for   ACS and WFPC2, respectively} and the procedure described in \citet{holtzman95} and \citet{sirianni05} for WFPC2 and ACS, respectively.

\section{The optical counterpart to IGR J18245-2452}\label{M28I_identification}
\begin{figure*}[t]
\begin{center}
\includegraphics[width=120mm]{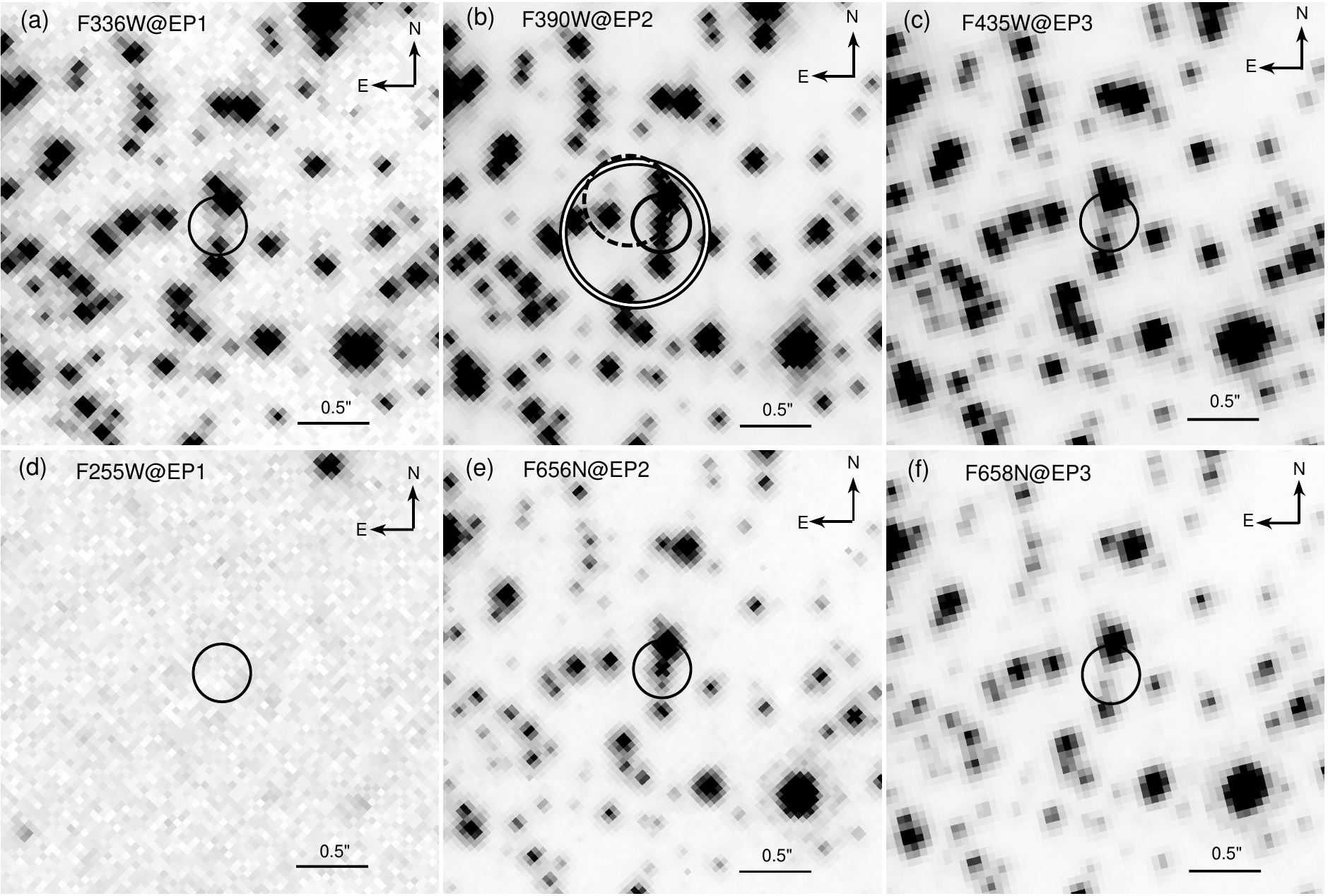}
\caption[Multi-epoch HST images of the optical counterpart to
  \igr]{HST images of the optical counterpart (solid circle) to
  \igr. The filters and epochs of observation are labelled in each
  panel (see Table \ref{M28I_dataset} for more details).  Clearly, the
  source is in a quiescent state in EP1 and EP3 (leftmost and
  rightmost panels), while it has been caught in outburst during EP2
  (central panels). In panel (b) the double and dashed circles mark,
  respectively, the position of the variable ATCA source detected by
  \citet{pavanAtel} and the Chandra X-ray source \# 23
  \citep{becker03}, with the radii corresponding to the quoted
  astrometric uncertainties.}
\label{M28I_map}
\end{center}
\end{figure*}

During a systematic study of the GC M28 aimed at searching for the companion stars to binary MSPs, we found a peculiar object (see Figure
\ref{M28I_map}) located at $\alpha_{2000}=18^{\rm h}24^{\rm m}32.50^{\rm   s}$ and $\delta_{2000}=-24^\circ52'07.8''$, in very good agreement
with the position of the X-ray source \#23 reported by \citet{becker03} and of the variable ATCA radio source discussed by \citet{pavanAtel}.

\begin{figure*}[p]
\begin{center}
\includegraphics[width=150mm]{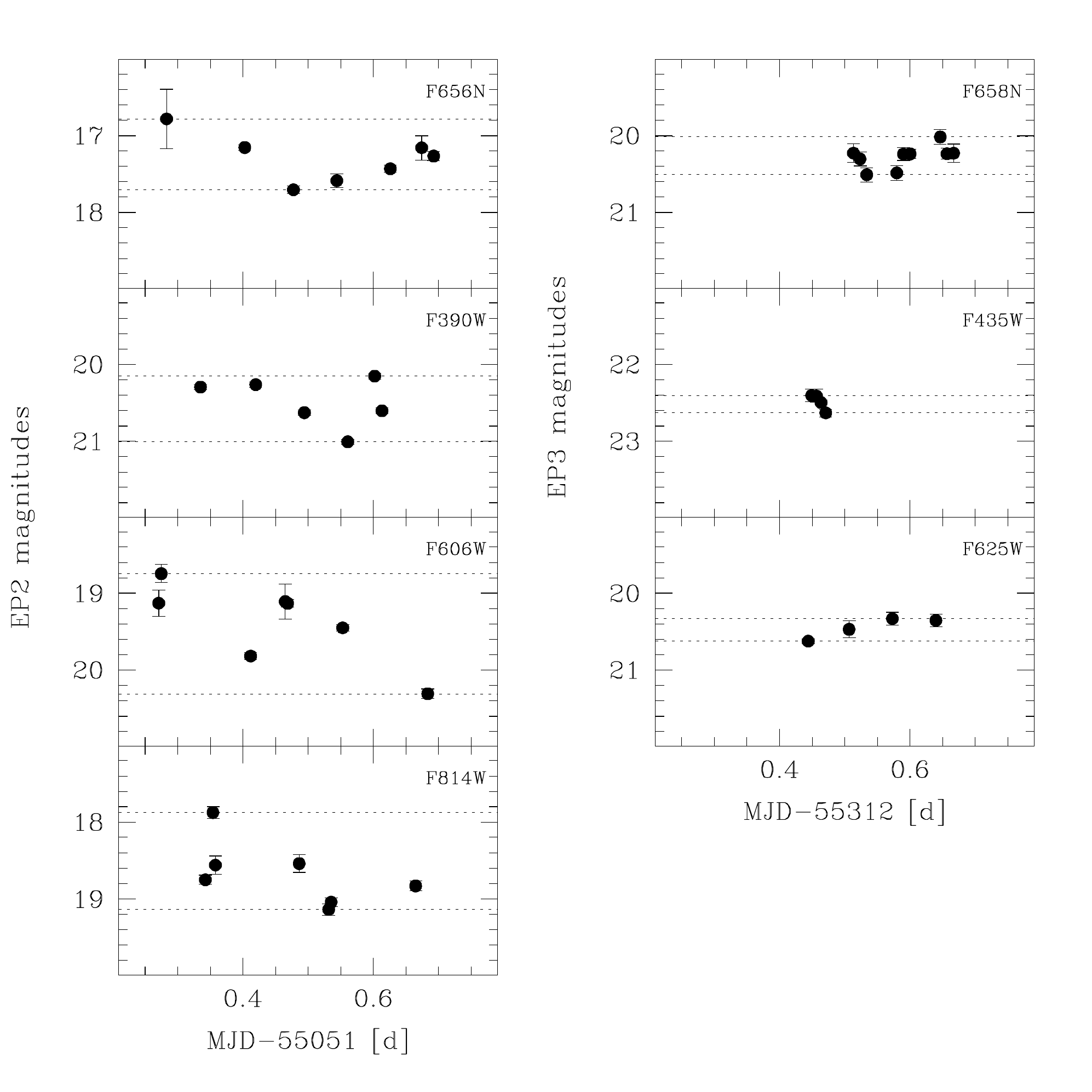}
  \caption[Variability of the optical counterpart to \igr.]{Light curves of the optical counterpart to \igr\ during the outburst state (left panels) and the  quiescent state (right panels). The dotted lines mark the maximum  range of variability detected in each set of observations. Photometric errors are reported, but in most
    cases are smaller than the point size.}
\label{M28I_time}
\end{center}
\end{figure*}

\begin{figure*}[t]
\begin{center}
\includegraphics[width=100mm]{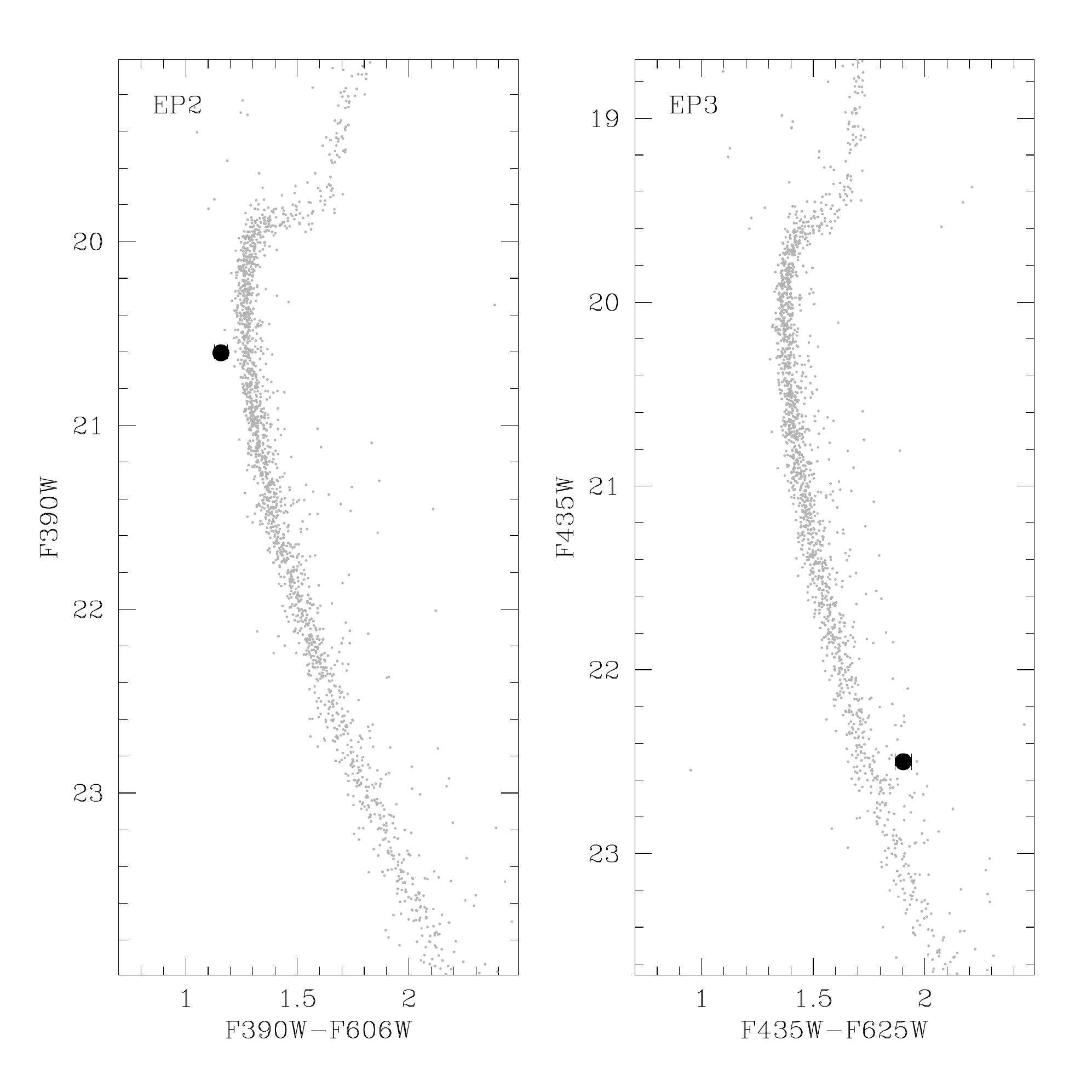}
  \caption[Multi-epoch CMDs of the optical counterpart to \igr. ]{CMDs obtained during outburst epoch (left panel) and quiescence epoch (right panel) for all  stars (gray points) located  within 10\arcsec\ from \igr.  The location of the optical counterpart to \igr, as obtained by averaging the observed light curves (see Figure \ref{M28I_time} and footnote 2), is shown as a large solid circle with error bars.}
\label{M28I_cmd}
\end{center}
\end{figure*}

In EP2 this star showed a strong and irregular variability in each filter, on a timescale of $\sim 10$ hours (Figure \ref{M28I_time}). Based on the mean magnitudes\footnote{It is important to note that, given   the  variability and an under-sampled time coverage, the mean   magnitudes  (and hence the  colors) derived here could not exactly correspond   to the true average luminosities of the star over the entire variability period.}  (\U$=20.61\pm0.01$, \V$ =19.45\pm0.02$, \I$ =18.83\pm0.03$ and \H$ = 17.42\pm0.02$), this star turns out to be about 0.5-1 magnitude fainter than the MS  TO and  bluer than the MS both in the (\U, \U-\V) and in the (\V, \V-\I)  CMDs (see Figure \ref{M28I_cmd}). Even more interesting is the comparison of the photometric properties among the three epochs of observations. Unlike the CMD location in EP2, the magnitudes derived for EP1 (\VV$=21.17\pm0.06$ and \UU$=23.04\pm0.21$) and for EP3 (\B$=22.50\pm0.03$, \R$=20.60\pm0.03$ and \Ha$=20.27\pm0.03$)  approximately locate the star onto the MS. Unfortunately, given the different instruments and filters, it is not possible to directly compare the magnitudes but, both from the visual inspection of images (see Figure \ref{M28I_map}) and from the CMD locations with respect to the TO point, it turns out that during EP1 and EP3 the star was about 2-3 magnitudes fainter than the TO, and hence $\sim 2$ mag fainter than in EP2.  This likely indicates that the observations during EP1 and EP3 sampled the object in quiescence, while EP2 data caught the star in an outburst state.  In addition, during each epoch a magnitude modulation is present, with an indication of a smaller amplitude in EP3 with respect to the variability detected during the EP2 outbursting state. In fact, the frame to frame magnitude scatter of the peculiar star during the outburst epoch (EP2) is $10-20\sigma$ larger than the scatter of normal stars in the same magnitude bin, while this value decreases to $\sim4\sigma$ in EP3.

\begin{figure*}[t]
\begin{center}
\includegraphics[width=120mm]{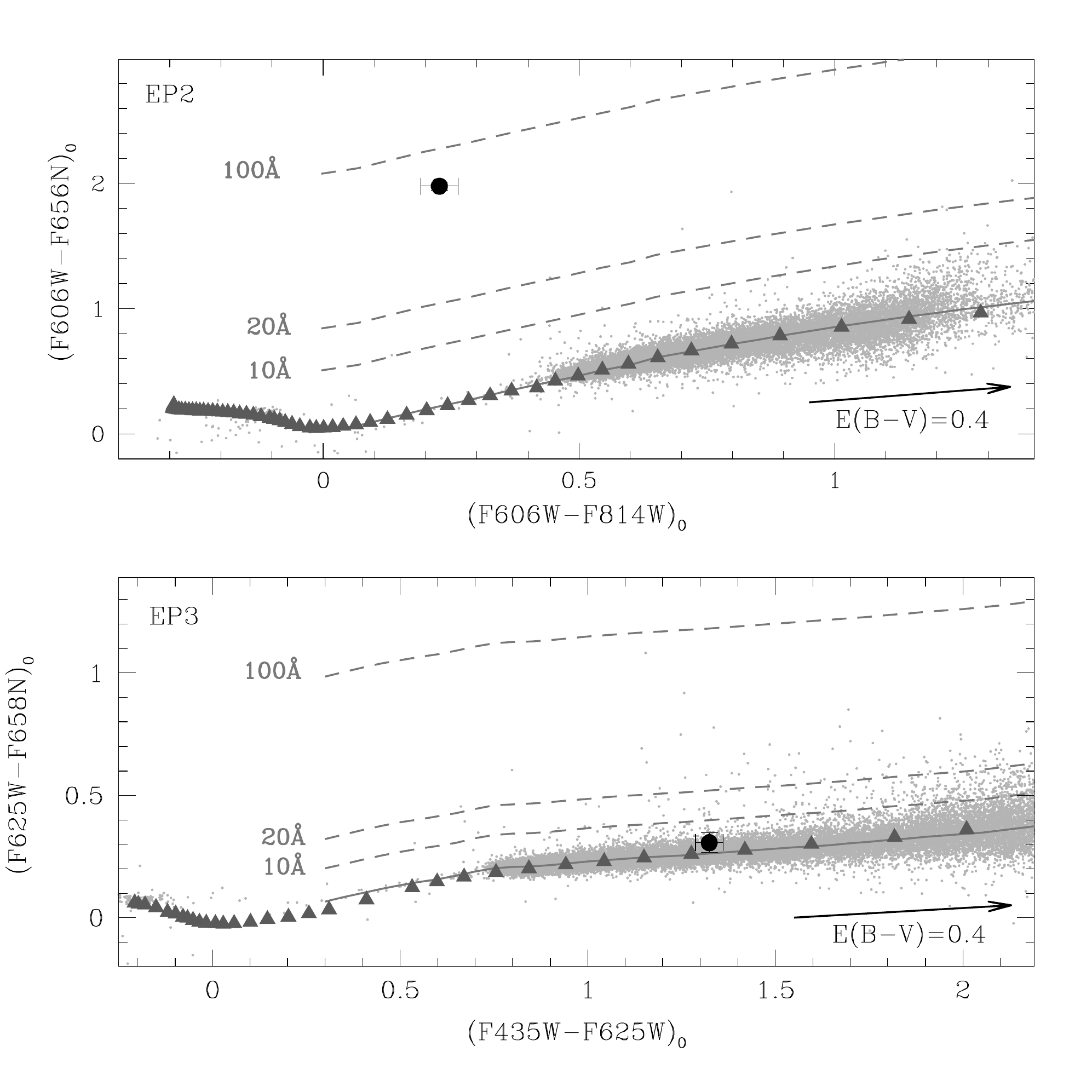}
\caption[H$\alpha$ excess of the optical counterpart to \igr.]{Reddening corrected color-color diagrams for both EP2 and EP3. In each panel the solid line is the median color of stars (gray dots) with no \halpha\  excess  and hence the location of stars with EW$_{{\rm H\alpha}}=0$. It well corresponds to the location (gray triangles) predicted from  atmospheric models \citep{bessel98}. Dashed lines show the expected position for stars with increasing levels of \halpha\  emission, with the corresponding  EW$_{\rm H\alpha}$ labelled. The black dots mark the positions of the optical counterpart to \igr\ in each epoch. During the outburst (upper panel) its H$\alpha$ emission corresponds to  EW$_{\rm H\alpha}=71.6^{+5.5}_{-5.1}$ \AA, while in the quiescent state (lower panel) the star is located on the continuum reference line of stars with no  \halpha\ excess.  }
\label{M28I_halpha}
\end{center}
\end{figure*}

 In principle, for actively accreting LMXBs,  \halpha\ emission is expected from the accretion disk, while the contribution from the heated 
companion star should be minimal or even absent. A visual inspection of EP2 images already suggests that this peculiar star also has \halpha\ excess: in fact, in the \H\ image (panel $(e)$ in Figure \ref{M28I_map}) it is significantly brighter than its southern neighbor, while these two objects show essentially the same magnitude in broad band filters (as the \U, see panel $(b)$ in Figure \ref{M28I_map}).   In order to quantify this excess we used a photometric technique based on the comparison between the magnitudes obtained from broad band  and  \halpha\ narrow filters  \citep[][]{cool95}. In particular, in this work we used a method  commonly applied to star forming regions \citep{demarchi10} and recently tested for the first time in the GC 47 Tucanae \citep[][]{beccari13}. First of all, we corrected all magnitudes for reddening by adopting $E(B-V)=0.4$ \citep{harris96}. Then we selected the peculiar star in the (\V-\H)$_0$ vs (\V-\I)$_0$ color-color diagram.  Note that this color combination well samples the continuum of stars with no \halpha\ emission for different spectral types through the (\V-\I)$_0$ color index, and it provides a good estimate of the \halpha\ emission through the (\V-\H)$_0$ color index, since the \halpha\ line contribution to the \V\ band is negligible.  The \halpha\ excess ($\Delta$\halpha) can be evaluated from the distance between the (\V-\H)$_0$ color index of the considered star and an empirical line\footnote{The reference line for the continuum has been   determined from the median (\V-\H)$_0$ color as function of   (\V-\I)$_0$ for stars with combined photometric error smaller than   0.05 magnitudes. As shown in Figure \ref{M28I_halpha}, this empirical   relation agrees very well with the theoretical one, obtained from   atmospheric models \citep{bessel98}. We also emphasize that, while   M28 may be affected by mild differential reddening, the reddening   vector is almost parallel to the empirical line tracing the   continuum (see Figure \ref{M28I_halpha}). This   means that even large fluctuations in the reddening   of individual   stars would not significantly affect the identification of objects   with \halpha\ excess \citep{be10}. } representative of the continuum.  In addition, the equivalent width (EW) of the \halpha\ emission can be quantitatively estimated from $\Delta$\halpha\ by applying equation (4) in \citet{demarchi10}: ${\rm   EW}_{{\rm H\alpha}}={\rm RW} \times [1-10^{-0.4\times\Delta{\rm       H}\alpha}]$, where RW is the rectangular width of the filter,  \citep[see Table 4 in][]{demarchi10}.  With such a method we estimated a \halpha\ excess $\Delta {\rm H\alpha}=  1.98\pm0.03$;   (upper panel in Figure \ref{M28I_halpha}) and an EW of the \halpha\ emission ${\rm EW_{H\alpha}}=71.6^{+5.5}_{-5.1}$ \AA\ (where the uncertainties take into account the errors in both colors) during the EP2 outburst state.  By applying an analogous method to EP3 data, making use of a suitable combination of \B, \R\ and \Ha\ filters, we find that the star is located on the continuum reference line (see the lower panel in Figure \ref{M28I_halpha}). Hence there is no indication of \halpha\ emission  during the quiescent state.

Finally, we tried to investigate the possible presence of UV emission by using the EP1 dataset in filters \nUV\ and \fUV. No source is detected at the location of the peculiar star, most probably because the images are not deep enough to reach its faint magnitudes.

\section{Discussion and conclusions}\label{M28I_conclusions}
The photometric analysis revealed the presence of a very peculiar star, which  underwent a strong luminosity increase and showed significant \halpha\ excess in EP2.  Even if this optical outburst occurred a few years before the INTEGRAL discovery, this evidence, combined with the positional coincidence with the ATCA variable source recently detected by \citet{pavanAtel} and with the Chandra X-ray source \#23 revealed by \citet{becker03} and firmly associated to \igr\ by \citet{homanAtel}, strongly suggests that we have identified the optical counterpart to \igr.  Indeed several outbursts separated by a few year delay are quite typical of LMXBs containing a NS \citep[e.g 4U 1608-52, Aquila X-1;][]{asai12}.

Unfortunately, the poor and irregular time coverage of our data prevented us to blindly determine the period of the magnitude modulation, which is expected to be correlated with the binary orbital motion.

During the quiescent state the companion star is approximately located on the MS, $\sim 3$ magnitudes fainter than the TO, while during the outburst it is $\sim 2$ magnitudes brighter and it is characterized by a bluer color. As known from the study of companions to MSPs and LMXBs, such an anomalous position is indicative of a perturbed state \citep[see,   e.g.][]{ferraro01com6397,cocozza08,pallanca10,testa12}.  In fact tidal deformations, heating processes and the presence of an accretion disk can significantly affect the magnitude and temperature of the star, thus also altering its position in the CMDs.  The main tool to discriminate between these effects is the determination of the light curve shape, but the available data-sets prevent us to perform this study.

Finally, the presence of strong \halpha\ emission (with ${\rm EW}_{\rm   H\alpha}=71.6^{+5.5}_{-5.1}$ \AA) during the outburst phase suggests
the presence of material accreting onto the NS.  On the other hand,  no \halpha\ excess has been detected in quiescence, in agreement with the
 fact that when the accretion rate is low, the disk is  much weaker and 
 the companion star contributes to the emission only through ``standard''  \halpha\ absorption.

\section{ Further preliminary works about \igr}

Very recently,  \citet{papitto13} claimed the association of \igr\ with \psrI\ in M28 on the basis of orbital parameters coincidence. In particular, they found that in the past decade
this source swung between rotation-powered and accretion-powered  states on timescales of a few days to a few months; this establishes the existence of an evolutionary phase during which a source can alternate between these two states over a timescale much shorter than the billion-year-long evolution of these binary systems, as they are spun-up by mass accretion to millisecond spin periods \citep{bhattvan91}.  This object therefore represents an  unambiguous tracer of both rotation-powered (``radio dominating'')  and accretion-powered (``X-ray dominating'') activity and it provides conclusive evidence for the evolutionary link between NSs in LMXBs and radio MSPs.

In the light of the \igr/\psrI\ association, we performed a preliminary variability analysis by adopting the orbital parameters of \psrI\ known both from radio and X-ray. Given the high crowding of the region and in particular the presence of a very close and significantly brighter star (see Figure \ref{M28I_map}),  we used the software {\sc romafot} \citep[that allows a visual check of the fit quality;][]{buonanno83} and we limited the analysis to  the EP2 dataset (because of the relative brightness of the target). The most reliable measures have been obtained in the \H\ and \U\ bands, where the optical counterpart is brighter and the closest star does not saturate.

\begin{figure*}[p]
\begin{center}
\includegraphics[width=140mm]{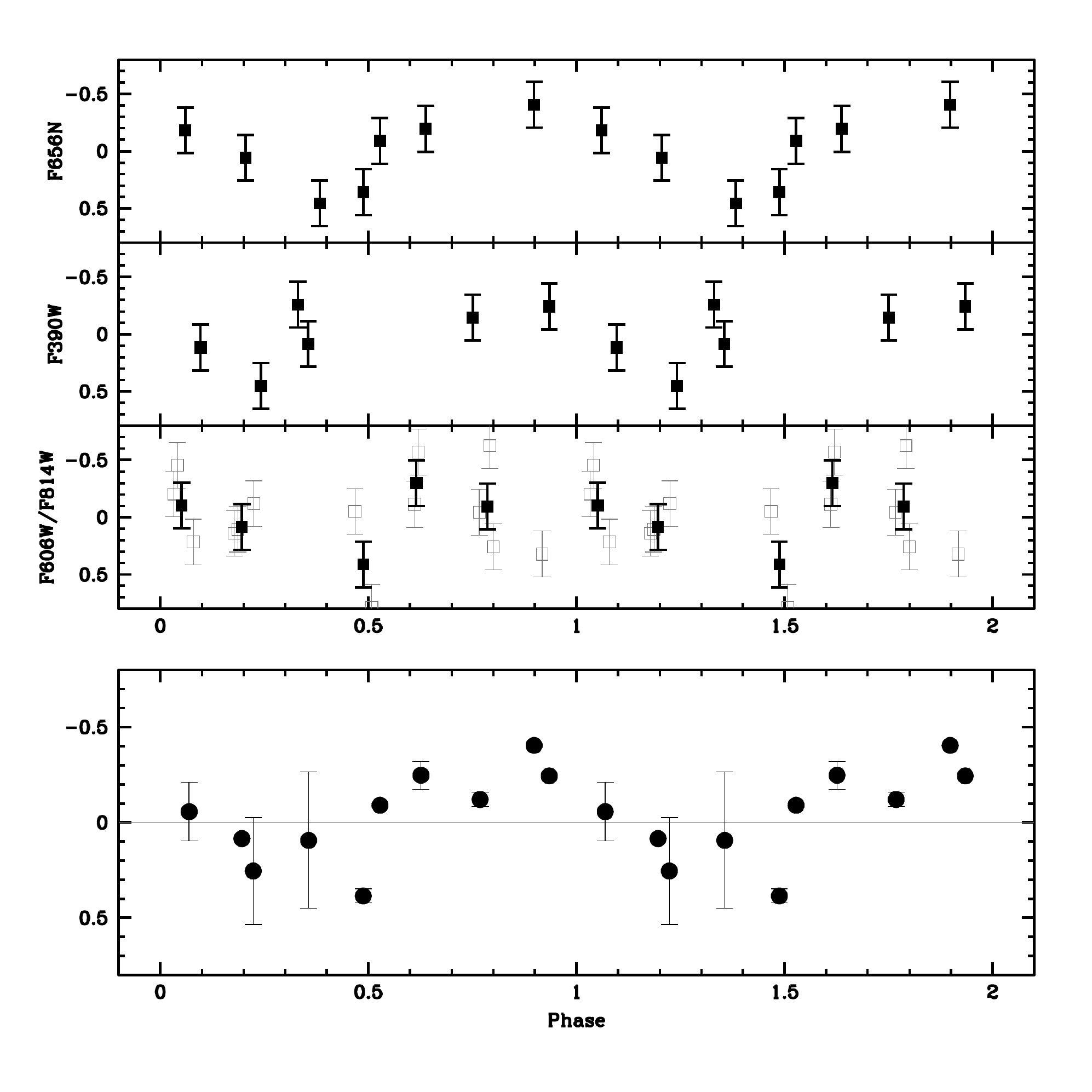}
\caption[Light curve of the optical counterpart to \igr.]{Preliminary light curve of the companion to \igr/\psrI\ phased with orbital parameters known from radio and X-ray. In all panels 
all magnitudes have been referred to mean value. In the third panel from top, the gray open squares are the measures obtained in the \V\ and \I\ bands, while the black filled squares are the combination of \V\ and \I\ measures at similar phases. The bottom panel shows the combined light curve obtained by averaging all available measures in interval of 0.1 in phase. }
\label{M28I_lc}
\end{center}
\end{figure*}

With the obtained measures, we built the light curve of IGR J18245-2452/PSR J1824-2452I  in all filters for the EP2 dataset. The star shows a clear variability in the \U\ and \H\ bands 
which seems to be well folded with the orbital motion of \psrI\  (two top panels of Figure \ref{M28I_lc}). 
Conversely in the \V\ and \I\ bands (gray open squares in the third panel from top) the magnitude modulation seems to be dominated by random scatter, likely caused by the contamination from the close saturated star. In order to reduce these problems we averaged the \V\ and \I\ magnitudes
(once referred to their respective mean values) measured at similar orbital phases. 
We thus obtained a light curve (black points of the third panel from top) with a shape potentially in agreement with the PSR orbital period. Lastly,  in the bottom panel of Figure \ref{M28I_lc}, we reported a combined light curve, obtained by averaging in bins of 0.1 in phase all the available \U, \H\ and \V / \I\  measures.

The analysis of the light curve seems to suggest that the detected variability is in agreement with the PSR orbital period, but unfortunately the poor sampling and the large photometric  uncertainties,  mainly due to  the contamination by the close star, prevented us to well constrain the light curve shape and hence to infer the source of variability. 
 An additional dataset, opportunely phase sampled and with optimized exposure times long enough to detect with a proper S/N the companion star and avoiding  the saturation of the close bright star is required to properly perform a variability analysis and possibly confirm with  good significance    the agreement between the magnitude modulation and the orbital motion periods. Moreover, from the  light curve shape it would be possible to infer some orbital and physical parameters of the system.

Also a spectroscopic analysis, that, given the high crowding  and the relative faint magnitude, is possible only during a  bright state, could help to characterize such system and the possible presence of an accretion disk through  the study  of the radial velocity curve, the chemical abundance patterns and UV emission lines.  However to properly derive the companion radial velocity curve, it is necessary to detect the spectral lines associated with the companion, avoiding those coming from the accretion disk.

\clearpage{\pagestyle{empty}\cleardoublepage}

\chapter[The companion to PSR 1518+0204C in M5]{The Black Widow PSR 1518+0204C in the Globular Cluster M5: optical photometry of the companion star. }\label{Chap:M5C}

In This Chapter we present the first identification of a companion to a BW in a GC.
The companion is a very faint star ($\Uc \gapp 24.8$, $\Vc \gapp 24.3$, $\Ic \gapp
 23.1$) located at 0.25\arcsec\ from the PSR nominal position. This star is visible only in a sub-sample of images, while it is below the detection threshold in the others, thus indicating that it is strongly variable. The resulting light curve shows  a maximum around the PSR inferior conjunction and a minimum around the PSR superior conjunction, with an amplitude of variation larger than 3 mags. We then speculate that such a light curve is shaped   by the heating processes due to the PSR flux. We constrain the reprocessing efficiency as a function of the inclination angle.  - {\it This chapter is mainly based on Pallanca et al. 2014, to be submitted to ApJ.}

\section{Introduction}\label{M5Cintro}
M5 (NGC 5904) is a Galactic GC with intermediate central density \citep[$\log \rho_0=4.0$ in units of $M_\odot/$pc$^3$;][]{pryormeylan93}  and relatively high metallicity \citep[{[Fe/H]} $\sim$ -1.3,][]{carretta09}   located at $\sim 7.5$ kpc from the Earth \citep[][ 2010  version]{ferraro99, harris96}. M5 harbors six MSPs \citep{hessels07,freire08}. Among them, \psrC\ deserves special attention since it is a 
BW system.

BWs are MSPs characterized by relatively small eccentricity and very small mass function \citep[thus indicating a companion mass smaller than $0.1\Msun$;][]{roberts13}.  See Section \ref{introbw}  for more details.

Up to now, 
only a few companions to BWs have been detected  and none of them is in GCs \citep[][]{fruchter88b, stappers96b,  stappers99, stappers00,     reynolds07,  vankerk11, pallanca12, romani12, breton13}.
The companions to BWs in the GF are found to be low mass stars likely ablated and with a filled RL.
Moreover, these objects show a light modulation of a few magnitudes correlated with the orbital motion.
The light curve of BW companions is usually characterized by a  maximum around PSR inferior conjunction ($\Phi\sim0.75$) and one minimum around PSR superior conjunction ($\Phi\sim0.25$).
Such a shape is thought to be due to the reprocessing of the PSR flux by the companion star.

\psrC,    has a spin period of 2.48 ms and it is in a 2.1 hr orbit with a companion of minimum mass  $\sim0.04 \Msun$. It shows regular eclipses for $15\%$ of its orbit, as well as eclipse delays at eclipse ingress and egress, which can last up to 0.2 ms, and are presumably due to dispersive delays as the PSR passes through the ionized wind of its companion \citep[][]{hessels07}. Moreover, the radio position of \psrC\  is coincident with a soft X-ray counterpart seen in a 45 ks Chandra ACIS-S observation of the cluster \citep[][]{hessels07}.

In this chapter we present  the  identification of the companion to \psrC, the first optical counterpart to a BW PSR in a GC. 

\section{Observations and data analysis}\label{M5CSec:dataan}
The photometric data-set used for this work consists of HST high-resolution images obtained with the UVIS channel of the WFC3.  
Our analysis has been focused only on  CHIP1, which contains the  PSR region. 

The analyzed images have been acquired through four filters, in two different epochs.
The first epoch images have been obtained on 2010 July 5 (Prop. 11615, P.I.  Ferraro).
The data-set consists of: 6 images  in the F390W filter  with an exposure time $t_{\rm exp}=500$ s each, 4 images in F606W  with $t_{\rm exp}=150$ s, 4 images in F814W  with $t_{\rm exp}=150$ s, and 6 images in F656N (a narrow filter corresponding to H$\alpha$) with exposure time ranging from $t_{\rm exp}=800$ s to $t_{\rm exp}=1100$ s.  
The second epoch images have been obtained during four visits between 2012  June 6 and 2012 June 9 (Prop. 12517, P.I. Ferraro) and by using   the same three wide filters of the first epoch. 
In particular, the data-set consists of: 4 images obtained through the F390W filter  with an exposure time  $t_{\rm exp}=735$ s each, 8 images in F606W  with $t_{\rm exp}=350$ s, and 12 images in F814W  with $t_{\rm exp}=230$  s and $240$ s.

 The photometric analysis  has been performed on the WFC3 ``flat fielded" (flt) images, once corrected for ``Pixel-Area-Map" (PAM) by using standard IRAF procedures,
 as described  in Chapter \ref{Chap:introCOM}. 
 In particular the master-frame has been produced combining all the F390W, F606W and F814W images.

The astrometry and the magnitude calibration have been performed by following the procedures described in Chapter \ref{Chap:introCOM}.
In particular we first corrected the instrumental positions of stars from geometric distortions within the FOV and we then reported the  WFC3 catalog onto the absolute astrometric system ($\alpha$, $\delta$) by using the stars in common with the HST WFPC2 catalog from \citet[][]{lanzoni07M5} as secondary astrometric standards.
The astrometric solution has an accuracy of $\sim0.2''$ in both $\alpha$ and $\delta$.

The WFC3 instrumental magnitudes have been calibrated to the VEGAMAG system by using the photometric zero-points and the procedures reported on the WFC3 web page.\footnote{http://www.stsci.edu/hst/wfc3/phot\_zp\_lbn}

\section{The companion to \psrC}
\begin{figure*}[t]
\begin{center}
\includegraphics[width=140mm]{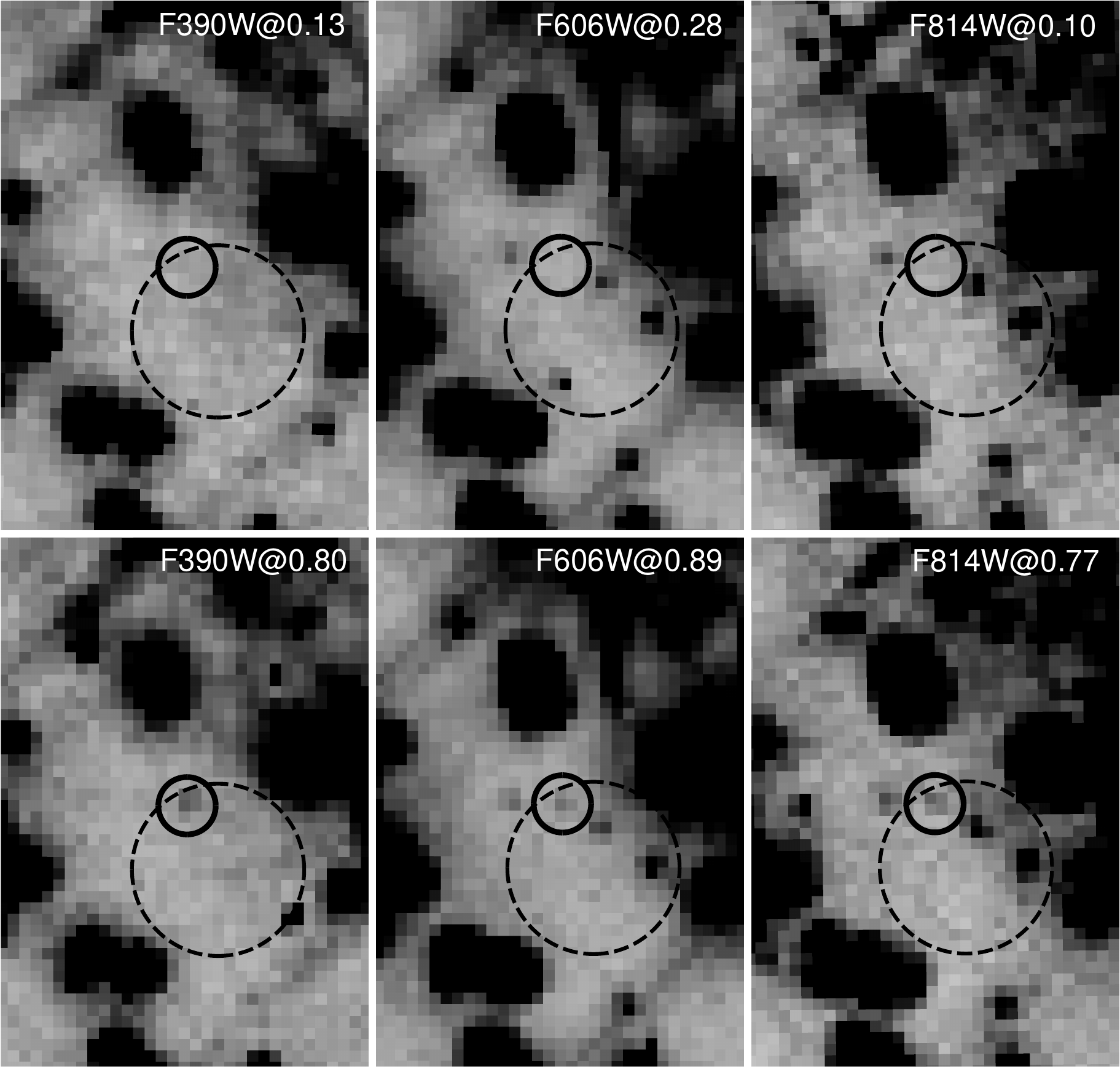}
  \caption[HST images of the region around \psrC.]{HST images of the \psrC\ region. The filters and orbital phases are labelled in each panel. The dashed circle (of radius $0.3\arcsec$) marks the PSR position. The solid circles indicate the identified companion star. It is clearly visible in the lower panels (around PSR inferior conjunction), while it disappears in the upper panels (around PSR superior conjunction). 
  }\label{M5Cfig:map}
\end{center}
\end{figure*}

In order to  identify the companion to \psrC\ we searched for peculiar objects located within $1\arcsec$ from the nominal PSR position (Ransom S., private communication). 
At a first visual inspection of the PSR region, it was possible to identify a star showing strong variability and  lying at $\sim 0.25\arcsec$  from the PSR nominal position.
This star is  very faint and  visible only  in a few images, while it is  below the detection limit in the other frames (see Figure \ref{M5Cfig:map}).

\begin{figure*}[t]
\begin{center}
\includegraphics[width=140mm]{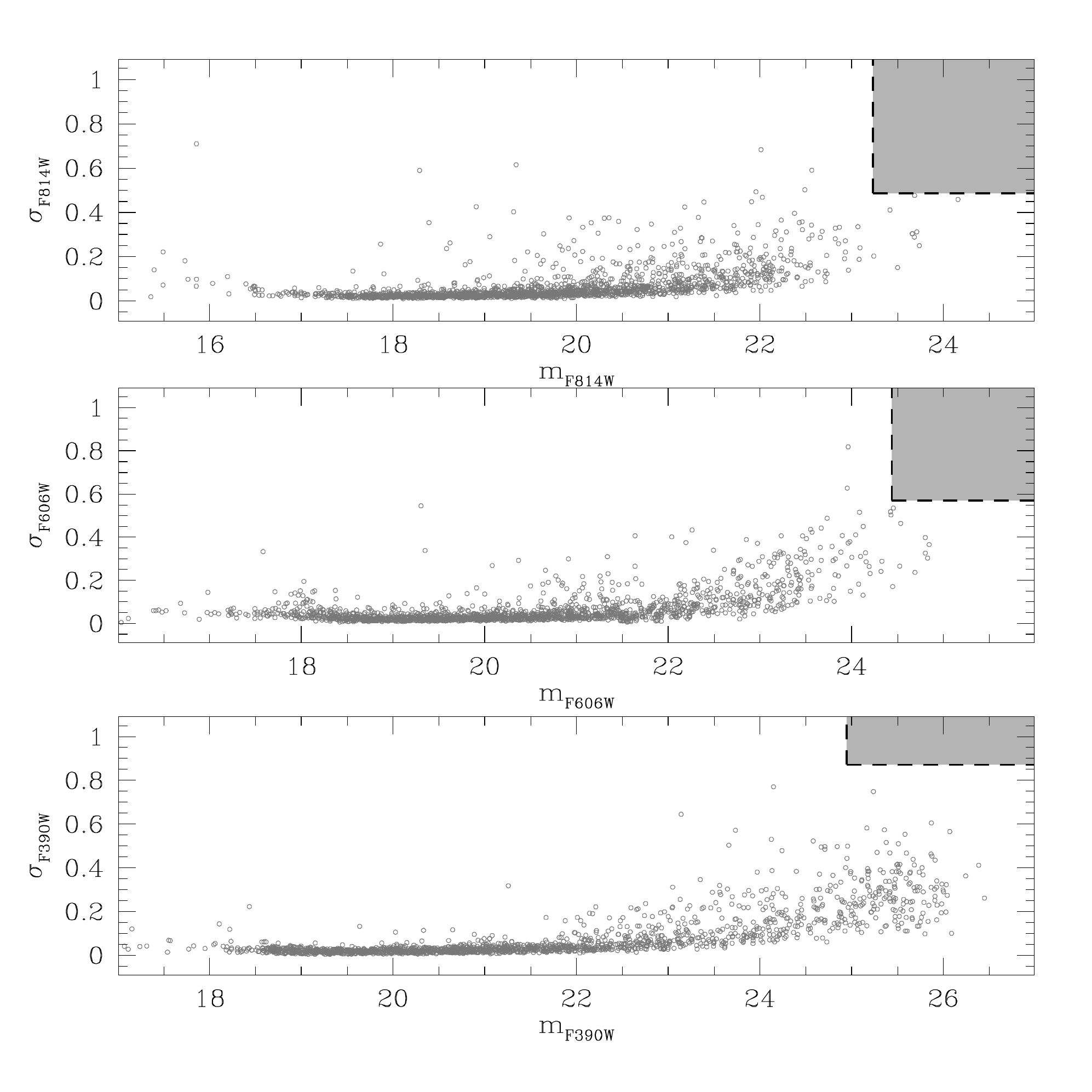}
  \caption[Frame to frame magnitude scatter in the region of \psrC.]{Frame to frame magnitude scatter (gray points) for the stars within $10\arcsec$ from the nominal PSR position, as a function of the magnitude. The standard deviations have been computed by using all the available images. 
The shaded gray areas mark the regions corresponding to the lower limits to the frame to frame scatters,
computed by considering the detection thresholds of the images  where the companion has not been detected.
}\label{M5Cfig:sigma}
\end{center}
\end{figure*}

Since the master-list consisted of the only  objects detected in at least 19 out of 38 images, this star was not included, because it was detectable in a smaller  number of  images.
Hence, we added by hand this star to the master-list and, by following the procedure described in Chapter \ref{Chap:introCOM}, we  measured its magnitude in as many images as possible. We found that its detectability is independent from exposures times, filters and the two epochs of observations.
We were able to measure its magnitude in only 14 images (4 in the F390W, 3 in the F606W and 7 in the F814W filters) and  we observed a significant magnitude variations among them: $\Delta \Uc \sim 0.32$ mag (from $\Uc=24.83 \pm .0.17$ to $\Uc=25.15 \pm 0.22$),  $\Delta\Vc \sim 0.62$ mag (from $\Vc=24.34 \pm 0.20$ to $\Vc=24.96 \pm 0.19$), and $\Delta\Ic \sim 0.87$ mag (from $\Ic=23.13\pm 0.07$ to $\Ic=24.00 \pm 0.28$).
The object is under the detection threshold in all F656N images, hence we do not have any $\Hc$ measure. 
In the other images this star is not detected likely because its flux is below the  detection threshold. For each band, we estimated the detection threshold  as the average value of the magnitudes of the five faintest detected stars within 20\arcsec\ from the PSR position obtaining ${\rm DT_{F390W}} \sim26.59\pm 0.13$, ${\rm DT_{F606W}}\sim25.67 \pm 0.10$ and ${\rm DT_{F814W}}\sim24.41 \pm 0.14$.
In turn, these values imply the following lower limits to the amplitudes of variation: $\Delta\Uc \ge 1.76$, $\Delta\Vc \ge 1.33$ and $\Delta\Uc \ge 1.28$.
Moreover, as shown in Figure \ref{M5Cfig:sigma} this star shows  a magnitude scatter larger than that computed for objects of similar luminosity. This confirms that it is a variable object with a magnitude modulation larger than what it is expected from the photometric errors.

\begin{figure*}[b]
\begin{center}
\includegraphics[width=140mm]{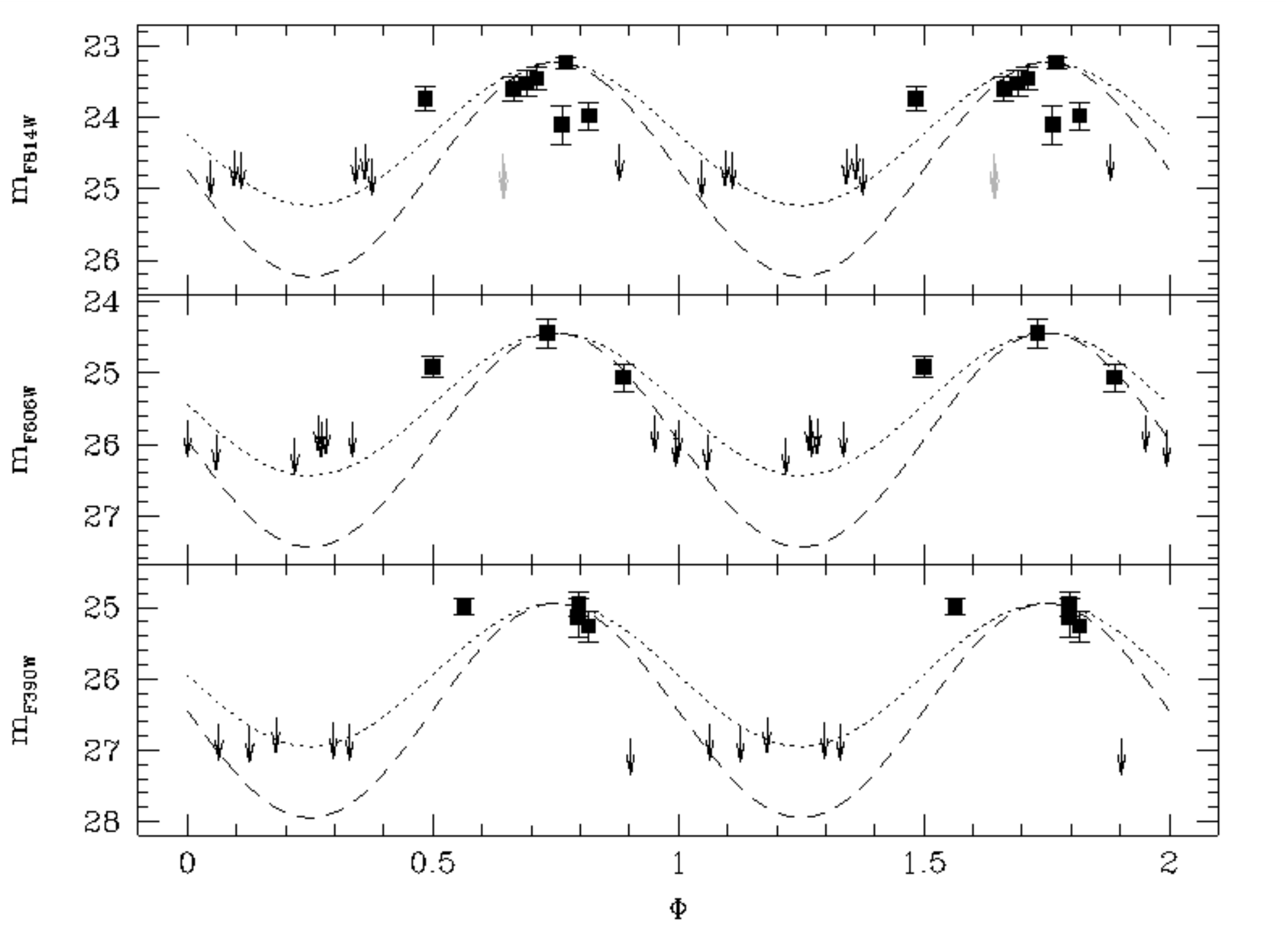}
  \caption[Light curve of the companion to \psrC.]{The observed light curve of the companion to \psrC\ folded with radio timing orbital parameters (Ransom S., private communication). The arrows are the estimated magnitude upper-limits for the images where the star is below the detection threshold. The gray  arrows correspond to
 the detection threshold  for  two images where a visual inspection suggests the possible presence of a star. The dotted and dashed   lines are sinusoidal first-guessed models of the light curve with amplitudes of two and three magnitudes, respectively.  Note that three mags is a lower limit to the amplitude of the magnitude modulation. 
 In fact the sinusoidal curves with  smaller amplitudes are brighter than the detection thresholds at $\Phi=1$ and the star should  therefore be detectable (in contrast with the observations). 
  }\label{M5Cfig:lc}
\end{center}
\end{figure*}

In order to establish if the magnitude variation is related to the PSR  orbital phase (and hence establish  a firm connection between this star and the PSR), we computed the light curve in the three bands  folding each measurement (or magnitude upper limits of the images where the star is not detected) with the orbital period and the  ascending node of the PSR (Ransom S., private communication).  
Although the available data do not allow a complete coverage of the orbital period, the  flux modulation of the star  nicely correlates with the PSR orbital phase  (see Figure \ref{M5Cfig:lc}). In fact the data are consistent with a luminosity maximum (in each band) around $\Phi=0.75$,  corresponding to the PSR inferior conjunction (when we are observing the companion side facing the PSR) and a luminosity minimum (at least a few mags fainter) around $\Phi=0.25$, corresponding 
to the PSR superior conjunction (when we are observing the back side of the companion).
Such a shape is the typical light curve expected when the surface of the companion is heated by the PSR flux.

For reference, we over-plotted  sinusoidal functions   with a maximum at $\Phi=0.75$ and a minimum at $\Phi=0.25$ onto the observed light curve. The first important point to note is that in order to account for most of the upper limits where the star is not detected, an amplitude variation of at least three magnitudes is required (see  Figure \ref{M5Cfig:lc}).
Such a large modulation ($\Delta$mag$\ge3$ mags) is in good agreement with what observed for similar objects  in the GF \citep[e.g.][]{stappers99, stappers01, pallanca12}.
Second, despite the low significance of the detection, there are some hints that the light curve could be asymmetric (e.g. the decrease to minimum is steeper than the increase to maximum), likely due to an asynchronously rotating companion as in the case of PSR 2051$-$0827 \citep[][]{doroshenko01,  stappers01}.

All these pieces of evidence suggest that the identified variable star is the companion to \psrC\ and  we name it \comC.
Since the available $\Uc$, $\Vc$ and $\Ic$  measurements are mainly clustered toward the maximum of the emission, they do not allow 
us to have a reliable measure of the mean magnitudes of this star. 
Therefore, in the following analysis we will use the values at maxima ($\Uc=24.83$,  $\Vc=24.34$ and $\Ic=23.13$) and a plausible range of magnitude variation ($\ge 3$ mags). 

\begin{figure*}[p]
\begin{center}
\includegraphics[width=140mm]{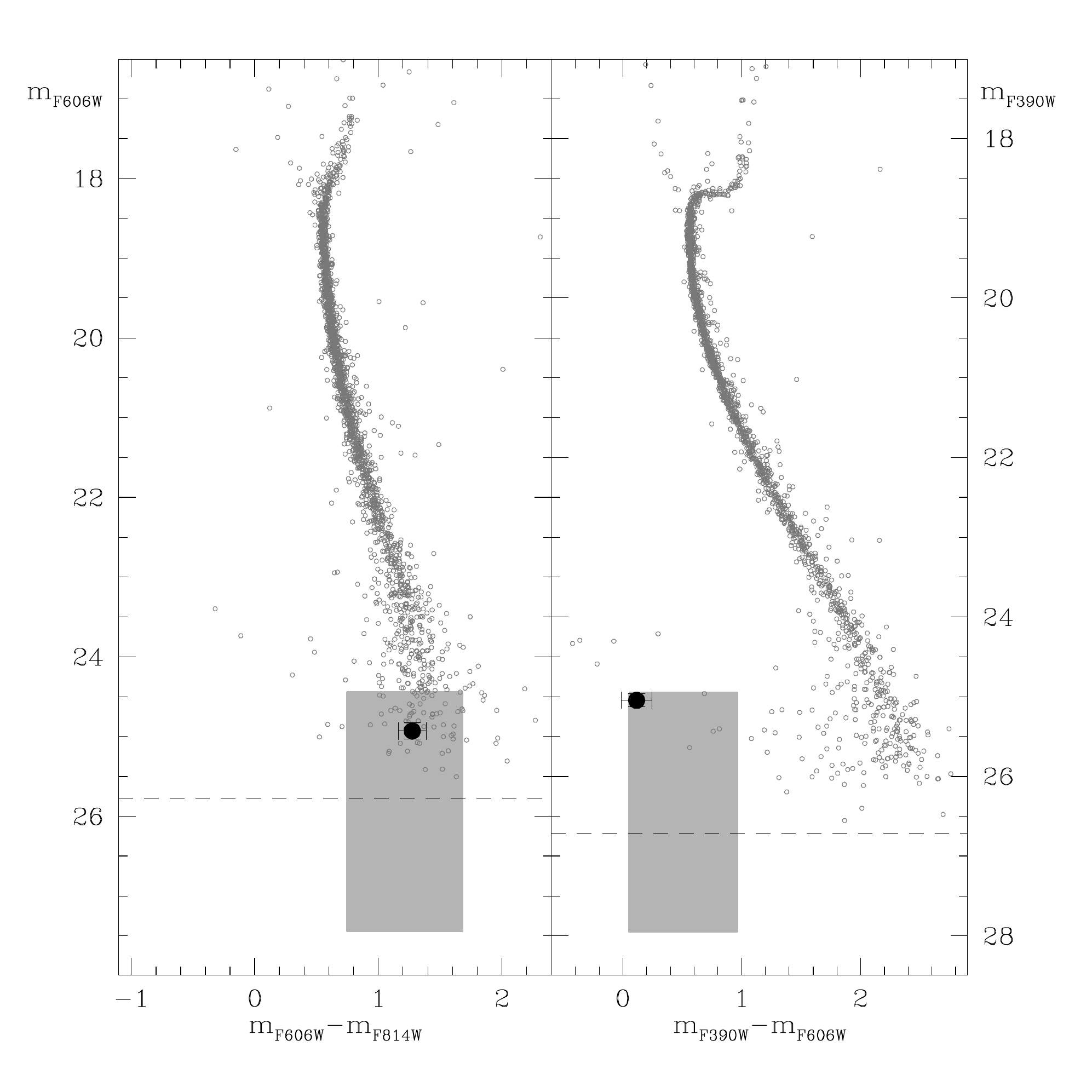}
  \caption[CMD position of the companion to \psrC.]{Location of  the companion to \psrC\ (black point) 
  in the optical CMDs (gray points represent all stars detected within $10\arcsec$ from the PSR nominal position). The companion is a faint star located between the MS  and the WD cooling sequence. 
  The dashed lines corresponds to the detection limits in the F606W and F390W bands (left and right panel, respectively).
  The  gray shaded regions mark the possible location of the companion star based on  the magnitude modulation between the measured maximum and a minimum assumed to be three magnitudes fainter. The color has been calculated as the color at maximum, with an uncertainty estimated to be of the order of the standard deviation of objects with similar magnitudes. }\label{M5Cfig:cmd}
  \end{center}
\end{figure*}

Figure \ref{M5Cfig:cmd} shows the position of \comC\  in the ($\Vc$, $\Vc - \Ic$) and in the  ($\Uc$, $\Uc - \Vc$)  CMDs.
As can be seen, \comC\ is located at faint magnitudes between the MS and the WD cooling sequence,  thus suggesting that it is probably a non-degenerate or a semi-degenerate, low mass, swollen star. Indeed similar objects have been previously identified in Galactic GCs \citep[see][] {ferraro01com6397, edmonds02, cocozza08, 
pallanca10, pallanca13comM28I}.

\section{Discussion}

We have constrained mass, luminosity and temperature of \comC\ by comparing  its position in the ($\Vc, \Vc-\Ic$) CMD (Figure \ref{M5Cfig:cmd}) with a reference isochrone \citep{girardi00,marigo08}.
We adopted a metallicity [Fe/H] $\sim$ -1.3 \citep{carretta09}, an age  $t=13$ Gyr, a distance modulus $(m-M)_0$=14.37 and a color excess $E(B-V)=0.03$ \citep{ferraro99}. 
In particular,  for the star magnitude we assumed the value at maximum ($\Vc=24.34$) as a lower limit and for its color we adopted the value at maximum  ($\Vc - \Ic = 1.21 \pm 0.46$) as a reference.
The color uncertainty has been estimated from the typical photometric error of stars of similar magnitudes (see the gray shaded region in Figure \ref{M5Cfig:cmd}). 
The resulting effective temperature, luminosity  and mass of the star are  $3440{\rm K} \lapp T_{\rm eff} \lapp 5250 {\rm K}$,  $L \lapp 1.19\times 10^{-3} \Lsun$ and  $\mcom \lapp 0.2 \Msun$.

Note that if we consider the largest derived value of the mass ($\mcom\sim0.2\Msun$)  and we combine it with the PSR mass function, 
 we can rule out very small inclination angles (e.g., $i \lapp 10^\circ$ for a $1.4 \Msun$ NS). 
 In addition, as 
found for other companion stars \citep[e.g.][]{ferraro03rv6397, pallanca10,   mucciarelli13}, masses derived from the position in CMDs  might be overestimated\footnote{
In fact, from the measured luminosity we can directly derive the mass through comparison with models of unperturbed MS stars, but 
if the star filled the RL  (and thus not follows anymore the hydrostatic  equilibrium law) such a mass value could be overestimated.}
and hence the limit to the inclination angle could be even more stringent. 
On the other hand, when we assume the mass lower limit to be a core hydrogen burning star ($\mcom \ge 0.08 \Msun$) we would obtain an upper limit to the inclination angle for the system 
 ($i\le30^\circ$). However, such inclinations are not compatible with the presence of radio eclipse.

Under the assumption that  the optical emission of \comC\ is well reproduced by a blackbody (BB), 
the stellar radius is $R_{\rm BB} \lapp 0.30\ \Rsun$.
However,  companions to BWs are expected to be affected by the tidal distortion exerted by the PSR and to have filled their RL or have even reached larger dimensions. 
In particular to justify the presence of eclipses of the radio signal, the size of the RL might be a more appropriate value  \citep[e.g., see PSR J2051$-$0827 and PSR J0610$-$2100][]{stappers96b, stappers00, pallanca12}. According to \citet[][]{eggleton83} the RL radius can be computed as: \begin{displaymath} \frac{\rrl}{a} \simeq \frac{0.49q^{\frac{2}{3}}}{0.6q^{\frac{2}{3}}+\ln \left(1+q^{\frac{1}{3}}\right)}  \end{displaymath} where $q$ is the ratio between the companion and the PSR masses ($\mcom$ and $\mpsr$, respectively). This relation can be combined with the PSR mass function $f_{\rm PSR}(i,\mpsr,\mcom)=(\mcom\sin i)^3/(\mcom+\mpsr)^2$. By assuming a  NS mass ranging from $\sim 1.2 \Msun$ to $2.5 \Msun$ \citep[][]{ozel12},
this yields  $\rrl\sim0.13-0.27\ \Rsun$, depending on the inclination angle ($i$) of the orbital system, ranging from $90^\circ$ to $1^\circ$, respectively.

Under the assumption that the optical variation shown in Figure \ref{M5Cfig:lc} is mainly due to irradiation from the MSP,  reprocessed by the surface of \comC, we can estimate how the re-processing efficiency depends on the inclination angle and, hence, infer the companion mass. To this end, one can compare the observed flux variation ($\Delta F_{obs}$)  between the maximum ($\Phi=0.75$) and the minimum ($\Phi=0.25$) of the light curve, with the expected flux variation ($\Delta F_{exp}$) computed from the rotational energy loss  rate ($\dot{E}$). 
Unfortunately, $\dot{E}$ is not measurable with the available radio observations. However we took as reference the
value measured for some BWs in the GF, that typically have $\dot{E}$ values ranging from $10^{34}$ to  $10^{35}$ erg s$^{-1}$ \citep{breton13}. 
Actually, since we do not observe the entire light curve, $\Delta F_{obs}$ can just put a lower limit to the reprocessing efficiency. Moreover, since these quantities depend on the inclination angle of the system (see below) we can just estimate the reprocessing efficiency as function of $i$. 

\begin{figure*}[p]
\begin{center}
\includegraphics[width=140mm]{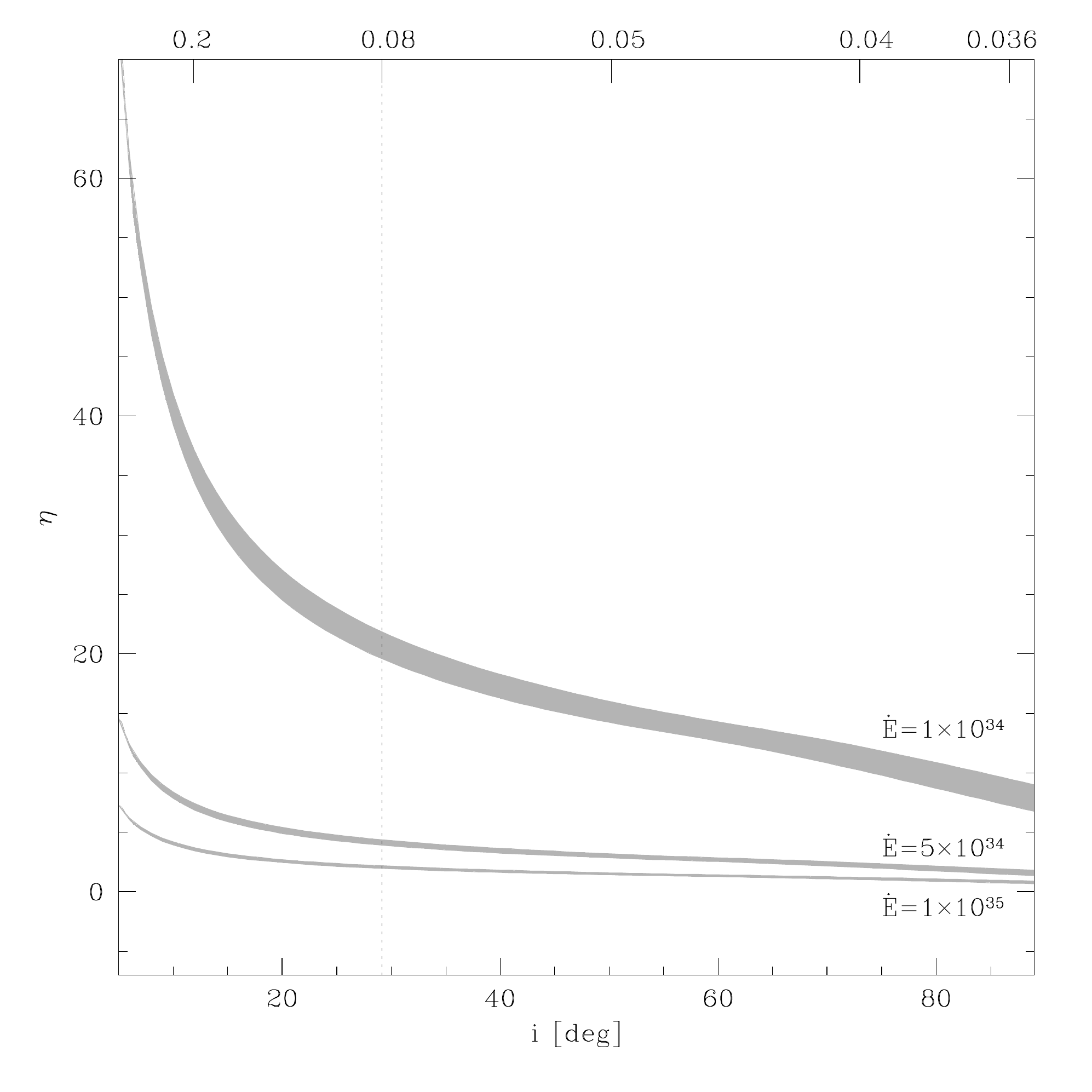}
  \caption[Reprocessing efficiency of the companion to \psrC.]{Lower limit to  the reprocessing efficiency for isotropic emission ($\eta$) calculated as a function of the inclination angle ($i$) and assuming the lower limit to the magnitude modulation ($\Delta$mag $= 3$ mags).
  The three gray strips correspond to different values of $\dot{E}$: $1.0\times10^{34}$, $5.0\times10^{34}$ and $1.0\times10^{35}$, from top to bottom.
  The thickness of each strip corresponds to a PSR  mass ranging from $1.24\Msun$ to $2.5\Msun$. 
  On the top axis, the companion masses in units of $\Msun$  calculated for a $\mpsr=1.4\Msun$ are reported. The dotted line marks the physical limit ($M\gapp0.08\Msun$) for core hydrogen burning stars.}\label{M5Cfig:repr}
\end{center}
\end{figure*}

At first we converted the observed  $\Delta \Vc$ modulation into a flux variation. 
We assumed the maximum measured magnitude $\Vc =24.34$ for $\Phi=0.75$ and an amplitude of variation $\Delta \Vc =3$ between $\Phi=0.75$ and $\Phi=0.25$, thus obtaining  $\Delta F_{obs} = 2.96 \times 10 ^{-15}$ erg s$^{-1}$ cm$^{-2}$.
On the other hand, the expected flux variation due to irradiation between $\Phi=0.75$ and $\Phi=0.25$ is given by \begin{displaymath} \Delta F_{exp}(i)= \eta \frac{\dot{E} }{a^2}  \rcom^2  \frac{1}{4\pi d^2_{\rm PSR}} \varepsilon (i) \end{displaymath} where $\eta$ is the re-processing efficiency under the assumption of isotropic emission, $a$ is semi-major axis of orbit, $\rcom$ is the radius of the companion star,  which we assumed  to be equal to $\rrl(i)$, $\dpsr$ is the distance of PSR, adopted to be equal to the GC distance \citep[$\dpsr=7.5$ kpc;][]{harris96,ferraro99} and $\varepsilon (i)$ parametrizes the difference of the re-emitting surface visible to the observer between maximum and minimum\footnote{ In the following we assume 
$\varepsilon (i)=(i/180)(1-\rcom/a)$.
Note that for a $\rcom \ll a$ the second term reduces to zero.
For the two limit configurations we find that in the case of a face-on system ($i=0^{\circ}$), the fraction of the heated surface visible to the observer is constant and hence  no flux variation is expected, while in the case of an edge-on system ($i=90^{\circ}$) the fraction of the heated surface that is visible to the observer varies between 0.5 (for $\Phi=0.75$) to 0 (for $\Phi=0.25$) and hence $\varepsilon=0.5$.}. By assuming $\Delta F_{obs}=\Delta F_{exp}(i)$ between $\Phi=0.75$ and $\Phi=0.25$, we can derive how $\eta$ varies as a function of $i$. The result is shown in Figure \ref{M5Cfig:repr}. 
Obviously, it is important to note that  all the calculations have been performed using a lower limit ($\Delta$mag $= 3$) for the magnitude modulation and hence also the estimated reprocessing efficiency has to be considered as a lower limit to the true value.

As an example,  the observed optical modulation can be reproduced considering a system seen at an inclination angle of about $60^{\circ}$,  with a very low mass companion ($\mcom\sim0.04-0.05\Msun$) that has filled its RL, and reprocess the PSR flux with an efficiency $\eta \gapp 1-15\%$. Similar results are obtained performing the same calculations in the F390W and F814W bands.

On the other hand, if we use $R_{\rm BB}$ instead of $\rrl$ for the stellar radius, the efficiency increases and for several configurations it becomes larger than 100\%.
In such  cases the only possible scenario would be that of an anisotropic PSR emission. However, given the presence of eclipses and the observed behavior of other similar objects,  $R_{ \rm BB}$ is likely too small to provide a good estimate of  the star physical size. However, future studies are needed to better constrain the system parameters.

Unfortunately, with the current generation of telescopes the star is too faint and located in a crowded region to perform a spectroscopical follow-up that would be furnish to uniquely characterize this object.
However, an optimized photometric follow-up aimed to detect the star at the minimum will give the opportunity to better constrain the light curve shape of the companion and hence to better characterize this binary  system.

\clearpage{\pagestyle{empty}\cleardoublepage}

\chapter[Spectroscopy of  \comA\ in NGC 6397]{New clues on the nature of the companion to PSR J1740$-$5340A in NGC 6397 from XSHOOTER spectroscopy}\label{Chap:6397}

By using XSHOOTER spectra acquired at the ESO VLT, we have studied the surface chemical composition of the companion star to the binary MSP J1740$-$5340A  in the GC NGC 6397. The measured abundances of Fe, Mg, Al and Na confirm that the star belongs to the cluster. On the other hand, the measured surface abundance of nitrogen ([N/Fe]$=+0.53\pm 0.15$ dex), combined with the carbon upper limit (${\rm [C/Fe]}<-2$ dex) previously obtained from UVES spectra, allows us to put severe constraints on the nature of this object, strongly suggesting that the PSR companion is a deeply peeled star.  In fact, the comparison with theoretical stellar models indicates that the matter currently observed at the surface of this star has been processed by the hydrogen-burning CN-cycle at equilibrium. In turn, this evidence suggests that the PSR companion is a low mass ($\sim 0.2 \Msun$) remnant star, descending from a $\sim 0.8 \Msun$ progenitor which
lost $\sim 70-80$\% of its original material because of mass transfer activity onto the PSR. - {\it This chapter is mainly based on \citealt{mucciarelli13}, ApJ, 772, L27.}

\section{Introduction}   
\label{6397intro}  

\begin{figure}[t]
\begin{center}
\includegraphics[width=140mm]{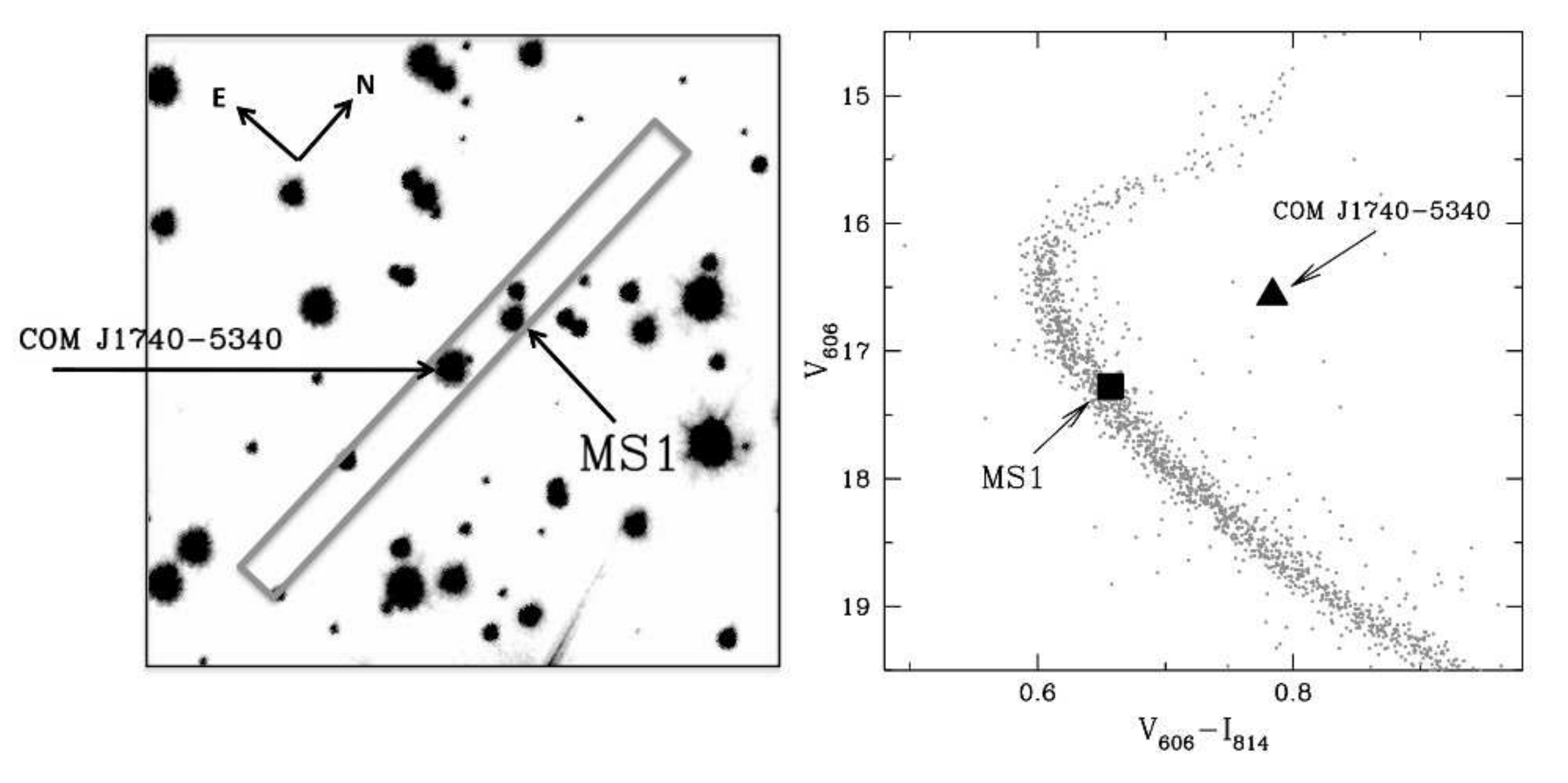}
\caption[Map and CMD position of  \comA.]{ 
{\sl Left panel}: 
HST-ACS archival $V_{606}$-band image with the position of the XSHOOTER slit and the identification of the two targets (the two other objects included in the slit are significantly fainter with $V_{606}>19.5$). {\sl Right panel}: 
CMD of the region at $R<40 \arcsec$ around \comA\, with marked the position of \comA\ and MS1.}
\label{6397loca}
\end{center}
\end{figure}

The MSP J1740$-$5340A in the GC NGC 6397  belongs to a binary system with an  orbital period of $\sim 1.35$ days \citep{damico01a}. At 1.4 GHz  radio frequency it shows eclipses for about 40\% of the orbital period, likely due to matter released from the companion \citep{damico01b}, which probably is also responsible for its X-ray emission \citep[see][]{grindlay01}. The companion star (hereafter, \comA) was identified by \citet{ferraro01com6397} as a variable object with a luminosity comparable to that of the MS  
TO and an anomalously redder color (see right panel in Figure \ref{6397loca}).  The shape of its light curve suggests that it is tidally distorted by the interaction with the PSR \citep{ferraro01com6397, kaluzny03}.

In virtue of its brightness ($V\sim16.6$), \comA\ represents one of the rare cases where spectroscopy of a GC MSP companion can be successfully performed (see Section \ref{specana}). Indeed, a detailed study of the companion radial velocity curve was performed by \citet{ferraro03rv6397} and, once combined with the PSR radial velocity curve, it allowed to derive the mass ratio of the system ($M_{\rm PSR}/M_{\rm COM}=5.85\pm 0.13$). In turn, this constrained the companion mass in the range $0.22 \Msun\le M_{\rm COM}\le 0.32 \Msun$ \citep{ferraro03rv6397,kaluzny03}. In addition, also a chemical analysis has been performed
on UVES@VLT high resolution spectra,  highlighting {\sl (a)} a complex structure of the H$\alpha$ profile, well reproduced by two different emission components \citep{sabbi03Ha}, {\sl   (b)} an unexpected detection of a He I line, suggesting the existence of a hot (T$>$10000 K) region located on the stellar hemisphere facing the MSP \citep{ferraro03rv6397}, and {\sl (c)} some anomalous chemical patterns (for Li, Ca and C) with respect to the  chemical composition of normal cluster stars, in particular a significant depletion of Carbon \citep{sabbi03C}.

Several hypotheses have been proposed to explain the nature of this system and find a coherent picture for the observational evidence collected \citep{possenti,burderi02,orosz03}.  In particular, two possible origins for COM J1740$-$ 5340A can be advanced: {\sl (1)} it is a low mass ($<0.3 \Msun$) MS star perturbed by the PSR; {\sl
  (2)} it is a {\sl normal} star (at the TO or slightly evolved, according to its luminosity), deeply peeled by mass loss processes.

Following the suggestions of \citet{ergma03}, the CN surface abundances are an ideal tool to discriminate between the two proposed scenarios: in fact, if it is a perturbed MS star, its C and N abundances should be unmodified with respect to the pristine cluster chemical composition. On the other hand, if it is a peeled star, its chemical composition is expected to show the signatures of  H-burning CN-cycle (in particular, a decrease of $^{12}$C  and an increase of $^{14}$N)\footnote{Note that this is also the   chemical signature used by \citet{ferraro06} to infer a   mass-transfer origin for a sub-sample of blue straggler stars in 47  Tucanae.}. A first evidence in favor of the latter scenario has been provided by the significant lack of C
\citep{sabbi03C}.  Unfortunately, however, those spectra do not allow to measure the N abundance. In this 
chapter we present XSHOOTER spectroscopic observations of \comA, focussing on the N abundance. 

\section{Observations}
Observations of \comA\ were secured with the XSHOOTER spectrograph at the ESO-VLT. A second target (hereafter MS1) was included in the same slit  (Figure \ref{6397loca}, left panel)  and used as comparison star. This is a MS star with $V_{606}=17.28$, located at $\sim 1.3\arcsec$ from the main target (see Figure \ref{6397loca}, right panel). 
A first observing run has been performed in June 2010, enabling simultaneously the UVB ($\sim$ 3300-5500 \AA) and the VIS
($\sim$5500-10000 \AA) channels of XSHOOTER.  The adopted slit width was $0.8\arcsec$ (R=6200) and $0.7\arcsec$ (R=11000) for the UBV and VIS channels, respectively, and the exposure time was 1200 s in both cases. To increase the S/N in the region around the NH band ($\sim$3360 \AA), a second observation has been secured in July 2011, using only the UVB channel, with the same slit width and with an exposure time of 2700 s.

The data reduction was performed with the XSHOOTER ESO pipeline, version 2.0.0, including bias subtraction, flat-fielding, wavelength calibration,  rectification and order merging.  Because the pipeline does not support efficiently the spectral extraction for many sources in the same slit, this task was performed manually with the IRAF package {\sc apall} in optimal extraction mode.  The final spectra have S/N=50-100 for COM J1740$-$5340A, and S/N=40-70 for MS1.

\section{Chemical analysis}

The chemical abundances of Fe, C, N, Na, Al and Mg have been derived through a $\chi^2$-minimization between the observed spectral features and a grid of synthetic spectra computed with different abundances for each species, following the procedure described in \citet{m12b}.  With respect to the traditional method of the equivalent widths, this approach reduces the difficulties in the continuum location, a critical task in the analysis of low-resolution spectra because of the severe line blanketing conditions.

For the analysis of \comA\ we adopted the atmospheric parameters $T_{eff}=5530$ K, $\log g = 3.46$ and $v_{turb}=1.0$ km/s derived by \citet{sabbi03C}.  For MS1 we derived  $T_{eff}$= 6459 K and $\log g=4.44$ by comparing the position  of the star in the CMD with a theoretical isochrone from the BaSTI dataset \citep{pietr06}, with Z=0.0003 (corresponding to [Fe/H]=--2.1), $\alpha$-enhanced chemical mixture, and an age of 12 Gyr, assuming the reddening and the distance modulus quoted by \citet{ferraro99}.  The photometric value of $T_{eff}$ for MS1 is confirmed by the analysis of the wings of the H$\alpha$ line.
Since the effective temperature of  \comA\ was also derived from the H$\alpha$ wings \citep{sabbi03C}, we can considered the values of $T_{eff}$ of the two objects on the same scale. For the microturbulent velocity 
we assumed 1 km s$^{-1}$, that is a reasonable value for unevolved low mass stars \citep{gratton01}.

The synthetic spectra have been computed with the SYNTHE code by R. L. Kurucz \citep{sbordone}, including all the atomic and molecular transitions listed in the Kurucz/Castelli line list\footnote{http://wwwuser.oat.ts.astro.it/castelli/linelists.html}. All the synthetic spectra have been convolved with a Gaussian profile to reproduce the appropriate spectral resolution. The synthetic spectra used for the analysis of COM J1740$-$5340A have been also convolved with a rotational profile with $v \sin i$=~50 km s$^{-1}$  \citep{sabbi03C}.  Instead, for MS1 no additional rotational velocity is added, according to the very low values ($<$3-4 km s$^{-1}$) typically measured in unevolved low mass stars \citep{lucatello03}.  The model atmospheres have been calculated with the {\sc atlas9} code \citep{castelli04},  assuming [M/H]$=-2.0$ dex and $\alpha$-enhanced chemical composition \citep[according to the analysis by][]{sabbi03C}.

The spectral lines for the analysis have been selected through the detailed inspection of the synthetic spectra, considering only those transitions that are unblended at the XSHOOTER resolution. A total of 15 and 13 Fe~I lines have been selected in COM~J1740$-$5340A and in MS1, respectively. The nitrogen abundances were derived by fitting the band-head of the A-X (0-0) and (1-1) transitions located at 3360 \AA\ and 3370 \AA, respectively.  The inspection of the solar-flux spectrum by \citet{neckel} suggested that we need to decrease by 0.5 dex the Kurucz value of $\log$ gf, in order to properly reproduce the solar N abundance. 
The carbon abundances were derived from the CH G-band at 4300 \AA. The Kurucz $\log$ gf for the CH transitions were decreased by 0.3 dex, in order to reproduce the G-band observed in the solar-flux spectrum by \citet{neckel}, as discussed in \citet{mucciarelli12}. 
Aluminum abundances were derived from the UV resonance line at 3961 \AA, applying a non-LTE correction of +0.7 dex for both targets, according to the calculations of \citet{andr08}. To derive the sodium abundance, we used the Na doublet at 8183-8194 \AA: these lines fall in a spectral region severely contaminated by telluric features. Despite the accuracy of the telluric subtraction (performed with the IRAF task {\tt telluric} by adopting as template the spectrum of an early-type star observed during the observing runs), 
the radial velocities of the two stars prevent a total deblending between the Na lines and the telluric features.  For both stars we therefore provide only upper
limits for the Na abundance, including the non-LTE corrections by \citet{lind}. 

Abundance uncertainties have been estimated by adding in quadrature the errors obtained from the fitting procedure and those arising from the atmospheric parameters.  The uncertainties in the fitting procedure have been estimated by resorting to Monte Carlo simulations. Uncertainties due to atmospheric parameters are calculated by varying one parameter at a time, while keeping the other ones fixed, and repeating the analysis.

\section{Results}
\begin{table}[t]
\begin{center}
\begin{tabular}{lll}
\hline
\hline
Ratio & COM J1740$-$5340A &MS1\\
\hline
\hline
${\rm [Fe/H]}$     & $-2.00 \pm 0.12$ & $-1.93 \pm 0.18$ \\
${\rm [Mg/Fe]}$    &  $+0.38 \pm 0.13$ &  $+0.30\pm0.15$ \\
${\rm [Al/Fe]}$    &  $+0.31 \pm 0.14$ &  $+0.35\pm0.20$ \\
${\rm [Na/Fe]}$    &  $<0.00$          &  $< 0.15$       \\
\hline
${\rm [N/Fe]}$     &  $+0.53 \pm 0.15$  &  $<+0.10$ \\
${\rm [C/Fe]}$     &  $<-1.0$          &  $+0.10\pm 0.20$   \\
\hline
\hline
\end{tabular}
\caption{Chemical abundances measured  for \comA\ and MS1. Reference solar abundances are by \citet{gs98}.}\label{6397table}
\end{center}
\end{table}

Table \ref{6397table} lists the chemical abundances derived for \comA\ and MS1, together with their total uncertainties.  The iron content of the MSP companion is [Fe/H]$=-2.00\pm 0.12$ dex, in agreement with both the iron abundance of MS1 ([Fe/H]$=-1.93\pm 0.18$ dex) and previous estimates of the cluster metallicity \citep[see   e.g.][]{carretta09,lovisi12}.  In addition, \comA\ and MS1 show very similar values of the Na, Mg and Al abundances.  These elements are involved in the chemical anomalies usually observed in GCs \citep{gratton12} and are explained as due to two or more bursts of star formation in the early phases of the cluster evolution \citep[see][]{dercole08}. In particular, given the ranges of values measured in unevolved stars both in NGC 6397 \citep{gratton01,carretta05,pasquini08} and in GCs of similar metallicity \citep{carretta09}, the [C/Fe], [Mg/Fe] and [Al/Fe] abundance ratios, as well as the upper limit of [Na/Fe] derived for MS1 suggest with this object belongs to the first generation of stars\footnote{ For TO stars in NGC~6397 \citet{lind09} derive  temperatures lower than those predicted by the BaSTI isochrone.  We repeated the analysis of MS1 decreasing $T_{eff}$ by 250 K, in order to match the $T_{eff}$ scale by \citet{lind09}. The differences in the derived [C/Fe] and [N/Fe] are smaller than 0.2 dex and do not change our conclusions about this star.}. In turn, the observed chemical similarity indicates that this is likely the case also for \comA.

\begin{figure}[t]
\begin{center}
\includegraphics[width=120mm]{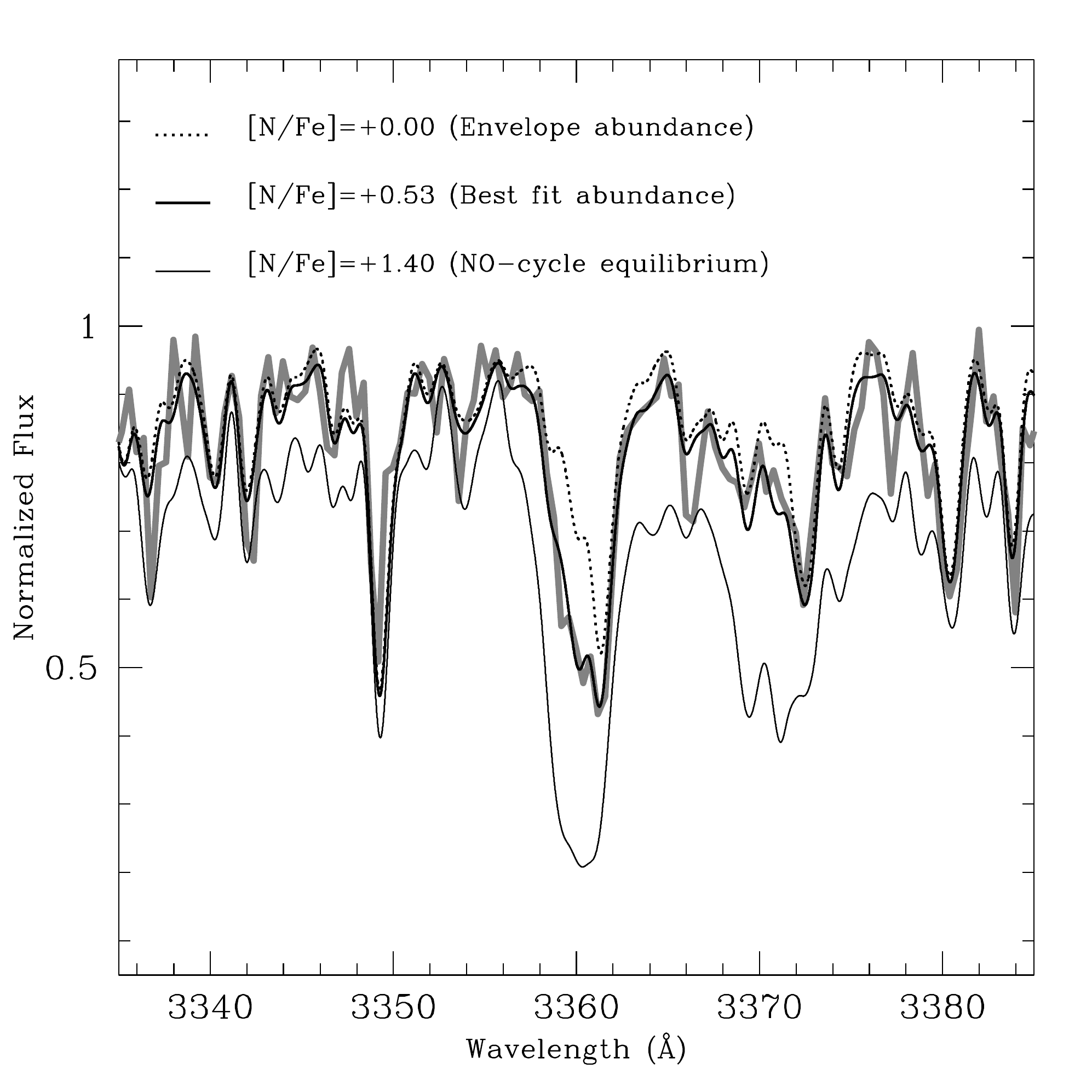}
\caption[Observed spectrum of \comA\ around the NH molecular band.]{Observed spectrum of \comA\ (thick gray line) in the  spectral region around the NH molecular band,   with over-imposed  synthetic spectra calculated with [N/Fe]=+0.0   (corresponding to the stellar envelope abundance, dotted line), +0.53 (best fit abundance, thick solid line), +1.40 (NO-cycle equilibrium abundance, thin solid line).}
\label{6397nh}
\end{center}
\end{figure}

The [N/Fe] upper limit obtained for MS1 is also consistent with what expected for the first stellar generation \citep[see][]{carretta05}, while the value measured for \comA\ ([N/Fe]$= +0.53\pm 0.15$ dex) is significantly larger. Figure \ref{6397nh} shows the observed spectrum of \comA\ in the region around the NH band, compared with synthetic spectra calculated with different values of [N/Fe].  Besides the best-fit synthetic spectrum (thick solid line), two other spectra are shown: one has been computed assuming [N/Fe]$=0$ dex, consistent with what expected on the surface of an unperturbed star and the results obtained for MS1; the other one has [N/Fe]$=+1.4$ dex, which is the value predicted for the NO-cycle equilibrium (see Section \ref{6397sec:disc}).  Clearly both these additional values are incompatible with the measured abundance. As discussed in
Section \ref{6397sec:disc}, this provides interesting constraints on the structure and the nature of this star.

Because of the weakness of the G-band only an upper limit for the C abundance of \comA\ is derived: at the XSHOOTER resolution, we measure [C/Fe]$<-1.0$ dex. This is compatible with the (more stringent) limit derived by \citet{sabbi03C} from high-resolution UVES spectra ([C/Fe]$<-2.0$ dex), which is therefore adopted in the
following discussion. Note that these upper limits are significantly smaller than any C abundance measured in NGC 6397 stars \citep[e.g.][]{carretta05}. 

\section{Discussion}
\label{6397sec:disc}
The abundances of C and N measured for \comA\ are incompatible with the values expected on the surface of an unperturbed MS star.  The very low [C/Fe] is also incompatible with the  C abundances range observed in the cluster \citep{carretta05}, thus excluding that \comA\ is an unperturbed second generation star. Hence, option $(1)$ discussed in the Introduction can be ruled out. 

In order to verify the possibility that  \comA\ is, instead, a deeply peeled TO/sub-giant star and to put new constraints on its nature, we compare the derived C and N abundances with the chemical gradients predicted in theoretical stellar models.
We have calculated the evolution of a 0.8 $\Msun$ stellar model, from the pre-MS to the red giant branch phase, using the same code and input physics of the BaSTI models \citep[see e.g.][]{pietr04,pietr06}. We adopted Z=0.0003, 
Y=0.245, and an $\alpha$-enhanced metal mixture ([$\alpha$/Fe]=+0.4),  which is appropriate for the first generation of stars in Galactic GCs (in particular, we assume [C/Fe]=0 and [N/Fe]=0, also consistently with the abundances measured for MS1).  The TO age of the model is 12 Gyr. This value, however, depends on the efficiency of atomic diffusion, which can be (partially or totally) inhibited by additional turbulent mixing (for which an adequate physical description is still lacking). We therefore calculated models both with and without atomic diffusion, finding that they are basically indistinguishable (in the following we therefore present only the results obtained from models without atomic diffusion).  Also the effects of radiative levitation are totally negligible at the metallicity of NGC 6397, because the radiative acceleration on the C and N atoms is always smaller than the gravitational acceleration \citep[see Figure 3 in][]{richard02} and we therefore do not include them in our models.

\begin{figure}[t]
\begin{center}
\includegraphics[width=120mm]{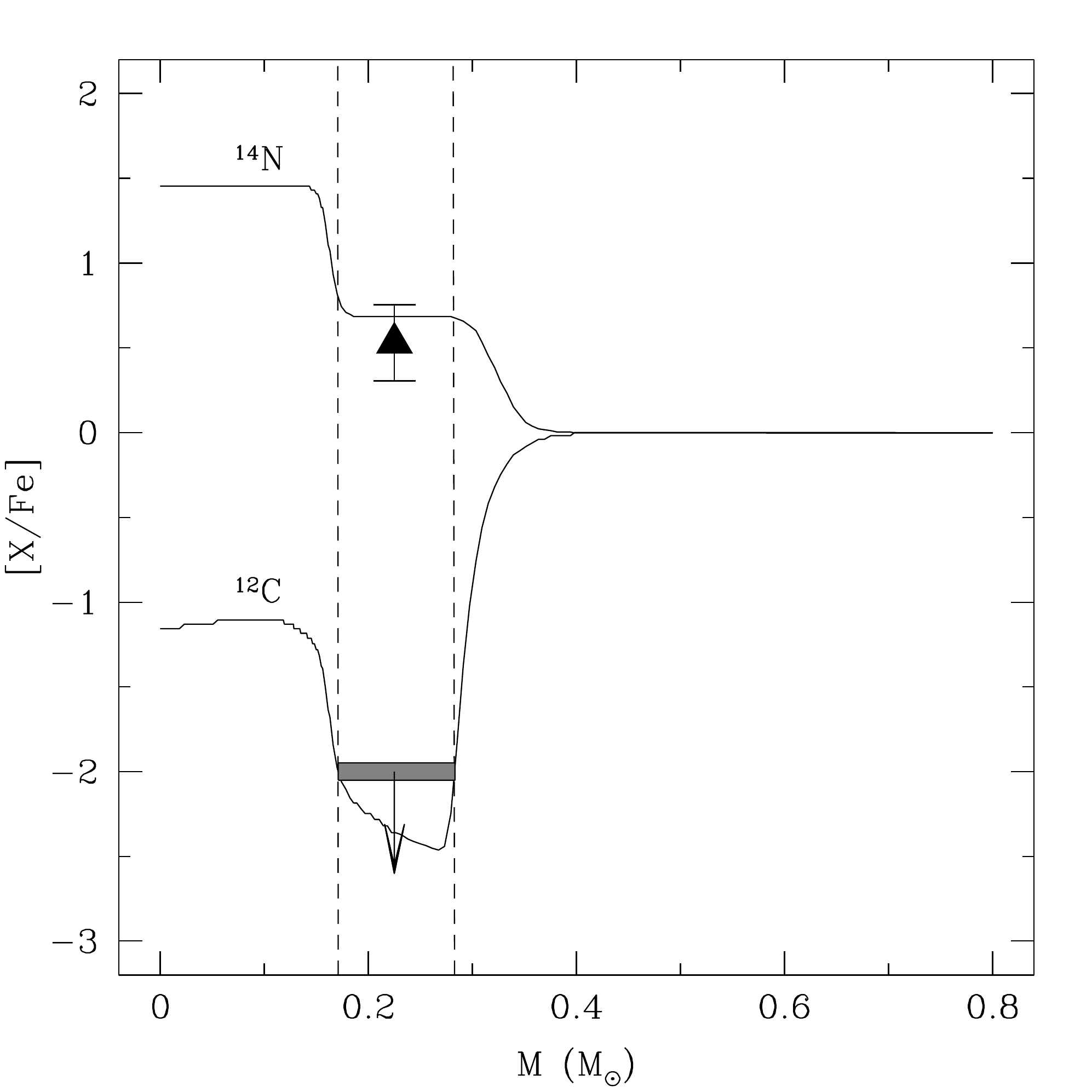}
\caption[ C and N  abundance ratios  as a function of the mass for a $0.8 \Msun$ sub-giant star.]{
Behavior of [X/Fe] abundance ratio for $^{12}$C and $^{14}$N as a function of the mass, for a stellar model of a sub-giant star with $M=~0.8 \Msun$, 
Z=~0.0003 and no atomic diffusion. The black triangle indicates the [N/Fe] measured in \comA, while the gray bar is the upper limit for [C/Fe]. The dashed vertical lines mark the mass range defined by the upper limit for [C/Fe].}
\label{6397cnm}
\end{center}
\end{figure}

Figure \ref{6397cnm} shows the gradients of the [$^{12}$C/Fe] and [$^{14}$N/Fe] abundance ratios in the interior of a sub-giant star\footnote{It is not easy to identify the evolutionary stage of the   star in the scenario of a peeled star.  Its position in the CMD   (Figure \ref{6397loca}) suggests that the object is a slightly evolved   star (see also \citealp{burderi02}), but we cannot exclude that it still belongs to   the MS. In any case, its luminosity seems to exclude that it is a   giant star.}, as a function of the stellar mass, from the center to the surface ($M=0.8 \Msun$).  As apparent, these abundances remain constant along the entire stellar envelope,  from the surface, down to the radius including half of the total mass.  The flat chemical profiles at [N/Fe]$=+1.4$ dex and [C/Fe]$\sim-1$ dex observed in the very central region ($M\lsim 0.15 \Msun$) are the consequence of the CNO-burning that occurred in the stellar core during the MS evolution.  The gradients observed in the intermediate region ($0.15 \Msun\lsim M\lsim 0.4 \Msun$), instead, are due to the ongoing hydrogen burning in a thick shell above the inactive core (with a mass of $\sim0.11 \Msun$).  In particular, the most external portion of the shell is mainly interested by the CN-cycle and therefore shows an increase of $^{14}$N and a drop of $^{12}$C.  At $0.18 \Msun\lsim M\lsim 0.38 \Msun$ the abundances of $^{12}$C and $^{14}$N reach the equilibrium: C displays its minimum value ([C/Fe]$=-2.45$ dex) and the N abundance profile shows a {\sl plateau} at [N/Fe]$=+0.68$ dex. In the innermost portion of the shell the NO-cycle is active, thus producing a further increase of both $^{14}$N and $^{12}$C, up to the most central values.

The dashed vertical lines in Figure \ref{6397cnm} mark the range of mass profile where the carbon abundance of the stellar model is in agreement with the upper limit ([C/Fe]$<-2$ dex) derived for \comA\ \citep{sabbi03C}.  In this same mass range also the N abundance shows a good agreement between the model prediction and the measured value (black triangle in Figure \ref{6397cnm}). Instead the [N/Fe] ratio observed in \comA\ is  incompatible with the abundance ratio predicted in any other region of the stellar model (consistently with what discussed above; see Figure \ref{6397nh}).  This evidence strongly suggests that \comA\ is a star peeled down to the region where the CN-cycle occurs, as a result of heavy mass transfer onto the NS.

\begin{figure}[b]
\begin{center}
\includegraphics[width=120mm]{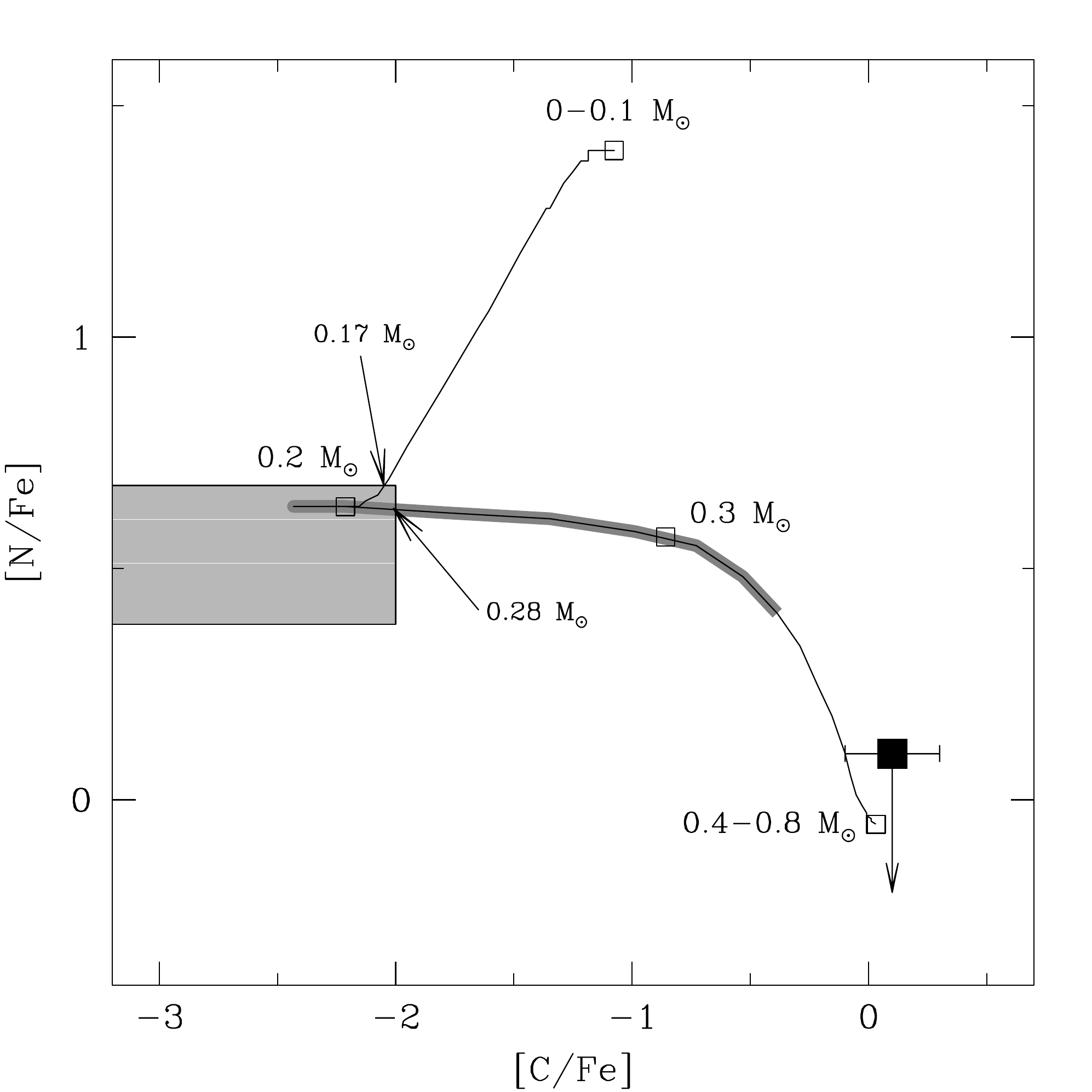}
\caption[N as a function of C for  a  $0.8 \Msun$ sub-giant star.]{
Behavior  of [N/Fe] as a function of [C/Fe] (thin solid lines) for the same stellar model shown in Figure~\ref{6397cnm}. Empty squares mark the position of the stellar masses at step of $0.1 \Msun$. The light gray box indicates the mean locus of [C/Fe] and [N/Fe] of COM J1740$-$5340A. The dark gray region indicates the mass range proposed by \citet{ferraro03rv6397}. The black square marks the position of MS1.}
\label{6397cn}
\end{center}
\end{figure}

The C and N abundances allow us to also identify a reasonable mass range for the MSP companion. Figure \ref{6397cn} shows the behavior of the discussed stellar model in the [C/Fe]--[N/Fe] plane. The light gray box indicates the locus corresponding to the abundances of \comA\ (taking into account the quoted uncertainties). Clearly, only the portion of the stellar model between 0.17 and $0.28 \Msun$ overlaps this region.  This mass range is in very good agreement with the value ($0.22-0.32 \Msun$; \citealp{ferraro03rv6397}) estimated for \comA\ from the binary system mass ratio (inferred from the radial velocity curve of the companion) and the orbital inclination angle (inferred from the optical light curve).  In order to quantify how much the results depend on the evolutionary phase of the star before the onset of heavy mass transfer, we repeated the analysis by using models on the MS, at the base of the red giant branch, before and after the occurrence of the first dredge-up.  Following the evolution from the TO to the base of the red giant branch, the region where the NO-cycle occurs increases in mass, thus reducing the mass range where the model C and N abundances match the observed ones. If the MSP companion was a TO star, \comA\ should now have a mass between 0.13 and $0.27 \Msun$, while the range decreases significantly in case of a red giant branch star: $0.22-0.28 \Msun$ for a star before the first dredge-up, and $0.25-0.28 \Msun$ for a star after the first dredge-up.  Note that in all the cases, the
upper mass limit remains basically unchanged, confirming a value smaller than $0.3 \Msun$, even if the upper limit for the C abundance inferred from the XSHOOTER spectra ([C/Fe]$\sim -1$) is assumed.    We finally stress that different assumptions about the initial   C and N abundances of the star have the effect of rigidly shifting   the model curves in both the planes shown in Figs. \ref{6397cnm} and   \ref{6397cn}.

The evidence presented here adds an important piece of information to properly characterize \comA.   The analysis of the C and N surface abundances
provides a diagnostic of the companion mass which is totally independent of other, commonly used methods, well confirming the previous estimates. In addition, the chemical patterns observed at the surface of \comA\ solidly confirm that this object is a deeply peeled star, and can be even used to obtain a quantitative evaluation of  the amount of mass lost by the star.  In fact, our analysis indicates that the entire envelope of the star has been completely removed and the peeling action has extended down to an interior layer where the CN-cycle approximately reached the equilibrium. By assuming an initial mass of $\sim 0.8 \Msun$, we estimate that \comA\ has lost $\sim 75\%$ of its initial mass during the interaction with the PSR.

\clearpage{\pagestyle{empty}\cleardoublepage}

\chapter[The companion to \psrBW\ in the Galactic Field]{The identification of the optical companion to the binary Millisecond Pulsar J0610$-$2100 in the Galactic Field}\label{Chap:0610}

In this study we have used deep $V$ and $R$ images acquired at the ESO VLT to identify the optical companion to the binary  \psrBW,  one of the BW MSP recently detected by the  Fermi Gamma-ray Telescope in the Galactic  plane. We found a  faint star ($V\sim26.7$) nearly coincident ($\delta r \sim0\arcsec.28$)  with the  PSR nominal position. This star is visible only in half of the available images, while it disappears in  the deepest ones (those acquired under the best seeing conditions), thus indicating that it is  variable. Although our observations do not sample the entire orbital period  ($P_b=0.28$ d) of the PSR, we found that the optical modulation  of the variable star nicely correlates with  the PSR orbital period and describes a well  defined peak  ($R\sim25.6$) at $\Phi=0.75$, suggesting a modulation due to the PSR heating. We tentatively conclude that the companion to \psrBW\ is a  heavily ablated, very low mass star  ($\approx 0.02\Msun$) that completely filled its RL. - {\it This chapter is mainly based on \citealt{pallanca12}, ApJ, 755, 180.}

\section{Introduction}

\psrBW\ is a MSP located in the GF with period $\Pspin=3.8$ ms and a radio flux $S_{\rm 1.4}= 0.4 \pm 0.2$ mJy at 1.4 GHz,  discovered during the Parkes High-Latitude pulsar survey \citep[][ hereafter B06]{burgay06}. The period derivative $\dot{P}=1.235\times10^{-20}$s s$^{-1}$ implies a characteristic age $\tau=5$ Gyr, a magnetic field $B=2.18 \times 10^8$ G, and a rotational energy loss rate  $\dot{E}= 2.3 \times 10^{33}$ erg s$^{-1}$, similar to the values measured for other  MSPs.  \psrBW\ is in a binary system, with an orbital period of $\sim 0.28$ d.  
It is located at a  distance of  $3.5\pm 1.5$ kpc, estimated from its dispersion measure (DM=60.666 pc cm$^{-3}$) and the Galactic electron density model of  \citet{cordeslazio02}. The PSR has a proper motion  $\mu_{\alpha} cos\delta = 7 \pm 3$ mas yr$^{-1}$ and $\mu_{\delta}=11\pm 3$ mas yr$^{-1}$ (B06), which implies a transverse velocity of  $228\pm53$ km s$^{-1}$, one of the highest measured for Galactic MSPs. The system mass function ($f_{\rm PSR}=5\times10^{-6}$) implies a lower limit of $0.02 \Msun$ for the mass of 
the companion, assuming $1.35 \Msun$ for the PSR (B06). Thus, in agreement with the definition presented in Section \ref{masstype}, \psrBW\ is probably a BW seen at a low inclination angle (in fact no eclipse is detected).

Until a few years ago just two other BWs were known in the GF, namely PSR B1957+20 \citep{fruchter88a} and PSR J2051$-$0827 \citep{stappers96a}. But very recently, thanks both to  dedicated surveys of $\gamma$-ray sources and to new blind searches,  new BWs have been discovered, most of  them having been detected in $\gamma$-rays \citep[see][ and references therein]{roberts11}. Also \psrBW\ has been detected in $\gamma$-rays by the Large Area Telescope (LAT) on board the {\em Fermi} Gamma-ray Space Telescope. Based upon positional coincidence with the LAT error box, it was  initially associated with the $\gamma$-ray source 1FGL\, J0610.7$-$2059 \citep{abdo09, abdo10}. The detection of $\gamma$-ray pulsations at the radio period of \psrBW\ has been recently reported and  used to confirm it as the $\gamma$-ray counterpart to the MSP \citep{espinoza13}. In the X-rays \psrBW\ has not been detected, neither by the {\em ROSAT} All Sky Survey \citep{voges99}, nor by {\em Swift} (Marelli, private communication).

Studying the optical emission properties of binary MSP companions is important to better constrain the orbital parameters and to clarify the evolutionary status of these systems and then to track back their history and characteristic timescales. In spite of their importance, only a few optical companions to BWs in the GF have been detected to date. 
The binary MSP B1957+20 is the first discovered BW and one of the best studied members of this class. The optical companion to PSR J1957+20 was identified by \citet{kulkarni88}, while subsequent observations found the companion to vary by  30\%-40\% in flux over  the orbital period \citep{callanan95}. \citet{reynolds07} modeled the light curve and thus constrained the system inclination ($63 ^{\circ}<i<67^{\circ}$) and the filling factor of the RL ($0.81<f<0.87$). Moreover, they ruled out the possibility that the companion is a WD, suggesting that most probably is a BD.  A recent spectroscopic analysis, combined with the knowledge of the inclination angle inferred from models of the light curve, suggested that the PSR B1957+20 is massive, with $M_{\rm PSR}=2.4\Msun$ \citep[$M_{\rm PSR}>1.66\Msun$ being a conservative limit;][]{vankerk11}. The optical companion to binary PSR J2051$-$0827 was identified by \citet{stappers96b}. They found that the amplitude of the companion's light curve was at least 1.2 mag, and that the variation was consistent with the companion rotating synchronously with the PSR and one side being heated by the impinging PSR flux. In  subsequent works it has been possible to study the entire lightcurve, measuring amplitudes of 3.3 and 1.9 magnitudes in the $R$-band and $I$-band respectively. The companion  has been modeled as a gravitationally distorted low mass star which is irradiated by the impinging PSR wind. The resulting best-fit model corresponds to a RL filling companion star, which converts approximately 30\% of the incident PSR spin down energy into optical flux \citep{stappers00}.
 
In this chapter we present  the  identification of  the optical companion to \psrBW, the third discovered counterpart to a BW system in the GF.

\section{Observations and data analysis}\label{Sec:analysis}
The used photometric data set consists of a series of ground-based optical images acquired with 
FORS2 mounted at the ESO-VLT. We used the Standard Resolution Collimator, with a pixel scale of 0.25\arcsec /pixel and a FOV of $6.8'\times6.8'$.  

Six short acquisition images (of 5 s each) and a total of 29 deep images in the $V_{BESSEL}$ and $R_{SPECIAL}$ bands ($V$ and $R$ hereafter)  were collected during six nights, from mid December 2004 to the beginning of January 2005 (see Table \ref{0610dataset}), under program $074.D-0371(A)$ (PI: Sabbi). These data allow us to sample  $\sim 25\%$ of the orbital period in $V$ and  $\sim 40\%$ of it in $R$ (see column 4 in Table \ref{0610dataset}).

\begin{table}
\begin{center}
\begin{tabular}{|l|l|l|l|c|}
\hline
FILTER &  N $\times$ $t_{exp}$ & Night & PSR-Phase & Detection \\
\hline
\hline
$V_{BESSEL}$ & 6 $\times$ 1010 s & 17/12/04 & 0.04-0.26 & NO\\ 
$V_{BESSEL}$ & 3 $\times$ 1010 s & 20/12/04 & 0.58-0.67 & YES$^1$\\ 
\hline
$R_{SPECIAL}$ & 5 $\times$ 590 s & 14/12/04 & 0.73-0.83 & YES\\ 
$R_{SPECIAL}$ & 5 $\times$ 590 s & 21/12/04 & 0.06-0.16 & NO\\ 
$R_{SPECIAL}$ & 5 $\times$ 590 s & 05/01/05 & 0.26-0.36 & NO\\ 
$R_{SPECIAL}$ & 5 $\times$ 590 s & 06/01/05 & 0.76-0.86 & YES\\ 
\hline
\end{tabular}
\end{center}
\caption{$^1$ Detected in only 2 images out of 3}
\label{0610dataset}
\end{table}

By following standard reduction procedures, we corrected the raw images for bias and flat-field. In particular, in order to obtain high-quality  master-bias and  master-flat images, we selected  a large number of  images obtained during each observing night  and, for each filter, we properly combined them by using the tasks  {\tt zerocombine} and {\tt flatcombine} in the IRAF package {\sc ccdred}. The calibration files thus obtained have been applied to the raw images by using the dedicated task  {\tt ccdproc}.

Based on the  Word Coordinate System (WCS) of the images, we approximately located the  PSR position and decided to limit the photometric analysis to a region of 500 pixel $\times$ 500 pixel ($\sim$ 125\arcsec $\times$ 125\arcsec) centered on it.

We carried out the photometric analysis by applying the PSF-fitting method (see Chapter \ref{Chap:introCOM}).
In particular, the master-frame for the object detection has been built combining  the three images obtained under the best seeing conditions ($\sim0.6\arcsec$) in both filters.

The photometric calibration has been performed by following the procedure described in Chapter \ref{Chap:introCOM}.
In particular, in this case we first derived the calibration equation for ten standard stars in the field PG0231 \citep{stetson00}, which has been observed with FORS2 during a photometric night (December 17, 2004) in both $V$ and $R$ under  a calibration program. To analyze the standard star field we performed an aperture photometry with a radius $r=14$ pixels and  we  compared the obtained magnitudes with the standard  Stetson catalog available on the  CADC web  site\footnote{http://cadcwww.dao.nrc.ca/community/STETSON/standards/}. 
The resulting calibration equations are $V=v-0.092(v-r)+27.79$ and $R=r-0.019(v-r)+27.97$, where $v$ and $r$ are the instrumental magnitudes. The color coefficient is very small for the $R$ band, while it could be not negligible in the $V$ band.

The astrometry procedure has been performed as described in Chapter \ref{Chap:introCOM}.
In particular, since most of the astrometric standard stars are saturated in our catalog, we used the stars detected in short $R$-band images as secondary astrometric standards. 
At the end of the procedure the typical accuracy of the astrometric solution was $\sim0\arcsec.2$ in both 
$\alpha$
and 
$\delta$.

\section{The companion to PSR J0610$-$2100}
In order to  identify the companion to \psrBW\ we  first searched for objects with coordinates compatible with the nominal PSR position: $\alpha= 06^h 10^m 13^s.59214(10)$, $\delta=-21^{\circ} 00' 28''.0158(17)$ at Epoch MJD=53100 (B06). Since the epoch of observations  is within less than one year from the epoch  of the reference radio position, we neglected the effect of proper motion, which is much smaller than the  accuracy of the astrometric solution of the FORS2 images.

\begin{figure*}[t]
\includegraphics[width=140mm]{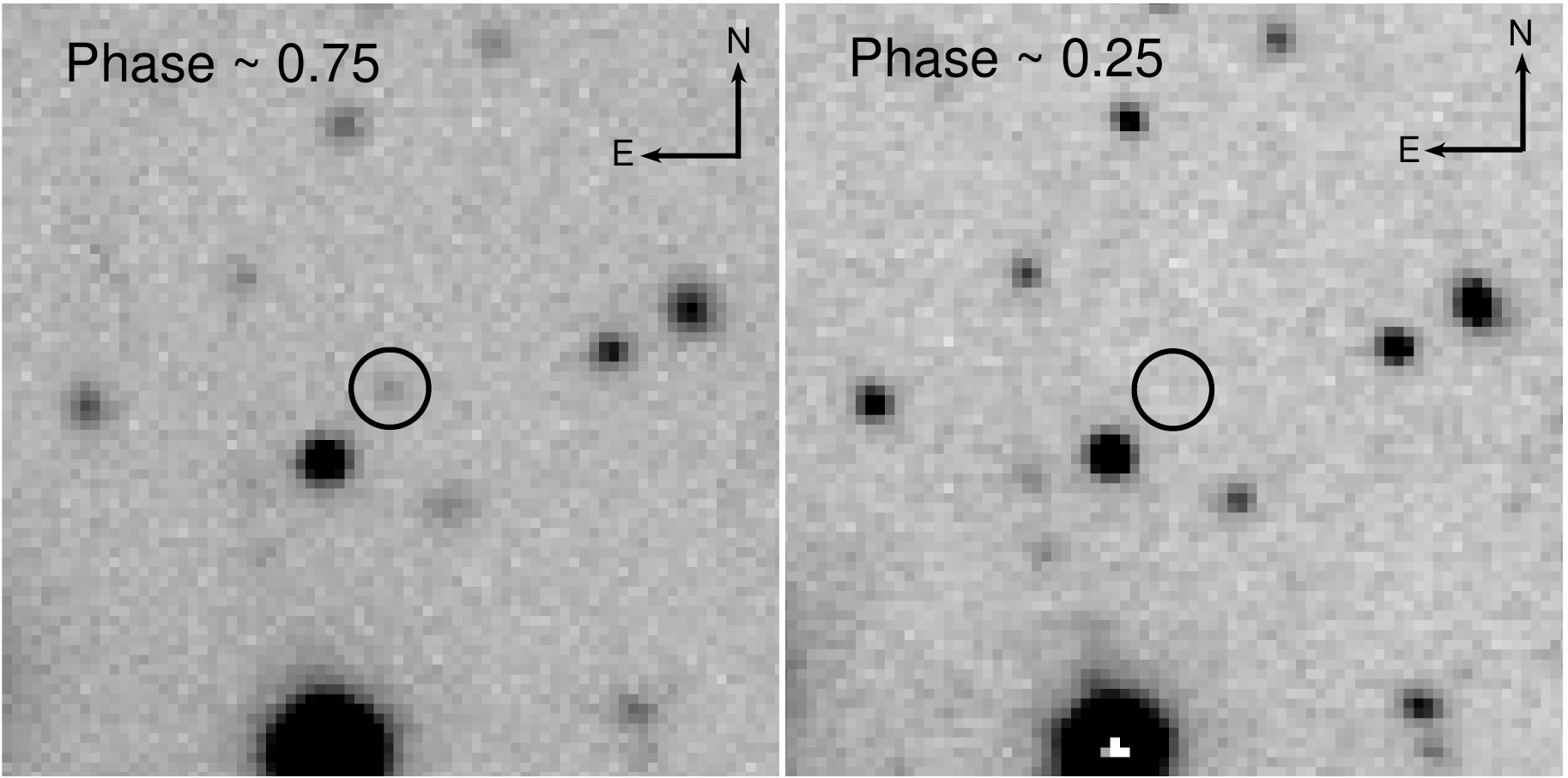}
  \caption[Map of the region around \psrBW.]{$R$-band images of  the $20\arcsec\times20\arcsec$ region around the nominal position of \psrBW, at two different epochs   corresponding to the orbital phases $\Phi=0.75$ (left panel) and $\Phi=0.25$ (right panel). Each image is the average of 3   images. The circle of radius $1\arcsec$ marks the system position. A star is clearly visible in the left panel,   while it vanishes in the right panel. Note that the images at $\Phi=0.25$ (where the star is undetected) have been obtained   under the best seeing conditions   (FWHM$=0.6\arcsec$).}
\label{0610map}
\end{figure*}

\begin{figure*}[t]
\begin{center}
\includegraphics[width=145mm]{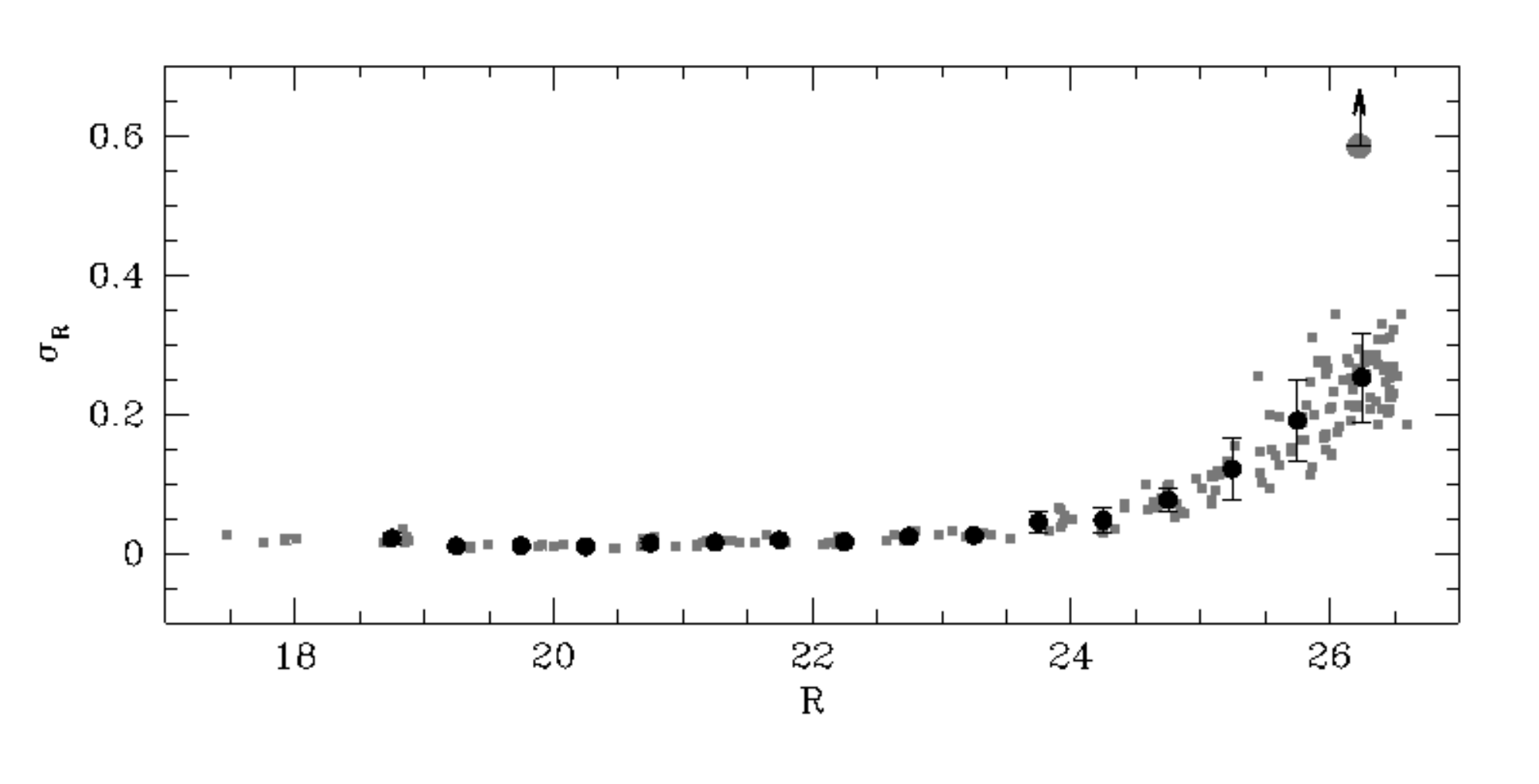}
  \caption[Frame to frame magnitude scatter  in the region of \psrBW.]{Frame to frame magnitude scatter (gray points) for the 173 stars identified in a $125 \arcsec \times125 \arcsec$ region   around the nominal PSR position,  as a function of the $R$ magnitude.    The standard deviation $\sigma _R$ has been computed by using all the available images.    Black circles and the corresponding error bars are the mean and the standard deviation values     in 0.5 magnitude bins, respectively.       The mean magnitude ($R\sim26.2$) and standard deviation ($\sigma_R\sim0.6$) of the companion star to PSR J0610$-$2100      (large grey circle) likely represent lower limits to the true values, because they have been computed adopting $R=27$      (the $R$-band detection limit) in all images where the star was not visible.  }
\label{0610sigma}
\end{center}
\end{figure*}

A first visual inspection of the PSR region clearly shows that only one star lies within a couple of arcseconds from the MSP radio position: it is located at $\alpha= 06^h 10^m 13^s.58$ $\delta=-21^{\circ} 00' 27''.83$, just $0.\arcsec28$ from \psrBW. Thus, from positional coincidence alone, we found a very good candidate  companion to \psrBW.  Note that the chance coincidence probability\footnote{The chance coincidence probability is calculated as $P=1-\exp(-\pi \sigma R^2)$ where $\sigma$ is the stellar density of stars with similar magnitude to the candidate companion and $R$ is the accuracy of the astrometric solution.} that a star is located at the PSR position is only $P=0.0007$. Hence, this star is the companion to \psrBW\ with a probability of $\sim 99\%$. Interestingly enough, this star was not present in the  master-list obtained from the stacking of the $R$-band images because it is visible in only half of the images, while it completely disappears in the others (see Figure \ref{0610map}). We performed a detailed photometric analysis of this star for measuring its magnitude in as many images as possible, and we found that it is not  detected in the  deepest $R$ images obtained under the best seeing conditions (FWHM$\sim0.6\arcsec$). In summary, we were able to measure the magnitude of the star in only 12 images (10 in the $R$-band and 2 in the $V$-band), finding significant variations: $\delta R \sim 1$ mag (from $R=25.3 \pm 0.1$ to $R=26.3 \pm 0.2$), and $\delta V \sim 0.5$ mag (from $V=26.7 \pm 0.2$ to $V=27.2 \pm 0.2$). In the remaining images the star magnitude is below the  detection thresholds ($R=27 \pm 0.3$ and $V=27.3 \pm 0.3$),  thus suggesting a very pronounced optical variation. Considering the entire data set,  the object's photometry  shows a quite large scatter, significantly ($>5\sigma$) larger than that computed for stars of similar magnitude in the same FOV (see Figure \ref{0610sigma}). These findings confirm that this is a variable object near the detection limit of our sample.

\begin{figure*}[b]
\begin{center}
\includegraphics[height=110mm]{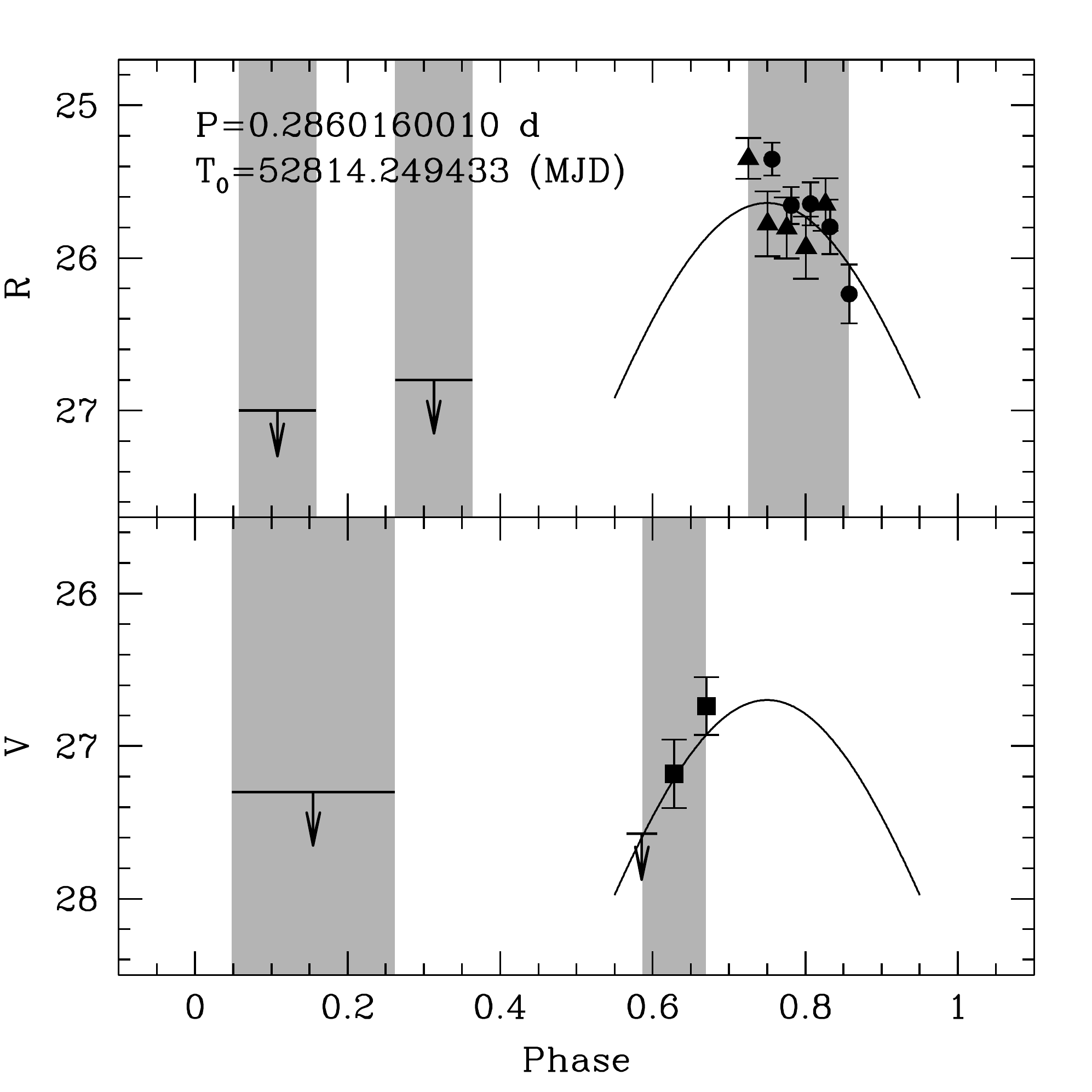}
  \caption[Light curve of the companion to \psrBW.]{The observed light curve of the companion to \psrBW\ folded with the orbital period ($P$) of the PSR using the    reference epoch ($T_0$) known from radio observations (B06).   The different symbols represent images obtained in different nights. The horizontal lines with arrows are the estimated   magnitude upper-limits for the images where   the star is  below the detection threshold. The gray regions correspond to the fraction of the orbital phase sampled by   each observations.   The solid black line is a first-guess modeling of the light curve by means of a simple sinusoidal mode.}
\label{0610lc}
\end{center}
\end{figure*}

In order to establish a firm connection between this star and the PSR, we computed the $V$ and $R$ light curves folding each  measurement with the orbital period ($P_b=0.2860160010$ d) and the  ascending node ($T_0=52814.249433$)  from the radio ephemeris (B06). As shown in Figure \ref{0610lc}, although the available data do not allow a complete coverage of the orbital period, the  flux modulation of the star  nicely correlates with the PSR orbital phase. The available data are consistent with  the rising (in the $V$ band) and the decreasing (in $R$-band) branches of a light curve with a peak at $\Phi=0.75$. This is the typical behavior expected when the surface of the companion is heated by the PSR flux  and the orbital plane  has a sufficiently high inclination angle. In fact, in this configuration a light curve with a maximum at $\Phi=0.75$ (corresponding to the  PSR inferior conjunction, when the companion surface faces the observer) and a minimum at $\Phi=0.25$ (corresponding to the  PSR superior conjunction) is expected.
Indeed  the star is not detectable at the  epochs corresponding to the orbital phases where the luminosity minimum is predicted. 

Based on all these pieces of evidence we propose the identified variable star as the companion to the PSR; following the nomenclature adopted in some papers of our group \citep[see][]{ferraro01com6397, ferraro03com6752, cocozza08, pallanca10}, we name it \comBW.

\begin{figure*}[t]
\begin{center}
\includegraphics[height=110mm]{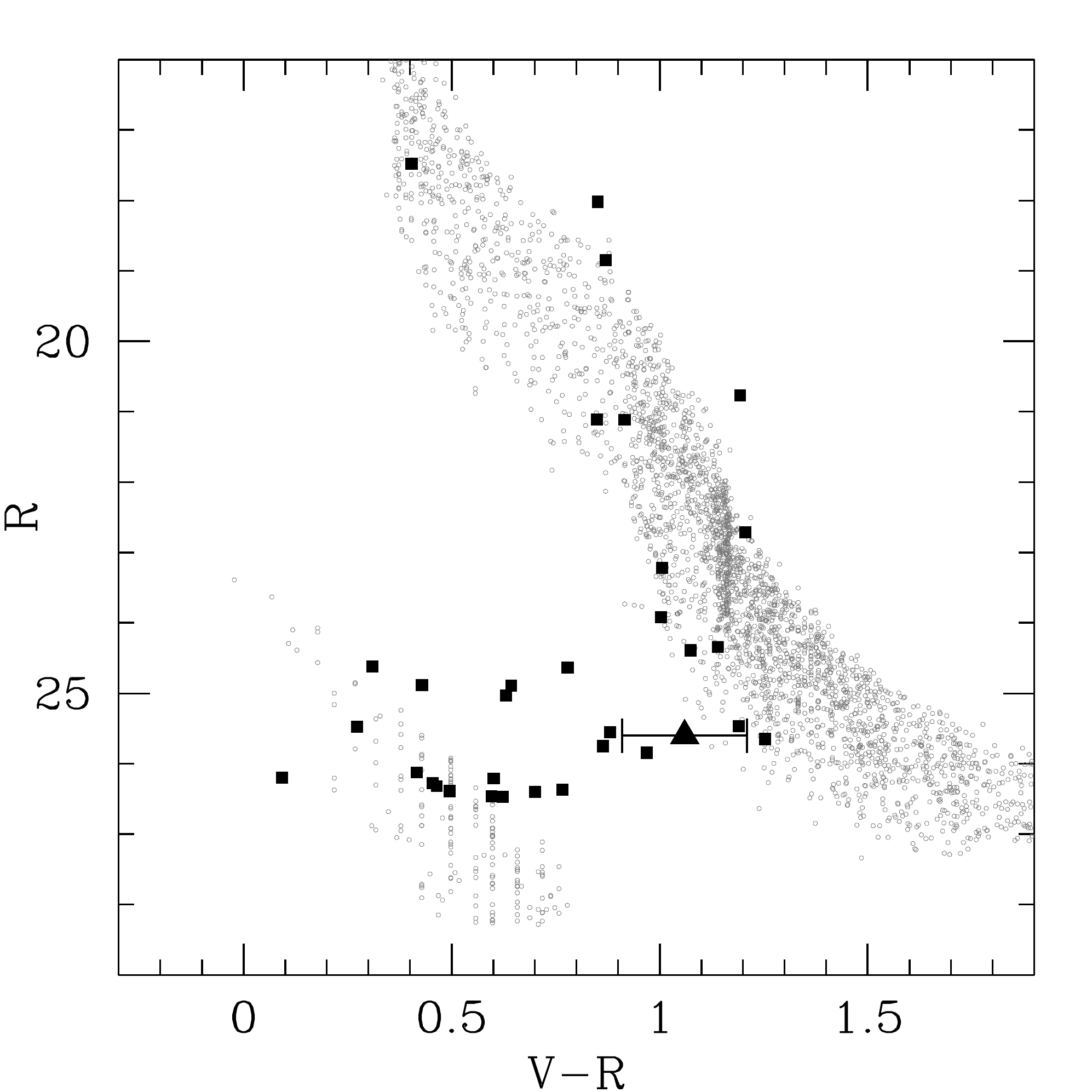}
  \caption[CMD position of the companion to \psrBW.]{($R$, $V-R$) CMD for \comBW\ (large black triangle) and for the objects detected in a   circle of about $30\arcsec$ around its nominal position (black squares).    \comBW\ is located at $V=26.7$ and $R=25.6$, corresponding to the values  estimated from the first-guess light-curve at its   maximum ($\Phi=0.75$). The error bar corresponds to the photometric  error at those magnitude values.   The open gray circles represent the Galaxy disk population obtained with the   Besancon model \citep{robin03} in the direction of the PSR and for a distance between 2 and 6 kpc.}
\label{0610cmd}
\end{center}
\end{figure*}

Since the available $V$ and $R$ measurements are mainly clustered toward the maximum of the emission, but do not allow to precisely determine it, we used  a simple  sinusoidal function\footnote{ Although this assumption is not supported by a physical reason, it provides a first estimate of the magnitude and color of COM J0610$-$2100 at maximum.} to obtain a first-guess modeling of the light curve. In the following analysis we will use these values, instead of the mean magnitudes averaged over the entire orbital period, because these latter are not available in this case. Also note that during the calibration procedure the color term entering the equations has been computed as the difference between the average value of the available $V$ and $R$ instrumental magnitudes. While this is strictly correct for non variable stars, in the case of \comBW\ it could have introduced an error in the estimated magnitudes.  However, since in the calibrating equations the coefficients of the color terms are very small, 
especially in the $R$-band, this uncertainty should be negligible. The resulting magnitudes of \comBW\ at maximum are $R=25.6$ and $V=26.7$. Figure \ref{0610cmd} shows the position in the ($R$, $V-R$)  CMD  of \comBW\     and the stars detected within  $30\arcsec$ from it. For the sake of clarity a simulation of the Galactic disk population in the direction of \psrBW\ computed with the Besancon Galaxy model \citep{robin03} is also shown. As can be seen, \comBW\ is located at a slightly bluer  color with respect to the reference MS, thus suggesting that it probably is a non-degenerate, low mass, swollen star. Indeed similar objects have been previously identified in Galactic GCs \citep[see][]{ferraro01com6397, edmonds02, cocozza08, pallanca10}.

\section{Discussion}
We have determined the physical parameters of \comBW\ from the comparison of its position in the CMD (Figure \ref{0610cmd}) with a reference zero age main sequence, assuming an interstellar extinction of $E(B-V)\sim0.074$\footnote{ from NED, Nasa/ipac Extragalactic Database - Galactic Extinction Calculator available at the web site {\it http://ned.ipac.caltech.edu/forms/calculator.html}.} and the typical metallicity of the Galactic disk ($Z=0.02$). The resulting effective temperature and bolometric luminosity of the star are $T_{eff}\sim3500$ K and $L_{bol}\sim 0.0017 L_{\odot}$ respectively, with a conservative uncertainty of $\pm 500$ K and $\pm0.0001L_{\odot}$. Under the assumption that  the optical emission of \comBW\ is well reproduced by a BB, it is possible to derive its radius: $R_{\rm BB}\sim 0.14R_{\odot}$. However, since the companion to a BW is expected to be affected by tidal distortions exerted by the PSR and to have filled its RL, the dimension of the RL might be a more appropriate value  \citep[i.e., see PSR J2051$-$0827;][]{stappers96b,stappers00}. According to \citet{eggleton83} we assume: 
\begin{displaymath}
\frac{\rrl}{a} \simeq \frac{0.49q^{\frac{2}{3}}}{0.6q^{\frac{2}{3}}+\ln \left(1+q^{\frac{1}{3}}\right)}  
\end{displaymath}
where $q$ is the ratio between the companion and the PSR masses ($\mcom$ and $\mpsr$, respectively). This relation can be combined with the PSR mass function $f_{\rm PSR}(i,\mpsr,\mcom)=(\mcom\sin i)^3/(\mcom+\mpsr)^2$ by assuming a  NS mass $\mpsr=1.5 \Msun$ (as recently estimated for recycled PSRs by \citealt{ozel12}, see also \citealt{zhang11} and \citealt{kiziltan11}), thus yielding  $\rrl(i)\sim0.24-0.47 R_{\odot}$, depending on the inclination angle ($i$) of the system. These values are about  1.7-3.4 times larger than $R_{\rm BB}$. In the following discussion we assume  the value of the RL as a measure of the size of \comBW\ and we discuss how the scenario would change by using $R_{\rm BB}$ instead of $\rrl$. While these assumptions trace two extreme possibilities, the situation is probably in the midway. In fact, in the case of a completely filled RL, the mass lost from the companion should produce some detectable signal in the radio band (unless for very small orbital inclinations) and ellipsoidal variations could be revealed in the light curve (unless the heating from the PSR is dominating).

Under the assumption that the optical variation shown in Figure \ref{0610lc} is mainly due to irradiation from the MSP, reprocessed by the surface of \comBW\, we can estimate how the re-processing efficiency depends on the inclination angle and, hence, on the companion mass. To this end, we compare the observed flux variation ($\Delta F_{obs}$)  between the maximum ($\Phi=0.75$) and the minimum ($\Phi=0.25$) of the light curve, with the expected flux variation ($\Delta F_{exp}$) computed from the rotational energy loss  rate ($\dot{E}$). Actually, since we do not observe the entire light curve, $\Delta F_{obs}$ can just put a lower limit to the reprocessing efficiency. Moreover, since these quantities depend on the inclination angle of the system (see below) we can just estimate the reprocessing efficiency as function of $i$.

At first we have to convert the observed magnitude variation into a flux. We limited our analysis to the $R$ band since we have more observations and a more reliable sampling of the light curve. At maximum ($\Phi=0.75$) we assume $R=25.6$, and between $\Phi=0.75$ and $\Phi=0.25$ we estimate an amplitude variation $\Delta R \gapp 1.5$. Hence we obtain $\Delta F_{obs} \sim 1.88 \times 10 ^{-30}$ erg s$^{-1}$ cm$^{-2}$ Hz$^{-1}$ and considering the filter width ($\Delta \lambda = 165$ nm) we have $\Delta F_{obs} \sim 3.4 \times 10 ^{-15}$ erg s$^{-1}$ cm$^{-2}$.

 \begin{figure*}[t]
\begin{center}
\includegraphics[height=110mm]{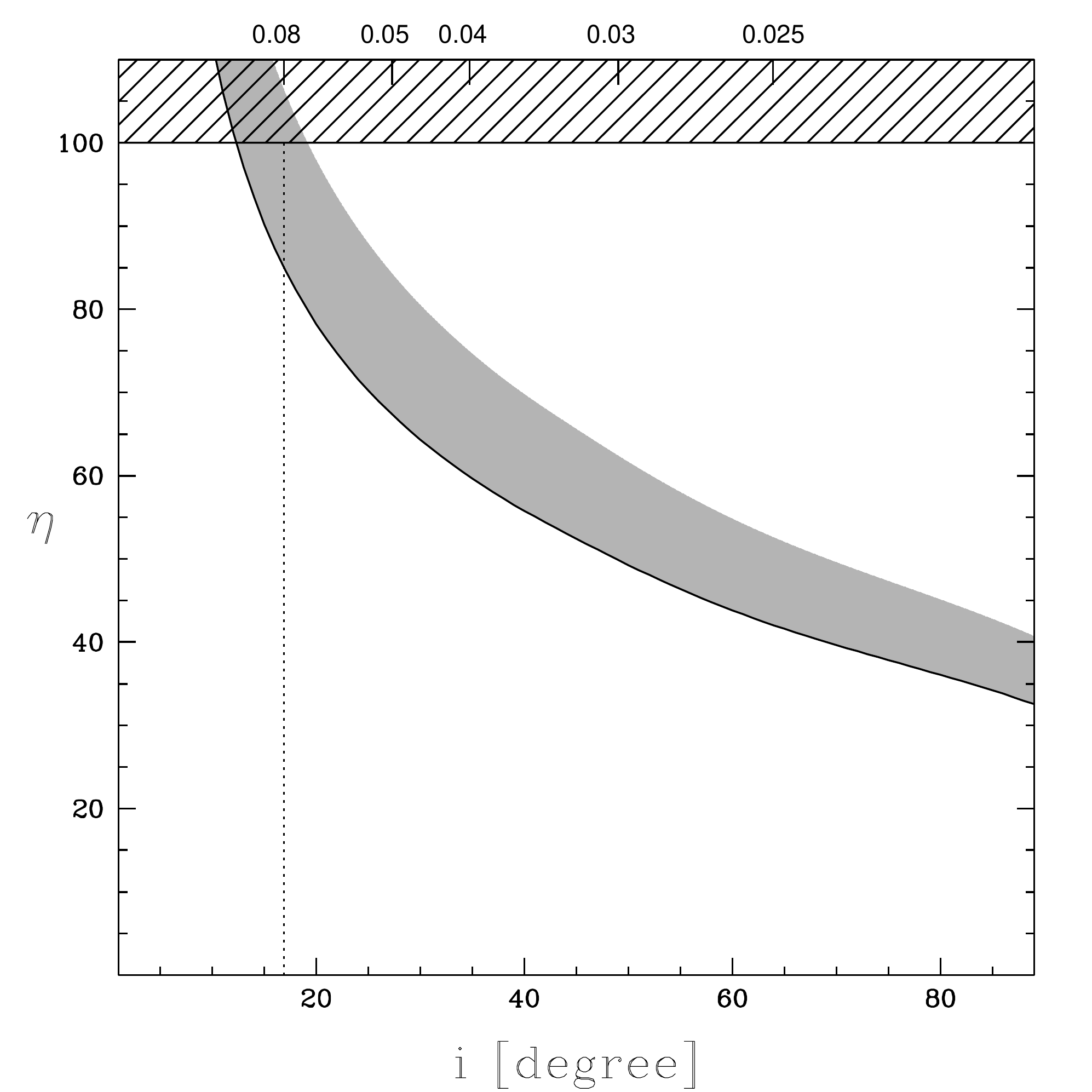}
  \caption[Reprocessing efficiency of the companion to \psrBW.]{Reprocessing efficiency for isotropic emission ($\eta$) as a function of the inclination angle ($i$) calculated assuming    $\mpsr=1.5 \Msun$ and $\rcom=\rrl(i)$. The corresponding values for the companion mass in units of $\Msun$ are reported on    the top axis.    The solid line marks  the lower limit of the amplitude of variation  ($\Delta R =1.5$), while the gray region corresponds  to    values of $\Delta R$ up to 3, that should be appropriate if considering the entire light curve \citep[i.e, see PSR J2051$-$0827;][]{stappers96b}.    The shaded area marks the region of the diagram where only anisotropic emission of the re-processed flux is admitted (in case of    isotropy, in fact, the efficiency would be unphysical: $\eta>100\%$).    The dotted line marks  the physical limit for core hydrogen burning  stars (i.e., objects with masses $\ge 0.08\Msun$).}
\label{0610repr}
\end{center}
\end{figure*}

On the other hand, the expected flux variation between $\Phi=0.75$ and $\Phi=0.25$ is given by 
\begin{displaymath}
\Delta F_{exp}(i)= \eta \frac{\dot{E} }{a^2}  \rcom^2  \frac{1}{4\pi d^2_{\rm PSR}} \varepsilon (i)
\end{displaymath}
where $\eta$ is the re-processing efficiency under the assumption of isotropic emission, $a$ is semi-major axis of the orbit which depends on the inclination angle, $\rcom$ is the radius of the companion star,  which we assumed  to be equal to $\rrl(i)$, $\dpsr$ is the distance of PSR  (3.5 kpc \footnote{In these calculations we adopted a distance of 3.5 kpc, while we discuss below how the scenario changes  by varying the distance between the range of values within the quoted uncertainty.}) and $\varepsilon (i)$ is the fraction of the re-emitting surface visible to the observer\footnote{ In the following we assume $\varepsilon (i)=i/180$. In fact, for a face-on configuration ($i=0^{\circ}$) no flux variations are expected, while for an edge-on system ($i=90^{\circ}$) the fraction of the heated surface that is visible to the observer varies between 0.5 (for $\Phi=0.75$) to zero (for $\Phi=0.25$).}. By assuming $\Delta F_{obs}=\Delta F_{exp}(i)$ between $\Phi=0.75$ and $\Phi=0.25$, we can derive a relation linking the re-processing efficiency and the inclination angle. The result is shown in Figure \ref{0610repr}. The absence of eclipses in the radio signal allows us to exclude very high inclination angles. As shown in Figure \ref{0610repr}, PSR companions  with stellar mass above the physical limit for core hydrogen burning star (i.e., with $M\ge0.08\Msun$) necessarily imply a non-isotropic emission mechanism of the PSR flux (otherwise a larger than $100\%$ physical efficiency would be required). On the contrary a re-processing efficiency between 40\% and 100\% is sufficient for less massive companions and intermediate inclination angles.

Taking into account the uncertainty on the PSR distance, only re-processing efficiencies larger than $\sim60\%$ for inclination angles in excess to $~50^{\circ}$ and companion masses lower than $\sim0.03\Msun$ are allowed in the case of the distance upper limit (5 kpc). Instead, in case of a closer distance, $\eta$ decreases for all inclination angles, thus making acceptable also companion stars with masses larger than $0.08 \Msun$. For instance, for companions masses between $0.08$ and $0.2 \Msun$ and a 2 kpc distant PSR, the re-processing efficiency ranges between $30\%$ and $60\%$ for any value of $i$.

 The observed optical modulation can be reproduced considering a system seen at an inclination angle of about $60^{\circ}$,  with a very low mass companion ($\mcom\sim0.02\Msun$) that has filled its RL, and a re-processing efficiency of about $50\%$. On the other hand, if we use $R_{\rm BB}$ instead of the $\rrl$, the efficiency becomes larger than 100\% for every inclination angle, and the only possible scenario would be that of an anisotropic PSR emission. However, even with this assumption it is very difficult to obtain an acceptable value for $\eta$. This seems to confirm that $R_{ \rm BB}$ is too small to provide a good estimate of the star physical size. Forthcoming studies will allow us to better constrain the system parameters.

\clearpage{\pagestyle{empty}\cleardoublepage}

\chapter[The companion to \psrIM\ in the Galactic Field]{The optical companion to the intermediate mass Millisecond Pulsar J1439$-$5501 in the Galactic Field}\label{Chap:1439}

In this chapter we describe the identification of the  companion star to the intermediate mass binary PSR J1439$-$5501 obtained by means of ground-based deep images in the $B$, $V$ and $I$ bands, acquired with FORS2 at the Paranal ESO observatory.   The companion is a massive WD with $B=23.57 \pm 0.02 $, $V=23.21 \pm 0.01$ and $I=22.96 \pm 0.01$, located at only $\sim0.05\arcsec$ from the PSR  radio  position.  Comparing the WD location in the ($B$, $B-V$) and ($V$, $V-I$) CMDs with theoretical cooling sequences we derived a range of plausible combinations 
of companion  masses ($1 \lapp \mcom \lapp 1.3~\Msun$), distances ($d \lapp 1200$ pc), radii ($\lapp7.8~10^{-3}~\Rsun$) and temperatures ($T_{eff}=31350^{+21500}_{-7400}$). From the PSR mass function and the estimated  mass range we also constrained the inclination angle ($i \gapp 55 ^{\circ}$) and the PSR mass ($\mpsr \lapp2.2 \Msun$).
The comparison between the WD cooling age and the spin down age suggests that the latter is overestimated by a factor of about  ten. - {\it This chapter is mainly based on \citealt{pallanca13com1439}, ApJ, 773, 127.}

\section{\psrIM}
\label{1439Sec:pulsar}
We have collected and analyzed a sample of deep multi band images in the direction of the IMM J1439$-$5501 in the GF, with the aim of identifying the companion star. 

The binary \psrIM\ was discovered during the first data reprocessing \citep{faulkner04}  of the Parkes Multibeam Pulsar Survey \citep{manchester01} and its accurate timing parameters were published by \citet[][hereafter L06]{lorimer06}. The spin period ($\Pspin\sim 29$ ms) and the surface magnetic field ($B_{s}\sim 2.05\times10^9$ G) suggest that \psrIM\ belongs to the class of the {\it mildly} recycled PSRs, i.e. the NS that underwent a relatively short phase of mass accretion from the companion star.

The distance of \psrIM\ can be estimated from its dispersion measure (DM$\sim 15$ pc cm$^{-3}$, L06) once a model for the
distribution of the electrons in the interstellar medium is adopted. In particular, the TC93 model \citep{taylorcordes93} predicts 760 pc,
the NE2001 model \citep{cordeslazio02} places the PSR at a  smaller distance (600 pc), whereas applying the updated Taylor \& Cordes model \citep{schnitzeler12} we obtained a larger distance, of 950 pc.   Considering  these uncertainties, a possible range\footnote{ However, for our detailed analysis below, the range of available WD models forces us to use a slightly reduced range, $d= 600-1200$ pc (but see Section \ref{1439Sec:Results}).} for the distance is $d= 500-1200$ pc.

From the observed orbital period, $P_b = 2.117942520(3)$ d, and the projected semi-major axis of the orbit, $\apsr\sin i=2.947980(3)\times 10^{11}$ cm $\sim 4.2$ $\Rsun$ (where $i$ is the orbital inclination), L06 derived a PSR mass function $f_{\rm PSR}=0.227597~\Msun$. This value implies that the companion to \psrIM\ has a mass $\gapp 1~\Msun$. In particular, adopting a NS mass $\mpsr = 1.4~\Msun$, the
minimum companion mass is $1.13~\Msun$ (L06), while assuming the minimum value for   radio PSR mass measured so far 
\citep[$\mpsr = 1.24~\Msun$,][]{faulkner05}, the minimum companion mass would be $1.07~\Msun.$
 
The minimum value for the companion mass leaves 3 options for the nature of this object: an ordinary non-degenerate star, a NS, or a very
massive CO- or ONe-WD.  A MS star with mass as large as $1.1~\Msun$ located at the system distance would have magnitude $R\lapp 15$. However, the inspection of archive images (ESO-DSS)  shows that no such a bright stars is observed close 
 to the nominal position of \psrIM. On the other hand the very small observed eccentricity of the orbit ($e = 5\times 10^{-5}$; L06) strongly argues against the hypothesis that the companion is a NS. In fact, in this case it should be the remnant of the massive star that first recycled \psrIM, and then exploded as a supernova. This should have probably left a significant eccentricity in the system, at odds with the observations.

In view of the considerations above, the companion is most likely a massive WD, and the system J1439$-$5501 belongs to the growing class of the so-called IMMs \citep{camilo96}. Given its orbital and spin period, the favored scenario, among those discussed by \citet[][see also Chapter \ref{Chap:introPSR}]{tauris12}, is that the system was originated from the evolution of an IMXB with an asymptotic giant branch companion, through a common envelope phase.

Of the seventeen candidate IMM systems currently known in the GF \citep{vankerk05, jacoby06, tauris12, burgay12}, only 5 companions to IMMs have been detected so far in the optical band: PSR B0655$+$64, PSR J1022$+$1001, PSR J1528$-$3146, PSR J1757$-$5322 and PSR J2145$-$0750 \citep{vanKerkKulk95, vankerk05, lundgren96a, lundgren96b, lohmer04,   jacoby06}.

\section{Observations and data analysis}
\label{1439Sec:analysis}

The photometric data set used for this work consists of a series of ground-based optical images acquired with 
FORS2.  We performed the observations in the {\it Standard Resolution mode}, with a pixel scale of $0.126\arcsec /pixel$ (adopting a binning of $1\times1$ pixels) and a 
FOV of $6'.8\times6'.8$.  All the brightest stars in the FOV have been covered with occulting masks in order to avoid artifacts produced by objects exceeding the detector saturation limit in long exposures, which would have significantly hampered the search for faint objects.

A total of 39 deep images in the $B_{HIGH}$, $V_{HIGH}$ and $I_{BESS}$ bands were collected during five nights in May 2009, under program $383.D-0406(A)$ (PI: Lanzoni).  Since the goal of this work is to identify the companion to \psrIM, only the chip containing the region around the nominal position of the PSR has been analyzed.

By following standard reduction procedures, we corrected the raw images for bias and flat-field as described in Chapter \ref{Chap:0610}.
We then performed the photometric analysis by applying the PSF fitting method (see  Chapter \ref{Chap:introCOM}).
At the end of the 
reduction procedure we obtained a catalog of about  4000 sources (see Table \ref{1439Table:cat} for a small sub-sample of the catalog).

\begin{table}
\begin{center}
\begin{tabular}{| l | c  c | c  c | c  c | c  c | }
\hline
\hline
ID &  R.A. (J2000) & Dec. (J2000) & $B$ & $eB$ & $V$ & $eV$ & $I$ & $eI$\\
\hline
     1$\star$ &   14:39:39.746   &  -55:01:23.66   &     23.57   &  0.02    &  23.21  &   0.01  &  22.96   &   0.02\\
     2   &   14:39:43.006   &  -55:00:47.58   &     22.24   &  0.02    &  21.00  &   0.01  &  19.66   &   0.01\\
     3   &   14:39:41.277   &  -55:01:10.81   &     22.16   &  0.02    &  20.51  &   0.01  &  18.82   &   0.01\\
     4   &   14:39:40.619   &  -55:01:14.66   &     21.46   &  0.02    &  19.96  &   0.01  &  18.43   &   0.01\\
     5   &   14:39:40.234   &  -55:00:55.79   &     22.70   &  0.02    &  21.21  &   0.01  &  19.71   &   0.01\\
     6   &   14:39:38.596   &  -55:00:40.84   &     21.34   &  0.02    &  19.98  &   0.01  &  18.60   &   0.01\\
     7   &   14:39:38.068   &  -55:00:45.54   &     23.09   &  0.02    &  21.75  &   0.01  &  20.32   &   0.01\\
     8   &   14:39:37.287   &  -55:00:50.09   &     23.53   &  0.02    &  21.98  &   0.01  &  20.28   &   0.01\\
     9   &   14:39:36.626   &  -55:01:24.52   &     21.39   &  0.02    &  20.20  &   0.01  &  18.88   &   0.01\\
    10  &   14:39:35.910   &  -55:01:29.44   &     23.68   &  0.02    &  22.14  &   0.01  &  20.49   &   0.01\\
\hline
\end{tabular}
\end{center}
\caption{Position and $B$, $V$ and $I$ magnitudes (with relative errors) of ten stars around \com\ to \psrIM. Symbol $\star$ marks the identified companion. }
\label{1439Table:cat}
\end{table}

The photometric calibration has been performed by following the procedure described in Chapter \ref{Chap:introCOM}. In particular, in this case we selected ten bright and isolated stars and, for each of them, we performed aperture photometry with
different radii ($r$) and we compared these magnitudes with those obtained with the PSF fitting.  The mean value of the differences between PSF and aperture magnitudes has been found to be constant for $r\ge13$ pixels.  Thus we used the value at $r=13$ pixels as aperture correction to be applied to all the stars in our catalog.  For a straightforward comparison with theoretical models, we decided to calibrate the instrumental magnitudes ($b$, $v$, $i$) to the standard Johnson photometric system ($B$, $V$, $I$).  To this aim, we first derived the calibration equations for ten standard stars in the field PG1323 \citep{stetson00}, which has been observed with FORS2  during  the observing run under photometric conditions. To analyze the standard star field we used the {\sc daophot} {\tt PHOT} task and we performed aperture photometry with the same radius used for the aperture correction.  We then compared the obtained magnitudes with the standard Stetson catalog available on the CADC web site\footnote{http://cadcwww.dao.nrc.ca/community/STETSON/standards/}. The comparison shows a clear dependence on color. Hence we performed a linear fit in order to derive the trend as a function of the color $(v-i)$ in the case of $V$ and $I$ bands and of the $(b-v)$ for the $B$-band.  The resulting calibration equations are $B=b+0.126(b-v)+27.16$, $V=v+0.051(v-i)+27.53$ and $I=i-0.002(v-i)+26.95$  and the final uncertainties on the calibrated magnitudes  are $\pm0.016$, $\pm0.004$ and $\pm0.010$ for $B$, $V$ and $I$, respectively.  We neglected the dependence on airmass since all exposures were taken at similar values.

The astrometry procedure has been performed as described in Chapter \ref{Chap:introCOM}.
In particular, since in the GF proper motions may not be negligible, we used as astrometric reference stars the objects in the catalog PPMXL \citep{roeser10}, where the proper motion of each star is listed. Since the number of objects in common with our dataset is
large enough (126 stars), we could derive appropriate coordinate transformations.  As first step of the procedure we derived the position of the astrometric stars at the epoch of the observations. Then we registered the pixel coordinates of the reference image onto the absolute coordinate system through the cross-correlation of the primary astrometric standards in common with our catalog, by using CataXcorr.  The root mean square of the adopted transformations is  $\sim0.3\arcsec$ in 
$\alpha$ and $\sim0.2\arcsec$ in 
$\delta$, while the typical uncertainty of the PPMXL stars in this field is $\sim0.1\arcsec$. These quantities give an accuracy 
of $\sim0.11\arcsec$ for our astrometric solution.

\section{The identification of the companion to PSR J1439$-$5501}
\label{1439Sec:Results}

In order to identify the companion to \psrIM\ we focused our attention to any object located close to the PSR nominal position, as derived
from the timing solution in the radio band: $\alpha_{2000}=14^{\rm h} 39^{\rm m} 39^{\rm s}.742(1)$ and $\delta_{2000}=-55^\circ 01' 23''.62(2)$ at the reference epoch MJD=53200 (L06). No determination of the PSR proper motion was reported in L06. However, assuming a conservative transverse velocity of $\sim 100$ km $s^{-1}$ \citep{hobbs05} the expected total positional shift would be $\sim
0.15\arcsec$ (for the system distance) over the $\sim 5$ yr interval between the reference epoch of the radio ephemeris and the
date of the optical observations. Therefore, the offset between the position of the system J1439$-$5001 in the optical images and the coordinates given by L06 would be at most of the order of the astrometric solution accuracy.  We also note that follow-up timing observations
of the  system provide additional support to a safe use of the coordinates reported in L06: in fact these observations indicate a transverse velocity for \psrIM\ well below $100$ km s$^{-1}$ (Burgay, private communication).

\begin{figure*}[b]
\begin{center}
\includegraphics[width=140mm]{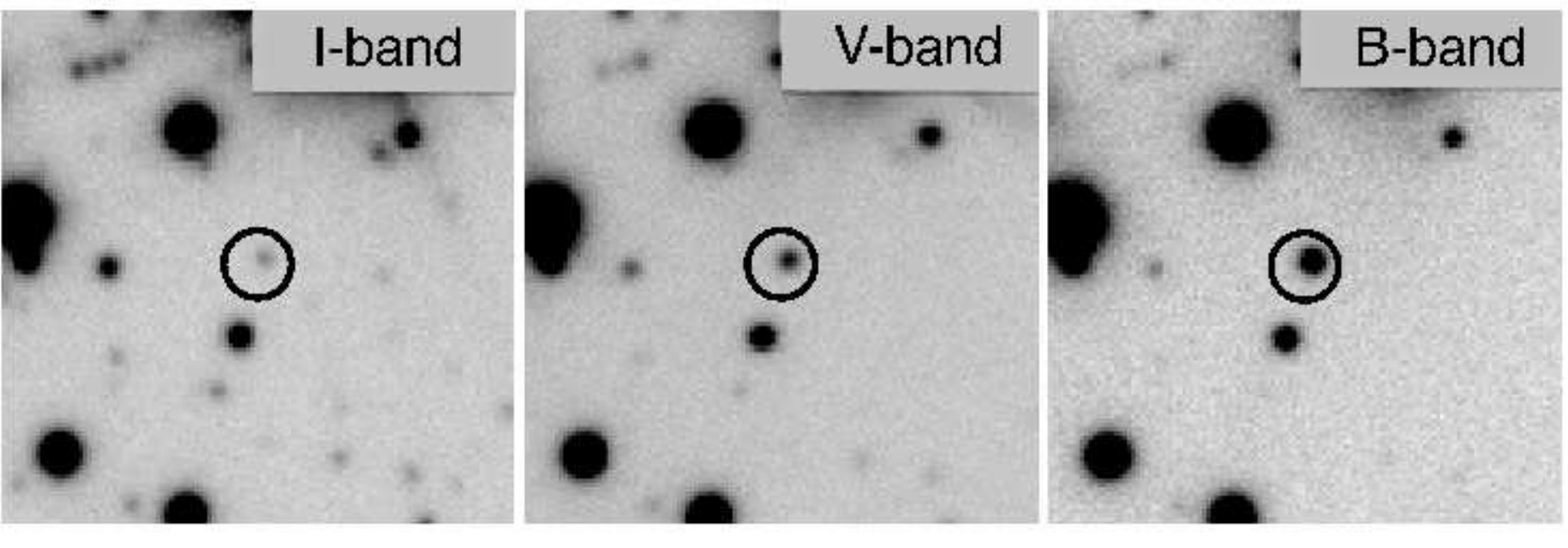}
  \caption[Image of the region around \psrIM.]{From left to right, $15\arcsec \times 15\arcsec$ maps in the $I_{BESS}$,  $V_{HIGH}$ and $B_{HIGH}$   bands around the PSR nominal position. The black circles are centered on \psrIM\   and have a radius of $1\arcsec$.}
\label{1439map}
\end{center}
\end{figure*}

A visual inspection of the deep images shows that there is a star located at only $0.05\arcsec$  from the nominal position of \psrIM\ (see Figure \ref{1439map}).  Moreover, no other star in our catalog is found within an error circle centered on the PSR position and having a radius of  several times the aforementioned uncertainty in the astrometric solution.

\begin{figure*}[b]
\begin{center}
\includegraphics[width=105mm]{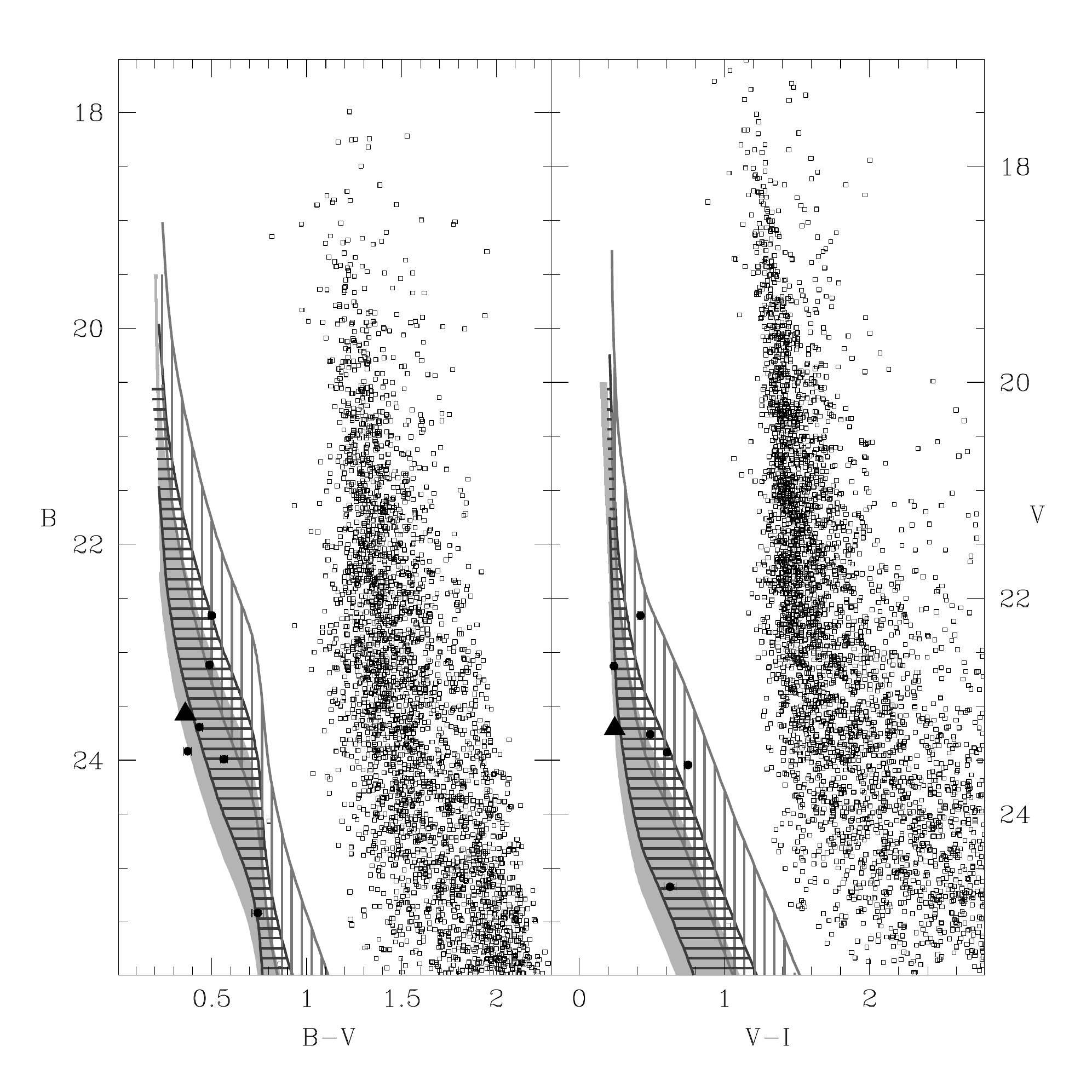}
  \caption[CMD positions of the companion to \psrIM.]{($B$, $B-V$) and ($V$, $V-I$) CMDs for   \com\ to \psrIM\  (black triangle) and for all the other objects observed in the detector FOV.  The companion is  located at $B=23.57\pm0.02$,   $V=23.21\pm0.01$ and   $I=22.96\pm 0.02$, corresponding to the region  of WD cooling sequences. The    vertically and the horizontally hatched bands correspond to the CO-WD cooling sequences,  for  $0.6$ and   $1.0~\Msun$, respectively, in a range of   distances between 600 and 1200 pc \citep[BaSTI database;][]{salaris10}, while the shaded light gray strip marks, in the same distance range, the location of a $1.0~\Msun$ ONe-WD \citep{althaus07}.
}
\label{1439cmd}
\end{center}
\end{figure*}

The comparison among the $I$, $V$ and $B$ images  shown in Figure \ref{1439map} clearly suggests that this star has a color bluer than most of the others objects in the field.  This feature is confirmed by the inspection of the 
CMDs, where the star  is located on the left side of the bulk of the detected stars, in a region compatible with  WD cooling sequences (see Figure \ref{1439cmd}). This is in nice agreement with the scenario proposed for the evolution of the system J1439$-$5501 discussed in Section~\ref{1439Sec:pulsar}. The probability that any star in our catalog falls at the PSR position by chance coincidence is low ($\sim 2\%$), and it further reduces to $\sim 0.003\%$ if only WDs are considered. The combination of all these  pieces of evidence strongly suggests that the detected star is the companion to \psrIM.

In principle, the companions to MSPs can show optical variability due to irradiation by the PSR or by rotational modulation, as observed in the case of PSR 0655+64 \citep{vanKerkKulk95, vanKerk97}. By comparing the flux of the PSR intercepted by the WD with the WD flux we estimated that the flux enhancement due to re-heating is negligible. Unfortunately we could not check if the system shows any variability due to rotation since the largest available number of observations is in the $I$ band, in which the object is very faint (see Figure \ref{1439map}), while on the $B$ and $V$ bands the orbital sampling is very poor and prevents any study of variability. Hence with the available data it is not possible to set a definitive conclusion about the presence of magnitude modulations. Therefore, in order to detect possible variability, and eventually study its nature, future phase resolved observations are required.

\begin{figure*}[t]
\begin{center}
\includegraphics[width=140mm]{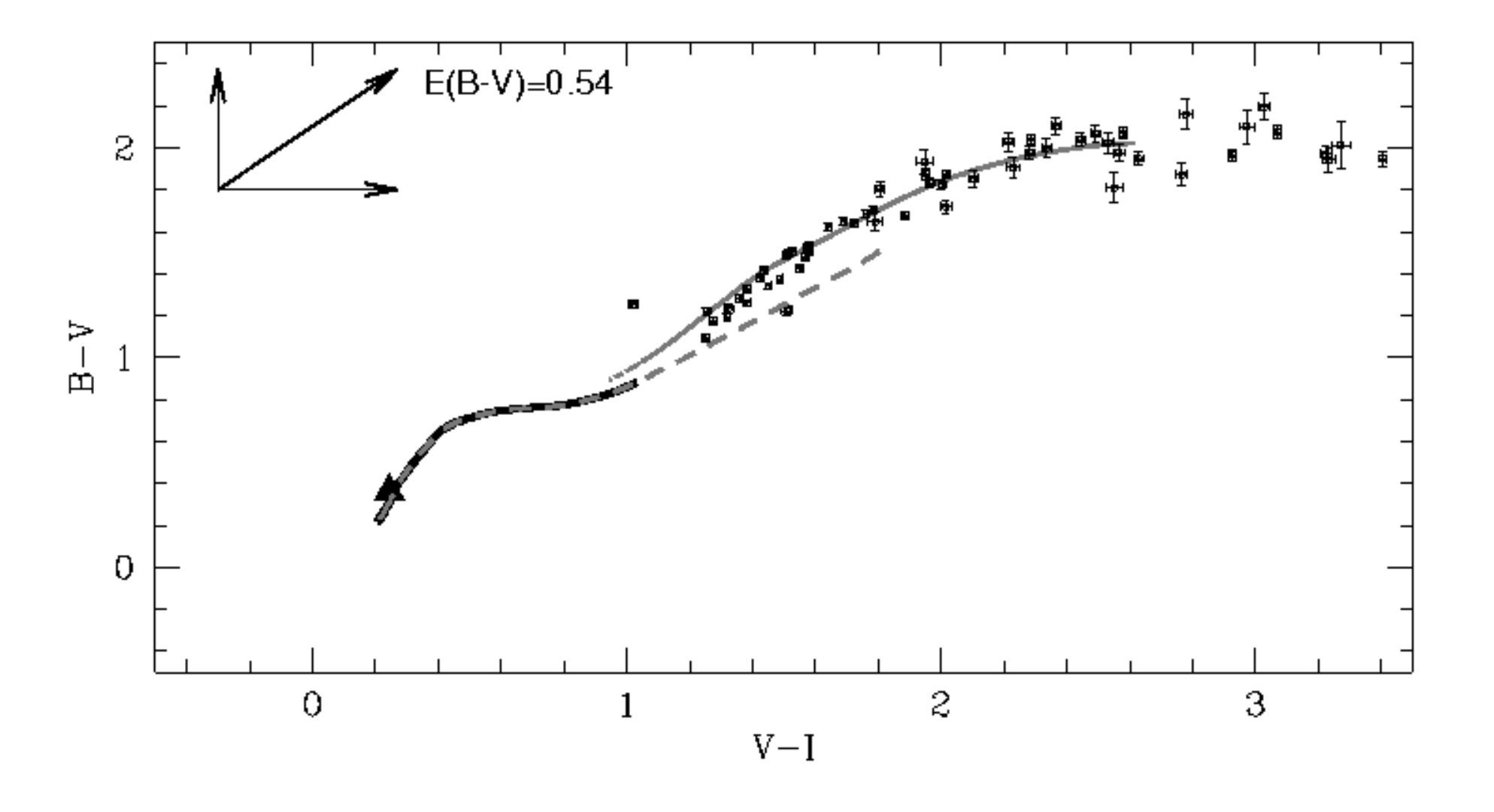}
  \caption[Color-color diagram for stars in the region of \psrIM.]{ ($V-I$, $B-V$) color-color diagram for all stars (open squares)  within $30\arcsec$ from \psrIM.    The companion   is highlighted with a large black triangle.   The gray  solid  region marks the location of MS populations having solar metallicity    and different ages, consistent with those observed in the   GF.   The  black solid and gray dashed  lines  correspond to the cooling sequences for  $1.2 \Msun$  CO- and ONe-WDs, respectively.    A color excess $E(B-V)=0.54$ is applied to the models (arrows mark the entity and direction of the applied reddening).
  }
\label{1439colcolred}
\end{center}
\end{figure*}

With the aim of deriving the mass and the age of \com, we compared its position  in the CMDs ($B=23.57 \pm 0.02 $, $V=23.21 \pm 0.01$ and $I=22.96 \pm 0.01$) with  a set  of theoretical CO- and ONe-WD cooling sequences of different masses \citep[BaSTI database,][]{salaris10, althaus07}.  In order to make consistent the ONe-WD models with the CO-WD cooling sequences we applied to the formers 
the same color transformations used for the latters.  As a first step, we calculated the extinction coefficient $E(B-V)$ by generating the  color-color  diagram  ($V-I,B-V$),  where  the dependence on distance disappears.  In particular, we compared the observed distribution of MS stars (within $30\arcsec$ from \psrIM) with the locus of theoretical models, with solar metallicity and different ages, typical of the GF population. Different reddening values,  ranging  between 0.4 and 0.7 with steps of $\delta E(B-V)=0.001$, have been iteratively applied to theoretical models. By performing a  $\chi^2$ test, we found that the best-fit value\footnote{The uncertainties on $E(B-V)$ have been estimated by accounting for the   magnitude errors of the observed population.} is $E(B-V)\sim0.54^{+0.06}_{-0.05}$ (see Figure~\ref{1439colcolred}).

In the following we will adopt $E(B-V)=0.54$, keeping in  mind that any estimate of the WD mass, distance and derived quantities depends on the $E(B-V)$ value. As can be seen from  Figure~\ref{1439colcolred},  if   $E(B-V)=0.54$   is assumed, the colors of \com\ nicely match the theoretical sequences  for WDs\footnote{ Note  that the  theoretical models shown in Figure~\ref{1439colcolred} are for  $1.2~\Msun$ WDs.
 However  the dependence of the WD cooling sequences on mass is negligible in the ($V-I,B-V$) color-color diagram \citep{bergeron95}.}.

 By taking into account  the absorption coefficients properly calculated for the effective wavelengths of the filters  \citep{cardelli89},
we derived the unabsorbed colors $(V-I)_0=-0.32^{+0.09}_{-0.07}$ and $(B-V)_0=-0.21^{+0.09}_{-0.07}$,    from which we estimated an effective  temperature of the star  $T_{eff}=31350^{+21500}_{-7400}$.

 By using the derived value of the reddening and  the adopted range of distances ($600-1200$ pc; see section {\ref{1439Sec:pulsar}}) 
we placed the WD cooling sequences for masses of $0.6$ and $1.0~\Msun$ selected from the BaSTI database \citep{salaris10}  in the ($V, V-I$) and the ($B, B-V$) CMDs.  As can be seen from Figure~\ref{1439cmd}, \com\ is clearly not compatible with low mass WDs, while it 
appears to be slightly more massive than $1.0~\Msun$.

\begin{figure*}[b]
\begin{center}
\includegraphics[width=100mm]{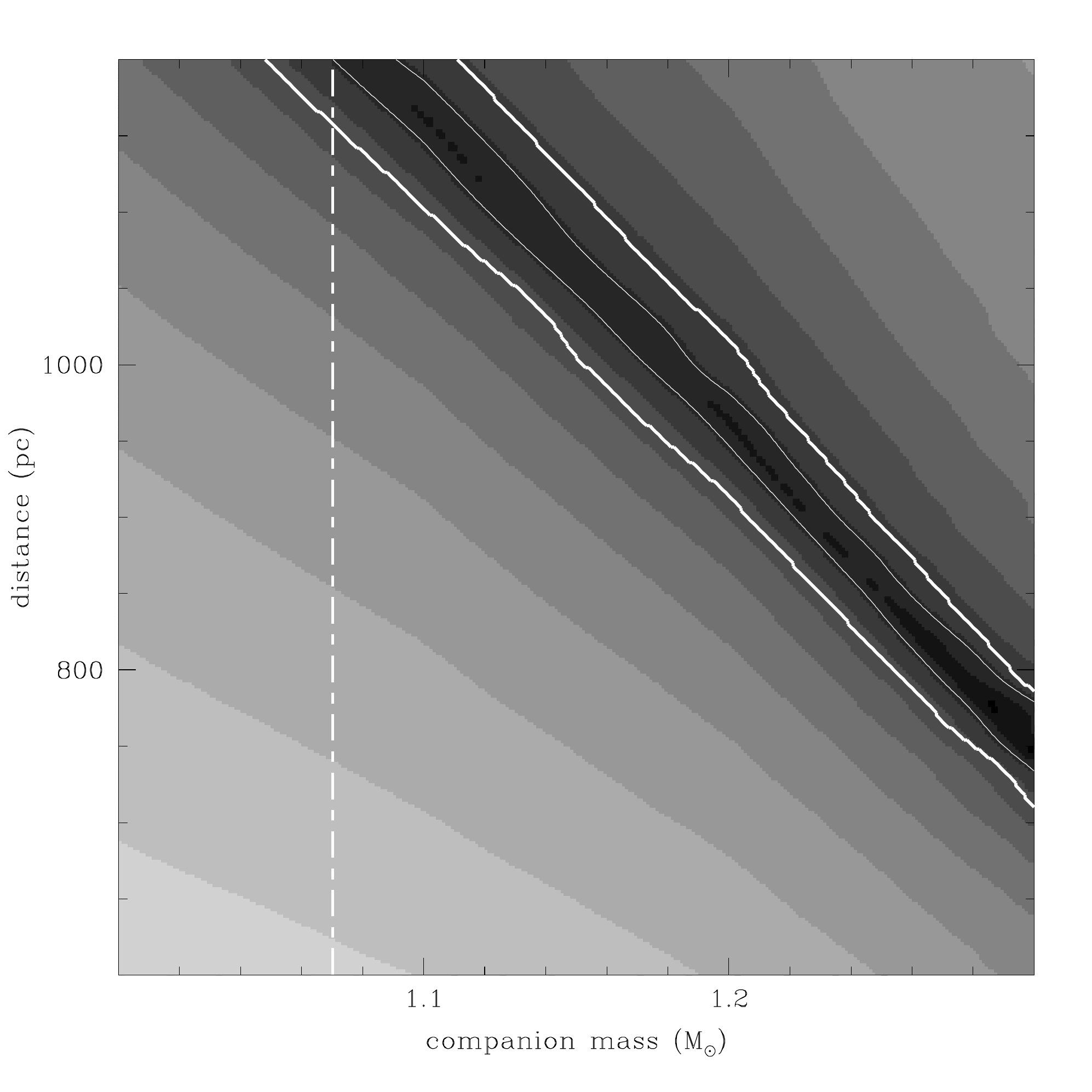}
  \caption[Distribution of probabilities  for a CO-WD companion in the  $m$-$d$ plane.]{Distribution of probabilities  for CO-WDs in the plane of companion mass ($m$) and system distance ($d$). Darker regions mean more probable configurations.    The thick white contour marks the region occupied by couples $m$-$d$ providing confidence values $\le1$ in both CMDs,   the thin white line encircles the configurations with confidence value $\le 1$.}
\label{1439mapmd}
\end{center}
\end{figure*}

In order to better constrain the mass of \com\ we needed a tight sampling in mass. Starting from the available tracks for CO- and ONe-WDs in the mass range $1-1.25~\Msun$ \citep{salaris10, althaus07} and assuming as a first approximation a linear relation between magnitude, color and mass in such a small range, we derived cooling sequences  between $1.0$ and $1.25~\Msun$ at regular steps.
By using the same linear relation we obtained tracks up to $1.3~\Msun$. For each mass ($m$) step we varied the PSR distance ($d$)  in the  range $600-1200$ pc, and we calculated the difference ($\Delta$ expressed in magnitude units) between the observed location of \com\ in the CMD and its perpendicular projection onto the cooling sequence. We applied this method in both the ($B,B-V$) and ($V,V-I$) CMDs,
thus obtaining $\Delta_1$ and $\Delta_2$, respectively, for any $m$-$d$ pair. For each value of $m$ we then selected the value of $d$ which minimizes the sum of  $|\Delta_1|+|\Delta_2|$ and we associated a confidence value to each $m$-$d$ pair. In each CMD the confidence value is calculated as $|\Delta|$ normalized to the photometric combined error $\sqrt{(e_{COL}^2+e_{MAG}^2)}$ of the star. In this way, the smaller is the confidence value, the larger is the probability associated to that configuration. In particular, a confidence value $\le 1$ means that the cooling sequence and the observed position are in agreement within the photometric errors. The resulting confidence value, when information from both CMDs is combined, is the sum of the confidence values derived from the two CMDs. The resulting distribution of confidence values in the $m$-$d$ plane is plotted in Figure~\ref{1439mapmd}. All the $m$-$d$  couples for which the confidence value is $\le1$ in both CMDs, have been selected as plausible combinations of $m$ and $d$ for \com\ (they are encircled by the white contour in Figure~\ref{1439mapmd}).

From this analysis we find that, for the considered upper limit to the distance ($d=1200$ pc), $M_{\rm COM} \gapp 1 \Msun$ (see Figure~\ref{1439mapmd}). Considering the constraint on the minimum companion mass from the PSR mass function ($M_{\rm COM}>1.07 \Msun$), we can safely confirm that $d\le1200$ pc. At small distances, the analysis is limited by the availability of WD models (see above). However, the adopted upper limit for the mass in our analysis ($M_{\rm COM}=1.3\Msun$) likely does not affect the results, even if we cannot definitely  rule out a more massive companion in the range $1.3<M_{\rm COM}<M_{\rm CH}$, where $M_{\rm CH}=1.44\Msun$ is the Chandrasekhar mass limit for a WD \citep{chandra35}. If \com\ is  a CO-WD  it would have a mass $1.07 \lapp M_{\rm COM}  \lapp 1.3~\Msun$, and it would be located at a distance $710\lapp d \lapp 1200$ pc (see Figure~\ref{1439mapmd}), which imply a radius in the range $4.5-7.7~10^{-3}~\Rsun$.  
In the case of ONe-WDs models, we found $m$-$d$  configurations as probable as those obtained for CO-WDs. Hence, \com\ to \psrIM\ could also be a ONe-WD with mass $1.07\lapp M_{\rm COM} \lapp 1.3~\Msun$, radius in the range $4.0-7.8~10^{-3}~\Rsun$ and located at distance  $640\lapp d \lapp1200$ pc. Note that, because of the larger molecular weight of the ONe-WDs with respect to the CO-WDs, the former have a smaller radius and hence, for a fixed mass, to fit the observed properties they should be located closer than CO-WDs.

\section{Discussion and conclusions}

\begin{figure*}[b]
\begin{center}
\includegraphics[width=100mm]{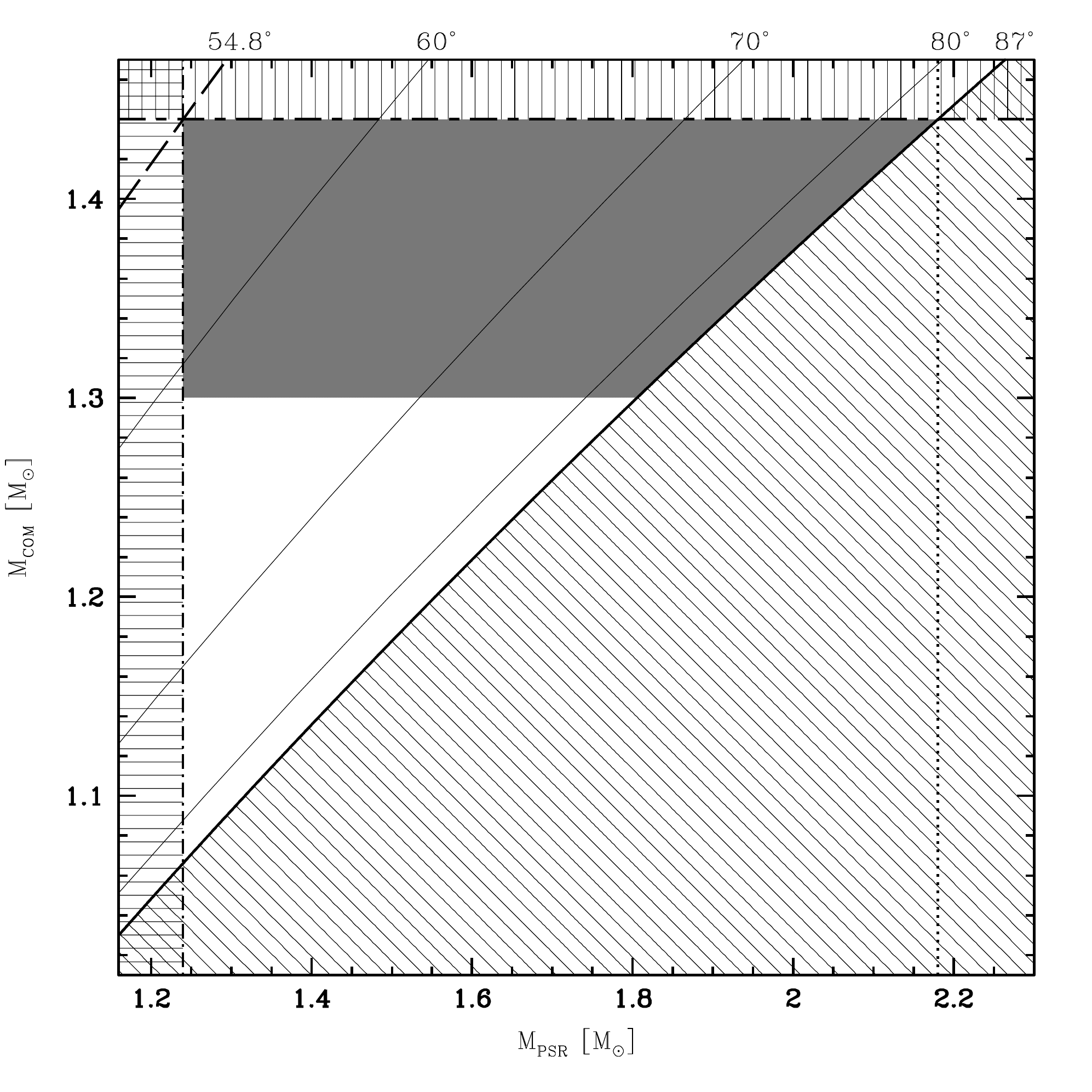}
  \caption[Mass of the companion to \psrIM\ as a function of the PSR mass.]{
  Companion mass as a function of the PSR mass. Solid lines indicate constant orbital inclinations from the PSR mass function. In particular the thick solid line marks the limit  of $i_{\rm max}=87^{\circ}$. The short-long dashed line marks the physical limit $M_{\rm CH}=1.44 \Msun$ for the WD mass. The dot dashed line marks the minimum limit $M_{\rm NS,min}=1.24\Msun$ ever measured  for a radio PSR mass. 
 The hatched regions are excluded because of the combination of the physical limits above. The dotted line indicates the upper limit to the PSR  mass $M_{\rm PSR}<2.18$ derived assuming $M_{\rm COM}=M_{\rm CH}$ and $i=i_{\rm max}$; while the long dashed line indicates the lower limit to the inclination angle $i>54.8^{\circ}$ calculated assuming $M_{\rm COM}=M_{\rm CH}$ and $M_{\rm PSR}=M_{\rm NS,min}$. The white region marks the investigated range of WD masses by comparison between optical photometry and theoretical models.  Note that we could not extend the analysis to  companions more massive than $1.3 \Msun$ (gray region) because models for WD with larger  masses are not available.}
 \label{1439mapmpmc}
\end{center}
\end{figure*}

From the PSR mass function  $f_{\rm PSR}=0.227597~\Msun$ (L06) and the range of permitted companion masses and radii derived in the previous Section, we constrained $\mpsr$ and the inclination of the system, as summarized in Figure \ref{1439mapmpmc}. First of all, given the relatively large orbital separation ($\sim 9.5~\Rsun$) and the small radius of the companion, the absence of observed eclipses of the radio signal along the orbit only constraints $i<89.4^\circ$.  The observation of the Shapiro delay effect could allow one to impose further constraints on the inclination. Unfortunately, the quality of the available radio timing (with a non uniform orbital coverage especially close to superior conjunction, where the effect of the Shapiro delay is maximum) does not allow at the moment to accurately determine the Shapiro delay parameters in this system. However, simulations show that, even with the few radio measures (times of arrival) obtained around orbital phase 0.25, a signature of an almost edge-on orbit would be detectable. We can hence constrain the inclination angle to be $\lapp 87^\circ$.
Simulating a data set with monthly observations and the current instrumentation, a determination of the Shapiro delay would take from $\sim 1$ to 3 decades, depending on the inclination of the source (in the range from 60 to 85 degrees). Instead, a lower limit of $54.8^\circ$ for $i$ results from the assumption that $\mcom=M_{\rm CH}$ and that the PSR has a mass larger than $1.24 \Msun$ (see Section \ref{1439Sec:pulsar}). Moreover, by adopting $\mcom=M_{\rm CH}$ and $i=87^{\circ}$, an upper limit of $2.18~\Msun$  can be inferred for the PSR mass.  

Of all the CO-WD companions to IMBPs identified so far,  \com\ to \psrIM\  is one of the few with a mass estimate. Moreover, comparing its estimated mass with the median  of companion masses of previously identified CO-WDs \citep[see][]{tauris12},  it turns out to be among the most massive.

Although the cooling age of a WD suffers from uncertainties, it is the only reliable age indicator of a PSR binary system \citep{tauris12}.
Hence, the optical identification of a WD companion to a binary MSP and its cooling age estimate are of crucial importance to constrain the spin-down theory 
through comparison with the PSR characteristic age (see Appendix).
\citet{kulkarni86} proposed for the first time a comparison between these two ages, finding a quite good agreement. However, there is an increasing body of evidence \citep[see e.g.][]{tauris12} that the characteristic age is a very poor  measure of the time that a recycled PSR spent since the completion of mass transfer.  
In fact, in most cases WDs are observed to be younger than the MSPs  \citep{hansenPhinney98a, hansenPhinney98b}, thus suggesting that the standard spin-down model may overestimate the PSR age \citep{jacoby06}.  For the parameters inferred above, the cooling age of \com\ is in the range   $ 0.1\lapp t_{cool} \lapp 0.2$ Gyr and $0.1\lapp t_{cool} \lapp 0.4$ Gyr for CO- and ONe-WD respectively (and up to $0.4-0.5 $ Gyr for the minimum reddening configuration) and it is several times smaller than its estimated spin-down age \citep[$3.2-4.5$ Gyr;][]{kiziltanThor10}. This discrepancy 
can be explained by considering that the characteristic age for PSRs in the millisecond regime could not be a proper estimate of the PSR age, since the hypothesis that $P_0$ is negligible with respect to the current $P$ is not applicable to recycled PSRs. Such a scenario is in agreement with the results of simulations of  an evolved synthetic population of MSPs \citep{kiziltanThor10}, which show that the characteristic ages could either over- or under-estimate the true age of MSPs by more than a factor of ten. Hence, we stress that  the identification of the companion and its age estimate (e.g. from the cooling sequences for WD companions) could be a powerful tool to derive the true age of a recycled PSR.

While a spectroscopic analysis, both for  the radial velocity curve and the chemical analysis, is hardly feasible with the current generation of instrumentation, a photometric follow-up could provide plenty of useful information. In particular, phase resolved data-sets could lead to the possibility of revealing any optical variability of the companion, while multi-band photometry could allow to better estimate the reddening and, by applying the same method used in this  work, to better constrain the mass and the distance of the companion.

\clearpage{\pagestyle{empty}\cleardoublepage}

\newpage
\markboth{CONCLUSIONS}{CONCLUSIONS}
\chapter{Conclusions}

MSPs are  fast rotating, highly  magnetized  NSs.
They are thought to form  in binary systems  containing a slowly rotating NS that, during a phase of heavy mass accretion from an evolving companion, is spun up to millisecond period.  
The final stage consists of a binary made of a fast rotating  PSR (a MSP) and a deeply peeled or even exhausted star (as a WD).
This theoretical scenario of MSP formation is known as {\it ``canonical recycling scenario''}.
In the last years, however  an increasing  number of systems deviating from the expectations of the {\it canonical recycling scenario} has been discovered and the formation and evolution of such exotic objects  still remain  unclear.

The identification of companion stars to MSPs is  useful  to fully characterize the binary systems and to constrain the formation and evolution of these objects.
In dense environments such as GCs, the identification of the companion stars is helpful also to understand the interplay between dynamics and the evolution of binary systems.

\subsubsection*{Summary of the results}
In this Thesis we report the  identification of 5 new companions to binary MSPs located both in GCs and in the GF: 
 
\begin{itemize}
\item[]{ \it  J1824$-$2452H in M28 -}
The identified companion to this RB is a RL filled, bloated star, highly perturbed by the NS tidal field. Likely, it is a $0.2 M_\odot$, orbiting a $1.4 M_\odot$ MSP, in a plane with an orbital inclination of $\sim 60-70^\circ$. 
The results obtained suggest that, it  is currently losing mass, and that the system is surrounded by large clouds of gas extending well beyond the RL which probably is constantly replenished.
Its location outside the cluster core suggests that the NS was recycled by another companion (that eventually became a very low mass, exhausted star, because of the heavy mass transfer). Then an exchange interaction occurred in the cluster core between the MSP binary and a MS star,  causing the ejection of the lightest star and kicking the newly-formed system away from the centre. The new companion started to suffer heavy perturbations (bloating, mass loss, etc.)  induced by the MSP. We currently observe it as the detected $0.2\Msun$, bloated companion, and  it eventually will become a He-WD.

\item[]{\it J1824$-$2452I in M28 -}
The identified companion to this RB X-ray transient is a star that showed a strong variability and  underwent a significant increase of optical magnitude a few years before the recently detected X-ray outburst. In particular, during the quiescent state the companion is approximately located along the MS, $\sim 3$ magnitudes fainter than the TO, while during the optical burst it is $\sim 2$ magnitudes brighter and it is characterized by a bluer color, indicating  a perturbed state. 
Moreover, the presence of strong \halpha\ emission 
during the optical outburst phase suggests the presence of  material accreting  onto the NS.  
Finally, on the light of the association of \igr\ with \psrI\  \citep{papitto13},  we performed a preliminary  light curve study finding a possible correlation between the orbital period and the magnitude modulation.

\item[]{\it J1518$+$0204C in M5 -}
Its companion is the first identified companion to a BW  in a GC.
It is a very  faint and strongly variable star, located between the MS and the WD cooling sequences in the CMD.
By comparing its position in the CMD with a reference isochrone we constrained the  effective temperature, luminosity  and mass of the star.
Considering the largest permitted mass value ($\mcom\sim0.2\Msun$) and the PSR mass function, it is possible to 
rule out very small inclination angles of the system (e.g. $i \lapp 10^\circ$ for a $1.4 \Msun$ NS).
The magnitudes of this star have been measured in 14 out of 44 analyzed images. In particular, the star has been detected around the PSR inferior conjunction, while it is under the detection limit at the PSR superior conjunction. A simple sinusoidal model suggests an amplitude of the light curve larger than 3 mags.
Since the  flux variation seems to be mainly  due to irradiation from the MSP, we  estimated a lower limit to the reprocessing efficiency as a function of $i$. 
As an example configuration, the observed optical modulation can be reproduced by considering a system seen at an inclination angle of about $60^{\circ}$,  with a very low mass RL filled companion ($\mcom\sim0.04-0.05\Msun$), that is reprocessing the PSR flux with an efficiency $\eta \gapp 1\%-15\%$.

\item[]{\it J0610$-$2100 in the GF -}
The companion to this BW is a faint, very low mass object.
It has been detected only in images obtained at the phase corresponding to the PSR inferior conjunction.
Its  magnitude modulation seems to be mainly due to irradiation from the MSP, reprocessed by its surface. For the lower limit to the observed flux variation, we thus estimated the reprocessing efficiency as function of  the inclination angle.   
As an example configuration,  the observed optical modulation can be reproduced by considering a system seen at an inclination angle of about $60^{\circ}$,  with a very low mass companion ($\mcom\sim0.02\Msun$) that has filled its RL, and with a reprocessing efficiency of about $50\%$.

\item[]{\it J1439$-$5501 in the GF -}
The identified companion to this IMM is a massive WD ($M>1\Msun$). 
 Because of the relatively large orbital separation, the small radius of the companion and the absence of observed eclipses of the radio signal along the orbit 
 allow us to derive only a weak  constraint to the inclination angle,  $i<89.4^\circ$.  
Moreover, from the non-determination of the Shapiro delay  the inclination angle is constrained to be $\lapp 87^\circ$.
 Instead, a lower limit of $54.8^\circ$ for $i$ results from the assumption that $\mcom=M_{\rm CH}$ and that the PSR has a mass larger than $1.24 \Msun$. On the other hand, by adopting $\mcom=M_{\rm CH}$ and $i=87^{\circ}$, an upper limit of $2.18~\Msun$  is obtained for the PSR mass.  
Of all the WD companions to IMMs identified so far,  \com\ to \psrIM\  is one of the few with a mass estimate. Moreover, comparing its estimated mass with the median  of companion masses of previously identified CO-WDs,  it turns out to be among the most massive.
 For the parameters inferred above, the cooling age of \com\ is in the range   $ 0.1\lapp t_{cool} \lapp 0.2$ Gyr and $0.1\lapp t_{cool} \lapp 0.4$ Gyr for CO- and ONe-WD respectively (and up to $0.4-0.5 $ Gyr for the minimum reddening configuration). This is several times smaller than the estimated spin-down age, $3.2-4.5$ Gyr.
 After other similar results in the literature, such a discrepancy further confirms that the characteristic ages could either over- or under-estimate the true age of MSPs by more than a factor of ten.
 
\end{itemize}
 
In this Thesis work we also performed a spectroscopical follow-up of 
{\it COM J1740-5340A in NGC 6397 -}
We measured the  abundances of C and N, finding that they are incompatible with the values expected on the surface of a normal MS star.
The 
[C/Fe] is also incompatible with the  C abundance range observed in the parent GC.
By comparing the  C and N abundances with the chemical gradients predicted by theoretical stellar models, we were able
to put new constraints on the nature of this object.
In particular, it seems to be a star peeled down to an interior layer where the CN-cycle approximately reached the equilibrium. 
The entire envelope of the star has been completely removed and, assuming an initial mass of $\sim 0.8 \Msun$, we estimated that it likely   lost $\sim 75\%$ of its initial mass during the interaction with the PSR.
The C and N abundances also allowed us to identify a reasonable mass range for the MSP companion: only the portion of the stellar model between 0.17 and $0.28 \Msun$ are in agreement with both the observed C and N abundances.  This confirms previous mass estimates, demonstrating that the analysis of the C and N surface abundances provides a powerful diagnostic of the companion mass which is totally independent of other, commonly used methods.

\subsubsection*{The used method}
The identification and characterization of MSP companions in this Thesis work are based on the careful photometric analysis of multi-wavelength and multi-epoch images (see Chapter \ref{Chap:introCOM}).
The identifications are based on the positional coincidence and the measurement of a flux modulation with the same period of the MSP orbital period.
Then,  the information from the CMD location is combined with that from  the light curve shape (in the case of a variable star) to derive the nature of the companion and to constrain the binary system parameters.

As discussed above, in this Thesis we performed a search for MSP companions in both GCs and the GF.
As emerged from the previous Chapters, in the case of MSPs located in the GF, the data reduction procedures can be easily performed, because of the low crowding conditions and the absence of cosmic rays. However, with respect to the case of GCs, the situation is slightly more complicated in the following steps of the analysis. For example, dedicated observations to perform the photometric calibration are needed. Moreover, once the companion has been identified there are still several uncertainties, which can be only partially constrained, such as the system distance, the reddening and the metallicity. This prevents a full characterization of the companion star.
On the other hand, in GCs  this kind of study is  complicated by the severe stellar crowding conditions
which make observations from the space mandatory.
However, after the first steps of the analysis, the advantage is that  quantities such as distance, reddening and metallicity are reasonably well known from previous detailed studies of the host cluster.

In general, however, it is important to keep in mind that the adopted method  suffers from a  selection effect. In fact,   non-degenerate MS companions are easier to detect than WDs because they are brighter at any given distance. 
Moreover, a magnitude variability in agreement with the orbital period  is typically present in the case of non-degenerate companions but see the notable exception of PSR J1911-5958A in NGC 6752 \citep{ferraro03com6752, cocozza06}.
Hence search for this signature preferentially leads to discover such kind of companion stars. 

We also used a spectroscopic approach for the chemical characterization of MSP companions, with the aim of better clarifying the evolutionary processes of these objects and to provide an
 alternative method for the mass estimate in the case of a peeled star.

\subsubsection*{Different classes of MSPs}
The results of this Thesis well fit in the context of the MSP classification outlined in Chapter \ref{Chap:introPSR}.
 In fact the studied companions to the RB systems are low mass ($\sim 0.2 \Msun$) stars, 
a few magnitudes fainter than the TO,  probably  peeled because of mass loss.
The companions to BWs are very low mass objects, likely below the mass limit for hydrogen burning, and they appear to be heated by the PSR flux.
Instead, the companion  to the IMM is a massive WD, as expected.

The companions to both the BWs discussed here  have been detected only at the PSR inferior conjunction, while they are under the detection limit at the PSR superior conjunction.
The few magnitude measures show  a maximum at $\Phi=0.75$,  thus suggesting  ``single-hill'' light curve shapes. 
Such a behavior is indicative of  reprocessing mechanisms of the PSR flux by a faint cold companion (as a BD).
Moreover the magnitude modulation between minimum and maximum  is larger than a few magnitudes for both objects,
thus suggesting a quite large inclination angle and a RL filled companion.

The two companions to RBs both appear to be perturbed non-degenerate objects. 
However, they show significantly different properties. In particular, the companion to M28H has physical properties stable in time, while the companion to M28I (in agreement with the X-ray/radio behavior) showed an optical outburst phase characterized by a high \halpha\ excess, likely due to an accretion event (e.g. characterized by the presence of a disk).
Moreover, also the location in the GC is different and seems to suggest a different scenario for the two objects. In fact the position of M28H out of the cluster core ($d\sim1.5 r_c$) has been interpreted as the result of a very recent exchange interaction, while M28I (at least in projection) is located in the innermost region ($d\sim0.6 r_c$) where it probably formed. Hence, they could be similar companions but looking different because they are undergoing  different evolutionary phases. Interestingly, also the companions to the two RBs in NGC 6397 and NGC 6266 are  non-degenerate stars, but they show
 different optical properties (e.g. the CMD location) with respect to those in M28. An explanation could be that all these objects are located along the same evolutionary path but at different phases.  
In addition, the spectroscopic study of \comA\  added a few pieces of information to the currently known scenario: the C and N abundances suggested that it is a star peeled up to the innermost region.

\subsubsection*{The state of the art}
Depending on the environment, the study of MSP companions has different applications.
Both in GCs and in the GF it is useful to understand the formation and evolution of MSPs descending from  primordial binaries. In addition, such an investigation in GCs is 
useful to clarify the role and the efficiency of dynamical  processes.

In this Thesis we identified 2 new companions belonging to two different mass classes of MSPs located in the GF:
\begin{itemize}
\item[-] 1 massive WD companion;
\item[-] 1  very low mass object. 
\end{itemize}

\begin{figure}[t]
\begin{center}
\includegraphics[width=120mm]{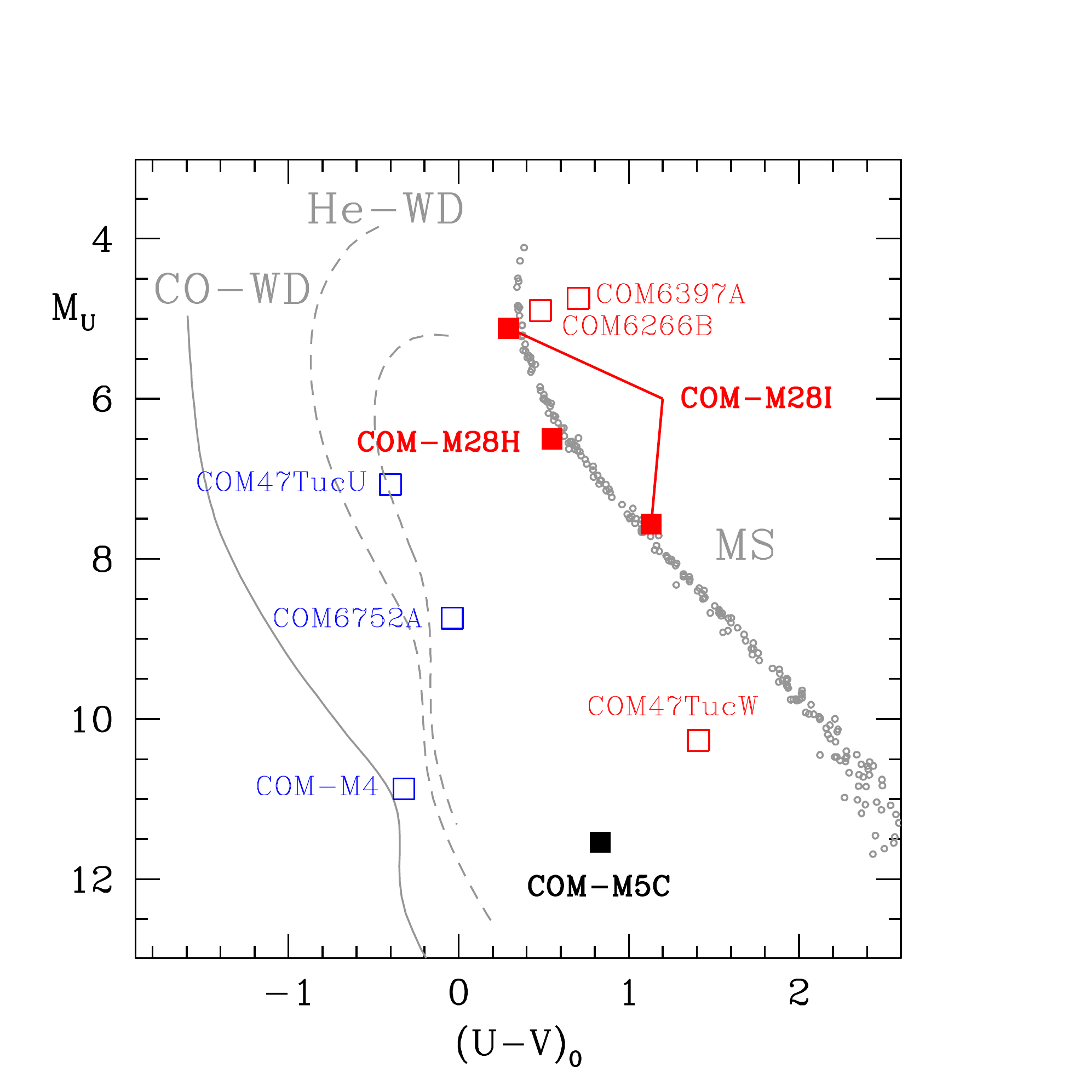}
\caption[CMD position of all MSP companions known to date in GCs.]{Location in the ($\rm M_U, (U-V)_0$) absolute plane of
all the currently  known optical companions to binary MSPs in GCs:
the six stars identified before this Thesis are marked as empty squares (same objects as in Figure \ref{stateArt}); the three newly discovered stars are shown as filled squares.
Blue, red and black colors mark the companions to ``canonical'', RB and BW systems, respectively. Note that the companion to \igr /\psrI\ is marked both in the quiescent and in the outburst state position.}\label{soaPOST}
\end{center}
\end{figure}

We also discovered 3 new companions in GCs. Thanks to this study, 
 the number of known MSP companions in GCs has increased by 50\%, bringing the 
 current total number to 9 (see Figure \ref{soaPOST}):
 \begin{itemize}
\item[-] 3 are He-WDs (blue points in Figure \ref{soaPOST});
\item[-] 5 are non-degenerate, likely peeled stars (red points in Figure \ref{soaPOST});
\item[-] 1 is a very low mass object (black point in Figure \ref{soaPOST}). 
\end{itemize}

 Therefore, only 3 (out of 9) are He-WD companions, the product expected by the {\it canonical recycling scenario}. This strongly indicates that exchange interactions are common events in the dense environments of GC cores and they are quite effective in modifying the ``natural" outcomes of the pulsar recycling processes. 
However, it is important to remember that this number could be affected by some observational biases, as discussed in the previous paragraph. 
  
On the other hand, the increasing number of non ``canonical''  MSPs in the GF (see the companion to \psrBW\ as a specific contribution of this Thesis) suggests that also the evolution of primordial binaries could lead to the formation of  such objects. Hence, despite the high probability of  dynamical interactions, also a non-degenerate companion in GCs could 
be the result of the evolution of primordial binaries.

\subsubsection*{The proposed evolutionary scenario}
\begin{figure}[b]
\begin{center}
\includegraphics[width=140mm]{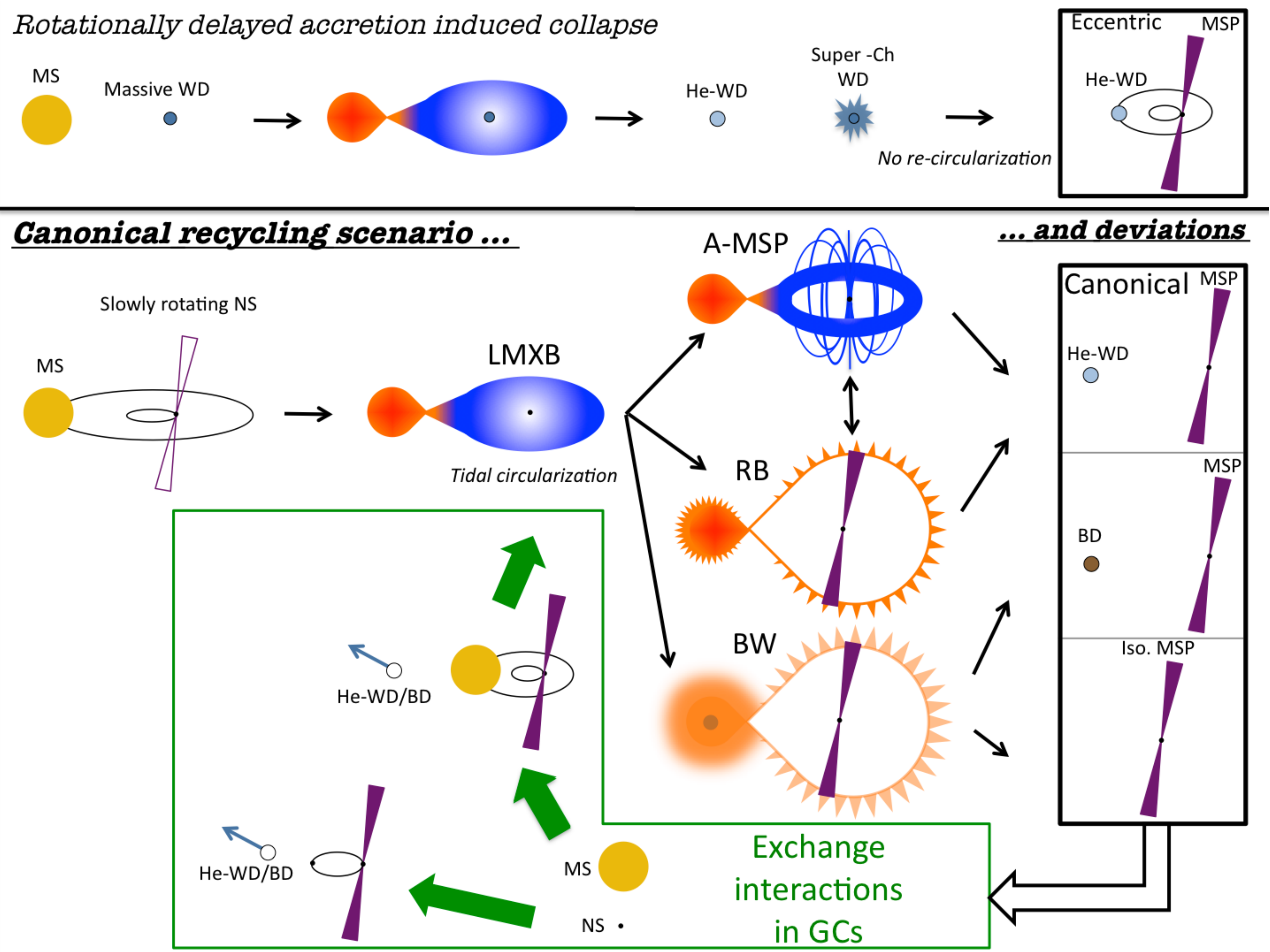}
\caption[Cartoon of the evolution of MSPs.]{Schematic view of the key phases of the  {\it canonical recycling scenario} and the deviations from it (as in the case of RB and BW systems),
also including  the possibility to produce exotic systems as the product of an exchange interaction in a GC.
In the top panel we represent the ``rotationally-delayed accretion-induced collapse'', proposed by    \citet{freiretauris13}  as an alternative formation mechanism for eccentric MSPs with He-WD companions. 
Note that all the cases in which the orbits of the two objects are drawn correspond to systems with non circular orbits.
}
\label{cartoon}
\end{center}
\end{figure}

In order to clarify the evolutionary history of binary low mass MSPs we tried to  sketch a possible scenario in the cartoon reported in Figure \ref{cartoon}.

In the top panel we schematically  report the ``rotationally-delayed accretion-induced collapse'' formation mechanism proposed by \citet{freiretauris13}, to account for a few eccentric binary systems made of  MSPs and  He-WDs recently detected in the GF (note that, following the standard expectations, these systems should have non eccentric orbits because of circularization during the mass transfer phase; see also Section \ref{Sec:ecc}).

The lower panel of this cartoon is an extension of  Figure \ref{recycling} (representative of the three key phases of the {\it canonical recycling scenario}).
In  the top-left part  we show a binary system made  of a slowly rotating NS and an unevolved MS companion star. Then the systems evolves through a LMXB phase characterized by X-ray emission, and the final stage consists of a binary made of a  MSP  and a He-WD (top square in the right part of the plot labelled by ``Canonical'').

However, the evolution between the LMXB phase and the final stage of the system is still shady and several pieces of evidence, both from radio and X-ray observations of the NSs and from the optical detection of the companions, suggest that the evolution could follow different channels.
We tentatively considered the  observed properties of different classes as representative of different evolutionary paths. From top to bottom we reported the Accreting-MSPs (A-MSPs), the RBs and the BWs.
In particular, from the LMXB phase,  MSPs can evolve through a RB phase, during which the MSP is active in radio band and the companion is bloated (thus producing the typical ``double-hill''  light curve) and likely it is losing material  causing radio eclipses (as in the case of \psrH). However, as observed for \psrI\ there are a few systems swinging between a radio dominated phase (as the RB phase) and a X-ray dominated phase (as the A-MSP phase), during which the NS is accreting material. The final stage of these systems is characterized by  what remains of the companion star: likely the He-core.
On the other hand, from the  LMXB phase, MSPs can evolve through BW systems, in which the companions are ablated and probably are semi-degenerate stars, like  BDs surrounded of lost material that produces the long lasting observed radio eclipses (as \psrC). Moreover given the relative faintness of these objects, the illumination by the PSR flux is responsible of the  ``single-hill'' light curve commonly observed for such systems. These systems likely evolve in  binaries containing a MSP and a very low mass object (a BD). If the companions have been completely ablated, they could also evolve in isolated MSPs.

Of course in GCs the scenario is even more complex because if an evolved MSPs (one of the three scenarios reported in the rightmost squares of the plot) run into another star (or even a binary system) an exchange interaction can happen. Likely, the MSP acquires a new (more massive) companion,  while the original light companion is ejected from the system. In particular, if the MSP acquires a NS, what remain is probably an eccentric double NS binary system, while if the MSP acquires a MS,  the recycling  process can start again. Moreover during the evolution, the orbit (originally eccentric as a consequence of the exchange interaction) becomes circular  due to tidal effects during the mass transfer.

Summarizing, 
the proposed scenario tentatively includes exotic RBs and BWs in the canonical evolutionary path. In the past years, the non-degenerate companions to such exotic objects 
were interpreted as the result of  exchange interactions. However, the increasing number of similar systems in the GF seems to suggest that MSPs with non-degenerate companions could be also a  transitory stage of  not fully recycled MSPs. 
Therefore a crucial step forward in our knowledge of MSP formation would be the definition of a diagnostic able to discriminate between the original companion of the NS and a star acquired because of an exchange interaction.
A possible clue could be obtained from the location in the GC and the three-dimensional motion in the cluster. In fact, in the case of an exchange interaction, the binary system could 
be expelled out of the core  and still have  a residual kick.  For example,  \psrH\ was suggested to be the product of a recent exchange interaction because it is located out of the core
and, in absence of perturbations, it would have already sunk in the innermost region under the effect of mass segregation.

Also the timescales of each phase are unknown. Hence catching a large sample of systems in these peculiar phases and determining their relative frequency 
would be crucial.

\subsubsection*{Future works}
 
 In the light of this Thesis results, we can outline future crucial steps in the  optical study of MSP companions.
 Spectroscopic data for the companion to \psrH\ have been already acquired and will be soon analyzed.
 Moreover, fine-tuned observations (in order to measure with a good S/N and  avoid the saturation of close stars, as in the particular case of \psrI) can be performed for most of the object studied in order to better constrain their variability. 
 In particular, for the two BWs (\psrC\ and \psrBW) it would be interesting to detect the companion flux at minimum, so to derive the exact amplitude of variation (instead of only a lower limit as in this Thesis). 
In the case of the companion to \psrIM, further observations could help to detect (if present) a magnitude modulation and, thanks to a multi-band analysis, to precisely  constrain the distance and thereby  the WD mass. 
In turn, this would allow us to  derive a more precise estimate of the age, to be compared with the characteristic age.  
Also radio follow-up could be useful to characterize even better the identified companions.
For example, in the case of the MSPs in the GF a more precise distance is useful to reduce the uncertainties on the physical parameters of the companion. 

 On the other hand, it will be crucial to  apply the same method used in this Thesis to 
 other MSPs in GCs, thus to  increase the number of identified companions (note that proper observations have been already secured for a few cases).
In particular, with a larger sample of companions it will be possible to understand how many are the product of  exchange interactions and to better characterize   recycling scenario.
This could help to indirectly quantify the occurrence of  dynamical interactions in GCs.

Moreover, the study of the companion to \igr /\psrI\ could be considered as a pilot case. In fact, objects similar to \igr /\psrI\ (likely precursors to transient  X-ray sources), might be identified in advance thanks to multi-epoch scheduled observations  in the most promising GCs.

\clearpage{\pagestyle{empty}\cleardoublepage}

\addcontentsline{toc}{chapter}{Appendix}
\newpage
\markboth{APPENDIX }{APPENDIX }
\setcounter{equation}{0}
\renewcommand{\theequation}{\Roman{equation}}
\chapter*{Appendix}
This appendix is denoted to report a few useful definitions (as the {\it characteristic age} and {\it characteristic magnetic field} of a PSR) and to define the formalism used to characterize MSPs in binary systems.
It is mainly based on the thorough  book {\it ``Handbbok of pulsar astronomy''} \citep[][and references therein]{handbook}.

\section*{Characteristic age of PSRs}
A PSR, during its evolution, spins-down with a rate $\dot P$ that is related to the loss of rotational kinetic energy.
In particular from the spin down theoretical model of a PSR, it is possible to express the PSR age  as the so called  {\it spin down age} 
\begin{equation}\label{spindownage} 
T_{\rm SD}=\frac{P}{(n-1)\dot P} \left[1-\left(\frac{P_0}{P}\right)^{n-1}\right],
\end{equation}
where $P_0$ is the spin period at birth and $n$ is the braking index.
Under the assumption that $P_0 \ll P$ and that the spin down is due to magnetic dipole radiation ($n=3$),  equation (\ref{spindownage}) simplifies to the {\it characteristic age}
\begin{equation}\label{carage}
\tau_c \equiv \frac{P}{2\dot P}\simeq 15.8 {\rm Myr} \left(\frac{P}{\rm s}\right) \left (\frac{\dot P}{10^{-15}} \right)^{-1}.
\end{equation}
Of course equation (\ref{carage}) depends on the adopted assumptions. 
Indeed, when an independent estimate of the age is available (as the cooling age of a WD companion) the comparison with the {\it characteristic age} can provide indirect constrain to the breaking index $n$.
For example a value $n=3$ typical of pure magnetic dipole could be not a proper assumption (as found for a few objects for which a direct measure of the braking index $n$  is available). Moreover, in the particular case of recycled PSRs, also  $P_0\ll P$ could  not be a proper assumption.

\section*{Characteristic magnetic field of PSRs}
As for the characteristic age, assuming that the spin down process is dominated by dipole braking, it is possible to estimate the magnetic field strength.
From a few theoretical equations and by assuming typical values for the main NS physical parameters, it is possible to estimate the  {\it characteristic magnetic field} as
\begin{equation}
B_S= 3.2\times10^{19} G \sqrt{ P\dot P} \simeq 10^{12} G \left (\frac{\dot P}{10^{-15}} \right)^{1/2}  \left(\frac{P}{\rm s}\right)^{1/2}. 
\end{equation}

\section*{Orbit description}
In the case of a non relativistic binary system, the orbit can be described by Kepler's laws, through
 six so called ``Keplerian parameters'':
\begin{itemize}
\item[-] the orbital period, $P_b$;
\item[-] the projected semi-major axis, $\apsr \sin i$;
\item[-] the orbital eccentricity, $e$;
\item[-] the longitude of periastron, $\omega$;
\item[-] the epoch of periastron passage, $T_0$;
\item[-] the position angle of the ascending node, $\Omega_{asc}$. 
\end{itemize}

The ``Keplerian parameters'' are related to the {\it eccentric anomaly} $E$, and the {\it true anomaly} $A_T(E)$ as follows:
\begin{equation}
E-e \sin E= \Omega_b\left[(t-T_0)-\frac{1}{2}\frac{\dot P_b}{P_b}(t-T_0)^2\right]
\end{equation}
\begin{equation}
A_T(E)=2\arctan \left[ \sqrt{\frac{1+e}{1-e}}\tan \frac{E}{2}\right]
\end{equation}
\begin{equation}
\omega=\omega_0+\frac{\dot \omega}{\Omega_b}A_T(E)
\end{equation}
where $\Omega_b \equiv 2\pi /P_b$ is  the mean angular velocity.
The quantity $M\equiv(2\pi /P_b)(t-T_0)$ is known as the {\it mean anomaly}.

Kepler' third law relates the size of the semi-major axis of the {\it relative} orbit ($a=\apsr+\acom$), to the mean angular velocity ($\Omega_b$), and the total mass of the system as follows
\begin{equation}
\Omega_b^2\left(\frac{a}{c}\right)^3=T_\odot(\mpsr+\mcom),
\end{equation}
where $\mpsr$ and $\mcom$ are the masses of the PSR and of the companion, respectively and the constant $T_\odot=G \Msun/c^3=4.925490947 \mu$s is used to express these masses in solar units. Moreover the semi-major axes of the orbits of the two objects  around the common centre of mass are 
\begin{equation}
\apsr=a\frac{\mcom}{\mpsr+\mcom}\ {\rm and} \  \acom=a\frac{\mpsr}{\mpsr+\mcom}
\end{equation}
from which the relation
\begin{equation}\label{rappa}
\frac{\apsr}{\acom}=\frac{\mcom}{\mpsr}
\end{equation}
can be derived.

It is important to remind that the orbital parameters of a PSR binary system are derived from  the variation of the observed pulse period, which changes in time  due to Doppler effect.
In particular the velocity along the line of sight  is:
\begin{equation}\label{velocity}
V(A_T)=\Omega_b \frac{\apsr \sin i}{\sqrt{1-e^2}}\left[\cos(\omega+A_T)+e \cos\omega\right]
\end{equation} 
and, if it is small compared to $c$, the observed Doppler-shifted pulse period can be written as
\begin{equation}
P(A_T)\simeq P_0 \left(1+  \frac{V(A_T)}{c}   \right),
\end{equation} 
where $P_0$  is the pulse period in the PSR rest frame.

One more important quantity, very helpful to constrain the companion mass, is the PSR mass function:
\begin{equation}
f_{\rm PSR}(\mpsr,\mcom,i)=\frac{(\mcom \sin i)^3}{(\mpsr+\mcom)^2}=\frac{4\pi^2c^3x^3}{G\Msun P_b^2},
\end{equation}
where $x$ is the measurable projected semi axis of the PSR orbit.
In particular, at a given PSR mass, solving this equation for an inclination angle of $90^\circ$ (corresponding to a  system seen edge-on) gives the minimum possible value for the companion mass.

\subsection*{Post keplerian parameters}
When a PSR  is in a close binary system with a WD or a NS or a black hole the relativistic effects are non negligible.
These systems are particularly important because, thanks to a parametrization of the timing model through ``post-Keplerian'' parameters, they  offer the opportunity to derive the masses of the two components of the binary system and to test the gravity theories.
In fact, in each gravity theory the ``post-Keplerian'' parameters should depends only on  the two masses of the binary  and on the Keplerian parameters. Hence, if the measure of at least two ``post-Keplerian'' parameters is available, the two masses are univocally determined,  while the measure of a third or more parameters provides a check of the assumed theory.

To date general relativity appears to be the best description of strong gravitational field regime. In this scenario the post Keplerian parameters are:
\begin{flalign}
\dot \omega=3T_\odot^{2/3} \left(\frac{P_b}{2\pi}\right)^{-5/3} \frac{1}{1-e^2}(\mpsr+\mcom)^{2/3}\\
\gamma=T_\odot^{2/3} \left(\frac{P_b}{2\pi}\right)^{1/3} e \frac{\mcom(\mpsr+2\mcom)}{(\mpsr+\mcom)^{4/3}}\\
r=T_\odot \mcom\\
s=\sin i = T_\odot^{-1/3} \left(\frac{P_b}{2\pi}\right)^{-2/3} x \frac{(\mpsr+\mcom)^{2/3}}{\mcom}\\
\dot P_b= -\frac{192\pi}{5}T_\odot^{5/3} \left(\frac{P_b}{2\pi}\right)^{-5/3}f(e)\frac{\mpsr \mcom}{(\mpsr+\mcom)^{1/3}}
\end{flalign}
where  all masses are expressed in solar units, G is the Newton gravitational constant, c the speed of light and
\begin{equation}
f(e)=\frac{(1+(73/24)e^2+(37/96)e^4)}{(1-e^2)^{7/2}}.
\end{equation}

In particular, $\dot\omega$ describes the relativistic advance of periastron and its measure  directly gives the total mass of the system.
The parameter $\dot P_b$ is the change of the orbital period and it represents the orbital decay due to gravitational-wave damping, while
$r$ and $s$ are the range and the shape of the Shapiro delay, which is a measure of the delay of the pulse due to the time needed to travel through the curved space-time.
Finally, $\gamma$ is the Einstein delay.

One of the most  exemplar cases in which a test of general relativity has been applied thanks to the measure of post Keplerian parameters is  the case of the  double PSR system \citep[e.g. see][]{kramer06}.

\section*{The case of circular orbits}
Most MSPs are observed in circular orbits. Hence all the equations above are simplified.
In particular,  $e=0$, $\omega=0$ and $T_0$ is assumed to be the time of passage at the ascending node $T_{asc}$.

In this Thesis the phase is defined as $\Phi=M/2\pi$, that for circular orbits is simply:
\begin{equation}
\Phi=\frac{t-T_{asc}}{P_b},
\end{equation}
ranging between 0 and 1. In particular with this phase definition, the orbital phase is a linear function of time and $\Phi=0$ corresponds to the PSR ascending node,  $\Phi=0.25$ to the PSR superior conjunction, $\Phi=0.5$ to the PSR descending node and $\Phi=0.75$ to the PSR inferior conjunction.

Also  equation (\ref{velocity}) is simpler in the case of circular orbits. In fact
\begin{equation}\label{velcirc}
V(\Phi)=\Omega_b\ \apsr \sin i\ \cos(\Phi).
\end{equation} 
It is important to note that, if the measure of the velocity along the line of sight is available for both the stars of the binary system, from  equations (\ref{rappa}) and (\ref{velcirc}) it is possible to directly obtain the mass ratio of the system from the ratio of the amplitudes of the two radial velocity curves:
\begin{equation}
\frac{\vert V_{\rm PSR} \vert}{\vert V_{\rm COM} \vert}=\frac{\apsr}{\acom}=\frac{\mcom}{\mpsr}.
\end{equation}

\clearpage{\pagestyle{empty}\cleardoublepage}

\clearpage{\pagestyle{empty}\cleardoublepage}

\markboth{ACKNOWLEDGEMENTS}{ACKNOWLEDGEMENTS}
\chapter*{Acknowledgements} 
First of all I would like to thank my supervisor Francesco Ferraro for giving me the opportunity to develop this project  and for teaching me what I needed to carry out this Thesis.

A simple thanks is not enough to express my gratitude to Emanuele Dalessandro for sharing his expertise and for answering (essentially every day) to all my questions.

I would also like to thank Barbara Lanzoni  for the useful suggestions  during these years and for her contribution to improve my manuscript.

I am grateful to the Cosmic-Lab project for providing financial support for conference attendance and to all the members of Cosmic-Lab group for the interesting meetings.

I  acknowledge Giacomo Beccari and Elena Valenti for everything that concerned my stay at ESO.

I also would like to thank Scott Ransom for the  great hospitality during my visit at NRAO.

I  acknowledge  Alessio Mucciarelli, Giacomo Beccari, Scott Ransom, Paulo Freire, Roberto Mignani, Andrea Possenti, Marta Burgay, Maurizio Salaris and Ingrid Stairs  for their contribution to the papers reported in this Thesis.

I thank also Gabriele Cocozza for the useful discussions that we had during these years.


\begin{thebibliography}{}
\bibitem[Abdo et al.(2009)]{abdo09} Abdo, A.~A., Ackermann, M., Ajello, M., et al.\ 2009, Science, 325, 848 

\bibitem[Abdo et al.(2010)]{abdo10} Abdo, A.~A., Ackermann, M., Ajello, M., et al.\ 2010, \apjs, 188, 405  

\bibitem[Alpar et al.(1982)]{alpar82} Alpar, M.~A., Cheng, A.~F., Ruderman, M.~A., \& Shaham, J.\ 1982, \nat, 300, 728 

\bibitem[Althaus et al.(2007)]{althaus07} Althaus, L.~G., Garc{\'{\i}}a-Berro, E., Isern, J., C{\'o}rsico, A.~H., \& Rohrmann, R.~D.\ 2007, \aap, 465, 249 

\bibitem[Anderson et al.(1990)]{anderson90} Anderson, S.~B., Gorham, P.~W., Kulkarni, S.~R., Prince, T.~A., \& Wolszczan, A.\ 1990, \nat, 346, 42 

\bibitem[Anderson(2003)]{anderson03} Anderson, J.\ 2003, HST Calibration Workshop : Hubble after the Installation of the ACS and the NICMOS Cooling System, 13 

\bibitem[Andrievsky et al.(2008)]{andr08}Andrievsky, S. M., Spite, M., Korotin, S. A., Spite, F., Bonifacio, P., Cayrel, R., Hill, V., \& Francois, P., 2008, A\&A, 481, 481

\bibitem[Antoniadis et al.(2013)]{antoniadis13} Antoniadis, J., Freire, P.~C.~C., Wex, N., et al.\ 2013, Science, 340, 448 

\bibitem[Archibald et al.(2009)]{archibald09} Archibald, A.~M., et al.\ 2009, Science, 324, 1411 

\bibitem[Archibald et al.(2013)]{archibald13} Archibald, A.~M., Kaspi, V.~M., Hessels, J.~W.~T., et al.\ 2013, arXiv:1311.5161 

\bibitem[Asai et al.(2012)]{asai12} Asai, K., Matsuoka, M., Mihara, T., et al.\ 2012, \pasj, 64, 128 

\bibitem[Backer et al.(1982)]{backer82} Backer, D.~C., Kulkarni, S.~R., Heiles, C., Davis, M.~M., \& Goss, W.~M.\ 1982, \nat, 300, 615 

\bibitem[Bagchi et al.(2011)]{bagchi11} Bagchi, M., Lorimer, D.~R., \& Chennamangalam, J.\ 2011, \mnras, 418, 477 

\bibitem[Bahramian et al.(2013)]{bahramian13} Bahramian, A., Heinke, C.~O., Sivakoff, G.~R., \& Gladstone, J.~C.\ 2013, \apj, 766, 136 

\bibitem[Bailyn(1992)]{bailyn92} Bailyn, C.~D.\ 1992, \apj, 392, 519

\bibitem[Bailyn(1995)]{bailyn95} Bailyn, C.~D.\ 1995, \araa, 33, 133

\bibitem[Barr et al.(2013)]{barr13} Barr, E.~D., Champion, D.~J., Kramer, M., et al.\ 2013, \mnras, 435, 2234 

\bibitem[Beccari et al.(2010)]{be10} Beccari, G., Spezzi, L., De   Marchi, G., et al.\ 2010, \apj, 720, 1108

\bibitem[Beccari et al.(2013)]{beccari13} Beccari, G., De Marchi, G., Panagia, N., \& Pasquini, L.\ 2013, \mnras, 2722 

\bibitem[Becker et al.(2003)]{becker03} Becker, W., Swartz, D.~A.,   Pavlov, G.~G., et al.\ 2003, \apj, 594, 798

\bibitem[B\'egin(2006)]{begin06} B\'egin, S., M.Sc. thesis,  Dept. of Physics and Astronomy, University of British Columbia   (2006).

\bibitem[Bekki \& Freeman(2003)]{bekkifreeman03} Bekki, K., \& Freeman, K.~C.\ 2003, \mnras, 346, L11 

\bibitem[Bellazzini et al.(1995)]{bellazzini95} Bellazzini, M., Pasquali, A., Federici, L., Ferraro, F.~R., \& Pecci, F.~F.\ 1995, \apj, 439, 687 

\bibitem[Bellini \& Bedin(2009)]{bellini09} Bellini, A., \& Bedin, L.~R.\ 2009, \pasp, 121, 1419 

\bibitem[Bellini et al.(2011)]{bellini11} Bellini, A., Anderson, J., \& Bedin, L.~R.\ 2011, \pasp, 123, 622 

\bibitem[Bellm et al.(2013)]{bellm13} Bellm, E., Djorgovski, S.~G., Drake, A.~J., et al.\ 2013, American Astronomical Society Meeting Abstracts, 221, \#154.10 

\bibitem[Bergeron et al.(1995)]{bergeron95} Bergeron, P., Wesemael, F., \& Beauchamp, A.\ 1995, \pasp, 107, 1047 

\bibitem[Bessell et al.(1998)]{bessel98} Bessell, M.~S., Castelli, F.,   \& Plez, B.\ 1998, \aap, 333, 231

\bibitem[Bhattacharya \& van den Heuvel(1991)]{bhattvan91} Bhattacharya, D., \& van den Heuvel, E.~P.~J.\ 1991, \physrep, 203, 1 

\bibitem[Bogdanov et al.(2005)]{bogdanov05} Bogdanov, S.,   Grindlay, J.~E., \& van den Berg, M.\ 2005, \apj, 630, 1029

\bibitem[Bogdanov et al.(2011)]{bogdanov11} Bogdanov, S., van den Berg, M., Servillat, M., et al.\ 2011, \apj, 730, 81 

\bibitem[Breton et al.(2013)]{breton13} Breton, R.~P., van Kerkwijk, M.~H., Roberts, M.~S.~E., et al.\ 2013, \apj, 769, 108 

\bibitem[Buonanno et al.(1983)]{buonanno83} Buonanno, R., Buscema,   G., Corsi, C.~E, Ferraro, I., \& Iannicola, G., 1983, A\&A, 126, 278

\bibitem[Burderi et al.(2002)]{burderi02} Burderi, L., D'Antona, F., \& Burgay, M., 2002, ApJ, 574, 325

\bibitem[Burgay et al.(2003)]{burgaypsr2} Burgay, M., D'Amico, N., Possenti, A., et al.\ 2003, \nat, 426, 531 

\bibitem[Burgay et al.(2006)]{burgay06} Burgay, M., Joshi, B.~C., D'Amico, N., et al.\ 2006, \mnras, 368, 283 

\bibitem[Burgay et al.(2012)]{burgay12} Burgay, M., Keith, M.~J., Lorimer, D.~R., et al.\ 2012, \mnras, 289 

\bibitem[Callanan et al.(1995)]{callanan95} Callanan, P.~J., van Paradijs, J., \& Rengelink, R.\ 1995, \apj, 439, 928 

\bibitem[Camilo et al.(1996)]{camilo96} Camilo, F., Nice, D.~J., Shrauner, J.~A., \& Taylor, J.~H.\ 1996, \apj, 469, 819 

\bibitem[Camilo et al.(2000)]{camilo00} Camilo, F., Lorimer, D.~R., Freire, P., Lyne, A.~G., \& Manchester, R.~N.\ 2000, \apj, 535, 975 

\bibitem[Cardelli et al.(1989)]{cardelli89} Cardelli, J.~A., Clayton, G.~C., \& Mathis, J.~S.\ 1989, \apj, 345, 245 

\bibitem[Carretta \& Gratton(1997)]{carrettagratton97} Carretta, E.,   \& Gratton, R.~G.\ 1997, \aaps, 121, 95

\bibitem[Carretta et al.(2005)]{carretta05}Carretta, E., Gratton, R. G., Lucatello, S., Bragaglia, A., \& Bonifacio, P., 2005, A\&A, 433, 597

\bibitem[Carretta et al.(2009)]{carretta09}Carretta, E., Bragaglia, A., Gratton, R. G., D'Orazi, V., \& Lucatello, S., 2009, A\&A, 508, 695

\bibitem[Castelli \& Kurucz(2004)]{castelli04}Castelli, F. \& Kurucz, R. L., 2004, arXiv:astro-ph/0405087v1

\bibitem[Chakrabarty \& Morgan(1998)]{chak98} Chakrabarty, D., \& Morgan, E.~H.\ 1998, \nat, 394, 346 

\bibitem[Chandrasekhar(1935)]{chandra35} Chandrasekhar, S.\ 1935, \mnras, 95, 207 

\bibitem[Chen et al.(2013)]{chen13} Chen, H.-L., Chen, X., Tauris, T.~M., \& Han, Z.\ 2013, \apj, 775, 27 

\bibitem[Clark(1975)]{clark75} Clark, G.~W.\ 1975, \apjl,  199, L143

\bibitem[Cocozza et al.(2006)]{cocozza06} Cocozza, G., Ferraro, F.~R., Possenti, A., \& D'Amico, N.\ 2006, \apjl, 641, L129 

\bibitem[Cocozza et al.(2008)]{cocozza08} Cocozza, G., Ferraro, F.~R., Possenti, A., et al.\ 2008, \apjl, 679, L105 

\bibitem[Cohn et al.(2013)]{cohnAtel} Cohn, H.~N., Lugger, P.~M., Bogdanov, S., et al.\ 2013, The Astronomer's Telegram, 5031, 1 

\bibitem[Cool et al.(1995)]{cool95} Cool, A.~M., Grindlay, J.~E., Cohn, H.~N., Lugger, P.~M., \& Slavin, S.~D.\ 1995, \apj, 439, 695 

\bibitem[Cordes \& Lazio(2002)]{cordeslazio02} Cordes, J.~M., \& Lazio, T.~J.~W.\ 2002, arXiv:astro-ph/0207156 

\bibitem[Cudworth \& Hanson(1993)]{cudworth93} Cudworth,   K.~M., \& Hanson, R.~B.\ 1993, \aj, 105, 168

\bibitem[Dalessandro et al.(2008a)]{dalessandro086388} Dalessandro, E., Lanzoni, B., Ferraro, F.~R., et al.\ 2008a, \apj, 677, 1069 

\bibitem[Dalessandro et al.(2008b)]{dalessandro082419} Dalessandro, E., Lanzoni, B., Ferraro, F.~R., et al.\ 2008b, \apj, 681, 311 

\bibitem[D'Amico et al.(2001a)]{damico01a} D'Amico, N., Lyne, A. G., Manchester, R. N., Possenti, A., \& Camilo, F., 2001a, ApJ, 548L, 171

\bibitem[D'Amico et al.(2001b)]{damico01b} D'Amico, N., Possenti, A., Manchester, R. N., Sarkissian, J. Lyne, A. G., \& Camilo, F., 2001b, 561L, 89

\bibitem[D'Antona \& Ergma(1993)]{dantona93} D'Antona, F.,   \& Ergma, E.\ 1993, \aap, 269, 219

\bibitem[D'Avanzo et al.(2009)]{d'avanzo09} D'Avanzo, P., Campana, S.,   Casares, J., et al.\ 2009, \aap, 508, 297

\bibitem[Davies \& Hansen(1998)]{davieshansen98} Davies, M.~B.,   \& Hansen, B.~M.~S.\ 1998, \mnras, 301, 15

\bibitem[De Marchi et al.(2010)]{demarchi10} De Marchi, G., Panagia,   N., \& Romaniello, M.\ 2010, \apj, 715, 1

\bibitem[Deneva et al.(2013)]{deneva13} Deneva, J.~S., Stovall, K., McLaughlin, M.~A., et al.\ 2013, \apj, 775, 51 

\bibitem[D'Ercole et al.(2008)]{dercole08} D'Ercole, A., Vesperini,   E., D'Antona, F., McMillan, S.~L.~W., \& Recchi, S.\ 2008, \mnras,   391, 825

\bibitem[Doroshenko et al.(2001)]{doroshenko01} Doroshenko, O., L{\"o}hmer, O., Kramer, M., et al.\ 2001, \aap, 379, 579 

\bibitem[Eckert et al.(2013)]{eckertAtel} Eckert, D., Del Santo, M., Bazzano, A., et al.\ 2013, The Astronomer's Telegram, 4925, 1 

\bibitem[Edmonds et al.(2001)]{edmonds01} Edmonds, P.~D., Gilliland, R.~L., Heinke, C.~O., Grindlay, J.~E., \& Camilo, F.\ 2001, \apjl, 557, L57  

\bibitem[Edmonds et al.(2002)]{edmonds02} Edmonds, P.~D., Gilliland, R.~L., Camilo, F., Heinke, C.~O., \& Grindlay, J.~E.\ 2002, \apj, 579, 741 

\bibitem[Eggleton(1983)]{eggleton83} Eggleton, P.~P.\ 1983, \apj, 268, 368 

\bibitem[Engel et al.(2012)]{engel12} Engel, M.~C., Heinke, C.~O.,   Sivakoff, G.~R., Elshamouty, K.~G., \& Edmonds, P.~D.\ 2012, \apj,   747, 119

\bibitem[Ergma \& Sarna(2003)]{ergma03} Ergma, E., \& Sarna, M. J., 2003, A\&A, 399, 237

\bibitem[Espinoza et al.(2013)]{espinoza13} Espinoza, C.~M., Guillemot, L., {\c C}elik, {\"O}., et al.\ 2013, \mnras, 430, 571 

\bibitem[Faulkner et al.(2004)]{faulkner04} Faulkner, A.~J., Stairs, I.~H., Kramer, M., et al.\ 2004, \mnras, 355, 147 

\bibitem[Faulkner et al.(2005)]{faulkner05} Faulkner, A.~J., Kramer, M., Lyne, A.~G., et al.\ 2005, \apjl, 618, L119 
  
 \bibitem[Ferraro et al.(1991)]{ferraro91} Ferraro, F.~R., Clementini, G., Fusi Pecci, F., \& Buonanno, R.\ 1991, \mnras, 252, 357 

\bibitem[Ferraro et al.(1992)]{ferraro92} Ferraro, F.~R., Fusi Pecci, F., \& Buonanno, R.\ 1992, \mnras, 256, 376 
  
\bibitem[Ferraro et al.(1995)]{ferraro95} Ferraro, F.~R.,   Fusi Pecci, F., \& Bellazzini, M.\ 1995, \aap, 294, 80
  
\bibitem[Ferraro et al.(1999)]{ferraro99} Ferraro, F.~R.,   Messineo, M., Fusi Pecci, F., de Palo, M.~A., Straniero, O.,   Chieffi, A., \& Limongi, M.\ 1999, \aj, 118, 1738

\bibitem[Ferraro et al.(2001a)]{ferraro0147tuc} Ferraro, F.~R., D'Amico,   N., Possenti, A., Mignani, R.P., \& Paltrinieri, B., 2001a, ApJ,   561, 337
  
\bibitem[Ferraro et al.(2001b)]{ferraro01com6397} Ferraro, F.~R., Possenti,   A., D'Amico, N., \& Sabbi, E., 2001b, ApJ, 561, L93

\bibitem[Ferraro et al.(2003a)]{ferraro03com6752} Ferraro, F.~R., Possenti, A., Sabbi, E., \& D'Amico, N.\ 2003a, \apjl, 596, L211 

\bibitem[Ferraro et al.(2003b)]{ferraro03dyn6752} Ferraro, F.~R., Possenti, A., Sabbi, E., et al.\ 2003b, \apj, 595, 179 

\bibitem[Ferraro et al.(2003c)]{ferraro03rv6397} Ferraro, F.~R., Sabbi, E., Gratton, R., et al.\ 2003c, \apjl, 584, L13 

\bibitem[Ferraro et al.(2004)]{ferraro04} Ferraro, F.~R., Sollima, A., Pancino, E., et al.\ 2004, \apjl, 603, L81 

\bibitem[Ferraro et al.(2006a)]{ferraro06omega} Ferraro, F.~R., Sollima, A., Rood, R.~T., et al.\ 2006a, \apj, 638, 433 

\bibitem[Ferraro et al.(2006b)]{ferraro06} Ferraro, F.~R., Sabbi, E., Gratton, R., et al.\ 2006b, \apjl, 647, L53 

\bibitem[Ferraro et al.(2009a)]{fe09ter} Ferraro, F.~R., Dalessandro, E., Mucciarelli, A., et al.\ 2009a, \nat, 462, 483 

\bibitem[Ferraro et al.(2009b)]{fe09m30} Ferraro, F.~R., Beccari, G.,   Dalessandro, E., et al.\ 2009b, \nat, 462, 1028

\bibitem[Ferraro et al.(2012)]{ferraro12} Ferraro, F.R.  et al., 2012,   \nat, 492, 393
  
\bibitem[Fregeau(2008)]{fregeau08} Fregeau, J.~M.\ 2008, \apjl, 673,   L25

\bibitem[Freire et al.(2003)]{freire03} Freire, P.~C., Camilo, F., Kramer, M., Lorimer, D.~R., Lyne, A.~G., Manchester, R.~N., \& D'Amico, N.\ 2003, \mnras, 340, 1359 

\bibitem[Freire(2005)]{freire05} Freire, P.~C.~C.\ 2005,   Binary Radio Pulsars, 328, 405

\bibitem[Freire et al.(2008)]{freire08} Freire, P.~C.~C., Wolszczan, A., van den Berg, M., \& Hessels, J.~W.~T.\ 2008, \apj, 679, 1433 

\bibitem[Freire \& Tauris(2013)]{freiretauris13} Freire, P.~C.~C., \& Tauris, T.~M.\ 2013, arXiv:1311.3478 

\bibitem[Fruchter et al.(1988a)]{fruchter88b} Fruchter, A.~S., Gunn, J.~E., Lauer, T.~R., \& Dressler, A.\ 1988a, \nat, 334, 686 

\bibitem[Fruchter et al.(1988b)]{fruchter88a} Fruchter, A.~S., Stinebring, D.~R., \& Taylor, J.~H.\ 1988b, \nat, 333, 237 

\bibitem[Girardi et al.(2000)]{girardi00} Girardi, L., Bressan, A., Bertelli, G., \& Chiosi, C.\ 2000, \aaps, 141, 371 

\bibitem[Goodman \& Hut(1989)]{goodmanhut89} Goodman, J., \& Hut,   P.\ 1989, \nat, 339, 40

\bibitem[Gratton et al.(2001)]{gratton01}Gratton, R. G., et al., 2001, A\&A, 369, 87

\bibitem[Gratton et al.(2012)]{gratton12}Gratton, R. G., Carretta, E., \& Bragalia, A., 2012, A\&AR, 20, 50

\bibitem[Grevesse \& Sauval(1998)]{gs98}Grevesse, N., \& Sauval, A. J., 1998, Space Science Reviews, 85, 161

\bibitem[Grindlay et al.(2001)]{grindlay01} Grindlay, J.~E., Heinke,   C.~O., Edmonds, P.~D., Murray, S.~S., \& Cool, A.~M.\ 2001, \apjl,   563, L53

\bibitem[Hansen \& Phinney(1998a)]{hansenPhinney98a} Hansen, B.~M.~S., \& Phinney, E.~S.\ 1998, \mnras, 294, 569 

\bibitem[Hansen \& Phinney(1998b)]{hansenPhinney98b} Hansen, B.~M.~S., \& Phinney, E.~S.\ 1998, \mnras, 294, 557 

\bibitem[Harris(1996)]{harris96} Harris, W.~E. 1996, AJ, 112, 1487

\bibitem[Heinke et al.(2013)]{heinkeAtel} Heinke, C.~O., Bahramian, A., Wijnands, R., \& Altamirano, D.\ 2013, The Astronomer's Telegram, 4927, 1 

\bibitem[Hessels et al.(2007)]{hessels07} Hessels, J.~W.~T., Ransom, S.~M., Stairs, I.~H., Kaspi, V.~M., \& Freire, P.~C.~C.\ 2007, \apj, 670, 363 

\bibitem[Hills \& Day(1976)]{hillsday76} Hills, J.~G., \&   Day, C.~A.\ 1976, \aplett, 17, 87

\bibitem[Hobbs et al.(2005)]{hobbs05} Hobbs, G., Lorimer, D.~R., Lyne, A.~G., \& Kramer, M.\ 2005, \mnras, 360, 974 

\bibitem[Holtzman et al.(1995)]{holtzman95} Holtzman, J.~A., Burrows, C.~J., Casertano, S., et al.\ 1995, \pasp, 107, 1065 

\bibitem[Homan \& Pooley(2013)]{homanAtel} Homan, J., \& Pooley, D.\ 2013, The Astronomer's Telegram, 5045, 1 

\bibitem[Hui et al.(2010)]{hui10} Hui, C.~Y., Cheng, K.~S., \& Taam, R.~E.\ 2010, \apj, 714, 1149 

\bibitem[Hut et al.(1992)]{hut92} Hut, P., et al.\ 1992, \pasp, 104,   981

\bibitem[Hut et al.(2003)]{hut03} Hut, P., Shara, M.~M.,  Aarseth, S.~J., et al.\ 2003, \na, 8, 337 

\bibitem[Ivanova et al.(2008)]{ivanova08} Ivanova, N.,   Heinke, C.~O., Rasio, F.~A., Belczynski, K., \& Fregeau,   J.~M.\ 2008, \mnras, 386, 553

\bibitem[Jacoby et al.(2006)]{jacoby06} Jacoby, B.~A., Chakrabarty, D., van Kerkwijk, M.~H., Kulkarni, S.~R., \& Kaplan, D.~L.\ 2006, \apjl, 640, L183 

\bibitem[Johnson \& Morgan(1953)]{johnson53} Johnson, H.~L., \& Morgan, W.~W.\ 1953, \apj, 117, 313 

\bibitem[Johnson et al.(1966)]{johnson66} Johnson, H.~L., Mitchell, R.~I., Iriarte, B., \& Wisniewski, W.~Z.\ 1966, Communications of the Lunar and Planetary Laboratory, 4, 99 

\bibitem[Johnson \& Pilachowski(2010)]{johnson10} Johnson, C.~I., \& Pilachowski, C.~A.\ 2010, \apj, 722, 1373 

\bibitem[Kaluzny et al.(2003)]{kaluzny03} Kaluzny, J., Rucinski, S. M., \& Thompson, I. B., 2003, AJ, 125, 1546

\bibitem[Kaplan et al.(2013)]{kaplan13} Kaplan, D.~L., Bhalerao, V.~B., van Kerkwijk, M.~H., et al.\ 2013, \apj, 765, 158 

\bibitem[Katz(1975)]{katz75} Katz, J.~I.\ 1975, \nat, 253, 698 

\bibitem[King et al.(2003)]{king03} King, A.~R., Davies, M.~B., \& Beer, M.~E.\ 2003, \mnras, 345, 678 

\bibitem[King \& Beer(2005)]{king05} King, A.~R., \& Beer, M.~E.\ 2005, Binary Radio Pulsars, 328, 429 

\bibitem[Kiziltan \& Thorsett(2010)]{kiziltanThor10} Kiziltan, B., \& Thorsett, S.~E.\ 2010, \apj, 715, 335 

\bibitem[K{\i}z{\i}ltan(2011)]{kiziltan11} K{\i}z{\i}ltan, B.\ 2011, in ``Astrophysics of neutro stars 2010: A conference in honor of M. Ali Alpar'' , AIP Conference Proceedings, Vol. 1379, pp. 41-47 

\bibitem[Kochanek(1996)]{kochanek96} Kochanek, C.~S.\ 1996, \apj, 457, 228 

\bibitem[Kramer et al.(2006)]{kramer06} Kramer, M., Stairs, I.~H., Manchester, R.~N., et al.\ 2006, Science, 314, 97 

\bibitem[Kulkarni(1986)]{kulkarni86} Kulkarni, S.~R.\ 1986, \apjl, 306, L85 

\bibitem[Kulkarni et al.(1988)]{kulkarni88} Kulkarni, S.~R., Djorgovski, S., \& Fruchter, A.~S.\ 1988, \nat, 334, 504 

\bibitem[Lanzoni et al.(2007a)]{lanzoni07M5} Lanzoni, B., Dalessandro, E., Ferraro, F.~R., et al.\ 2007a, \apj, 663, 267 

\bibitem[Lanzoni et al.(2007b)]{lanzoni071904} Lanzoni, B., Sanna, N., Ferraro, F.~R., et al.\ 2007b, \apj, 663, 1040 

\bibitem[Lanzoni et al.(2007c)]{lanzoni076388} Lanzoni, B., Dalessandro, E., Ferraro, F.~R., et al.\ 2007c, \apjl, 668, L139 

\bibitem[Lanzoni et al.(2010)]{lanzoni10} Lanzoni, B., Ferraro, F.~R., Dalessandro, E., et al.\ 2010, \apj, 717, 653 

\bibitem[Lee et al.(1999)]{lee99} Lee, Y.-W., Joo, J.-M., Sohn, Y.-J., et al.\ 1999, \nat, 402, 55 

\bibitem[Linares(2013)]{linaresAtel} Linares, M.\ 2013, The Astronomer's Telegram, 4960, 1 

\bibitem[Lind et al.(2009)]{lind09}Lind, K., Primas, F., Charbonnel, C., Grundhal, F., \& Asplund, M., 2009, A\&A, 503, 545

\bibitem[Lind et al.(2011)]{lind}Lind, K., Asplund, M., Barklem, P. S., \& Belyaev, A. K., 2011, A\&A, 528, 103

\bibitem[Little \& Tremaine(1987)]{little87} Little, B., \& Tremaine, S.\ 1987, \apj, 320, 493 

\bibitem[L{\"o}hmer et al.(2004)]{lohmer04} L{\"o}hmer, O., Kramer, M., Driebe, T., et al.\ 2004, \aap, 426, 631 

\bibitem[Lorimer \& Kramer(2005)]{handbook} Lorimer, D.~R., Kramer, M.\ 2005, {\it Hanbook of pulsar astronomy}, Cambridge University Press 

\bibitem[Lorimer et al.(2006)]{lorimer06} Lorimer, D.~R., Faulkner, A.~J., Lyne, A.~G., et al.\ 2006, \mnras, 372, 777 

\bibitem[Lovisi et al.(2012)]{lovisi12}Lovisi, L., Mucciarelli, A., Lanzoni, B., Ferraro, F. R., Gratton, R. G., Dalessandro, E., \& Contreras Ramos, R., 2012, ApJ, 754, 91

\bibitem[Lucatello \& Gratton(2003)]{lucatello03}Lucatello, S., \& Gratton, R. G., 2003, A\&A, 406, 691

\bibitem[Lundgren et al.(1996a)]{lundgren96a} Lundgren, S.~C., Ergma, E., \& Cordes, J.~M.\ 1996a, IAU Colloq.~160: Pulsars: Problems and Progress, 105, 521 

\bibitem[Lundgren et al.(1996b)]{lundgren96b} Lundgren, S.~C., Cordes, J.~M., Foster, R.~S., Wolszczan, A., \& Camilo, F.\ 1996b, \apjl, 458, L33 

\bibitem[Lynch et al.(2012)]{lynch12} Lynch, R.~S., Freire, P.~C.~C., Ransom, S.~M., \& Jacoby, B.~A.\ 2012, \apj, 745, 109 

\bibitem[Lyne et al.(1987)]{lyne87} Lyne, A.~G., Brinklow, A.,   Middleditch, J., Kulkarni, S.~R., \& Backer, D.~C.\ 1987, \nat, 328,   399

\bibitem[Lyne et al.(2004)]{lyne04} Lyne, A.~G., Burgay, M., Kramer, M., et al.\ 2004, Science, 303, 1153 

\bibitem[Manchester et al.(2001)]{manchester01} Manchester, R.~N., Lyne, A.~G., Camilo, F., et al.\ 2001, \mnras, 328, 17 

\bibitem[Marigo et al.(2008)]{marigo08} Marigo, P.,   Girardi, L., Bressan, A., Groenewegen, M.~A.~T., Silva, L., \&   Granato, G.~L.\ 2008, \aap, 482, 883

\bibitem[Meylan \& Heggie(1997)]{meylanheggie97} Meylan, G., \&   Heggie, D.~C.\ 1997, \aapr, 8, 1

\bibitem[Milone et al.(2012)]{milone12} Milone, A.~P., Piotto, G., Bedin, L.~R., et al.\ 2012, \aap, 540, A16 

\bibitem[Moffat(1969)]{moffat69} Moffat, A.~F.~J.\ 1969, \aap, 3, 455   

\bibitem[Montegriffo et al.(1995)]{montegriffo95} Montegriffo, P., Ferraro, F.~R., Fusi Pecci, F., \& Origlia, L.\ 1995, \mnras, 276, 739 

\bibitem[Mucciarelli, Salaris \& Bonifacio(2012)]{mucciarelli12}Mucciarelli, A., Salaris, M., \& Bonifacio, P., 2012, MNRAS, 419, 2195

\bibitem[Mucciarelli et al.(2012)]{m12b} Mucciarelli, A., Bellazzini, M., Ibata, R., Merle, T., Chapman, S. C. Dalessandro, E., \& Sollima, A., 2012, MNRAS, 426, 2889

\bibitem[Mucciarelli et al.(2013)]{mucciarelli13} Mucciarelli, A., Salaris, M., Lanzoni, B., et al.\ 2013, \apjl, 772, L27 

\bibitem[Neckel \& Labs(1984)]{neckel}Neckel, H., \& Labs, D., 1984, SoPh, 90, 205

\bibitem[Norris \& Da Costa(1995)]{norris95} Norris, J.~E., \& Da Costa, G.~S.\ 1995, \apj, 447, 680 

\bibitem[Origlia et al.(2003)]{origlia03} Origlia, L., Ferraro, F.~R., Bellazzini, M., \& Pancino, E.\ 2003, \apj, 591, 916 

\bibitem[Origlia et al.(2011)]{origlia11} Origlia, L., Rich, R.~M., Ferraro, F.~R., et al.\ 2011, \apjl, 726, L20 

\bibitem[Origlia et al.(2013)]{origlia13} Origlia, L., Massari, D., Rich, R.~M., et al.\ 2013, \apjl, 779, L5 

\bibitem[Orosz \& van Kerkwijk(2003)]{orosz03} Orosz, J. A. \& van Kerkwijk, M. H., 2003, A\&A, 397, 237

\bibitem[{\"O}zel et al.(2012)]{ozel12} {\"O}zel, F., Psaltis, D., Narayan, R., \& Santos Villarreal, A.\ 2012, \apj, 757, 55 

\bibitem[Paczy{\'n}ski(1971)]{paczynski71} Paczy{\'n}ski,   B.\ 1971, \araa, 9, 183

\bibitem[Pallanca et al.(2010)]{pallanca10} Pallanca, C., Dalessandro, E., Ferraro, F.~R., et al.\ 2010, \apj, 725, 1165 

\bibitem[Pallanca et al.(2012)]{pallanca12} Pallanca, C., Mignani, R.~P., Dalessandro, E., et al.\ 2012, \apj, 755, 180 

\bibitem[Pallanca et al.(2013a)]{pallanca13com1439} Pallanca, C., Lanzoni, B., Dalessandro, E., et al.\ 2013a, \apj, 773, 127 

\bibitem[Pallanca et al.(2013b)]{pallanca13comM28I} Pallanca, C., Dalessandro, E., Ferraro, F.~R., Lanzoni, B., \& Beccari, G.\ 2013b, \apj, 773, 122 

\bibitem[Pallanca et al.(2013c)]{pallancaAtel} Pallanca, C., Dalessandro, E., Ferraro, R.~F., Lanzoni, B., \& Beccari, G.\ 2013c, The Astronomer's Telegram, 5003, 1 

\bibitem[Pancino et al.(2002)]{pancino02} Pancino, E., Pasquini, L., Hill, V., Ferraro, F.~R., \& Bellazzini, M.\ 2002, \apjl, 568, L101 

\bibitem[Papitto et al.(2013)]{papitto13} Papitto, A., Ferrigno, C., Bozzo, E., et al.\ 2013, \nat, 501, 517 

\bibitem[Paresce et al.(1992)]{paresce92} Paresce, F., de Marchi, G., \& Ferraro, F.~R.\ 1992, \nat, 360, 46 

\bibitem[Pasquini et al.(2008)]{pasquini08}Pasquini, L., Ecuvillon, A., Bonifacio, P., \& Wolff, B., 2008, A\&A, 489, 315

\bibitem[Pavan et al.(2013)]{pavanAtel} Pavan, L., Wong, G., Wieringa, M.~H., et al.\ 2013, The Astronomer's Telegram, 4981, 1 

\bibitem[Phinney(1992)]{phinney92} Phinney, E.~S.\ 1992, Royal Society   of London Philosophical Transactions Series A, 341, 39

\bibitem[Phinney \& Kulkarni(1994)]{phinneykulk94} Phinney, E.~S., \& Kulkarni, S.~R.\ 1994, \araa, 32, 591 

\bibitem[Pietrinferni et al.(2004)]{pietr04}Pietrinferni, A., Cassisi, S., Salaris, M., \& Castelli, F. 2004, ApJ, 612, 168

\bibitem[Pietrinferni et al.(2006)]{pietr06} Pietrinferni, A., Cassisi, S., Salaris, M., \& Castelli, F. 2006, ApJ, 642, 797

\bibitem[Pooley et al.(2003)]{pooley03} Pooley, D., et al.\ 2003,   \apjl, 591, L131

\bibitem[Possenti(2002)]{possenti} Possenti, A.\ 2002, Neutron Stars, Pulsars, and Supernova Remnants, 183 

\bibitem[Possenti et al.(2003)]{possenti03} Possenti, A., D'Amico, N.,   Manchester, R.~N., Camilo, F., Lyne, A.~G., Sarkissian, J., \&   Corongiu, A.\ 2003, \apj, 599, 475

\bibitem[Pryor \& Meylan(1993)]{pryormeylan93} Pryor, C., \&   Meylan, G.\ 1993, Structure and Dynamics of Globular Clusters, 50,   357

\bibitem[Ransom et al.(2005)]{ransom05} Ransom, S.~M.,   Hessels, J.~W.~T., Stairs, I.~H., Freire, P.~C.~C., Camilo, F.,   Kaspi, V.~M., \& Kaplan, D.~L.\ 2005, Science, 307, 892

\bibitem[Ransom(2008a)]{ransom07} Ransom, S.~M.\ 2008a, IAU Symposium, 246, 291 

\bibitem[Ransom(2008b)]{ransom08} Ransom, S.~M.\ 2008b, 40 Years of Pulsars: Millisecond Pulsars, Magnetars and More, 983, 415 

\bibitem[Reynolds et al.(2007)]{reynolds07} Reynolds, M.~T., Callanan, P.~J., Fruchter, A.~S., et al.\ 2007, \mnras, 379, 1117 
 
\bibitem[Richard, Michaud \& Richer(2002)]{richard02}Richard, O., Michaud, G., \& Richer, J., 2002, ApJ, 580, 1100 
 
\bibitem[Roberts(2011)]{roberts11} Roberts, M.~S.~E.\ 2011, American Institute of Physics Conference Series, 1357, 127 

\bibitem[Roberts(2013)]{roberts13} Roberts, M.~S.~E.\ 2013, IAU Symposium, 291, 127 

\bibitem[Robin et al.(2003)]{robin03} Robin, A.~C., Reyl{\'e}, C., Derri{\`e}re, S., \& Picaud, S.\ 2003, \aap, 409, 523  
 
\bibitem[Roeser et al.(2010)]{roeser10} Roeser, S., Demleitner, M., \& Schilbach, E.\ 2010, \aj, 139, 2440 
 
\bibitem[Romani(2012)]{romani12} Romani, R.~W.\ 2012, \apjl, 754, L25  

\bibitem[Romano et al.(2007)]{romano07} Romano, D., Matteucci, F., Tosi, M., et al.\ 2007, \mnras, 376, 405 

\bibitem[Romano et al.(2013)]{romanoAtel} Romano, P., Barthelmy, S.~D., Burrows, D.~N., et al.\ 2013, The Astronomer's Telegram, 4929, 1 

\bibitem[Sabbi et al.(2003a)]{sabbi03Ha} Sabbi, E, Gratton, R. G.,  Ferraro, F. R., Bragaglia, A., Possenti, A., D'Amico, N., \& Camilo, F., 2003a, ApJ, 589, L41

\bibitem[Sabbi et al.(2003b)]{sabbi03C} Sabbi, E, Gratton, R. G., Bragaglia, A., Ferraro, F. R., Possenti, A., Camilo, F., \& D'Amico, N., 2003b, A\&A, 412, 829

\bibitem[Salaris et al.(2010)]{salaris10} Salaris, M., Cassisi, S., Pietrinferni, A., Kowalski, P.~M., \& Isern, J.\ 2010, \apj, 716, 1241 

\bibitem[Sbordone et al.(2004)]{sbordone}Sbordone, L., Bonifacio, P., Castelli, F., \& Kurucz, R. L., 2004, MSAIS, 5, 93 

\bibitem[Schnitzeler(2012)]{schnitzeler12} Schnitzeler, D.~H.~F.~M.\ 2012, \mnras, 427, 664 

\bibitem[Serino et al.(2013)]{serinoAtel} Serino, M., Takagi, T., Negoro, H., et al.\ 2013, The Astronomer's Telegram, 4961, 1 

\bibitem[Sigurdsson et al.(2003)]{sigurdsson03} Sigurdsson, S.,   Richer, H.~B., Hansen, B.~M., Stairs, I.~H., \& Thorsett,   S.~E.\ 2003, Science, 301, 193 
 
\bibitem[Sills et al.(2003)]{sills03} Sills, A., Deiters, S., Eggleton, P., et al.\ 2003, \na, 8, 605 

\bibitem[Sirianni et al.(2005)]{sirianni05} Sirianni, M., Jee, M.~J., Ben{\'{\i}}tez, N., et al.\ 2005, \pasp, 117, 1049 

\bibitem[Skrutskie et al.(2006)]{2mass06} Skrutskie, M.~F., Cutri, R.~M., Stiening, R., et al.\ 2006, \aj, 131, 1163 

\bibitem[Sollima et al.(2005)]{sollima05} Sollima, A., Pancino, E., Ferraro, F.~R., et al.\ 2005, \apj, 634, 332 

\bibitem[Sollima et al.(2007a)]{sollima07} Sollima, A., Ferraro, F.~R., Bellazzini, M., et al.\ 2007a, \apj, 654, 915 

\bibitem[Sollima et al.(2007b)]{sollima07bin} Sollima, A., Beccari, G., Ferraro, F.~R., Fusi Pecci, F., \& Sarajedini, A.\ 2007b, \mnras, 380, 781 

\bibitem[Sollima et al.(2008)]{sollima08} Sollima, A., Lanzoni, B., Beccari, G., Ferraro, F.~R., \& Fusi Pecci, F.\ 2008, \aap, 481, 701 

\bibitem[Stairs(2004)]{stairs04} Stairs, I.~H.\ 2004, Science, 304, 547 

\bibitem[Stairs et al.(2006)]{stairs06} Stairs, I.~H.,   B\'egin, S., Ransom, S., Freire, P., Hessels, J., Katz, J., Kaspi,   V., \& Camilo, F.\ 2006, Bulletin of the American Astronomical   Society, 38, 1118

\bibitem[Stappers et al.(1996a)]{stappers96b} Stappers, B.~W., Bessell, M.~S., \& Bailes, M.\ 1996a, \apjl, 473, L119 

\bibitem[Stappers et al.(1996b)]{stappers96a} Stappers, B.~W., Bailes, M., Lyne, A.~G., et al.\ 1996b, \apjl, 465, L119 
 
\bibitem[Stappers et al.(1999)]{stappers99} Stappers, B.~W., van Kerkwijk, M.~H., Lane, B., \& Kulkarni, S.~R.\ 1999, \apjl, 510, L45 

\bibitem[Stappers et al.(2000)]{stappers00} Stappers, B.~W., van Kerkwijk, M., \& Bell, J.~F.\ 2000, IAU Colloq.~177: Pulsar Astronomy - 2000 and Beyond, 202, 627 
 
\bibitem[Stappers et al.(2001)]{stappers01} Stappers, B.~W., van Kerkwijk, M.~H., Bell, J.~F., \& Kulkarni, S.~R.\ 2001, \apjl, 548, L183 
 
\bibitem[Stetson(1987)]{stetson87} Stetson, P.~B.\ 1987,   \pasp, 99, 191

\bibitem[Stetson(1994)]{stetson94} Stetson, P.~B.\ 1994, \pasp, 106,  250

\bibitem[Stetson(2000)]{stetson00} Stetson, P.~B.\ 2000, \pasp, 112, 925 

\bibitem[Taam et al.(2000)]{taam00} Taam, R.~E., King, A.~R., \& Ritter, H.\ 2000, \apj, 541, 329 
 
\bibitem[Tauris \& Savonije(1999)]{taurissav99} Tauris, T.~M., \& Savonije, G.~J.\ 1999, \aap, 350, 928 
 
\bibitem[Tauris et al.(2000)]{tauris00} Tauris, T.~M., van den Heuvel, E.~P.~J., \& Savonije, G.~J.\ 2000, \apjl, 530, L93  
 
\bibitem[Tauris(2011)]{tauris11} Tauris, T.~M.\ 2011, Evolution of Compact Binaries, 447, 285  
 
\bibitem[Tauris et al.(2012)]{tauris12} Tauris, T.~M., Langer, N., \& Kramer, M.\ 2012, \mnras, 425, 1601  
 
\bibitem[Taylor \& Cordes(1993)]{taylorcordes93} Taylor, J.~H., \& Cordes, J.~M.\ 1993, \apj, 411, 674 
 
 \bibitem[Testa et al.(2012)]{testa12} Testa, V., di Salvo, T., D'Antona, F., et al.\ 2012, \aap, 547, A28 
 
\bibitem[Tonry \& Davis(1979)]{tonry79} Tonry, J., \& Davis, M.\ 1979, \aj, 84, 1511 
 
\bibitem[van den Heuvel(1994)]{vandenheuvel94} van den Heuvel, E.~P.~J.\ 1994, \aap, 291, L39 

\bibitem[van Kerkwijk \& Kulkarni(1995)]{vanKerkKulk95} van Kerkwijk, M.~H., \& Kulkarni, S.~R.\ 1995, \apjl, 454, L141 

\bibitem[van Kerkwijk(1997)]{vanKerk97} van Kerkwijk, M.~H.\ 1997, White dwarfs, 214, 383 

\bibitem[van Kerkwijk et al.(2005)]{vankerk05} van Kerkwijk, M.~H., Bassa, C.~G., Jacoby, B.~A., \& Jonker, P.~G.\ 2005, Binary Radio Pulsars, 328, 357

\bibitem[van Kerkwijk et al.(2011)]{vankerk11} van Kerkwijk, M.~H., Breton, R.~P., \& Kulkarni, S.~R.\ 2011, \apj, 728, 95 

\bibitem[Verbunt \& Hut(1987)]{verbunthut87} Verbunt, F., \& Hut, P.\ 1987, The Origin and Evolution of Neutron Stars, 125, 187 

\bibitem[Verbunt et al.(1997)]{verbunt97} Verbunt, F., Bunk, W.~H.,   Ritter, H., \& Pfeffermann, E.\ 1997, \aap, 327, 602

\bibitem[Verbunt \& Freire(2013)]{verbfrei13} Verbunt, F., \& Freire, P.~C.~C.\ 2013, arXiv:1310.4669 

\bibitem[Voges et al.(1999)]{voges99} Voges, W., Aschenbach, B., Boller, T., et al.\ 1999, \aap, 349, 389 

\bibitem[von Zeipel(1924)]{vonzeipel24} von Zeipel, H.\ 1924, \mnras, 84, 684 

\bibitem[White et al.(1984)]{white84} White, N.~E., Kaluzienski, J.~L., \& Swank, J.~H.\ 1984, American Institute of Physics Conference Series, 115, 31 

\bibitem[Zhang et al.(2011)]{zhang11} Zhang, C.~M., Wang, J., Zhao, Y.~H., et al.\ 2011, \aap, 527, A83 

\bibitem[Zinn(1980)]{zinn80} Zinn, R.\ 1980, \apjs, 42, 19 













\end{thebibliography}
\end{document}